\documentclass[11pt,a4paper]{article}
\usepackage[backend=bibtex, style=authoryear, maxbibnames=9]{biblatex}
\usepackage{pdfpages}
\usepackage{color}
\usepackage{diagbox}
\usepackage{amsthm, amsmath, amsfonts,graphicx,xr,url}
\usepackage{amssymb}
\usepackage{enumitem}
\usepackage{float}
\usepackage{graphicx}
\usepackage{bm, bbm}
\usepackage{blkarray}
\usepackage{authblk}
\usepackage{multirow}
\usepackage[margin=1in]{geometry}
\usepackage{changepage}
\usepackage[font=normalsize,labelfont=bf]{caption}
\usepackage[font=footnotesize,labelfont=bf]{subcaption}
\numberwithin{equation}{section}
\usepackage{hyperref}
\hypersetup{
	colorlinks,
	citecolor=blue,
	linkcolor=cornflowerblue,
	filecolor=yaleblue,      
	urlcolor=cornflowerblue
}
\usepackage{cleveref} 
\usepackage{footnotebackref}
\crefformat{footnote}{#2\footnotemark[#1]#3}
\usepackage{easyReview}
\allowdisplaybreaks

\usepackage{chngcntr}
\counterwithin{figure}{section}
\counterwithin{table}{section}
\DeclareMathOperator*{\VAR}{VAR}
\DeclareMathOperator*{\Var}{Var}
\DeclareMathOperator*{\Cov}{Cov}
\DeclareMathOperator*{\AR}{AR}

\DeclareMathOperator*{\Bin}{Binomial}
\DeclareMathOperator*{\PSBP}{PSBP}
\DeclareMathOperator*{\GP}{GP}
\DeclareMathOperator*{\NNGP}{NNGP}
\DeclareMathOperator*{\PG}{PG}
\DeclareMathOperator*{\determ}{det}
\DeclareMathOperator*{\diag}{Diag}
\DeclareMathOperator*{\trace}{tr}
\DeclareMathOperator*{\unif}{Unif}
\DeclareMathOperator*{\Ga}{Gamma}

\DeclareMathOperator*{\iid}{iid}
\DeclareMathOperator*{\ind}{ind}
\DeclareMathOperator*{\supth}{th}
\DeclareMathOperator*{\Chol}{Chol}
\DeclareMathOperator*{\rooti}{Rooti}
\DeclareMathOperator*{\vect}{vec}
\DeclareMathOperator*{\MN}{\mathcal{M}\mathcal{N}}
\def\chol{\mathrm{Chol}}
\def\NN{\mathbb{N}}
\def\RR{\mathbb{R}}
\def\T{{\mathrm{\scriptscriptstyle T}}}
\def\Ocal{\mathcal{O}}

\usepackage{xr}
\makeatletter

\usepackage{pifont}
\newcommand{\cmark}{\ding{52}}%
\newcommand{\xmark}{\ding{56}}%

\definecolor{cornflowerblue}{rgb}{0.39, 0.58, 0.93}

\title{\textbf{Enhancing Scalability in Bayesian Nonparametric Factor Analysis of Spatiotemporal Data}}
\author{\small Yifan Cheng \thanks{y.cheng@u.nus.edu}  }
\author{\small Cheng Li \thanks{stalic@nus.edu.sg} }
\affil{\footnotesize Department of Statistics and Data Science, National University of Singapore}
\date{}
\begin{document}
	\maketitle
	\vspace{-6.5mm}
	\begin{abstract}
		This article introduces novel and practicable Bayesian factor analysis frameworks that are computationally feasible for moderate to large spatiotemporal data. Previous Bayesian analysis of spatiotemporal data has utilized a Bayesian factor model with separable temporal latent factors and spatial factor loadings, along with stick-breaking process priors on the loadings to enable clustering of spatial locations. Such a flexible Bayesian model, however, faces a prohibitively high computational cost in posterior sampling when the spatial and temporal dimensions increase to a couple hundred. We address this computational challenge with several speed-up proposals. We integrate a new slice sampling algorithm that permits varying numbers of spatial mixture components across all latent factors and guarantees them to be non-increasing through the posterior sampling iterations, thus effectively reducing the number of mixture parameters. Additionally, we introduce a spatial latent nearest-neighbor Gaussian process prior and new sequential updating algorithms for the spatially varying latent variables in the stick-breaking process prior. Our new models and sampling algorithms exhibit significantly enhanced computational scalability and storage efficiency and possess powerful inferential capabilities for both spatiotemporal prediction and clustering of spatial locations with similar temporal trajectories. The improvement in computational efficiency and inferential performance is substantiated by extensive simulation experiments.
	\end{abstract}
	\tableofcontents

		\section{Introduction}\label{sect1} 
	One pivotal task frequently occurring in spatiotemporal applications such as infectious disease control and natural disaster prevention involves identifying spatial clusters with similar temporal trajectories and picking out the regions that may require early special attention, which requires modeling the covariance structure with dependence along both the spatial and temporal dimensions. In general, there are several primary goals a spatiotemporal model would want to accomplish. 
	First, the model should effectively characterize the major sources of variation across both dimensions while retaining spatial and temporal dimension reduction properties and hence overall interpretability to a satisfactory extent, given that one typically observes only a single trajectory of data across both space and time with no independent replicates. Second, the model should properly account for spatiotemporal heterogeneity and nonstationarity prevalent in various data types such as imaging data in neurosciences and geosciences, as decently addressing these aspects is crucial for better prediction performances. Third, the model should entail a parsimonious structure that facilitates efficient computation on data with moderate to high spatial and temporal dimensions for practicable implementation, especially when Bayesian inference, which often involves computationally intensive posterior sampling, is concerned.
	
	This article focuses on the Bayesian spatiotemporal latent factor model developed in \textcite{Berchuck2021}, which achieves the first two goals described above. Factor models can effectively reduce dimension by producing a handful of informative latent factors from a host of raw variables. Previous works on Bayesian spatial factor analysis often assumed independent replicates (\cite{Christensen2002}, \cite{Wall2009}, \cite{Ren2013}) and conducted dimension reduction on a large number of observation types at each spatial location, which has been successfully generalized to nonnegative spatial factorization with applications in spatial transcriptomics (\cite{TowEng22}, \cite{MaZho22}). The separable factor models with temporally dependent latent factors and spatially dependent factor loadings used in recent works such as \cite{Braetal15} and \textcite{Berchuck2021}, in contrast, effectively reduce dimension both spatially and temporally, thus offering convenient interpretability. The general framework in \textcite{Berchuck2021} also enables clustering spatial locations into distinct groups, a commonly adopted modeling strategy typically achieved by placing Bayesian mixture priors. Examples of mixture priors employed in recent Bayesian spatiotemporal literature include the mixture of finite mixtures \parencite{Huetal23}, the Dirichlet process \parencite{Mozetal22}, the temporal dependent Dirichlet process \parencite{DeIetal23}, and the Indian buffet process \parencite{Yangetal22}. More spatially structured priors like the random spanning tree prior \parencite{Zhangetal23} may also be considered to enhance the spatial contiguity of clusters. In line with these, the model proposed by \textcite{Berchuck2021} incorporates clustering across spatial locations via a probit stick breaking process (PSBP) prior \parencite{Rodriguez2011} coupled with a spatial latent neighborhood structure from a Gaussian process, thereby capturing both spatial dependence and heterogeneity.
	
	We make three key contributions to enhance the computational scalability of this baseline Bayesian spatiotemporal factor model. 
	First, we follow \textcite{Walker2007} and develop a new slice sampling algorithm specifically tailored to mixture weights determined by the spatially varying latent variables in the PSBP prior, which ensures the correct conditional posterior distributions and differs from the infinite mixture algorithm presented in \textcite{Berchuck2021}. Under our version of slice sampling, the numbers of spatial mixture components, which do not need to be pre-specified and are permitted to vary across all factors, are ensured non-increasing through the Markov chain Monte Carlo (MCMC) iterations, thus leading to notably accelerated computation for model fitting and spatial prediction as well as significantly alleviated storage requirements for spatial prediction and clustering.
	Second, we impose the Nearest-Neighbor Gaussian Process (NNGP) prior \parencite{Datta2016} in place of the full GP prior and further propose novel sequential updating algorithms for the spatially varying latent variables. Our approach successfully attains high spatial scalability by reducing the corresponding posterior sampling computational complexity from cubic to linear in the number of locations. Under our spatial NNGP prior, the computational speed for making predictions at new locations is markedly increased as well, and storage with respect to the spatial covariance matrix and its inverse can be reduced from quadratic to linear if needed.
	Third, we devise and implement two critical inferential procedures--outcome predictions at new spatial locations and spatial clustering of temporal trends. Possibly due to the prohibitive storage requirements that ensue under the baseline model when the numbers of spatial mixture components and location points are large, both inferential procedures are not currently supported by \textcite{Berchuck2021}'s R package \texttt{spBFA}. 
	Consequently, our improved algorithm manifests considerably superior scalability for MCMC-based posterior sampling as well as subsequent vital inferential tasks on datasets with a few hundred to a couple thousand spatial locations and time points. Detailed scalability summaries for our novel approaches are presented in \Cref{appenH}.
	
	This paper is organized as follows. 
	\Cref{sect2} depicts the baseline Bayesian spatiotemporal Gaussian factor model in \textcite{Berchuck2021} and illustrates its weaknesses in computational scalability through detailed analysis of a motivating example, demonstrating our techniques' strong acceleration capabilities. 
	\Cref{sect3} introduces our novel slice sampling algorithm. 
	\Cref{sect4} derives a new spatial latent NNGP prior and sequential updating schemes with theoretical justifications for the improved computational and storage complexity. 
	\Cref{sect5} elaborates on spatial prediction and temporal trends clustering. \Cref{sect6} presents simulation experiments justifying scalability improvements and a real data example. 
	\Cref{sect7} discusses potential future work.
	Finally, the Supplementary Material details full posterior sampling steps, offers additional acceleration solutions and imposes more complicated temporal structures on equispaced time points,
	complements \Cref{sect2,sect3,sect4,sect5,sect6}, and provides extensions to model non-normal observations.

	\section{Bayesian Spatiotemporal Gaussian Factor Model and Three Acceleration Techniques} \label{sect2}
	
	We first describe the baseline latent factor model for spatiotemporal data in \textcite{Berchuck2021}. 
	For an outcome $y_t^o(\bm{s}_{i})$ of type $o$ ($o=1,\ldots,O$) observed at time $t$ ($t=1,\ldots,T$) and spatial location $\bm{s}_{i}$ ($i=1,\ldots,m$), we assume the separable latent factor model
	\begin{align} \label{eq:single.obs}
		y_t^o(\bm{s}_{i}) &= \bm{x}_t^o(\bm{s}_{i})^\T \bm{\beta} + \sum_{j=1}^k \lambda_j^o(\bm{s}_{i}) \eta_{tj} + \epsilon_t^o(\bm{s}_{i}),
	\end{align}
	where $\bm{x}_t^o(\bm{s}_{i})\in \RR^p$ is the spatiotemporal covariate vector, $\bm{\beta}\in \RR^p$ is the vector of coefficients, $\eta_{tj}$'s are the latent temporal factors, $\lambda_j^o(\bm{s}_{i})$'s are the spatial-specific factor loadings, and $\epsilon_t^o(\bm{s}_{i})$ is the error term. Define the stacked quantities $\bm{y}_t=\big(y_t^1(\bm{s}_{1}),\ldots,y_t^1(\bm{s}_{m}),\ldots,\\ y_t^O(\bm{s}_{1}),
	\ldots,y_t^O(\bm{s}_{m}) \big)^\T \in \RR^{mO\times 1}$, $\bm{X}_t = \big(\bm{x}_t^1(\bm{s}_{1}),\ldots,\bm{x}_t^1(\bm{s}_{m}),\ldots,\bm{x}_t^O(\bm{s}_{1}),\ldots,\bm{x}_t^O(\bm{s}_{m})\big)^\T \in \RR^{mO\times p}$, $\Lambda=(\bm{\lambda}_1,\ldots,\bm{\lambda}_k) \in \RR^{mO\times k}$ with $\bm{\lambda}_j = \big(\lambda_j^1(\bm{s}_{1}),\ldots,\lambda_j^1(\bm{s}_{m}),\ldots,\lambda_j^O(\bm{s}_{1}),\ldots,
	\lambda_j^O(\bm{s}_{m}) \big)^\T \\ \in \RR^{mO\times 1}$, $\bm{\eta}_t = (\eta_{t1},\ldots,\eta_{tk})^{\T} \in \RR^{k\times 1}$, and $\bm{\epsilon}_t = \big(\epsilon^1_t(\bm{s}_{1}),\ldots,\epsilon^1_t(\bm{s}_{m}),\ldots,\epsilon^O_t(\bm{s}_{1}),\ldots\epsilon_t^O(\bm{s}_{m})\big)^\T \in \RR^{mO\times 1}$. Then \eqref{eq:single.obs} is equivalent to the matrix-form $\bm{y}_t = \bm{X}_t\bm{\beta} + \Lambda\bm{\eta}_t + \bm{\epsilon}_t$, for $t=1,\dots,T$. 
	We assume that the error vector is distributed as $\bm{\epsilon}_t\sim N_{mO}\left(\bm{0}, \Xi\right)$ independently across time $t=1,\ldots,T$, where $\Xi=\diag\left((\sigma^2)^1(\bm{s}_{1}),\dots,(\sigma^2)^1(\bm{s}_{m}),\dots,(\sigma^2)^O(\bm{s}_{1}),\dots,(\sigma^2)^O(\bm{s}_{m})\right)$. \Cref{appenA1} details prior specification and Gibbs sampler steps for this basic model. 
	
	To further integrate spatial clustering capabilities, we opt to fully specify the factor loadings matrix and introduce spatial dependency by adapting the Probit Stick Breaking Process (PSBP, \cite{Rodriguez2011}), an extension of the stick-breaking construction of the Dirichlet Process (DP, \cite{Sethuraman1994}), the workhorse of Bayesian nonparametric clustering. 
	For any given $(j,i,o)$, we let the latent mixture distribution $G_j^{i,o}$ follow $\PSBP\big((G_{j})_0,\bm{\alpha}_{j}^{i,o}\big)$ with base measure $(G_{j})_0$ and concentration parameter vector $\bm{\alpha}_{j}^{i,o}$, whose scalar components are $\alpha_{jl_j}^o(\bm{s}_{i}),\:l_j\in \NN,\: l_j \geq 1$, if 
	\begin{align}\label{PSBP} 
		& G_j^{i,o}=\sum_{l_j=1}^{\infty} w_{jl_j}^o(\bm{s}_{i})\delta_{\theta_{jl_j}},\quad\theta_{jl_j}\overset{\iid}{\sim}(G_{j})_0,\; w_{jl_j}^o(\bm{s}_{i})=\Phi(\alpha_{jl_j}^o(\bm{s}_{i})) \prod_{r_j=1}^{l_j-1}[1-\Phi(\alpha_{jr_j}^o(\bm{s}_{i}))]
	\end{align}
	The above construction is indeed adequate since $\sum_{l_j=1}^{\infty}w_{jl_j}^o(\bm{s}_{i})=1$ almost surely for all $(j,i,o)$ \parencite{Rodriguez2011}. We can often truncate the infinite sequence at a large number of terms to obtain a finite mixture for practical implementation of posterior sampling. Let a pre-specified fixed $L\in\mathbb{N}$, $1<L<\infty$ be the number of clusters for all factors and let $w_{jL}^o(\bm{s}_{i})=\prod_{r_j=1}^{L-1}[1-\Phi(\alpha_{jr_j}^o(\bm{s}_{i}))]$ be the last mixture weight for all $(j,i,o)$ in \eqref{PSBP}. We then impose the PSBP in \eqref{PSBP} on $\bm{\lambda}_j$'s independently across all $(j,i,o)$,
	\begin{align}\label{PSBP2}
		&\lambda_j^o(\bm{s}_{i})=\theta_{j\xi_j^o(\bm{s}_{i})} ~|~ G_j^{i,o} \sim G_j^{i,o},~ \text{ where }
		\xi_j^o(\bm{s}_{i})\sim \text{Multinomial}\left(w_{j1}^o(\bm{s}_{i}),\dots,w_{jL}^o(\bm{s}_{i})\right) \nonumber \\
		&\text{ such that }
		\mathbb{P}(\xi_j^o(\bm{s}_{i})=l_j)=w_{jl_j}^o(\bm{s}_{i}), \text{ for any } l_j=1,\dots,L.
	\end{align} 
	
	With the PSBP model delineated in \Cref{PSBP,PSBP2}, spatial neighborhood proximity can now be incorporated into the columns $\bm{\lambda}_j,j\in\{1,\dots,k\}$ of the factor loadings matrix $\Lambda_{mO\times k}$ by jointly modeling the latent variables $\bm{\alpha}_{jl_j}$'s that dictate the weights $\bm{w}_{jl_j}$'s. Let $\bm{\alpha}_{jl_j}(\bm{s}_i)^\T = \big(\alpha_{jl_j}^1(\bm{s}_i),\alpha_{jl_j}^2(\bm{s}_i),\ldots,\alpha_{jl_j}^O(\bm{s}_i)\big)$ for all $i=1,2,\ldots,m$. We assign the prior
	\begin{equation}\label{alpha prior}
		\bm{\alpha}_{jl_j}=\big(\bm{\alpha}_{jl_j}(\bm{s}_1)^\T,\dots,\bm{\alpha}_{jl_j}(\bm{s}_m)^\T\big)^\T \sim N_{mO}\left(\bm{0},  F(\rho)_{m\times m}\otimes \kappa_{O\times O}\right),
	\end{equation}
	independently across all $(j,l_j)$ for $j=1,\dots,k$ and $l_j=1,\dots,L-1$. In \eqref{alpha prior}, the covariance structure of $\bm{\alpha}_{jl_j}$'s is modeled as the Kronecker product of two matrices. Specifically, the covariance matrix $F(\rho)$ stems from a spatial Gaussian process over the $m$ spatial locations with an exponential covariance function parameterized by $\rho$, a length-scale parameter specifying the level of spatial correlation. $\rho$ is assigned a uniform prior and requires Metropolis moves in posterior sampling. The other covariance matrix $\kappa_{O\times O}$ depicts the covariance between the $O$ observation types. It is left fully unstructured and imposed a standard conjugate prior $\mathcal{IW}(\nu,\Theta)$. Posterior sampling of $\bm{\alpha}_{jl_j}$'s typically relies on introducing latent normal variables $z_{jl_j}^o\left(\bm{s}_{i}\right)$'s for probit models to bring about conjugacy for $\bm{\alpha}_{jl_j}$'s (see \Cref{appenD}). 
	Finally, for the atoms $\theta_{jl_j}$'s in \Cref{PSBP2}, we follow \textcite{Bhattacharya2011} to assign a multiplicative gamma process shrinkage prior
	\begin{align} \label{eq:shrinkage}
		& \theta_{jl_j}\overset{\ind}{\sim}N(0, \tau_j^{-1}),\text{ where } \nonumber \\
		& 
		\tau_j=\prod_{h=1}^j\delta_h\text{ with }\delta_1\sim \text{Gamma}(a_1,1)
		\text{ and } \delta_h\sim \text{Gamma}(a_2,1), ~\text{ for all } h\geq2.
	\end{align} 
	
	Despite its capability and flexibility in model fitting, the standard MCMC algorithm derived in \textcite{Berchuck2021} and their R package \texttt{spBFA} become prohibitively slow even when $m$ and $T$ increase to a couple hundred. We identify within this algorithm the following principal causes of the computational burdens resulting from a large $m$. 
	\vspace{1.5mm}\\
	\noindent (i) Posterior sampling of the spatial covariance parameters $\rho$ and $\kappa$ requires $F(\rho)^{-1}$ and $\determ(F(\rho))$. A computational complexity of $\Ocal(m^3)$ is thus needed.\\
	\noindent (ii) Posterior sampling of the spatially varying latent variables ${\alpha}_{jl_j}^o\left(\bm{s}_{i}\right)$'s in the PSBP prior involves the notorious cubic computational complexity from fitting Gaussian processes. \\
	\noindent (iii) Posterior sampling of the mixture labels $\xi_j^o\left(\bm{s}_{i}\right)$'s and introduced latent normal $z_{jl_j}^o\left(\bm{s}_{i}\right)$'s across all spatial locations for a moderately large $L$ also incurs computational costs.\\
	\noindent (iv) A major spatial prediction step obtaining predicted spatially varying latent variables $\hat{\bm{\alpha}}\left(\bm{s}_{(m+1):(m+r)}\right)$ at $r$ new locations requires a computational complexity quadratic in $m$.
	
	We tackle these challenges by proposing three new techniques detailed in Sections \ref{sect3} and \ref{sect4}. In particular, slice sampling in \Cref{sect3} addresses the issues (iii) and (iv) above; the spatial NNGP prior in \Cref{sect4} addresses the issues (i) and (iv) above; the new sequential updating algorithm in \Cref{sect4d} addresses the issue (ii) above.
	
	We first present a small-scale toy example to illustrate posterior sampling time reduction attributable to our three novelties. We consider four methods. \texttt{fullGPfixedL} assumes a common number of spatial mixture components $L$ for all $k$ latent factors and imposes the full GP prior on $\alpha_{jl_j}^o\left(\bm{s}_{i}\right)$'s; \texttt{NNGPblockFixedL} assumes a common fixed $L$ and adopts our spatial NNGP prior with the standard block-wise updating scheme; \texttt{NNGPsequenFixedL} assumes a common fixed $L$ and adopts the spatial NNGP prior with our new sequential updating scheme in \Cref{sect4d}; \texttt{NNGPsequenVaryLj} adopts the spatial NNGP prior, the new sequential updating scheme, and our slice sampling algorithm in \Cref{sect3} with varying numbers of spatial mixture components $L_{1:k}$. For comparison, we also ran two variations of algorithms in the R package \texttt{spBFA} by \textcite{Berchuck2021}. \texttt{spBFAL10} specifies a fixed $L=10$ mixture components and \texttt{spBFALInf} is under \texttt{spBFA}'s infinite mixture implementation. Both adopt the full spatial GP prior. 
	
	\begin{figure}[h]
		\centering
		\includegraphics[width=.8\textwidth]{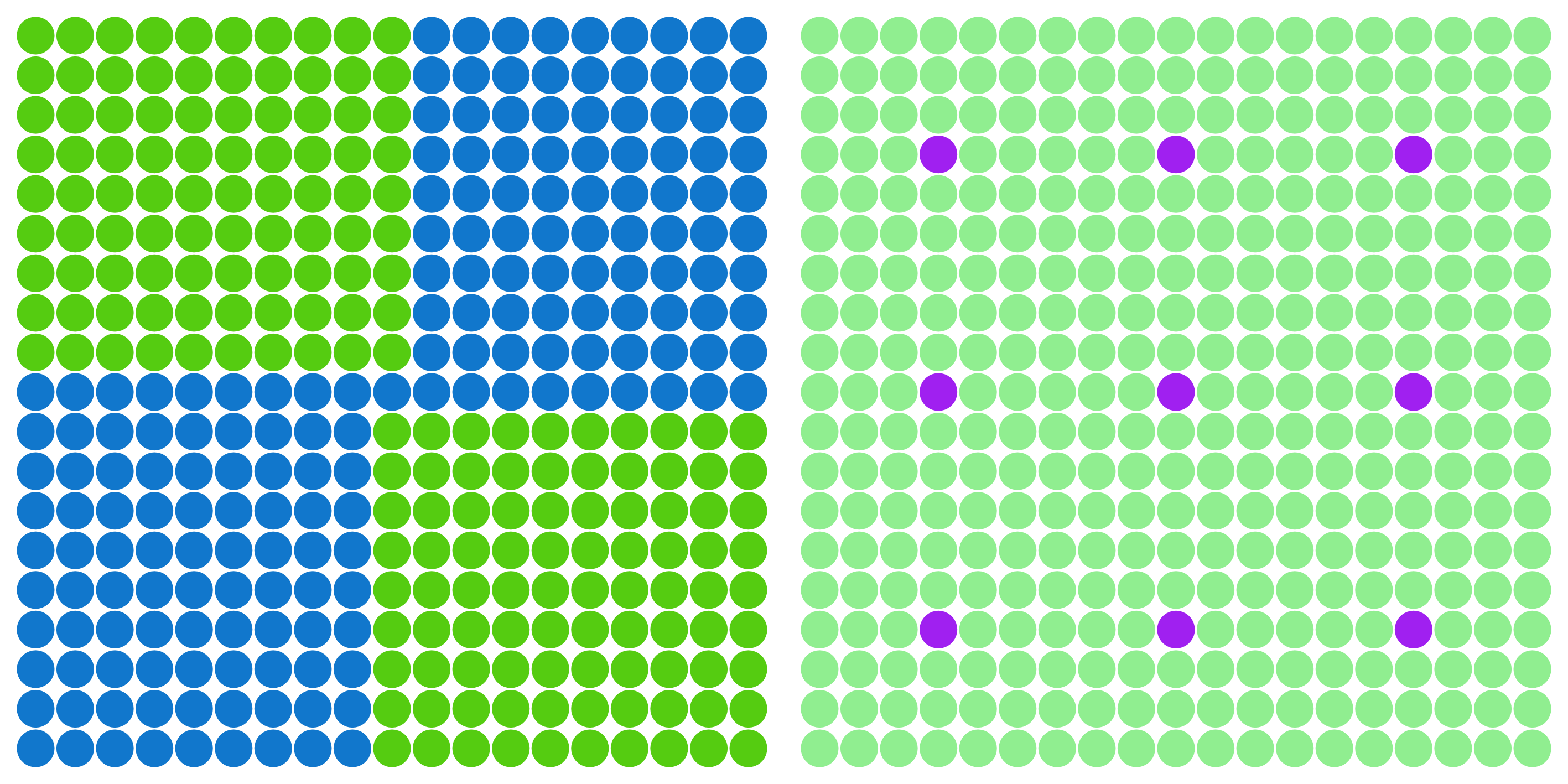}
		\caption{The actual spatial grouping (left) for the toy example. The two colors represent the two groups for $m=19^2=361$ spatial locations. In the right panel, the 9 purple locations are testing locations, and the 352 light green locations are training ones.} 
		\label{toyExSpatialPlots}
	\end{figure}
	
	We generated data following the model \eqref{eq:single.obs} without the regression term $\bm{x}_t^o(\bm{s}_{i})^\T \bm{\beta}$. The number of factors was set to $k=2$, and the actual numbers of clusters for both factors were also set to $2$. We specified $\theta_{11}=\theta_{12}=5$, $\theta_{21}=10$, and $\theta_{22}=-10$ for the atoms $\theta_{jl_j}$'s. We considered $m=19^2=361$ spatial locations $\bm{s}_i=(i_1,i_2)$ for $i=1,2,\ldots,361$ on an equispaced $2$-dimensional grid, where $i_1,i_2\in\{1,2,\ldots,19\}$ satisfy $19\cdot(i_1-1)+i_2=i$ for each $i$. These 361 locations were assigned to two actual groups as in \Cref{toyExSpatialPlots}\textbf{A}. One group corresponds to $(\theta_{11},\theta_{21})=(5,10)$ and the other corresponds to $(\theta_{12},\theta_{22})=(5,-10)$. We set $O=1$, $\psi=2.3$, $\sigma^2\left(\bm{s}_{i}\right)=0.01$ for all $i$ and specified $T=310$ time points. When fitting the models, we utilized a training data set that consists of $y_t(\bm{s}_i)$'s at 352 out of the 361 spatial locations (\Cref{toyExSpatialPlots}\textbf{B}) and the first 300 of the 310 time points. We used 10 for both the value of $L$ and an upper bound to $L_{1:k}$. A burn-in length of 80000 and a post-burn-in length of 20000 (thinned to 5000) were used throughout. Simulated $y_t(\bm{s}_i)$'s on the remaining 9 testing locations for the first 300 time points are used to assess spatial prediction performances. Similarly, $y_t(\bm{s}_i)$'s on the 352 training locations for the last 10 time points are used to assess temporal prediction performances.
	
	The overall model fitting time for \texttt{fullGPfixedL}, \texttt{NNGPblockFixedL}, \texttt{NNGPsequenFixedL}, \texttt{NNGPsequenVaryLj}, \texttt{spBFAL10}, and \texttt{spBFALInf} are 15.77 hours, 15.46 hours, 15.19 hours, 14.58 hours, 7.24 days, and 3.08 days, respectively. The first four methods based on our implementation are several times faster than the \texttt{spBFA} package, as we have optimized several places in \texttt{spBFA}. 
	Among the first four methods, \texttt{NNGPsequenVaryLj} is the fastest, followed by \texttt{NNGPsequenFixedL}, \texttt{NNGPblockFixedL}, and finally \texttt{fullGPfixedL}, as our three novelties significantly accelerate the posterior sampling of spatial-related parameters. 
	\begin{table}[h]
		\centering
		\scalebox{0.85}{
			\centering
			\begin{tabular}{|*{5}{c|}}
				\hline
				\backslashbox{\textbf{Parameter}}{\textbf{Method}} & \texttt{fullGPfixedL} & \texttt{NNGPblockFixedL} & \texttt{NNGPsequenFixedL} & \texttt{NNGPsequenVaryLj}\\
				\hline
				$\rho$ & 11.389 &  4.961 & 4.965 & 4.986 \\
				\hline
				$\kappa$ & 2.006 & 1.004 & 1.007 & 1.001 \\
				\hline
				$\alpha_{jl_j}^o(\bm{s}_{i})$'s & 19.829 & 17.728 &  7.14 &   21.003\\
				\hline
				$z_{jl_j}^o(\bm{s}_{i})$'s or $u_j^o(\bm{s}_{i})$'s & 24.408 &  24.158 & 24.079 &  $\approx 0$ \\
				\hline
				$\xi_j^o(\bm{s}_{i})$'s & 13.164 & 13.235 & 13.129 & 3.279\\
				\hline
				$\theta_{jl_j}$'s & 243.285 & 242.981 & 243.138 & 242.254\\
				\hline
				$\delta_{1:k}$ & 0.986 &  0.977 &  0.986 & 0.934\\
				\hline
			\end{tabular}         
		}
		\caption{Average sampling time per MCMC iteration (in milliseconds) corresponding to the spatial-related parameters $\rho$, $\kappa$, $\alpha_{jl_j}^o(\bm{s}_{i})$'s, $z_{jl_j}^o(\bm{s}_{i})$'s or $u_j^o(\bm{s}_{i})$'s, $\xi_j^o(\bm{s}_{i})$'s, $\theta_{jl_j}$'s, and $\delta_{1:k}$ across 5000 post-burn-in MCMC iterations for our four methods in the toy example.}
		\label{sect6dGibbsStepTimeTable}
	\end{table}  
	We further provide a detailed breakdown of the posterior sampling time for different sets of parameters and report the average time per MCMC iteration in \Cref{sect6dGibbsStepTimeTable}. The NNGP prior has significantly reduced computation time for covariance parameters $\rho$ and $\kappa$. On top of that, \texttt{NNGPsequenFixedL} using our sequential updating scheme has further reduced sampling time for the spatially varying latent variables $\alpha_{jl_j}^o\left(\bm{s}_{i}\right)$'s. \texttt{NNGPsequenVaryLj} based on our slice sampling algorithm bypasses sampling latent variables $z_{jl_j}^o(\bm{s}_{i})$'s and samples instead the auxiliary variables $u_j^o(\bm{s}_{i})$'s (see \Cref{sect3}), whose computation is much faster. The Gibbs sampler step for $\xi_j^o(\bm{s}_{i})$'s has also been significantly accelerated. 
	Overall, the method \texttt{NNGPsequenVaryLj}, which utilizes all our three new techniques, exhibits the best computational efficiency. In this toy example, the four methods are still close in overall computation speed, as sampling loadings parameters $\theta_{jl_j}$'s took the most model fitting time. As $m$ increases, however, the acceleration in sampling spatial-related parameters quickly becomes substantial, as we shall see in \Cref{sect6a} and \Cref{appenIb}. $m=3600$, for instance, leads to an overall computational speed increase of more than 10 times.
	
	\begin{table}[h]
		\centering
		\scalebox{0.72}{
			\begin{tabular}{|*{7}{c|}}
				\hline
				\textbf{Method} & \texttt{fullGPfixedL} & \texttt{NNGPblockFixedL} & \texttt{NNGPsequenFixedL} & \texttt{NNGPsequenVaryLj} & \texttt{spBFAL10} & \texttt{spBFALInf}\\
				\hline
				Spatial Prediction &  14.123149 &  14.003086 & 14.341286 & 12.764447 & N.A. & N.A. \\
				\hline
				Temporal Prediction & 18.518662 & 18.754345 & 18.299211 & 19.340736
				&  19.409493 & 164.42639\\
				\hline
		\end{tabular} }                                                             \caption{Toy example: posterior mean squared errors for spatial and temporal predictions.}
		\label{table:toy.prediction}
	\end{table}
	\Cref{appenIa} elaborates on all methods' performances in terms of posterior deviances, diagnostics statistics, temporal prediction metrics, spatial prediction time and metrics, and spatial clustering outcomes, from which we can see that adopting our three novel techniques can indeed significantly accelerate the default MCMC algorithm for the Bayesian spatiotemporal factor model in \textcite{Berchuck2021} without sacrificing inferential performances in model fitting, prediction, and spatial clustering. All four methods’ diagnostics, converged posterior deviances, and temporal prediction results are close to their \texttt{spBFAL10} counterparts and significantly better than their \texttt{spBFALInf} counterparts. Moreover, all four methods deliver close satisfactory spatial prediction performances and achieve 100\% accuracy for clustering based on recorded posterior samples the $m_0=352$ training locations into the two actual spatial groups. The \texttt{spBFA} package currently does not support spatial prediction or clustering. See \Cref{table:toy.prediction} and \Cref{appenIa} for the results we obtained.

	\section{Slice Sampling for Bayesian Spatial Clustering} \label{sect3}
	
	We propose a new slice sampling algorithm to address the computational burden incurred by using a common number of mixture components $L\in\NN, \,L>1$ for all latent factors in the PSBP model \eqref{PSBP2}. For one thing, pre-specifying a fixed common $L$ is likely restrictive since the numbers of spatial mixtures are unknown and may well differ across the $k$ latent factors. An at least moderately large $L$ would typically be required for fitting many mixture models across $m$ spatial locations, especially when $m$ becomes large. This leads to computational inefficiency since we need to perform more complicated calculations when updating $\xi_j^o\left(\bm{s}_{i}\right)$'s, $\delta_h$'s, $\kappa$, $\rho$ and sample more spatial clustering parameters $\theta_{jl_j}$'s, $\alpha_{jl_j}^o\left(\bm{s}_{i}\right)$'s in each MCMC iteration. 
	Moreover, when both $L$ and the number of spatial locations $m$ are large, storage requirements for either spatial prediction at new locations or clustering of temporal trends can become highly prohibitive, as one needs to store posterior samples of $\theta_{jl_j}$'s, $\bm{\alpha}_{jl_j}$'s, and $\bm{w}_{jl_j}$'s from all kept post-burn-in MCMC iterations. Last but not least, posterior sampling of the introduced latent normal variables $z_{jl_j}^o\left(\bm{s}_{i}\right)$'s (\Cref{appenD}) is required when a common fixed $L$ is assumed and is computationally demanding when $L$ is large. 
	
	These concerns can be effectively addressed by assuming a varying number of clusters $L_j$ for the $j$th latent factor loadings $\bm{\lambda}_j$ and adapting pertinent slice sampling ideas in \textcite{Walker2007}. Our novel algorithm follows the original version of slice sampling \parencite{Walker2007} and essentially differs from \textcite{Berchuck2021}'s flawed implementation in their supplementary material and R package \texttt{spBFA}. Some further notes are in \Cref{appenE}. 
	
	Let us fix a pair $(i,o)\in\{1,\dots,m\}\times\{1,\dots,O\}$. Consider the spatial PSBP prior under the finite-mixture version of \eqref{PSBP} with $L_j$ mixture components for the $j$th latent factor loadings. For $t=1,\dots,T$, the observation $y_t^o\left(\bm{s}_{i}\right)$ has density 
	\begin{align}\label{fyio}
		&f \left(y_t^o \left(\bm{s}_{i}\right) ~|~ \bm{\beta},\bm{\eta}_t,(\sigma^2)^o(\bm{s}_{i}), L_j\text{ for all }j,\bm{\alpha}^o\left(\bm{s}_{i}\right),\bm{\theta} \right) \\
		= &
		\sum_{l_1=1}^{L_1}\ldots \sum_{l_k=1}^{L_k}\left\lbrace \left[\prod_{j=1}^k w_{jl_j}^o\left(\bm{s}_{i}\right)  \right]
		\times f\left(y_t^o (\bm{s}_{i}) ~|~ \bm{\beta},\bm{\eta}_t,(\sigma^2)^o(\bm{s}_{i}),\theta_{jl_j}\text{ for all }j=1,\dots,k \right)\right\rbrace,\nonumber
	\end{align}	
	where $\bm{\alpha}^o\big(\bm{s}_{i}\big) = \{\alpha_{jl_j}^o(\bm{s}_i):j=1,\ldots,k,\,l_j=1,\ldots,L_j-1\}$. To incorporate slice sampling \parencite{Walker2007}, we further introduce latent variables $u_j^o\left(\bm{s}_{i}\right)$, $j\in\{1,\dots,k\}$, such that
	\begin{align}\label{fyuio}
		&f\left({y}_t^o\left(\bm{s}_{i}\right),u_{1:k}^o\left(\bm{s}_{i}\right)  ~|~ \bm{\beta},\bm{\eta}_t,(\sigma^2)^o(\bm{s}_{i}), L_j\text{ for all }j,\bm{\alpha}^o\left(\bm{s}_{i}\right),\bm{\theta} \right) \\
		= &
		\sum_{l_1=1}^{L_1}\ldots \sum_{l_k=1}^{L_k} \left\lbrace\left[\prod_{j=1}^k \mathbbm{1}_{\left\lbrace u_j^o\left(\bm{s}_{i}\right) < w_{jl_j}^o\left(\bm{s}_{i}\right)\right\rbrace}\right]
		\times f\left(y_t^o(\bm{s}_{i})~|~\bm{\beta},\bm{\eta}_t,(\sigma^2)^o(\bm{s}_{i}),\theta_{1l_{1}},\ldots, \theta_{kl_{k}}\right)\right\rbrace,\nonumber
	\end{align}
	where $\mathbbm{1}_E$ denotes the indicator function of an event $E$.
	Since integrating over $u_j^o\left(\bm{s}_{i}\right)$'s in \eqref{fyuio} with respect to the Lebesgue measure on the measurable space $\left((0,1]^k, \mathcal{B}\left((0,1]^k\right)\right)$ yields \eqref{fyio}, the joint density in \Cref{fyuio} with the corresponding marginal density in \Cref{fyio} indeed exists. Hence, the complete data likelihood is
	\begin{align}\label{completeLik1}
		\ell_{\text{complete}} &= \prod_{o=1}^O \prod_{i=1}^m\prod_{t=1}^T \left[\prod_{j=1}^k\mathbbm{1}_{\left\lbrace u_j^o(\bm{s}_{i})<w^o_{j\xi_j^o(\bm{s}_{i})}(\bm{s}_{i})\right\rbrace}\right] \\
		&\times f\left(y_t^o(\bm{s}_{i})~|~\bm{\beta},\bm{\eta}_t,(\sigma^2)^o(\bm{s}_{i}),\xi_{j}^o(\bm{s}_{i}),\theta_{j\xi_{j}^o(\bm{s}_{i})}\text{ for all }j=1,\ldots,k \right) , \nonumber
	\end{align}
	where for any $(i,o,j)\in\{1,\dots,m\}\times\{1,\dots,O\}\times\{1,\dots,k\}$ and $l_j\in\{1,\dots,L_j\}$,
	\begin{equation}\label{weightsFormula1}
		w_{jl_j}^o(\bm{s}_{i})=
		\begin{cases}
			\Phi\left(\alpha_{jl_j}^o(\bm{s}_{i})\right)\prod_{r_j<l_j}\left[1-\Phi\left(\alpha_{jr_j}^o(\bm{s}_{i})\right) \right],&\text{ if }l_j\in\{1,\dots,L_j-1\} \\
			\prod_{r_j<L_j}\left[1-\Phi\left(\alpha_{jr_j}^o(\bm{s}_{i})\right) \right],&\text{ if }l_j=L_j
		\end{cases}.	  
	\end{equation}
	
	Under this model specification, we no longer need to introduce and sample the latent normal $z_{jl_j}^o\left(\bm{s}_{i}\right)$'s (\Cref{sect2}) from their full conditional distributions (\Cref{postZ1} in \Cref{appenD}) to bring about conjugacy for $\alpha_{jl_j}^o(\bm{s}_i)$'s, which requires a total of $\Ocal(kLmO)$ floating point operations (flops) per MCMC iteration and is thus computationally inefficient for a large $L$. Instead, we only need to sample the latent uniform $u_j^o\left(\bm{s}_{i}\right)$'s from their full conditional distributions below. 
	\begin{align}\label{postU}
		&\text{For any }(i,o,j),\quad f\left(u_j^o(\bm{s}_{i})~|~ \cdot\right) \propto \mathbbm{1}_{\left\lbrace u_j^o(\bm{s}_{i})<w^o_{j\xi_j^o(\bm{s}_{i})}(\bm{s}_{i}) \right\rbrace}\sim\unif\left(0, w^o_{j\xi_j^o(\bm{s}_{i})}(\bm{s}_{i})\right) .
	\end{align}
	The corresponding computational complexity per MCMC iteration is reduced to $\Ocal(kmO)$. 
	
	The pivotal role played by the introduced parameters $L_j$'s and $u_j^o\left(\bm{s}_{i}\right)$'s is reflected in the step updating $L_j$'s and sampling cluster labels $\xi_j^o\left(\bm{s}_{i}\right)$'s from their full conditional distributions. Fix any arbitrary $j\in\{1,\dots,k\}$. For all $l_j\in\{1,\dots,L_j^{\text{old}}\}$ and for any $(i,o)$,
	\begin{align}\label{postXi1}
		\mathbb{P}\left(\xi_j^o(\bm{s}_{i})=l_j~|~\cdot\right) &\propto \mathbbm{1}_{\left\lbrace w_{jl_j}^o(\bm{s}_{i})>u_j^o(\bm{s}_{i})\right\rbrace } \nonumber \\
		& \times \prod_{t=1}^{T} f\left(y_t^o(\bm{s}_{i})|\bm{\beta},\bm{\eta}_t,(\sigma^2)^o(\bm{s}_{i}),\xi_j^o(\bm{s}_{i})=l_j,\bm{\xi}_{-j}^o(\bm{s}_{i}),\bm{\theta}_{\bm{\xi}^o(\bm{s}_{i})}\right),
	\end{align}	
	where for any $(i,o,j,l_j)$ such that $w_{jl_j}^o(\bm{s}_{i})>u_j^o(\bm{s}_{i})$, for all $t\in\{1,\dots,T\}$,
	\begin{align} &f\left(y_t^o(\bm{s}_{i})~|~\bm{\beta},\bm{\eta}_t,(\sigma^2)^o(\bm{s}_{i}),\xi_j^o(\bm{s}_{i})=l_j,\bm{\xi}_{-j}^o(\bm{s}_{i}),\bm{\theta}_{\bm{\xi}^o(\bm{s}_{i})}\right) 
		\nonumber \\
		=\;&(2\pi)^{-\frac{1}{2}}(\sigma^{-1})^o(\bm{s}_{i})\exp\left\lbrace -\frac{1}{2}(\sigma^{-2})^o(\bm{s}_{i})\Big[y_t^o(\bm{s}_{i})-\bm{x}_t^o(\bm{s}_{i})^{\T}\bm{\beta}-\sum_{h\neq j}^{1\leq h\leq k}\theta_{h\xi_h^o(\bm{s}_{i})}\eta_{th}-\theta_{jl_j}\eta_{tj} \Big]^2\right\rbrace. \nonumber
	\end{align}
	For any $(i,o)\in\{1,\ldots,m\}\times\{1,\ldots,O\}$, we let
	\begin{align}
		&L_j^{i,o}=\text{ the smallest positive integer such that }\sum\nolimits_{l_j=1}^{L_j^{i,o}}w_{jl_j}^o(\bm{s}_{i}) > 1-u_j^o(\bm{s}_{i}) \nonumber\\
		\text{ and }~ &L_j^{\text{new}}=\max\left\lbrace L_j^{i,o}:i=1,\dots,m,\,o=1,\dots,O\right\rbrace. \label{LjNewFormula1}
	\end{align}
	It is clear from \Cref{postXi1,LjNewFormula1} that for any $l_j>L_j^{\text{new}}$, $\mathbb{P}\left(\xi_j^o(\bm{s}_{i})=l_j|\cdot\right)=0$ for all $(i,o)$. Hence, we can update the parameter estimate for $L_j\in\mathbb{N}\backslash\{0\}$ from $L_j^{\text{old}}$ to $L_j^{\text{new}}\leq L_j^{\text{old}}<\infty$ in this MCMC iteration and only consider $l_j\in\{1,\dots,L_j^{\text{new}}\}$ to be the possible values for $\xi_j^o(\bm{s}_{i})$ for all $(i,o)$. 
	After obtaining $\{L_j^{\text{new}}:j=1,\dots,k\}$ by \Cref{LjNewFormula1} at the start of this Gibbs sampler step, we immediately update estimates for $\bm{\alpha}_{jl_j}$'s by only keeping the $mO\times 1$ vectors corresponding to $l_j\in\{1,2,\dots,L_j^{\text{new}}-1\}$ and discarding the ones corresponding to $l_j\in\{L_j^{\text{new}},L_j^{\text{new}}+1,\dots,L_j^{\text{old}}-1\}$. We then formulate our new weights $\bm{w}_{jl_j}$ for $l_j=1,\ldots,L_j^{\text{new}}$ (as specified by \Cref{weightsFormula1}) by discarding the previous $\bm{w}_{jl_j}$ estimates with $l_j\in\{L_j^{\text{new}}+1,\dots,L_j^{\text{old}}\}$ and adjusting the ${w}^o_{jL_j^{\text{new}}}(\bm{s}_{i})$ estimates if $L_j^{\text{new}}$ is strictly less than $L_j^{\text{old}}$. New cluster labels $\xi_j^o(\bm{s}_{i})$'s can then be sampled according to \Cref{postXi1} based on these new weights. Among all kept weights parameters, only the ${w}^o_{jL_j^{\text{new}}}(\bm{s}_{i})$ estimates are possibly altered (increased) in this step. Since all updated weights are ensured greater than or equal to their old values, it is indeed guaranteed that $\xi_j^o\left(\bm{s}_{i}\right)^{\text{old}}\leq L_j^{i,o}\leq L_j^{\text{new}}\leq L_j^{\text{old}}$ is still in the support for $\xi_j^o\left(\bm{s}_{i}\right)^{\text{new}}$ and $u_j^o(\bm{s}_{i})<w^o_{j\xi_j^o(\bm{s}_{i})^{\text{new}}}(\bm{s}_{i})^{\text{new}}$ for each $(j,i,o)$.
	
	Thanks to this non-increasing property of posterior samples for $L_j$'s through the MCMC iterations, we can perform simplified calculations when updating $\xi_j^o\left(\bm{s}_{i}\right)$'s, $\delta_h$'s, $\kappa$, $\rho$ and sample fewer posterior $\theta_{jl_j}$'s and $\bm{\alpha}_{jl_j}$'s from their full conditional distributions given in \eqref{postTheta} and below. In particular, the posterior sampling time of $\xi_j^o\left(\bm{s}_{i}\right)$'s is reduced most considerably (\Cref{sect6dGibbsStepTimeTable,table:summary.complexity}, \Cref{sect6a}, \Cref{appenIb}). 
	As each $L_j$ is usually much smaller than its initial upper bound in practice, storage concerns related to $\theta_{jl_j}$'s, $\bm{\alpha}_{jl_j}$'s, and $\bm{w}_{jl_j}$'s can be significantly alleviated as well. Spatial prediction can also be considerably accelerated (\Cref{sect5a,sect6b}, \Cref{appenG2b,appenIa}). See \Cref{appenH1} for theoretical analysis of complexities.
	
	We do not need to sample any $\bm{\alpha}_{jl_j}$'s for $j$ values with $L_j=1$, as the corresponding $w_{j1}^o(\bm{s}_{i})$'s have already been set to $1$ in the earlier Gibbs sampling step for $\xi_j^o(\bm{s}_{i})$'s.
	Fix any arbitrary $(i,o,j)$ where $L_j>1$ and fix an $l_j\in\{1,\dots,L_j-1\}$. The only term in the complete data likelihood \eqref{completeLik1} possibly involving $\alpha_{jl_j}^o(\bm{s}_{i})$ is $\mathbbm{1}_{\left\lbrace u_j^o(\bm{s}_{i})<w_{j\xi_j^o(\bm{s}_{i})}^o(\bm{s}_{i})\right\rbrace}$. Hence for any $(j, l_j)$ where $j\in\{1,\dots,k\}$ and $l_j\in\{1,\dots,L_j-1\}$,
	\begin{align}
		\bm{\alpha}_{jl_j}|\cdot 
		& \sim N_{mO}\left(\bm{0}, F(\rho)\otimes\kappa\right)\times\prod_{i=1}^m\prod_{o=1}^O \mathbbm{1}_{\left\lbrace l_j>\xi_j^o(\bm{s}_{i})\text{ or }  u_j^o(\bm{s}_{i})<w^o_{j\xi_j^o(\bm{s}_{i})}(\bm{s}_{i})\right\rbrace}\label{postAlphaVaryLj12}\\
		&\sim N_{mO}\left(\bm{0}, F(\rho)\otimes\kappa\right)\times\prod_{i=1}^m\prod_{o=1}^O
		\begin{cases}
			1,  &\text{ if }l_j>\xi_j^o(\bm{s}_{i}) \\
			\mathbbm{1}_{\left\lbrace {\alpha}_{jl_j}^o(\bm{s}_{i})^{\text{new}}>\text{lowerBound}_{(j,l_j,i,o)} \right\rbrace},  &\text{ if }l_j=\xi_j^o(\bm{s}_{i}) \\
			\mathbbm{1}_{\left\lbrace {\alpha}_{jl_j}^o(\bm{s}_{i})^{\text{new}}<\text{upperBound}_{(j,l_j,i,o)}\right\rbrace},  &\text{ if }l_j<\xi_j^o(\bm{s}_{i}) \\
		\end{cases},\nonumber
	\end{align}
	where $\bm{\alpha}_{jl_j}=\big(\bm{\alpha}_{jl_j}(\bm{s}_1)^\T,\dots,\bm{\alpha}_{jl_j}(\bm{s}_m)^\T\big)^\T$ with $\bm{\alpha}_{jl_j}(\bm{s}_i)^\T = \big(\alpha_{jl_j}^1(\bm{s}_i),\alpha_{jl_j}^2(\bm{s}_i),\ldots,\alpha_{jl_j}^O(\bm{s}_i)\big)$ for all $i=1,2,\ldots,m$, and for any $(j,l_j,i,o)$, 
	\begin{align}
		&\text{lowerBound}_{(j,l_j,i,o)}=\Phi^{-1}\left( \frac{u_j^o(\bm{s}_{i})\cdot\Phi\left({\alpha}_{jl_j}^o(\bm{s}_{i})^{\text{old}}\right) }{w^o_{j\xi_j^o(\bm{s}_{i})}(\bm{s}_{i})} \right) \text{ and }\label{alphaLowerBound}\\ 
		& \text{upperBound}_{(j,l_j,i,o)}=\Phi^{-1}\left(1 - \frac{u_j^o(\bm{s}_{i})\cdot\left[ 1-\Phi\left({\alpha}_{jl_j}^o(\bm{s}_{i})^{\text{old}}\right)\right]  }{w^o_{j\xi_j^o(\bm{s}_{i})}(\bm{s}_{i})} \right).\label{alphaUpperBound}
	\end{align}
	These truncated multivariate normal posterior distributions of the $\bm{\alpha}_{jl_j}$'s can be directly sampled via multivariate rejection sampling, as implemented by for instance the \texttt{rtmvnorm()} function in the R package \texttt{tmvtnorm}. When $m$ is large, we will further enhance computational efficiency in Section \ref{sect4d} by replacing this possibly high-dimensional rejection sampling by a sequential updating algorithm that uses low-dimensional rejection samplings. After obtaining new estimates for $\alpha_{jl_j}^o(\bm{s}_i)$'s, we formulate posterior samples of $w_{jl_j}^o(\bm{s}_i)$'s via \Cref{weightsFormula1} correspondingly. The complete Gibbs sampler based on this slice sampling algorithm can be found in \Cref{appenA3}. 
	
	We briefly discuss three computational complexity sources of this slice sampling algorithm. One is the posterior sampling of $u_j^o(\bm{s}_{i})$'s from \Cref{postU}, which has a smaller complexity $\Ocal(kmO)$ over the indices $j,i,o$. The standard Gibbs sampler for PSBP prior, on the other hand, requires sampling the latent normal variables $z_{jl_j}^o(\bm{s}_{i})$'s from their conditional posteriors that are possibly truncated normal distributions (\Cref{postZ}), which has complexity $\Ocal(kLmO)$ over all indices $j,l_j,i,o$. One concerns the posterior sampling of $\xi_j^o(\bm{s}_{i})$'s, which has computational complexity $\Ocal\left(\sum_{j=1}^k L_j\cdot mOT(p+k)\right)$. Another stems from the calculation of lower and upper bounds in \eqref{alphaLowerBound} and \eqref{alphaUpperBound} and the cost of rejection sampling from \eqref{postAlphaVaryLj12}. We provide a more detailed analysis of and considerably reduce the complexity of this part in \Cref{sect4d}. In practice, benefits from replacing the sampling of $z_{jl_j}^o(\bm{s}_{i})$'s by $u_j^o(\bm{s}_{i})$'s and accelerating the sampling of $\xi_j^o(\bm{s}_{i})$'s outweigh the additional cost from rejection sampling of $\bm{\alpha}_{jl_j}(\bm{s}_i)$'s, especially when $m$ and $L$ are large; see our numerical results in \Cref{sect6a} and \Cref{appenIb}.
	
	\section{Spatial Scalability via Nearest-Neighbor Gaussian Process and Sequential Updates}\label{sect4} 
	
	\subsection{Computational Burdens from a Large $F(\rho)_{m\times m}$}\label{sect4a} 
	
	For \textcite{Berchuck2021}'s Bayesian spatiotemporal factor model in \Cref{sect2}, one major source of computational bottleneck is the GP prior in \Cref{alpha prior}, where the covariance matrix has a large dimension of $mO\times mO$. For one, sampling $\rho$ and $\kappa_{O\times O}$ from their conditional posterior distributions will inevitably involve computation related to the inversion and determinant of a large matrix $F(\rho)_{m\times m}$. The computational complexity for inversion and determinant calculation are both typically $\Ocal(m^3)$ for $m\times m$ matrices without any special structure, and the storage cost of the entire $F(\rho)$ and $F(\rho)^{-1}$ is $\Ocal(m^2)$, both of which become prohibitive for a large $m$. For another, the posterior sampling of $\bm{\alpha}_{jl_j}$'s is also heavily affected. As deduced in \Cref{postAlphaFixedL1}, for any $(j,l_j)\in\{1,\dots,k\}\times\{1,\dots,L-1\}$, $\bm{\alpha}_{jl_j}|\cdot \sim N_{mO}\left( \left[ I_{mO}+F(\rho)^{-1}\otimes\kappa^{-1} \right]^{-1}\bm{z}_{jl_j}, \left[ 
	I_{mO}+F(\rho)^{-1}\otimes\kappa^{-1} \right]^{-1} \right)$. We thus need to evaluate $\chol\left( \left[ I_{mO}+F(\rho)^{-1}\otimes\kappa^{-1} \right]^{-1}  \right)$, the Cholesky factor of an $mO\times mO$ matrix, which requires $\Ocal(m^3)$ flops using standard methods. Furthermore, if we adopt the slice sampling algorithm in \Cref{sect3}, then the full conditional distribution of each $\bm{\alpha}_{jl_j}$ is likely a truncated $mO$-variate normal distribution given by \eqref{postAlphaVaryLj12}, which often has a very low acceptance rate when $m$ is large. As such, the slow rejection sampling step will likely result in a computational complexity even larger than $\Ocal(m^3)$ for updating all $\bm{\alpha}_{jl_j}$'s.
	
	We overcome these computational issues from large-scale GP using two techniques. First, by treating the $\bm{\alpha}_{jl_j}\big(\bm{s}_i\big)$'s as the location-specific latent random effects $\bm{w}(\bm{s}_i)$'s in \textcite{Datta2016} and \textcite{Finley2022}, we can straightforwardly apply the latent Nearest-Neighbor Gaussian Process (NNGP) model framework to significantly reduce the computational and storage complexities related to the inversion and determinant of large $m\times m$ matrices. Second, we propose a sequential updating scheme to bypass the inefficient posterior sampling of $\bm{\alpha}_{jl_j}$'s. Computation and storage efficiency can hence be greatly enhanced (\Cref{appenH2,appenH3}). \Cref{appenF} gives some further comments.
	
	\subsection{Incorporating the Spatial Latent NNGP}\label{sect4b}
	We derive an NNGP prior $\tilde \pi(\bm{\alpha}_{jl_j}|\kappa,\rho)=N_{mO}\left(\bm{0},\tilde{F}(\rho)\otimes\kappa\right) $ for all $(j,l_j)$, where $\tilde{F}(\rho)^{-1}_{m\times m}$ is sparse and can be calculated in $\Ocal(m)$ flops together with $\determ\big( \tilde{F}(\rho)\big)$. 
	Fix any arbitrary $(j,l_j)$ and consider the zero-mean $O$-variate Gaussian process $\bm{\alpha}_{jl_j}|\kappa,\rho\sim N_{mO}\left( \bm{0}, F(\rho)\otimes\kappa\right)$ over the spatial domain $\mathcal{D}\subset\mathbb{R}^d,d\in\mathbb{N}$, with $\bm{\alpha}_{jl_j}=\left(\bm{\alpha}_{jl_j}(\bm{s}_1)^\T,\dots,\bm{\alpha}_{jl_j}(\bm{s}_m)^\T\right)^\T$, where each $\bm{\alpha}_{jl_j}(\bm{s}_i),i\in\{1,\dots,m\}$, conceived as a realization of the latent process $\{\bm{\alpha}_{jl_j}(\bm{s})|\bm{s}\in\mathcal{D}\}$, is an $O\times 1$ vector corresponding to the observation at location $\bm{s}_i$ on the $O$ variables. We take the location reference set to be the same as the set of observed locations $\mathcal{S} =\{\bm{s}_1,\dots,\bm{s}_m\}\subset \mathcal{D}$, a simple yet effective choice that delivers superb inferences \parencite{Datta2016}.    
	Pick a positive integer $h\ll m$ and consider the Euclidean distance \parencite{Vecchia1988}. Let $N(\bm{s}_1)=\emptyset$ and define for each $i\in\{2,3,\dots,m\}$ location point $\bm{s}_i$'s neighbor set $N(\bm{s}_i) $ as the $\min\{h,i-1\}$ nearest neighbors of $\bm{s}_i$ in $\{\bm{s}_1,\dots,\bm{s}_{i-1}\}$. As shown by \textcite{Datta2016}, a quite small number of $h$ (between 10 to 15 for instance) can already yield highly competitive performance, and the predictive performance of NNGP is sensitive to neither the choice of reference set nor the ordering of locations. 
	
	The GP prior $\bm{\alpha}_{jl_j}|\kappa,\rho\sim N\left( \bm{0}, F(\rho)\otimes\kappa \right)$ gives the density 
	\begin{equation*}
		\pi(\bm{\alpha}_{jl_j}|\kappa,\rho) 
		= \pi(\bm{\alpha}_{jl_j}(\bm{s}_1)|\kappa,\rho)\times\prod_{i=2}^m f(\bm{\alpha}_{jl_j}(\bm{s}_i)|\bm{\alpha}_{jl_j}\left(\bm{s}_{1:(i-1)}\right),\kappa,\rho), 
	\end{equation*}
	which, under the latent NNGP prior, can be approximated by
	\begin{equation}\label{approxJointDensity}
		\tilde \pi(\bm{\alpha}_{jl_j}|\kappa,\rho) =\pi(\bm{\alpha}_{jl_j}(\bm{s}_1)|\kappa,\rho)\times\prod_{i=2}^m f(\bm{\alpha}_{jl_j}(\bm{s}_i)|\bm{\alpha}_{jl_j,N(\bm{s}_i)},\kappa,\rho),
	\end{equation}
	where  $\bm{\alpha}_{jl_j,N(\bm{s}_i)}$ is at most of length $hO$ for each $i\in\{2,3,\dots,m\}$.		
	It is clear that $\bm{\alpha}_{jl_j}(\bm{s}_1)|\kappa,\rho\sim N_O\left(\bm{0},\kappa\right)$. Since $\bm{\alpha}_{jl_j}|\kappa,\rho\sim N_{mO}\left( \bm{0}, F(\rho)\otimes\kappa\right) $, by properties of conditional distributions of jointly multivariate normals, we have
	\begin{equation}\label{compoCondiDist}
		\bm{\alpha}_{jl_j}(\bm{s}_i)|\bm{\alpha}_{jl_j,N(\bm{s}_i)},\kappa,\rho\sim N_{O}\left( B_{\bm{s}_i}\bm{\alpha}_{jl_j,N(\bm{s}_i)}, F_{\bm{s}_i} \right), \text{ for all }i\in\{2,\dots,m\},
	\end{equation}
	where the $O\times \left( \left| N(\bm{s}_i)\right|\cdot O \right)$ matrix $B_{\bm{s}_i}=F(\rho)_{\bm{s}_i, N(\bm{s}_i)}F(\rho)^{-1}_{N(\bm{s}_i)}\otimes I_{O\times O}$ and the $O\times O$ matrix $F_{\bm{s}_i} = \left[ F(\rho)_{\bm{s}_i, \bm{s}_i} - F(\rho)_{\bm{s}_i, N(\bm{s}_i)}F(\rho)^{-1}_{N(\bm{s}_i)}F(\rho)_{N(\bm{s}_i), \bm{s}_i}\right]\otimes\kappa$ with $F(\rho)_{\bm{s}_i, \bm{s}_i}=1$, $ F(\rho)_{\bm{s}_i, N(\bm{s}_i)}$, $F(\rho)_{N(\bm{s}_i), \bm{s}_i}, F(\rho)_{ N(\bm{s}_i)}$ being the corresponding $1\times 1, 1\times \left| N(\bm{s}_i)\right|, \left| N(\bm{s}_i)\right|\times 1, \left| N(\bm{s}_i)\right|\times \left| N(\bm{s}_i)\right|$ sub-matrices of $F(\rho)_{m\times m}$.
	Hence following \Cref{approxJointDensity}, we have
	\begin{equation}\label{approxMultiDens}
		\tilde \pi(\bm{\alpha}_{jl_j}|\kappa,\rho) \sim N_O\left(\bm{0},\kappa\right)\times\prod_{i=2}^m N_{O}\left( B_{\bm{s}_i}\bm{\alpha}_{jl_j,N(\bm{s}_i)}, F_{\bm{s}_i} \right).
	\end{equation}
	We thus no longer need to store the entire $F(\rho)$ and only need to store the sub-matrices $F(\rho)_{\bm{s}_i, N(\bm{s}_i)}, F(\rho)_{N(\bm{s}_i), \bm{s}_i}, F(\rho)_{ N(\bm{s}_i)}$ for each $i\in\{2,\dots,m\}$. The storage cost of $F(\rho)$ can hence be reduced from $\Ocal(m^2)$ to $\Ocal(m(h+1)^2)$, i.e., quadratic to linear in $m$ since $|N\left(\bm{s}_i\right)|\leq h\ll m$ for all $i\in\{2,\dots,m\}$. It is known that \Cref{approxMultiDens} is indeed the joint density from a valid stochastic process (Appendix A in \cite{Datta2016}), as its corresponding construction ensures a directed acyclic graph $\mathcal{G}=\{\mathcal{S},N_{\mathcal{S}}\}$. Also, for any $i\in\{2,\dots,m\},\;|N\left(\bm{s}_i\right)|\leq h \text{ and }B_{\bm{s}_i}, F_{\bm{s}_i}$ can be calculated by at most $\Ocal(h^3)$ flops, which result from evaluating $F(\rho)^{-1}_{N(\bm{s}_{i})}$. We thus need $\Ocal(mh^3)$ flops evaluating all $B_{\bm{s}_i}$'s and $F_{\bm{s}_i}$'s.
	
	\subsection{Approximating $F(\rho)^{-1}$ and $\determ(F(\rho))$}\label{sect4c}
	Let $F_{\bm{s}_1}=\kappa_{O\times O}$ and $B_{\bm{s}_1}^*=(1,0,\dots,0)_{1\times m}\otimes I_{O\times O}$. Then by \Cref{approxMultiDens}, 
	\begin{align} 
		\tilde \pi(\bm{\alpha}_{jl_j}|\kappa,\rho)
		&\propto \frac{1}{\prod_{i=1}^m\sqrt{\determ(F_{\bm{s}_i})}}\times \exp\left\lbrace -\frac{1}{2}\sum_{i=1}^m\bm{\alpha}_{jl_j}^{\T}(B^*_{\bm{s}_i})^{\T}F_{\bm{s}_i}^{-1}B^*_{\bm{s}_i}\bm{\alpha}_{jl_j} \right\rbrace\nonumber\\
		&=\frac{1}{\sqrt{\prod_{i=1}^m\determ(F_{\bm{s}_i})}}\times \exp\left\lbrace -\frac{1}{2}\bm{\alpha}_{jl_j}^{\T}B_{\mathcal{S}}^{\T}F_{\mathcal{S}}^{-1}B_{\mathcal{S}}\bm{\alpha}_{jl_j} \right\rbrace,\label{compactDens}
	\end{align}
	where the $mO\times mO$ matrix $F_{\mathcal{S}}=\diag(F_{\bm{s}_1}, \dots, F_{\bm{s}_m})$ and $B_{\mathcal{S}} = \left(\left[B^*_{\bm{s}_1}\right]^{\T}, \left[B^*_{\bm{s}_2}\right]^{\T},\ldots,\left[B^*_{\bm{s}_m}\right]^{\T}\right)^{\T}$. 
	For each $i$, the $O\times mO$ matrix $B^*_{\bm{s}_i}=\left(B^*_{\bm{s}_i,1},B^*_{\bm{s}_i,2},\dots,B^*_{\bm{s}_i,m} \right)$ is sparse with at most $h+1$ non-zero $O\times O$ blocks out of the $m$ blocks
	\begin{align}
		B^*_{\bm{s}_i,j}=
		\begin{cases}
			I_{O\times O},&\text{ if }j=i \\
			-B_{\bm{s}_i}[,(l-1)O+1:lO],&\text{ if }\exists\; l\in\{1,\dots,\min(h,i-1)\} \text{ such that } \bm{s}_j=N(\bm{s}_i)[l] \\
			\bm{0}_{O\times O},&\text{ otherwise}
		\end{cases},\nonumber
	\end{align}
	for $j\in\{1,\dots,m\}$, where $N(\bm{s}_i)[l]$ denotes the $l$-th point in $N(\bm{s}_i)$. Since for any $i\in\{2,3,\dots,m\}$, $N(\bm{s}_i)\subset\{\bm{s}_1,\dots,\bm{s}_{i-1}\}$, $B_{\mathcal{S}}$ is lower triangular whose main diagonal entries are all $1$. This implies that $\determ(B_{\mathcal{S}})=1$ and that $B_{\mathcal{S}}^{\T}F_{\mathcal{S}}^{-1}B_{\mathcal{S}}$ is invertible. Hence, we can let $\tilde F(\rho)^{-1}\otimes\kappa^{-1}=B_{\mathcal{S}}^{\T}F_{\mathcal{S}}^{-1}B_{\mathcal{S}}=\sum_{i=1}^m(B^*_{\bm{s}_i})^{\T}F^{-1}_{\bm{s}_i}B^*_{\bm{s}_i}$, whose computational complexity given $F_{\bm{s}_i}$'s and $B_{\bm{s}_i}$'s is $\Ocal(O^3)+\Ocal(m(h+1)^2)=\Ocal(m)$ since $h\ll m$ and $O\in\mathbb{N}$ is also assumed very small. 
	The $\tilde F(\rho)^{-1}_{m\times m}$ defined this way simplifies \Cref{compactDens} to
	\begin{equation}\label{approxJoint}
		\tilde \pi(\bm{\alpha}_{jl_j}|\kappa,\rho)\propto \left\{\determ\left(\tilde F(\rho)\otimes\kappa\right)\right\}^{-1/2} \times \exp\left\lbrace -\frac{1}{2}\bm{\alpha}_{jl_j}^{\T} [\tilde F(\rho)^{-1}\otimes\kappa^{-1}]\bm{\alpha}_{jl_j}\right\rbrace. 
	\end{equation}
	Hence $\tilde \pi(\bm{\alpha}_{jl_j}|\kappa,\rho)=N_{mO}\left(\bm{0},\tilde F(\rho)\otimes\kappa\right)$, which is our NNGP prior.
	As $\determ(B_{\mathcal{S}})=1$, we have $\determ\big(\tilde F(\rho)\otimes\kappa\big)=\determ\left( [B_{\mathcal{S}}^{\T}F_{\mathcal{S}}^{-1}B_{\mathcal{S}}]^{-1}\right)=\determ(F_{\mathcal{S}})=\prod_{i=1}^m\determ(F_{\bm{s}_i}) $, which can be calculated in $\Ocal(m)+\Ocal(O^3)=\Ocal(m)$ flops with evaluated $F_{\bm{s}_i}$'s. The total computational complexity calculating $\tilde F(\rho)^{-1}\otimes\kappa^{-1}$ and $\determ\big( \tilde F(\rho)\otimes\kappa\big)$ is thus linear in $m$. A full GP with a dense precision matrix $F(\rho)\otimes\kappa$, on the other hand, needs $\Ocal(m^3)$ flops updating the counterpart quantities $F(\rho)^{-1}\otimes\kappa^{-1}$ and $\determ\left(  F(\rho)\otimes\kappa\right)$ in each MCMC iteration. 
	
	We now consider elements of the precision matrix  $\tilde F(\rho)^{-1}_{m\times m}$. For any pair $(i,j)\in\{1,\dots,m\}^2$ and $i<j$, let $\tilde{F}_{ij}^{-1}$ be the $(i,j)$-th entry of $\tilde F(\rho)^{-1}$. Then $\tilde F(\rho)^{-1}\otimes\kappa^{-1}=\sum_{l=1}^m(B^*_{\bm{s}_l})^{\T}F^{-1}_{\bm{s}_l}B^*_{\bm{s}_l}$ implies  $\tilde{F}_{ij}^{-1}\otimes\kappa^{-1}=\sum_{l=j}^m (B^*_{\bm{s}_l,i})^{\T}F_{\bm{s}_l}^{-1}B^*_{\bm{s}_l,j}$. Hence $\tilde{F}_{ij}^{-1}\neq 0$ implies that there exists $l\in\{j,j+1,\dots,m\}$ such that $\bm{s}_i\in N(\bm{s}_l)$ and $\bm{s}_j\in N(\bm{s}_l)$ or $\bm{s}_j=\bm{s}_l$. We can thus obtain an upper bound, $m\cdot\frac{h(h+1)}{2}$, of the maximum number of pairs $(i,j),i<j$ leading to a non-zero $\tilde{F}_{ij}^{-1}$ by looping over $l\in\{1,\dots,m\}$. Further taking the lower triangular and diagonal entries of $\tilde F(\rho)^{-1}$ into consideration, we get that $\tilde F(\rho)^{-1}_{m\times m}$ is sparse with at most $2\times m\cdot\frac{h(h+1)}{2}+m = m\cdot[h(h+1)+1]$ non-zero entries. 
	Hence, the storage complexity for $\tilde F(\rho)^{-1}_{m\times m}$ is $\Ocal(mh^2)$, including the nonzero entries and their indices. The spatial NNGP's capability of reducing storage for $F(\rho)^{-1}$ from quadratic to linear in $m$ can be tremendously useful for fitting data with a very large $m$ value.

	\subsection{Sequentially Updating \boldmath{$\alpha$}$_{jl_j}$'s Under Spatial NNGP Prior}\label{sect4d}  
	
	Although the NNGP prior can reduce sampling costs of $\rho$ and $\kappa_{O\times O}$ (\Cref{appenA2}), the inefficiency in the posterior sampling of $\bm{\alpha}_{jl_j}$'s for a large $m$ as described in \Cref{sect4a} cannot be addressed by adopting the spatial NNGP prior alone, as we explain below. Therefore, we further propose a sequential updating scheme to tackle these issues.  
	
	Under the model in \Cref{sect2}, for any arbitrary $(j,l_j)\in\{1,\dots,k\}\times\{1,\dots,L-1\}$,	the full conditional posterior of $\bm{\alpha}_{jl_j}$ under the NNGP prior is	
	\begin{equation}\label{blockAlphaFixedL}
		\bm{\alpha}_{jl_j}|\cdot 
		\sim N_{mO}\left( \left[ I_{mO}+\tilde{F}(\rho)^{-1}\otimes\kappa^{-1} \right]^{-1}\bm{z}_{jl_j}, \left[ I_{mO}+\tilde{F}(\rho)^{-1}\otimes\kappa^{-1} \right]^{-1} \right) ,
	\end{equation}
	where $\bm{z}_{jl_j}$ is the introduced latent normal vector (\Cref{appenD}). We thus need to evaluate $\chol\big( \big[ I_{mO}+\tilde F(\rho)^{-1}\otimes\kappa^{-1} \big]^{-1} \big) $ and $\big[ I_{mO}+\tilde F(\rho)^{-1}\otimes\kappa^{-1} \big]^{-1}$, 
	which requires $\Ocal(m^3)$ flops via standard methods. Hence, directly proceeding with block updating $\bm{\alpha}_{jl_j}$'s according to \Cref{blockAlphaFixedL} for all $(j,l_j)$ still requires $\Ocal(m^3)+\Ocal(k(L-1)\cdot (mO)^2)=\Ocal(m^3)$ flops in each MCMC iteration and thus incurs a high computational cost.
	
	For the slice sampling algorithm in \Cref{sect3}, for any $(j, l_j)$ where $j\in\{1,\dots,k\}$ and $l_j\in\{1,\dots,L_j-1\}$, the full conditional posterior of $\bm{\alpha}_{jl_j}$ under the NNGP prior is
	\begin{align}\label{blockAlphaVaryLj}
		\bm{\alpha}_{jl_j}|\cdot
		&\sim N_{mO}\left(\bm{0}, \tilde{F}(\rho)\otimes\kappa\right)\times\prod_{i=1}^m\prod_{o=1}^O
		\begin{cases}
			1,&\text{ if }l_j>\xi_j(\bm{s}_{i,o})\\
			\mathbbm{1}_{\left\lbrace {\alpha}_{jl_j}^o(\bm{s}_{i})^{\text{new}}>\text{lowerBound}_{(j,l_j,i,o)} \right\rbrace},&\text{ if }l_j=\xi_j^o(\bm{s}_{i})\\
			\mathbbm{1}_{\left\lbrace {\alpha}_{jl_j}^o(\bm{s}_{i})^{\text{new}}<\text{upperBound}_{(j,l_j,i,o)}\right\rbrace},&\text{ if }l_j<\xi_j^o(\bm{s}_{i})\\
		\end{cases},
	\end{align}
	where $\text{lowerBound}_{(j,l_j,i,o)}$'s and $\text{upperBound}_{(j,l_j,i,o)}$'s are given by \Cref{alphaLowerBound,alphaUpperBound}.
	Therefore, we need to evaluate $\chol\big(\tilde{F}(\rho)\otimes\kappa\big)$, which generally still requires $\Ocal(m^3)$ flops. Finding a faster numerical algorithm to evaluate $\chol\left(\tilde{F}(\rho)\otimes\kappa\right)$ is challenging. More seriously, rejection sampling from a high-dimensional truncated $mO-$variate normal distribution in \Cref{blockAlphaVaryLj} may become extremely slow due to the low acceptance rate.
	
	We propose a novel algorithm to bypass these computational burdens under both models by sequentially updating components of $\bm{\alpha}_{jl_j}=\left(\bm{\alpha}_{jl_j}(\bm{s}_1)^\T,\dots,\bm{\alpha}_{jl_j}(\bm{s}_m)^\T \right)^{\T} $ for each $(j,l_j)$ instead. Computational complexity orders for updating all $\bm{\alpha}_{jl_j}$'s in each MCMC iteration are presented in \Cref{sectH3summaryTable}. Let $F_{\bm{s}_1}=\kappa_{O\times O}$ and $ F^{-1}_{\bm{s}_1}B_{\bm{s}_1}\bm{\alpha}_{jl_j,N(\bm{s}_1)}=\bm{0}_{O\times 1}$ since $N(\bm{s}_1)=\emptyset$.
	
	Under the model in \Cref{sect2} and the NNGP prior $\tilde{\pi}\left(\bm{\alpha}_{jl_j}|\kappa,\rho\right)=N_{mO}\left(\bm{0},\tilde{F}(\rho)\otimes\kappa\right)$ for all $(j,l_j)$, for any fixed $(j,l_j,i)\in\{1,\dots,k\}\times\{1,\dots,L-1\}\times\{1,\dots,m\}$,
	\begin{align}
		f(\bm{\alpha}_{jl_j}(\bm{s}_i)|\cdot)
		&\propto f(\bm{z}_{jl_j}(\bm{s}_i)|\bm{\alpha}_{jl_j}(\bm{s}_i))\times  {f}(\bm{\alpha}_{jl_j}(\bm{s}_i)|\bm{\alpha}_{jl_j,N(\bm{s}_i)},\kappa,\rho)\times\prod_{r:\bm{s}_i\in N(\bm{s}_r)}{f}(\bm{\alpha}_{jl_j}(\bm{s}_r)|\bm{\alpha}_{jl_j,N(\bm{s}_r)},\kappa,\rho) \nonumber\\
		&\sim N_O\left(V_{\bm{s}_i}\bm{\mu}_{jl_j,\bm{s}_i}, V_{\bm{s}_i} \right),\label{seqAlphaIndPostDensFixedL1}
	\end{align}
	where
	\begin{align*}
		V_{\bm{s}_i} &= \left[I_O+F^{-1}_{\bm{s}_i} + \sum_{r:\bm{s}_i\in N(\bm{s}_r)}B_{\bm{s}_r,k_r(\bm{s}_i)}^{\T}F^{-1}_{\bm{s}_r}B_{\bm{s}_r,k_r(\bm{s}_i)} \right]^{-1}\text{ and}\\ 
		\bm{\mu}_{jl_j,\bm{s}_i} &= \bm{z}_{jl_j}(\bm{s}_i) + F^{-1}_{\bm{s}_i}B_{\bm{s}_i}\bm{\alpha}_{jl_j,N(\bm{s}_i)} + \sum_{r:\bm{s}_i\in N(\bm{s}_r)}B_{\bm{s}_r,k_r(\bm{s}_i)}^{\T}F^{-1}_{\bm{s}_r}\left\lbrace \bm{\alpha}_{jl_j}(\bm{s}_r)-\sum^{1\leq k\leq |N(\bm{s}_r)|}_{k\neq k_r(\bm{s}_i)}B_{\bm{s}_r,k}\bm{\alpha}_{jl_j,N(\bm{s}_r)[k]} \right\rbrace. 
	\end{align*}

	Under the model in \Cref{sect3} and the NNGP prior $\tilde{\pi}\left(\bm{\alpha}_{jl_j}|\kappa,\rho\right)=N_{mO}\left(\bm{0},\tilde{F}(\rho)\otimes\kappa\right)$ for all $(j,l_j)$, for any fixed $(j,l_j,i)$ where $j\in\{1,\dots,k\}$, $l_j\in\{1,\dots,L_j-1\}$, and $i\in\{1,\dots,m\}$,
	\begin{align}\label{seqAlphaIndPostDens2}
		f\left(\bm{\alpha}_{jl_j}(\bm{s}_i)|\cdot\right) 
		&\propto  \prod_{r:\;r=i\text{ or }\bm{s}_i\in N(\bm{s}_r)}{f}(\bm{\alpha}_{jl_j}(\bm{s}_r)|\bm{\alpha}_{jl_j,N(\bm{s}_r)},\kappa,\rho)\times \prod_{o=1}^O \mathbbm{1}_{\left\lbrace l_j>\xi_j^o(\bm{s}_{i})\text{ or }  u_j^o(\bm{s}_{i})<w^o_{j\xi_j^o(\bm{s}_{i})}(\bm{s}_{i})\right\rbrace} \nonumber\\
		&\sim N_O\left(V_{\bm{s}_i}\bm{\mu}_{jl_j,\bm{s}_i}, V_{\bm{s}_i} \right)\times\prod_{o=1}^O \mathbbm{1}_{\left\lbrace l_j>\xi_j^o(\bm{s}_{i})\text{ or }  u_j^o(\bm{s}_{i})<w^o_{j\xi_j^o(\bm{s}_{i})}(\bm{s}_{i})\right\rbrace},
	\end{align}
	where
	\begin{align}
		V_{\bm{s}_i} & = \left[F^{-1}_{\bm{s}_i} + \sum_{r:\bm{s}_i\in N(\bm{s}_r)}B_{\bm{s}_r,k_r(\bm{s}_i)}^{\T}F^{-1}_{\bm{s}_r}B_{\bm{s}_r,k_r(\bm{s}_i)} \right]^{-1}\text{ and } \nonumber \\
		\bm{\mu}_{jl_j,\bm{s}_i} &=  F^{-1}_{\bm{s}_i}B_{\bm{s}_i}\bm{\alpha}_{jl_j,N(\bm{s}_i)} + \sum_{r:\bm{s}_i\in N(\bm{s}_r)}B_{\bm{s}_r,k_r(\bm{s}_i)}^{\T}F^{-1}_{\bm{s}_r}\left\lbrace \bm{\alpha}_{jl_j}(\bm{s}_r)-\sum^{1\leq k\leq |N(\bm{s}_r)|}_{k\neq k_r(\bm{s}_i)}B_{\bm{s}_r,k}\bm{\alpha}_{jl_j,N(\bm{s}_r)[k]} \right\rbrace. \nonumber
	\end{align}
	
	Full deduction details pertaining to the above are in \Cref{appenA2,appenA3}. For each $r$ such that  $\bm{s}_i\in N(\bm{s}_r) $, $k_r(\bm{s}_i)$ in the expressions above refers to the positive integer less than or equal to $|N(\bm{s}_r)|$ such that $\bm{s}_i=N(\bm{s}_r)[k_r(\bm{s}_i)]$. We have written the $O\times (|N(\bm{s}_r)|\cdot O)$ matrix $B_{\bm{s}_r}$ as $\big(B_{\bm{s}_r,1},\ldots,B_{\bm{s}_r,|N(\bm{s}_r)|}\big)$, where each sub-matrix is $O\times O$, and written the $\left(|N(\bm{s}_r)|\cdot O\right)\times 1$ vector $\bm{\alpha}_{jl_j,N(\bm{s}_r)}$ as
	$\big(\bm{\alpha}_{jl_j,N(\bm{s}_r)[1]}^\T, \ldots,\bm{\alpha}_{jl_j,N(\bm{s}_r)[|N(\bm{s}_r)|]}^\T\big)^\T$ so that $B_{\bm{s}_r}\bm{\alpha}_{jl_j,N(\bm{s}_r)}=\sum_{k=1}^{|N(\bm{s}_r)|}B_{\bm{s}_r,k}\bm{\alpha}_{jl_j,N(\bm{s}_r)[k]}$. Note also that we only need to consider $r=i+1,\dots,m$ for each $i<m$ in order to find out the set $\{r\in\{1,\dots,m\}:\bm{s}_i\in N(\bm{s}_r)\}$.
	
	Suppose that we have already calculated $B_{\bm{s}_i}$'s and $F_{\bm{s}_i}$'s in the Gibbs sampler step updating $\rho$. Then, sequentially sampling the elements $\bm{\alpha}_{jl_j}(\bm{s}_i),i\in\{1,\dots,m\},$ of $\bm{\alpha}_{jl_j}$ one by one according to \Cref{seqAlphaIndPostDensFixedL1} incurs a computational complexity of $\Ocal\big(m(O^3+h)\big) + \Ocal\big(k(L-1)mh(hO+O^2)\big) + \Ocal\big(mk(L-1)O^2\big)
	= \Ocal\big(m\big[k(L-1)hO(h+O)+O^3\big]\big)=\Ocal(m)$ if $L\ll m$. If we adopt the slice sampling framework in \Cref{sect3} and proceed with \Cref{seqAlphaIndPostDens2} instead, then the corresponding computational complexity for updating all $\bm{\alpha}_{jl_j}\big(\bm{s}_i\big)$'s is 
	\begin{align*}
		&\Ocal\big(O^3+mh\big) + \Ocal\Bigg(\sum_{j=1}^k (L_j-1) mh(hO+O^2) \Bigg) 
		+\Ocal\Bigg(\sum_{j=1}^k \Big[ mO + \sum_{l_j=1}^{L_j-1}\sum_{i=1}^m \big(k_{jl_ji}^* \cdot O + n_{jl_ji}^* \cdot O^2\big) \Big]\Bigg) \nonumber\\
		=\;&\Ocal\left(O\left\lbrace \sum_{j=1}^k \Bigg[(L_j-1)mh(h+O) + \sum_{l_j=1}^{L_j-1}\sum_{i=1}^m \left(k_{jl_ji}^*+n_{jl_ji}^*\cdot O\right)\Bigg] + O^2 \right\rbrace  \right)
	\end{align*}
	where $\Ocal\left(O\cdot\sum_{j=1}^k \left[m + \sum_{l_j=1}^{L_j-1}\sum_{i=1}^m k_{jl_ji}^*\right]\right)$ denote the overall computational complexity required for calculation of the upper and lower bounds, and $n_{jl_ji}^*$'s represent the numbers of draws from $N_O\left(V_{\bm{s}_i}\bm{\mu}_{jl_j,\bm{s}_i}, V_{\bm{s}_i} \right)$ required per sample. 
	Sequentially updating $\bm{\alpha}_{jl_j}$'s as above successfully replaces the inefficient rejection sampling from a high-dimensional truncated $mO$-variate normal distribution by the more efficient rejection sampling from low-dimensional truncated $O$-variate normal distributions. As the number of observation types $O\in\mathbb{N}$ is very small and often equals $1$, the rejection sampling from truncated $O$-variate normals (\Cref{seqAlphaIndPostDens2}) concerned with sequential updating would have adequate acceptance rates with small $n_{jl_ji}^*$'s. 
	If the largest number of mixture components $\max_{j\in\{1,\ldots,k\}} L_j$ is of constant order or increases only logarithmically with $m$ as in Dirichlet processes, then given $k,h,O\ll m$ and $\sum_{j=1}^k L_j \leq k\cdot \max_{j\in\{1,\ldots,k\}} L_j$, sequential updates in our Gibbs samplers reduce the corresponding computational complexity from likely larger than $\Ocal(m^3)$ to about linear in $m$ in each MCMC iteration (\Cref{appenH3}). 
	
	\section{Spatial Prediction and Temporal Trends Clustering}\label{sect5}
	\subsection{Predictions at New Spatial Locations}\label{sect5a}
	
	Bayesian nonparametric predictions at any arbitrary future time points or new spatial locations under our modeling frameworks in \Cref{sect2,sect3} are straightforward. We can decompose the integral representation of the posterior predictive distribution (PPD) into several known densities and then obtain the PPD via composition sampling. 
	Here in the main text, we elaborate on new-location predictions under our model in \Cref{sect3}. Details pertaining to the other prediction cases are given in \Cref{appenG}. 
	
	For Bayesian predictions at $r$ ($r\in\mathbb{N},r\geq 1$) new locations $\bm{s}_{(m+1):(m+r)}$ given their corresponding covariates matrix $X(\bm{s}_{(m+1):(m+r)})_{rTO\times p}$, if any, we can write the PPD as
	\begin{align*}
		&f(\bm{y}(\bm{s}_{(m+1):(m+r)})|\bm{y}(\bm{s}_{1:m}),X(\bm{s}_{(m+1):(m+r)}))\\
		={}&\int_{\Theta}f\left(\bm{y}(\bm{s}_{(m+1):(m+r)})|\Theta,\bm{y}(\bm{s}_{1:m}),X(\bm{s}_{(m+1):(m+r)}) \right)\pi\left(\Theta|\bm{y}(\bm{s}_{1:m})\right)d\Theta,
	\end{align*} 
	where $\Theta=\left(\bm{\eta},\bm{\beta},\bm{\sigma^2}(\bm{s}_{(m+1):(m+r)}),\bm{\theta},\bm{\xi}(\bm{s}_{(m+1):(m+r)}),\bm{\alpha}(\bm{s}_{(m+1):(m+r)}),\bm{\alpha},\kappa,\rho
	\right)$ and $\bm{\alpha}$ denotes $\bm{\alpha}(\bm{s}_{1:m})$, and then partition the integral into 
	\begin{align}\label{PPDnewLoc}		
		&\int_{\Theta}\underbrace{f\left(\bm{y}(\bm{s}_{(m+1):(m+r)})|\bm{\theta},\bm{\xi}(\bm{s}_{(m+1):(m+r)}),\bm{\eta},\bm{\beta},\bm{\sigma^2}(\bm{s}_{(m+1):(m+r)}),X(\bm{s}_{(m+1):(m+r)})\right) }_{T_1}\nonumber\\
		&\qquad\times\underbrace{f\left(\bm{\xi}(\bm{s}_{(m+1):(m+r)})|\bm{\alpha}(\bm{s}_{(m+1):(m+r)})\right) }_{T_2}
		\underbrace{f\left(\bm{\alpha}(\bm{s}_{(m+1):(m+r)})|\bm{\alpha},\kappa,\rho\right) }_{T_3}\nonumber\\
		&\qquad\times\underbrace{\pi\left(\bm{\eta},\bm{\beta},\bm{\theta},\bm{\alpha},\kappa,\rho |\bm{y}(\bm{s}_{1:m})\right)}_{T_4}
		\underbrace{\pi\left(\bm{\sigma^2}(\bm{s}_{(m+1):(m+r)})\right)}_{T_5}d\Theta
	\end{align}
	since the posterior density $\pi\left(\bm{\sigma^2}(\bm{s}_{(m+1):(m+r)})|\bm{y}(\bm{s}_{1:m})\right)$ equals the prior $\pi\left(\bm{\sigma^2}(\bm{s}_{(m+1):(m+r)})\right)$. 
	$\bm{\alpha}$ and $\bm{\alpha}\left(\bm{s}_{(m+1):(m+r)}\right)$ already incorporate the cluster number parameters $L_j$ for all $j$.
	
	In \Cref{PPDnewLoc}, $T_1$ is the likelihood; $T_4$ is the parameters' posterior distribution obtained from the original model fit’s MCMC sampler; $T_5$ denotes the prior density for $\bm{\sigma^2}(\bm{s}_{(m+1):(m+r)})=\left( (\sigma^2)^1\left(\bm{s}_{m+1}\right),\dots,(\sigma^2)^O\left(\bm{s}_{m+1}\right),\dots\dots,(\sigma^2)^1\left(\bm{s}_{m+r}\right),\dots,(\sigma^2)^O\left(\bm{s}_{m+r}\right) \right)$ with 
	$(\sigma^2)^o\left(\bm{s}_{m+i_r}\right)\overset{\ind}{\sim}\mathcal{IG}(a,b),\,i_r\in\{1,\dots,r\},\,o\in\{1,\dots,O\}$; $T_2$ is the density of the multinomial distribution described in \Cref{sect2}; and $T_3$ can be written as  
	\begin{equation}\label{spatpredCompoVaryLj} f\left(\bm{\alpha}(\bm{s}_{(m+1):(m+r)})|\bm{\alpha}(\bm{s}_{1:m}),\kappa,\rho\right)=\prod_{j=1}^k\prod_{l_j=1}^{L_j-1}f\left(\bm{\alpha}_{jl_j}(\bm{s}_{(m+1):(m+r)})|\bm{\alpha}_{jl_j}(\bm{s}_{1:m}),\kappa,\rho\right).
	\end{equation}	
	
	From the full spatial GP prior on the combined vector $\bm{\alpha}_{jl_j}(\bm{s}_{1:(m+r)})$, for any $(j,l_j)$,
	\begin{align}\label{spatpredFixedLCL2}
		\bm{\alpha}_{jl_j}(\bm{s}_{(m+1):(m+r)})|\bm{\alpha}_{jl_j}(\bm{s}_{1:m}),\kappa,\rho \sim N_{rO}\left(B_{\bm{s}_{(m+1):(m+r)}}\bm{\alpha}_{jl_j}(\bm{s}_{1:m}), F_{\bm{s}_{(m+1):(m+r)}}\right),
	\end{align}
	where $B_{\bm{s}_{(m+1):(m+r)}}=\left\lbrace F(\rho)_{\bm{s}_{(m+1):(m+r)}, \bm{s}_{1:m}}\left[F(\rho)_{ \bm{s}_{1:m}}\right]^{-1}\right\rbrace \otimes I_{O\times O}$ is $rO\times mO$ and $F_{\bm{s}_{(m+1):(m+r)}} = \left\lbrace F(\rho)_{\bm{s}_{(m+1):(m+r)}} - F(\rho)_{\bm{s}_{(m+1):(m+r)}, \bm{s}_{1:m}}\left[F(\rho)_{ \bm{s}_{1:m}}\right]^{-1}F(\rho)_{\bm{s}_{1:m}, \bm{s}_{(m+1):(m+r)}}\right\rbrace \otimes \kappa_{O\times O}$ is $rO\times rO$ with $F(\rho)_{ \bm{s}_{1:m}}, F(\rho)_{\bm{s}_{(m+1):(m+r)}, \bm{s}_{1:m}}, F(\rho)_{\bm{s}_{1:m}, \bm{s}_{(m+1):(m+r)}}, F(\rho)_{\bm{s}_{(m+1):(m+r)}}$ being the corresponding $m\times m$, $r\times m$, $m\times r$, $r\times r$ sub-matrices of the $(m+r)\times (m+r)$ matrix $F(\rho)_{\bm{s}_{1:(m+r)}}$.
	
	If we use instead the latent NNGP prior $\tilde{\pi}(\bm{\alpha}_{jl_j}(\bm{s}_{1:(m+r)})|\kappa,\rho)$, then one primary advantage of NNGP is that it is a valid Gaussian process on the entire spatial domain $\mathcal{D}\subset \mathbb{R}^d$ after extending $\tilde \pi\left(\bm{\alpha}_{jl_j}\left(\bm{s}_{1:m}\right)|\kappa,\rho\right)$ in \Cref{approxJointDensity}. For all $\bm{s}\in\mathcal{D}\backslash\mathcal{S}$, we let $N(\bm{s})$ consist of the $h$ nearest neighbors of $\bm{s}$ in ${\mathcal{S}}=\{\bm{s}_1,\bm{s}_2,\dots,\bm{s}_m\}$. We then specify the NNGP derived from the parent GP as
	\begin{align}\label{fullNNGP}
		\bm{\alpha}_{jl_j}\left(\bm{s}_{1:m}\right) | \kappa,\rho &\sim \pi(\bm{\alpha}_{jl_j}(\bm{s}_1)|\kappa,\rho) \times \prod_{i=2}^m f(\bm{\alpha}_{jl_j}(\bm{s}_i)|\bm{\alpha}_{jl_j,N(\bm{s}_i)},\kappa,\rho),\nonumber\\ 
		\bm{\alpha}_{jl_j}(\bm{s})|\bm{\alpha}_{jl_j}\left(\bm{s}_{1:m}\right),\kappa,\rho&\overset{\ind}{\sim}f\left(\bm{\alpha}_{jl_j}(\bm{s})|\bm{\alpha}_{jl_j,N(\bm{s})},\kappa,\rho\right),\text{ for all }\bm{s}\in\mathcal{D}\backslash\mathcal{S}. 
	\end{align}
	This generalized directed graph on $\mathcal{S} \cup \{\bm{s}\}$ is still ensured acyclic with a well-defined NNGP process constructed as above for any parent GP and any fixed reference set $\mathcal{S}$; see Section 2.2 and Appendix D \& E in \textcite{Datta2016}. Extension to any location $\bm{s}\in\mathcal{D}$ via nearest-neighbor kriging as in \Cref{fullNNGP} enables efficient hierarchical predictions (\Cref{appenG2b,appenH2}). In particular, for any $(j,l_j)$, at the $r$ new locations $\bm{s}_{(m+1):(m+r)}$,
	\begin{align}\label{spatpredFixedLCL3}
		&\;\tilde{f}\left(\bm{\alpha}_{jl_j}(\bm{s}_{(m+1):(m+r)})|\bm{\alpha}_{jl_j}(\bm{s}_{1:m}),\kappa,\rho\right)
		= \prod_{i_r=1}^r \tilde{f}\left(\bm{\alpha}_{jl_j}(\bm{s}_{(m+i_r)})|\bm{\alpha}_{jl_j}(\bm{s}_{1:m}),\kappa,\rho\right) \\
		= & \prod_{i_r=1}^r f\left(\bm{\alpha}_{jl_j}(\bm{s}_{(m+i_r)})|\bm{\alpha}_{jl_j,N(\bm{s}_{(m+i_r)})},\kappa,\rho\right) \sim \prod_{i_r=1}^r N_O\left(B_{\bm{s}_{(m+i_r)}}\bm{\alpha}_{jl_j,N(\bm{s}_{(m+i_r)})}, F_{\bm{s}_{(m+i_r)}}\right).\nonumber
	\end{align}
	For each $i_r$, $B_{\bm{s}_{(m+i_r)}}=\Big\lbrace F(\rho)_{\bm{s}_{(m+i_r)}, N(\bm{s}_{(m+i_r)})}\left[F(\rho)_{ N(\bm{s}_{(m+i_r)})}\right]^{-1}\Big\rbrace \otimes I_{O\times O}$ is $O\times hO$, and $F_{\bm{s}_{(m+i_r)}} =\Big\lbrace F(\rho)_{\bm{s}_{(m+i_r)}} - F(\rho)_{\bm{s}_{(m+i_r)}, N(\bm{s}_{(m+i_r)})}\left[F(\rho)_{ N(\bm{s}_{(m+i_r)})}\right]^{-1}F(\rho)_{N(\bm{s}_{(m+i_r)}), \bm{s}_{(m+i_r)}}\Big\rbrace \otimes \kappa_{O\times O}$ is $O\times O$ with $F(\rho)_{\bm{s}_{(m+i_r)}}=1$, $F(\rho)_{\bm{s}_{(m+i_r)}, N(\bm{s}_{(m+i_r)})}$, $F(\rho)_{N(\bm{s}_{(m+i_r)}), \bm{s}_{(m+i_r)}}$, $F(\rho)_{ N(\bm{s}_{(m+i_r)})}$ being the corresponding $1\times 1$, $1\times h$, $h\times 1$, $h\times h$ sub-matrices of $F(\rho)_{\bm{s}_{1:(m+r)}}$.
	
	\Cref{table:summary.complexity} summarizes computational complexities associated with the posterior sampling of some key model parameters and a major spatial prediction step. The detailed derivation of these computational complexities can be found in \Cref{appenH}.
	\begin{table}[h]
		{
			\begin{adjustwidth}{-.4cm}{-0.5cm}
				\begin{center}
					\scalebox{0.68}{\begin{tabular}{|*{6}{c|}}
							\hline
							\multicolumn{2}{|c|}{\textbf{Method}} & \texttt{fullGPfixedL} & \texttt{NNGPblockFixedL} & \texttt{NNGPsequenFixedL} & \texttt{NNGPsequenVaryLj}\\
							\hline 
							\hline
							\multicolumn{2}{|c|}{\textbf{NNGP Prior}} & \xmark & \cmark & \cmark & \cmark\\
							\hline
							\multicolumn{2}{|c|}{\textbf{Sequential Updates}} & \xmark & \xmark & \cmark & \cmark\\
							\hline
							\multicolumn{2}{|c|}{\textbf{Slice Sampling}} & \xmark & \xmark & \xmark & \cmark\\
							\hline \hline 
							
							\multicolumn{2}{|c|}{$\rho$ and $\kappa_{O\times O}$} & $\Ocal(m^3)$ &  \multicolumn{2}{|c|}{$\Ocal\left(\left\lbrace k(L-1)O^2[h^2+h+1]+h^3\right\rbrace m\right)$}  & $\Ocal\left(\left\lbrace\sum_{j=1}^k(L_j-1)\cdot O^2[h^2+h+1]+h^3\right\rbrace m\right)$\\
							\hline
							\multicolumn{2}{|c|}{\multirow{2}{*}{$\alpha^o_{jl_j}\big(\bm{s}_{i}\big)$'s}}
							& \multicolumn{2}{|c|}{\multirow{2}{*} {$\Ocal(m^3)$}} & \multirow{2}{*} {$\Ocal\big(m\big[k(L-1)hO(h+O)+O^3\big]\big)$}  & $\Ocal\big(O\big( \sum_{j=1}^k \big[(L_j-1)mh(h+O) +$ \\
							\multicolumn{2}{|c|}{} & \multicolumn{2}{|c|}{} & & $ \sum_{l_j=1}^{L_j-1}\sum_{i=1}^m \big(k_{jl_ji}^*+n_{jl_ji}^*\cdot O\big)\big] + O^2 \big)  \big)$ \\
							\hline
							\multicolumn{2}{|c|}{$z_{jl_j}^o\big(\bm{s}_{i}\big)$'s or $u_j^o\big(\bm{s}_{i}\big)$'s} & \multicolumn{3}{|c|}{$\Ocal(kLmO)$} & $\Ocal(kmO)$ \\
							\hline
							\multicolumn{2}{|c|}{$\delta_{1:k}$, $\theta_{jl_j}$'s,} & \multicolumn{3}{|c|}{\multirow{2}{*}{$\Ocal\big(kLmOT(p+k)\big)$}} & \multirow{2}{*}{$\Ocal\left(\sum_{j=1}^k L_j\cdot mOT(p+k)\right)$}  \\
							\multicolumn{2}{|c|}{and $\xi_j^o\big(\bm{s}_{i}\big)$'s} & \multicolumn{3}{|c|}{} & \\ 
							
							\hline \hline
							\multicolumn{2}{|c|}{\textbf{Spatial Prediction}} & $\Ocal\big(r[m^2+r^2+$ &  \multicolumn{2}{|c|}{\multirow{2}{*}{$\Ocal\Big(r\big[h^3+k(L-1)O(h+O)\big]+O^3\Big)$}}  & \multirow{2}{*}{$\Ocal\left(r\left[h^3+\sum_{j=1}^{k}(L_j-1)\cdot O(h+O)\right]+O^3\right)$}\\
							\multicolumn{2}{|c|}{\bf at $r$ Locations} & $k(L-1)mO]\big)$ &  \multicolumn{2}{|c|}{}  & \\
							\hline
					\end{tabular}}
				\end{center}
			\end{adjustwidth}
			\caption{Summary of computational complexities for the posterior sampling of model parameters and a major step obtaining $\hat{\bm{\alpha}}\left(\bm{s}_{(m+1):(m+r)}\right)$ for spatial prediction at $r$ new locations in each MCMC iteration, based on our 3 new techniques in Sections \ref{sect3} and \ref{sect4}.}
			\label{table:summary.complexity}
		}
	\end{table}
	
	\subsection{Summarizing Spatial Clusters with Similar Temporal Trends}\label{sect5b}	
	For our spatiotemporal factor models specified in \Cref{sect2,sect3}, clustering information for temporal trajectories $\{\bm{y}_t:t=1,\ldots,T\}$ is contained in all $k$ columns of the factor loadings matrix $\Lambda$. 
	The posterior distribution of the mixture weights $w_{jl_j}^o(\bm{s}_{i})$'s, which can be readily computed from posterior samples of the spatially varying latent variables $\alpha_{jl_j}^o(\bm{s}_{i})$'s, contains pertinent spatial clustering information for $\bm{\lambda}_{1:k}$. 
	Fix an arbitrary $o\in\{1,\dots,O\}$. Let $\bm{w}_o(\bm{s}_{i})=\{w_{jl_j}^o(\bm{s}_{i}):j=1,\dots,k,\,l_j=1,\dots,L_j\}$ for all $i$, where $L_j$ can be replaced by a common $L$ under our model in \Cref{sect2}. Then spatial proximity information has been encoded into $\bm{{w}}_o=
	\left[{\bm{w}}_o(\bm{s}_{1}),
	\ldots,
	{\bm{w}}_o(\bm{s}_{m})
	\right]^{\T}$, which is of dimension $m\times \sum_{j=1}^k L_j$. Any two locations $\bm{s}_{i}$ and $\bm{s}_{i'}$ ($i,i'\in\{1,\dots,m\}$) would have similar temporal trajectories for observation type $o$ if the two vectors $\bm{w}_o(\bm{s}_{i})$ and $\bm{w}_o(\bm{s}_{i'})$ are close to each other. Therefore, we can apply k-means clustering to the $m\times \sum_{i=1}^n\sum_{j=1}^k L_j^{(t_i)}$ matrix $\left(\bm{\hat{w}}_o^{(t_1)}, \bm{\hat{w}}_o^{(t_2)},\dots,\bm{\hat{w}}_o^{(t_n)}\right)$, where each $L_j^{(t_i)}$ represents the estimated number of clusters for factor $j$ in the $t_i$-th MCMC iteration, and $\bm{\hat{w}}_o^{(t_1)},\dots,\bm{\hat{w}}_o^{(t_n)}$ denote the corresponding posterior samples of weights parameters for all $m$ locations from $n\in\mathbb{N}$ selected kept post-burn-in MCMC iterations. Under the model in \Cref{sect2}, the matrix $\left(\bm{\hat{w}}_o^{(t_1)},\bm{\hat{w}}_o^{(t_2)},\ldots,\bm{\hat{w}}_o^{(t_n)}\right)$ is of dimension $m\times nkL$ instead.

	\section{Simulation Experiments}\label{sect6}
	
	This section concentrates on simulation studies that demonstrate the enhancement in computational scalability and ends with a real data example for sea surface temperature data. Additional numerical results for multiple observation types, more complex temporal structure under the vector autoregressive model, and sensitivity analysis to hyperparameters are presented in \Cref{appenI} of the Supplementary Material.
	
	\subsection{Computational Efficiency for Key Gibbs Sampler Steps}\label{sect6a} 
	We first corroborate the computational acceleration capabilities in Bayesian posterior sampling of our three techniques: slice sampling (\Cref{sect3}), spatial latent NNGP prior (\Cref{sect4}), and sequential updating algorithms (\Cref{sect4d}). 
	As described in \Cref{sect2} and \Cref{table:summary.complexity}, the four methods we implemented are encoded as \texttt{fullGPfixedL}, \texttt{NNGPblockFixedL}, \texttt{NNGPsequenFixedL}, and \texttt{NNGPsequenVaryLj}. 
	We simulated data from the model in \Cref{sect2} with no covariate $\bm{x}$, with exponential temporal covariance $[H(\psi)]_{tt'}=\exp\{-\psi|t-t'|\}$ for all $t,t'\in\{1,\ldots,T\}$ and spatial dependence under a GP with exponential covariance function, such that $F(\rho)=\exp(-\rho D)$, where $D_{m\times m}$ denotes the Euclidean distance matrix of $m$ locations. The true clustering mechanism was generated from the spatial PSBP prior. A unanimous cluster number upper bound of 10 was set, and the actual numbers of clusters for the $k$ factors were randomly sampled between 1 and 10. We set the number of observation types $O=1$, the number of latent factors $k=5$, the temporal tuning parameter $\psi=2.3$, the spatial tuning parameter $\rho=0.8$, and $\sigma^2\left(\bm{s}_{i}\right)=0.01$ for all $i\in\{1,\dots,m\}$. We specified $T$ equal-distanced time points $t=1,\ldots,T$ and $m$ spatial locations $\bm{s}_i=(i_1,i_2)$ for $i=1,\ldots,m$ on an equispaced $2$-dimensional grid, where $(i_1,i_2)\in\{1,\ldots,\sqrt{m}\}\times\{1,\ldots,\sqrt{m}\}$. We picked 9 different sets of $(m,T)$ pairs: $(m,T) = (400, 30), (400, 50), (900, 30), (900, 50), (1600, 30), (1600, 50), (1600, 100), (3600, 50), (3600, 100)$. $m$ is set large to highlight differences in the computation time of different methods. For the three methods with spatial latent NNGP priors, we set $h=15$ for the neighborhood size. The fixed number of clusters $L$ is specified to be $50$ for the first three methods, and a starting value and hence upper bound of $50$ is assigned to all $L_j$'s for \texttt{NNGPsequenVaryLj}. We ran each MCMC chain for $2\times 10^4$ burn-in iterations and $10^4$ post-burn-in iterations, which were thinned to $5000$ samples for analysis.  	
	
	\Cref{sect6aModelRuntime} reports the total model fitting time under the 9 $(m,T)$ pairs of our four methods, which are all markedly faster than their \texttt{spBFA} counterparts in \Cref{spBFAsect6aModelRuntime}. For most of the 9 cases,  the slowest to the fastest are \texttt{fullGPfixedL}, \texttt{NNGPblockFixedL}, \texttt{NNGPsequenFixedL}, and \texttt{NNGPsequenVaryLj}. The computational gain from our three new techniques becomes more evident for larger values of $m$ and $T$. When $m=3600$ and $T=100$, \texttt{NNGPsequenVaryLj} is more than 10 times faster than the full GP prior based \texttt{fullGPfixedL}, demonstrating a substantial enhancement in computational scalability.
	
	\begin{table}[h]
		\centering
		\scalebox{0.85}{
			\centering
			\begin{tabular}{|*{6}{c|}}
				\hline
				\multicolumn{2}{|c|}{\backslashbox{\textbf{Setting}}{\textbf{Model}}} & \texttt{fullGPfixedL} & \texttt{NNGPblockFixedL} & \texttt{NNGPsequenFixedL} & \texttt{NNGPsequenVaryLj}\\
				\hline
				\multirow{3}{*}{$T=30$} & $m=400$ & 3.3 hours & 2.17 hours & 2.41 hours & 1.68 hours\\
				\cline{2-6}
				& $m=900$ & 10.66 hours & 7.3 hours & 5.42 hours & 2.9 hours\\
				\cline{2-6}
				& $m=1600$ & 1.68 days & 1.01 days & 10.18 hours & 6.51 hours\\
				\hline
				\multirow{4}{*}{$T=50$} & $m=400$ & 3.68 hours & 3.01 hours & 3.21 hours & 2.17 hours\\
				\cline{2-6}
				& $m=900$ & 12.56 hours & 10.11 hours & 7.95 hours & 4.52 hours\\
				\cline{2-6}
				& $m=1600$ & 1.79 days & 1.12 days & 12.86 hours & 9.05 hours \\
				\cline{2-6}
				& $m=3600$ & 19.41 days & 9.42 days & 2.03 days & 1.39 days\\
				\hline
				\multirow{2}{*}{$T=100$} & $m=1600$ & 2.3 days & 1.5 days & 21.24 hours & 15.09 hours\\
				\cline{2-6}
				& $m=3600$ & 20.09 days & 9.9 days & 2.51 days & 1.89 days\\
				\hline
			\end{tabular}         
		}
		\caption{Total model fitting time for the four methods with $2\times 10^4$ burn-in and $10^4$ post-burn-in MCMC iterations under 9 different settings of $(m,T)$ values.}
		\label{sect6aModelRuntime}
	\end{table}
	\begin{figure}[ph!]
		\centering
		\includegraphics[width=\textwidth]{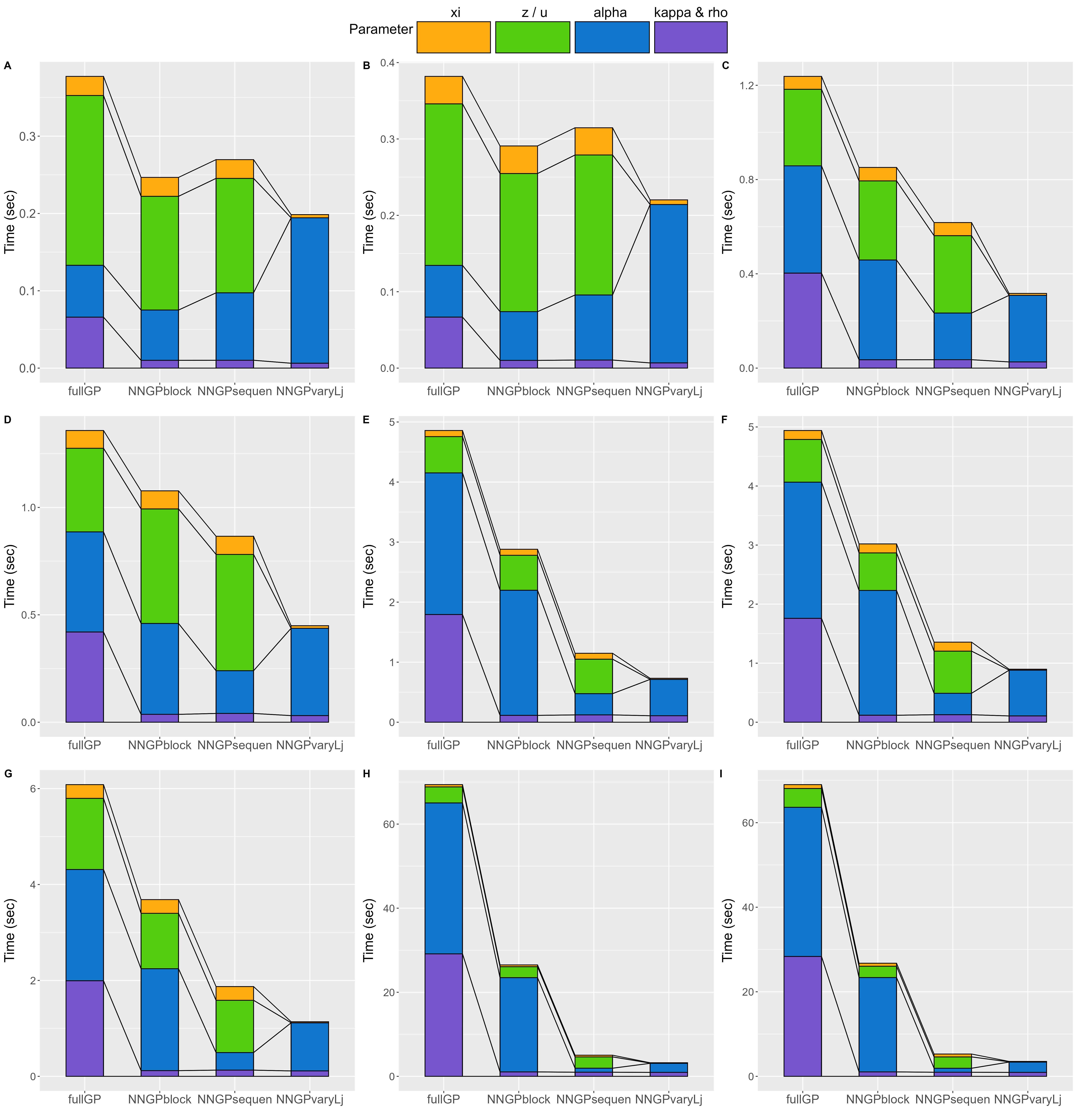}      
		\caption{Average sampling time per MCMC iteration (in seconds) corresponding to  $\rho$ and $\kappa$ (purple), ${\alpha}_{jl_j}^o(\bm{s}_{i})$'s (blue),  $\xi_j^o(\bm{s}_{i})$'s (orange), and $z_{jl_j}^o(\bm{s}_{i})$'s or $u_j^o(\bm{s}_{i})$'s (green) across all 5000 kept post-burn-in MCMC iterations for the four methods \texttt{fullGPfixedL} (\texttt{fullGP}), \texttt{NNGPblockFixedL} (\texttt{NNGPblock}), \texttt{NNGPsequenFixedL} (\texttt{NNGPsequen}), and \texttt{NNGPsequenVaryLj} (\texttt{NNGPvaryLj}), across 9 different $(m,T)$ settings - $(400, 30)$ (\textbf{A}), $(400, 50)$ (\textbf{B}), $(900, 30)$ (\textbf{C}), $(900, 50)$ (\textbf{D}), $(1600, 30)$ (\textbf{E}), $(1600, 50)$ (\textbf{F}), $(1600, 100)$ (\textbf{G}), $(3600, 50)$ (\textbf{H}), and $(3600, 100)$ (\textbf{I}).}
		\label{sect6aStackedPlots}
	\end{figure}
	Next, we provide a detailed breakdown of the posterior sampling time of different model parameters in each MCMC iteration. \Cref{sect6aStackedPlots} presents a detailed comparison across all 9 pairs of $(m,T)$ values, and \Cref{m400T30sect6aGibbsStepTimeTable,m400T50sect6aGibbsStepTimeTable,m900T30sect6aGibbsStepTimeTable,m900T50sect6aGibbsStepTimeTable,m1600T30sect6aGibbsStepTimeTable,m1600T50sect6aGibbsStepTimeTable,m1600T100sect6aGibbsStepTimeTable,m3600T50sect6aGibbsStepTimeTable,m3600T100sect6aGibbsStepTimeTable} reports the average sampling time for all 9 cases. The obtained results for large values of $m$ align with our theoretical computational complexities in \Cref{table:summary.complexity}. \texttt{NNGPblockFixedL} is evidently faster than \texttt{fullGPfixedL}, which is mainly attributable to the Gibbs sampler steps for $\rho,\,\kappa$ and results from replacing $F(\rho)_{m\times m}$ under the full GP prior by a sparse $\tilde{F}(\rho)_{m\times m}$ under our spatial latent NNGP prior (\Cref{sect4c}). 
	The only Gibbs sampler step whose run time significantly differs between \texttt{NNGPsequenFixedL} and \texttt{NNGPblockFixedL} is the step updating all $\alpha_{jl_j}^o(\bm{s}_{i})$'s, where \texttt{NNGPsequenFixedL} uses the sequential updating algorithm in \Cref{sect4d}. As $m$ gets larger, the sequential updates lead to more significant acceleration for sampling $\alpha_{jl_j}^o(\bm{s}_{i})$'s. When $m=3600$, for instance, sampling $\alpha_{jl_j}^o(\bm{s}_{i})$'s in \texttt{NNGPsequenFixedL} is more than 20 times faster than that in \texttt{NNGPblockFixedL}. 
	\texttt{NNGPsequenVaryLj} further accelerates the posterior sampling corresponding to $\xi_j^o(\bm{s}_{i})$'s and $z_{jl_j}^o(\bm{s}_{i})$'s (which are replaced by $u_j^o(\bm{s}_{i})$'s) by using the slice sampling algorithm in \Cref{sect3}. Consequently, \texttt{NNGPsequenVaryLj} stays as the conspicuous fastest algorithm, as it incorporates the non-increasing property for the numbers of spatial clusters $L_j$'s guaranteed by our slice sampling construction (\Cref{sect3}), the enhanced spatial scalability from our latent NNGP prior (\Cref{sect4c}), and the efficient sequential updates (\Cref{seqAlphaIndPostDens2} in \Cref{sect4d}). 
	
	\subsection{Computational Efficiency for Spatial Prediction}\label{sect6b} 
	
\begin{figure}[h]
		\centering
		\includegraphics[width=\textwidth]{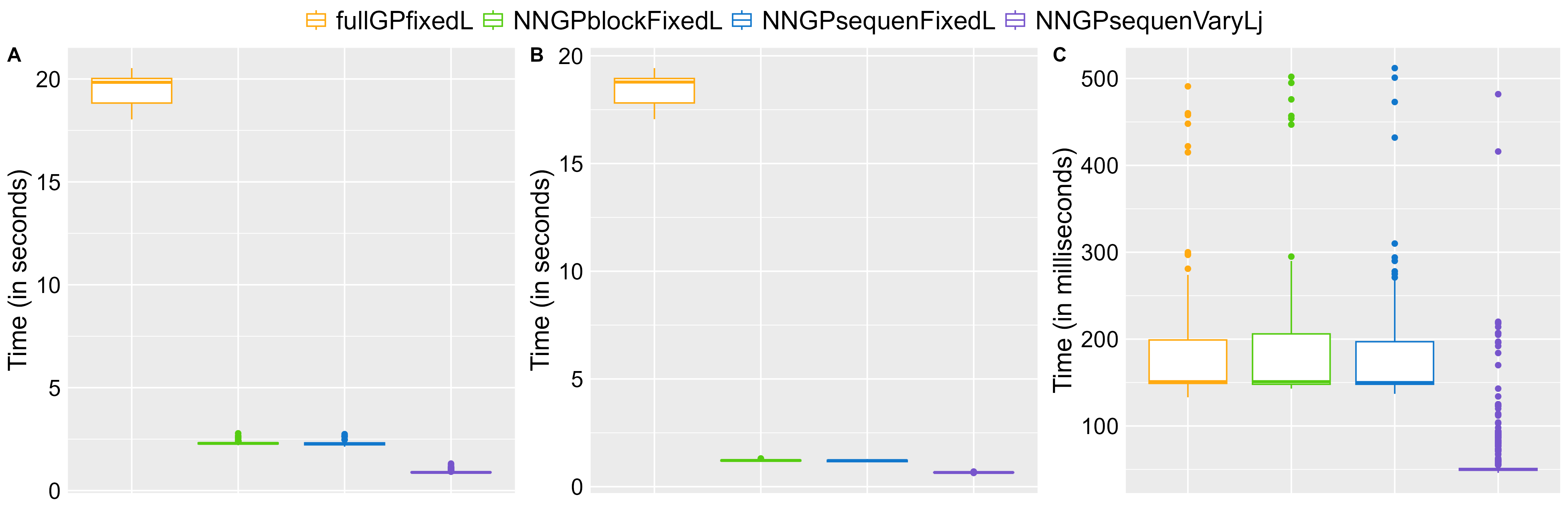}
		\caption{\small Boxplots of the $N\cdot 9 \cdot n$ time realizations for making predictions at the 9 testing locations for our four methods. We have recorded not only the overall spatial prediction time in \textbf{A}, but also the time needed for obtaining $\hat{\bm{\alpha}}\left(\bm{s}_{(m+1):(m+r)}\right)$ in \textbf{B} and $\hat{\bm{w}}\left(\bm{s}_{(m+1):(m+r)}\right)$, $\hat{\bm{\xi}}\left(\bm{s}_{(m+1):(m+r)}\right)$, and $\hat{{\Lambda}}\left(\bm{s}_{(m+1):(m+r)}\right)$ in \textbf{C}, at all kept post-burn-in MCMC iterations.}
		\label{spatpredtimeBoxPlots}
	\end{figure}

	We validate the significant computational acceleration in spatial prediction attributable to our spatial latent NNGP prior (\Cref{sect4}) and slice sampling techniques (\Cref{sect3}). We simulated $N=3$ datasets from the model in \Cref{sect2} with exponential temporal covariance, spatial dependence under a GP with the exponential covariance function, and the PSBP clustering mechanism. We set the number of latent factors $k=4$, a unanimous upper bound of 10 on the number of clusters for all spatial mixture components, the number of observation types $O=1$,  the spatial tuning parameter $\rho=0.8$, the temporal tuning parameter $\psi=2.3$, and $\sigma^2\left(\bm{s}_{i}\right)=0.01$ for all $i\in\{1,\dots,m\}$. We specified $T=300$ equal-distanced time points $t=1,2,\ldots,300$ and $m=20^2+3^2=409$ spatial locations $\bm{s}_i=(i_1,i_2)$ for $i=1,2,\ldots,m$ on a $2$-dimensional grid, where $(i_1,i_2)\in\{1,2,\ldots,20\}\times\{1,2,\ldots,20\}\cup \{9.5,10.5,11.5\}\times\{9.5,10.5,11.5\}$.
	Each simulated dataset is divided into a training set and a testing set, where the training set consists of data on $m_0=20^2=400$ spatial locations $\bm{s}_i=(i_1,i_2)$ for $i=1,2,\ldots,m_0$ with $(i_1,i_2)\in\{1,2,\ldots,20\}\times\{1,2,\ldots,20\}$, and the testing set consists of data on the remaining $r=3^2=9$ spatial locations $\bm{s}_{i_r}=(i_{r_1},i_{r_2})$ for $i_r=1,2,\ldots,r$ with $(i_{r_1},i_{r_2})\in\{9.5,10.5,11.5\}\times\{9.5,10.5,11.5\}$. 
	We set 40 as both the fixed cluster number $L$ in the first three methods and the upper bound for all $L_j$'s in the last method, and $h=15$ for the three methods with spatial latent NNGP priors. We run $3\times 10^4$ burn-in iterations and keep $W= 1000$ out of $10^4$ post-burn-in iterations' posterior samples for our MCMC chains. 
	From each of our $4\times N$ fitted results, we obtained $n=50$ spatial prediction instances using each of our 9 pre-specified random seeds. We thus have four sets of $N\cdot 9 \cdot n$ recorded total (\Cref{spatpredtimeBoxPlots}\textbf{A}) and two-major-step (\Cref{spatpredtimeBoxPlots}\textbf{B} and \textbf{C}) spatial prediction time realizations, which are summarized by the boxplots in \Cref{spatpredtimeBoxPlots}.
	
	We have derived in \Cref{appenH2} and summarized in \Cref{table:summary.complexity} theoretical computational complexities of a major spatial prediction step obtaining $\hat{\bm{\alpha}}_{jl_j}\big(\bm{s}_{(m+1):(m+r)}\big)$, which indicate that \texttt{fullGPfixedL} should take markedly longer time to predict outcomes at new locations than the other three methods for large $m$. Since there are no differences between the spatial prediction procedures for \texttt{NNGPblockFixedL} and \texttt{NNGPsequenFixedL}, these two methods should be comparably fast. \Cref{table:summary.complexity} and how we obtain $\hat{\Lambda}\left(\bm{s}_{(m+1):(m+r)}\right)$ from $\hat{\bm{\alpha}}\left(\bm{s}_{(m+1):(m+r)}\right)$ suggest that \texttt{NNGPsequenVaryLj} should be noticeably faster than the other three methods in spatial prediction, especially when the number of spatial mixture components is moderate to large (\Cref{appenH1}). 
	These theoretical conclusions correspond well to the computation time in \Cref{spatpredtimeBoxPlots}. 
	\texttt{NNGPsequenVaryLj} does turn out to be the fastest method in both major steps and thus overall for spatial prediction. 
	\texttt{fullGPfixedL} is also the evident slowest, which is caused by the step computing $\hat{\bm{\alpha}}\left(\bm{s}_{(m+1):(m+r)}\right)$ (\Cref{spatpredtimeBoxPlots}\textbf{B}). \texttt{NNGPblockFixedL} and \texttt{NNGPsequenFixedL} indeed have quite close computation times in spatial prediction both step-wise and overall.
	
	\subsection{Sea Surface Temperature Data}\label{sect6c}
	
	\begin{figure}[h]
		\centering
		\includegraphics[width=\textwidth]{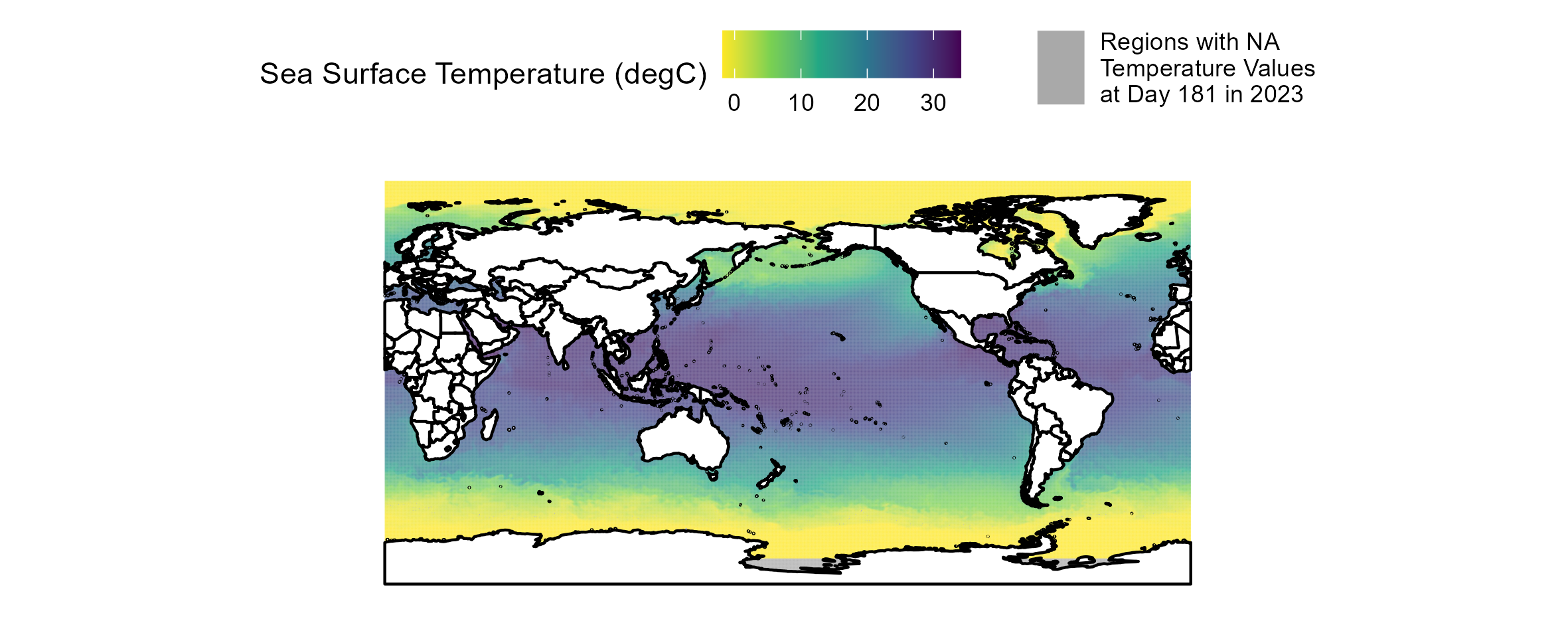}      
		\caption{Actual Sea Surface Temperature (in degrees Celsius) at all $1440\times 720=1036800$ spatial locations on Day 181 in 2023. World map from \textcite{ggmap}.}
		\label{realExDay181all}
	\end{figure}
\begin{figure}[h]
		\centering
		\includegraphics[width=\textwidth]{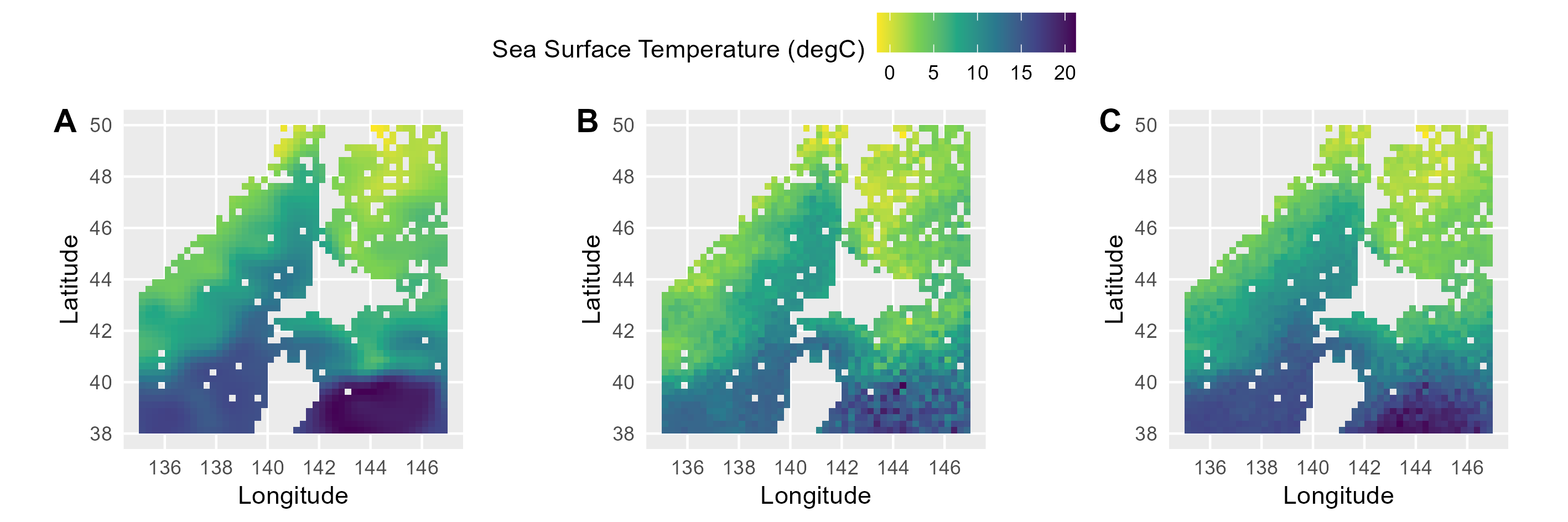}      
		\caption{Actual (\textbf{A}) and predicted (by \texttt{NNGPsequenFixedL} (\textbf{B}) and \texttt{NNGPsequenVaryLj} (\textbf{C})) Sea Surface Temperature (in degrees Celsius) at the 1577 training locations on Day 351 in 2023.}
		\label{realExDay351}
	\end{figure}

\begin{figure}[h]
		\centering
		\includegraphics[width=\textwidth]{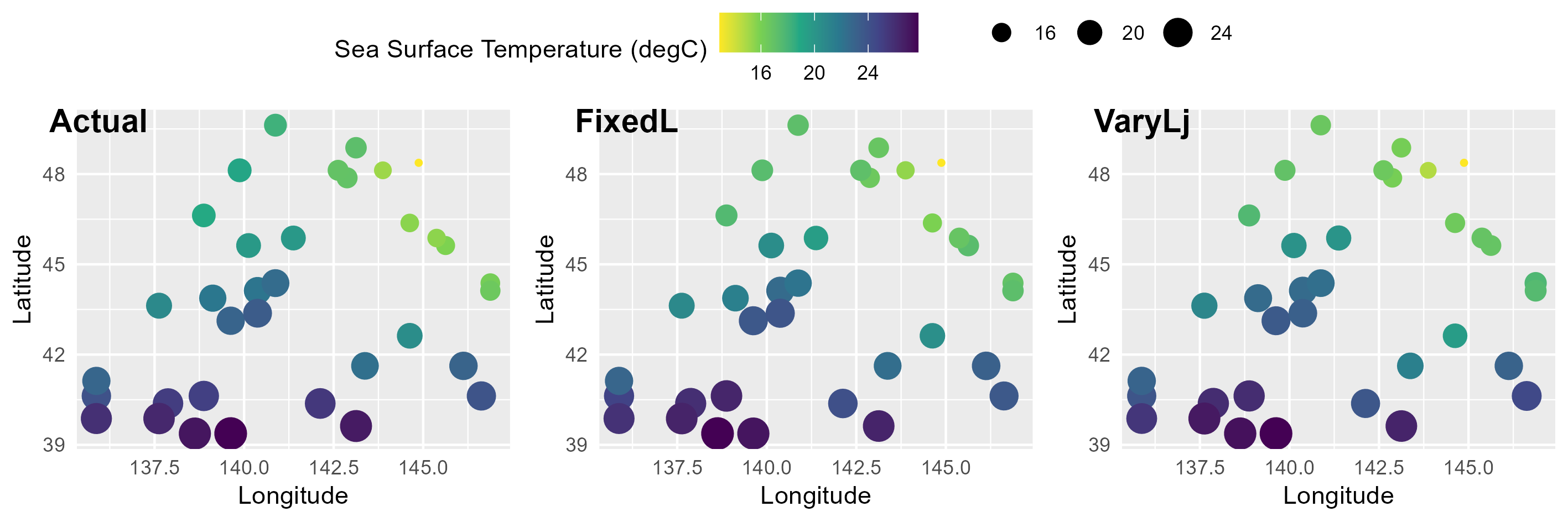}      
		\caption{Actual and predicted (by \texttt{NNGPsequenFixedL} and \texttt{NNGPsequenVaryLj}) Sea Surface Temperature (in degrees Celsius) at the 35 testing locations on Day 230 in 2023.}
		\label{realExDay230}
	\end{figure}

 
	We downloaded Daily Sea Surface Temperature (SST) data in 2023 from \href{https://psl.noaa.gov/data/gridded/data.noaa.oisst.v2.highres.html}{the NOAA website}. The complete data set consists of observations $y_t(\bm{s}_i)$'s, which include a large number of NAs, at $1440\times 720=1036800$ spatial locations specified by longitude-latitude pairs for all 365 days in 2023 (\Cref{realExDay181all}). We only retained locations with no NA observations at all 365 days in the year. Observations at the $m=1612$ spatial points located between $38^\circ$--$50^\circ$ north latitudes and $135^\circ$--$147^\circ$ east longitudes were used for analysis. We fitted \texttt{NNGPsequenFixedL} and \texttt{NNGPsequenVaryLj} with a training data set that consists of $y_t(\bm{s}_i)$'s at a randomly selected $m_0=1577$ out of the 1612 spatial locations and the first 350 days in 2023. We specified the temporal correlation structure and spatial neighborhood structure matrices as $[H(\psi)]_{tt'}=\exp\{-\psi|t-t'|\}$ for all $t,t'\in\{1,\ldots,T\}$ and $F(\rho)=\exp\{-\rho D\}$, respectively, where $D_{m_0\times m_0}$ is a distance matrix for our $m_0$ spatial locations. We also assumed no additional covariates $\bm{x}_t\left(\bm{s}_{i}\right)$'s. We set $h=15$, $L=50$ for \texttt{NNGPsequenFixedL}, and a starting value and hence upper bound of $50$ to all $L_j$'s for \texttt{NNGPsequenVaryLj}. We ran each MCMC chain for $10^5$ burn-in iterations and $2\times 10^4$ post-burn-in iterations, which were thinned to $1000$ samples for analysis. Figures \ref{realExDay351} and \ref{realExDay230} demonstrate these two methods' satisfactory temporal prediction and spatial prediction results, respectively. 

	\section{Discussion}\label{sect7}
	
	We discuss several potential extensions for our current work. 
	First, although we have adopted the same PSBP prior as in \textcite{Berchuck2021}, slice sampling for stick-breaking processes with non-probit links can be similarly derived. 
	Second, we have incorporated the NNGP prior mainly for its computational advantage as a sparse GP approximation. Since NNGP falls in the broader class of Vecchia approximation (\cite{Vecchia1988}, \cite{KatGui21}), one can also integrate other versions of Vecchia approximation that better characterize specific spatial dependencies. 
	Third, our treatment of latent temporal trajectories assumes simple homoscedastic and stationary processes for $\{\bm{\eta}_t\}_{t=1}^T$, which may be further extended to more complex models for temporal structure, such as Bayesian GARCH (\cite{Jensen2013}, \cite{Virbickaite2015}).      
	Finally, the Gaussian response model in the main article is inapplicable to non-normal observed data such as count data. In the Supplementary Material, we have accommodated the binary response model with a logit link by utilizing a P\'olya-Gamma data augmentation technique. We can also extend this framework to generalized linear models with non-Gaussian likelihoods such as Poisson regression, which will be particularly useful for modeling spatiotemporal count data.

	
	

	\appendix	
	\vfill
	\noindent This Supplementary Material is structured as follows. \Cref{appenA} provides full Gibbs sampling details for our spatiotemporal Bayesian Gaussian factor analysis models. \Cref{appenB} elaborates on temporal-related computational burdens for a large number of time points $T\in\mathbb{N},\,T>1$ and our corresponding powerful acceleration solutions when all time points are equally spaced. \Cref{appenC} models the latent temporal process $\{\bm{\eta}_t\}_{t=1}^T$ using a $\VAR(1)$ structure for evenly dispersed time points. \Cref{appenD,appenE,appenF,appenG,appenH,appenI} complement \Cref{sect2,sect3,sect4,sect5,sect6}. Specifically, \Cref{appenD} presents detailed methodology and justification with respect to introducing latent normal variables $z_{jl_j}^o(\bm{s}_{i})\overset{\ind}{\sim}N(\alpha_{jl_j}^o(\bm{s}_{i}),1)$ to bring about conjugacy for $\bm{\alpha}_{jl_j}$'s and thus enabling computationally feasible posterior sampling for \Cref{sect2}'s model; \Cref{appenE} offers some pivotal notes regarding our novel approach adapting slice sampling \parencite{Walker2007} for computationally and storage-wise efficient Bayesian spatial clustering not mentioned in \Cref{sect3}; \Cref{appenF} gives some general comments on the Nearest-Neighbor Gaussian Process (NNGP); \Cref{appenG} discusses detailed procedures for predictions at future time points and cases pertaining to spatial prediction not covered in \Cref{sect5a}; \Cref{appenH} summarizes the computational complexity and memory improvements (with respect to MCMC-based posterior sampling and subsequent vital inferential tasks including spatial prediction and clustering) corresponding to each of our three novelties, i.e., slice sampling, spatial NNGP prior, and sequential updates; \Cref{appenI} displays extensive complementary simulation results with detailed procedures and explanations. Finally, \Cref{appenJ} extends our models to address three non-normal observation data types that still correspond to Gaussian kernels, possibly after some shrewd transformations.
	\newpage
	
	\section{Gibbs Samplers for Our Spatiotemporal Bayesian Gaussian Factor Models} \label{appenA}	
	\par We first delineate prior specification regarding $\bm{\eta}_t$'s, $\Upsilon_{k\times k}$, $\psi$, $\bm{\beta}_{p\times 1}$, and $(\sigma^2)^o\left(\bm{s}_{i}\right)$'s for the model \eqref{eq:single.obs}. Each error variance $(\sigma^2)^o\left(\bm{s}_{i}\right)$ is assigned independent inverse gamma prior $\mathcal{IG}(a,b)$, and the coefficient vector $\bm{\beta}_{p\times 1}$ is imposed a standard conjugate normal prior $N_p\left(\bm{\mu}_{0\bm{\beta}},\Sigma_{0\bm{\beta}}\right)$ in the presence of covariates $\bm{x}_t^o(\bm{s}_{i})$'s. For the temporal latent factors $\bm{\eta}_t$'s, we consider in the main text a general Kronecker-type covariance structure on the prior $\bm{\eta}\sim N_{Tk}\left(\bm{0}, H(\psi)_{T\times T}\otimes\Upsilon \right)$, where $\bm{\eta}=(\bm{\eta}_1^\T,\dots,\bm{\eta}_T^\T)^\T \in \RR^{Tk\times 1}$, $H(\psi) \in \RR^{T\times T}$ captures the temporal correlation structure on $\bm{\eta}_1,\ldots,\bm{\eta}_T$ with tuning parameter $\psi$, and $\Upsilon\in \RR^{k\times k}$ describes the covariance between the $k$ latent components. Examples of $H(\psi)$ are the $\AR(1)$ process $[H(\psi)]_{tt'}=\psi^{|v_t-v_{t'}|}$ and the Gaussian process with exponential covariance function $[H(\psi)]_{tt'}=\exp\{-\psi|v_t-v_{t'}|\}$ for $t,t'\in\{1,\ldots,T\}$, where $v_t$ denotes the time point indexed by $t$ and permits irregularly spaced time points. We also consider the seasonal counterparts of $\AR(1)$ and Gaussian processes with an arbitrary temporal seasonal period $d\in\mathbb{N}, d\geq1$ for modeling the latent process $\{\bm{\eta}_t\}_{t=1}^T$. The tuning parameter $\psi$ is given a uniform prior on a bounded interval, which lacks conjugacy and requires a Metropolis step in posterior sampling. The covariance matrix $\Upsilon_{k\times k}$ is assumed to be unstructured and assigned a conjugate inverse Wishart prior $\Upsilon\sim \mathcal{IW}\left(\zeta, \Omega\right)$. 
	When all time points are equally spaced, a frequently encountered scenario in numerous practical applications, we can considerably accelerate the Gibbs sampling steps for temporal-related parameters $\bm{\eta}_t$'s, $\Upsilon_{k\times k}$, and $\psi$ (to be elaborated in \Cref{appenB}) and adopt a more complicated $\VAR(1)$ model for $\{\bm{\eta}_t\}_{t=1}^T$ (see \Cref{appenC}). 
	\par Throughout this section, \texttt{include.time = TRUE} and \texttt{include.space = TRUE} refer to the scenarios when temporal correlation and spatial dependence are taken into consideration, respectively, i.e., when the temporal correlation structure matrix $H(\psi)_{T\times T}$ and spatial neighborhood structure matrix $F(\rho)_{m\times m}$ are not fixed as identity matrices. The four specifications of \texttt{temporal.structure} - \texttt{`ar1'}, \texttt{`exponential'}, \texttt{`sar1'}, and \texttt{`sexponential'}, denote $\AR(1)$, the exponential process, and their seasonal counterparts, respectively. In almost all circumstances, \texttt{spatial.structure = `continuous'}, i.e., we set $F(\rho)=\exp\{-\rho D\}$, where $D_{m\times m}$ represents the Euclidean distance matrix for our $m$ spatial locations. In the very rare situation when \texttt{spatial.structure = `discrete'}, we consider $F(\rho)^{-1} = D_{\omega}-\rho W$, which corresponds to a Gaussian Markov random field for discrete spatial data. Finally, \texttt{spatApprox} is a logical indicating whether the spatial NNGP prior (\Cref{sect4}) is adopted in place of the full GP prior. If \texttt{spatApprox = TRUE}, then \texttt{spatial.structure} must be \texttt{`continuous'}.   
	\par In all our models, the Gibbs sampling steps with respect to $\bm{\eta}_t$'s, $\Upsilon_{k\times k}$, $\psi$, $\bm{\beta}_{p\times 1}$, $(\sigma^2)^o\left(\bm{s}_{i}\right)$'s are exactly the same and given as below.
	\begin{enumerate}
		\item Sampling from the full conditional distributions of $\bm{\eta}_t$'s:\par
		Fix any arbitrary $t\in\{1,\dots,T\}$. Denote $\bm{\eta}_{-t}$ as the vector $\left(\bm{\eta}_1^{\T},\dots,\bm{\eta}_{t-1}^{\T},\bm{\eta}_{t+1}^{\T},\dots,\bm{\eta}_T^{\T}\right)^{\T}$, which is of length $k(T-1)$. Let $H_{(1:T)_{-t},(1:T)_{-t}}, H_{(1:T)_{-t}, t}, H_{t,(1:T)_{-t}}, H_{t,t}$ be the corresponding $(T-1)\times (T-1)$, $(T-1)\times 1$, $1\times (T-1)$, $1\times 1$ sub-matrices of the temporal correlation structure matrix $H(\psi)_{T\times T}$. Then by properties of conditional distributions of jointly multivariate normals, $f\left(\bm{\eta}_t|\bm{\eta}_{-t},\Upsilon,\psi\right)\sim N_k\left(\mathbb{E}_{0\bm{\eta}_t}, \mathbb{C}_{0\bm{\eta}_t}\right)$, where $\mathbb{E}_{0\bm{\eta}_t}=\left(H_t^+\otimes I_{k\times k}\right)\bm{\eta}_{-t}$ and $\mathbb{C}_{0\bm{\eta}_t}=H_t^*\otimes {\Upsilon}_{k\times k}$ with $H_t^+=H_{t,(1:T)_{-t}}H_{(1:T)_{-t},(1:T)_{-t}}^{-1}$ and $H_t^*=H_{t,t}-H_t^+H_{(1:T)_{-t}, t}=H_{t,t}-H_{t,(1:T)_{-t}}H_{(1:T)_{-t},(1:T)_{-t}}^{-1}H_{(1:T)_{-t}, t}$. Hence
		\begin{align}\label{postEta}
			f\left(\bm{\eta}_t|\cdot\right)
			&\propto f\left(\bm{y}_t|\Lambda,\bm{\eta}_{t},\bm{\beta},\Xi\right) \times f\left(\bm{\eta}_{1:T}|\Upsilon,\psi\right) \\
			& = f\left(\bm{y}_t|\Lambda,\bm{\eta}_t,\bm{\beta},\Xi\right) \times f\left(\bm{\eta}_t|\bm{\eta}_{-t},\Upsilon,\psi\right) \times f\left(\bm{\eta}_{-t}|\Upsilon,\psi\right) \nonumber\\
			&\propto f\left(\bm{y}_t|\Lambda,\bm{\eta}_t,\bm{\beta},\Xi\right) \times f\left(\bm{\eta}_t|\bm{\eta}_{-t},\Upsilon,\psi\right) \nonumber\\
			&\propto \exp\left\lbrace-\frac{1}{2}\left(\bm{y}_t-\Lambda\bm{\eta}_t-X_t\bm{\beta}\right)^{\T}\Xi^{-1}\left(\bm{y}_t-\Lambda\bm{\eta}_t-X_t\bm{\beta}\right) \right\rbrace \nonumber \times\nonumber\\ &\;\;\exp\left\lbrace-\frac{1}{2}\left(\bm{\eta}_t-\mathbb{E}_{0\bm{\eta}_t}\right)^{\T}\mathbb{C}_{0\bm{\eta}_t}^{-1}\left(\bm{\eta}_t-\mathbb{E}_{0\bm{\eta}_t}\right)\right\rbrace\nonumber\\
			&\propto \exp\left\lbrace-\frac{1}{2}\bm{\eta}_t^{\T}\left[\mathbb{C}_{0\bm{\eta}_t}^{-1}+\Lambda^{\T}\Xi^{-1}\Lambda\right]\bm{\eta}_t+\bm{\eta}_t^{\T}\left[\mathbb{C}_{0\bm{\eta}_t}^{-1}\mathbb{E}_{0\bm{\eta}_t}+\Lambda^{\T}\Xi^{-1}\left(\bm{y}_t-X_t\bm{\beta}\right)\right]  \right\rbrace \nonumber\\
			&\sim N_k\left(\mathbb{C}_{\bm{\eta}_t}\bm{\mu}_{\bm{\eta}_t}, \mathbb{C}_{\bm{\eta}_t}\right), 
			\text{ where}\nonumber\\
			&\mathbb{C}_{\bm{\eta}_t}=\left[\mathbb{C}_{0\bm{\eta}_t}^{-1}+\Lambda^{\T}\Xi^{-1}\Lambda\right]^{-1}\text{ and } \bm{\mu}_{\bm{\eta}_t}=\mathbb{C}_{0\bm{\eta}_t}^{-1}\mathbb{E}_{0\bm{\eta}_t}+\Lambda^{\T}\Xi^{-1}\left(\bm{y}_t-X_t\bm{\beta}\right).\nonumber
			& \nonumber
		\end{align}
		\item Sampling from the full conditional distribution of $\Upsilon_{k\times k}$:
		\begin{align}\label{postUpsilon}
			&f(\Upsilon|\cdot)\propto f\left(\bm{\eta}_{1:T}|\Upsilon,\psi\right) \times f_0(\Upsilon) \sim \mathcal{IW}\left(T+\zeta, \Phi H(\psi)^{-1}\Phi^{\T} + \Omega \right),  \\
			&\text{where }\Phi_{k\times T} = \left(\bm{\eta}_1,\dots,\bm{\eta}_T\right). \nonumber
		\end{align}
		When the number of factors $k=1$, the above Inverse-Wishart distribution simplifies to an Inverse-Gamma distribution $\mathcal{IG}\left(\frac{T+\zeta}{2}, \frac{\Phi H(\psi)^{-1}\Phi^{\T} + \Omega}{2} \right)$.
		\item Sampling $\psi$ via a Metropolis step when \texttt{include.time = TRUE}:
		\begin{align}\label{postPsi}
			f(\psi|\cdot)\propto f\left(\bm{\eta}_{1:T}|\Upsilon,\psi\right) \times f_0(\psi) \times \left| \frac{\partial}{\partial\Delta}h^{-1}(\Delta) \right| \text{ with }\bm{\eta}_{1:T}|\Upsilon,\psi \sim N_{Tk}\left(\bm{0}, H(\psi)\otimes\Upsilon\right). 
		\end{align}
		In \Cref{postPsi}, $f_0(\psi)$ refers to the uniform prior $\unif\left(a_{\psi}, b_{\psi}\right)$ when \texttt{temporal.structure = "exponential"} or \texttt{"sexponential"} and the transformed Beta prior with p.d.f $f_0\left(\psi\right) \propto (1+\psi)^{\gamma-1}\times (1-\psi)^{\beta-1}$ when \texttt{temporal.structure = "ar1"} or \texttt{"sar1"}. $\Delta$ is a new parameter defined to be $h(\psi)=\ln\left(\frac{\psi-a_{\psi}}{b_{\psi}-\psi}\right)$. Hence $\psi=h^{-1}(\Delta)=\frac{b_{\psi}\exp(\Delta)+a_{\psi}}{1+\exp(\Delta)}$ and $\lvert\frac{\partial}{\partial\Delta}h^{-1}(\Delta)\rvert\propto\frac{\exp(\Delta)}{\left[1+\exp(\Delta)\right]^2}$. 
		\par At each MCMC iteration $s\in\mathbb{N}$, let the current parameter estimate for $\psi$ be $\psi^{(s)}$. We propose a new parameter value $\psi^*=h^{-1}(\Delta^*)$ with $\Delta^*$ sampled from a symmetric kernel $N\left(\Delta^{(s)},\delta\right)$, where $\Delta^{(s)}=h\left(\psi^{(s)}\right)$ and $\delta>0$ is a tuning parameter decided after the burn-in MCMC iterations based on the acceptance rate for $\psi^*$, and set
		\begin{align*}
			{\psi}^{(s+1)}=
			\begin{cases}
				{\psi}^*,&\text{ w.p. }\alpha\left({\psi}^{(s)},{\psi}^*\right)=\min\left\lbrace\frac{f({\psi}^*|\cdot)}{f({\psi}^{(s)}|\cdot)},1\right\rbrace \\ 
				{\psi}^{(s)},&\text{ w.p. }1-\alpha\left({\psi}^{(s)},{\psi}^*\right)
			\end{cases}.
		\end{align*}
		\item Sampling from the full conditional distribution of $\bm{\beta}_{p\times 1}$ when there are additional covariates $\bm{x}$:
		\begin{align}\label{postBeta}
			f\left(\bm{\beta}|\cdot\right) &\propto \prod_{t=1}^T f\left(\bm{y}_t|\Lambda,\bm{\eta}_t,\bm{\beta},\Xi\right)\times f_0(\bm{\beta}) \\
			&\propto  \exp\{-\frac{1}{2}\left(\bm{\beta}-\bm{\mu}_{0\bm{\beta}}\right)^{\T}\Sigma_{0\bm{\beta}}^{-1}\left(\bm{\beta}-\bm{\mu}_{0\bm{\beta}}\right)\} \times \nonumber\\
			&\;\prod_{t=1}^T \exp\left\lbrace-\frac{1}{2}\left(\bm{y}_t-\Lambda\bm{\eta}_t-X_t\bm{\beta}\right)^{\T}\Xi^{-1}\left(\bm{y}_t-\Lambda\bm{\eta}_t-X_t\bm{\beta}\right) \right\rbrace           \nonumber\\
			&\propto \exp\left\lbrace-\frac{1}{2}\bm{\beta}^{\T}\left[\Sigma_{0\bm{\beta}}^{-1} + \sum_{t=1}^TX_t^{\T}\Xi^{-1}X_t\right]\bm{\beta}+\bm{\beta}^{\T}\left[\Sigma_{0\bm{\beta}}^{-1}\bm{\mu}_{0\bm{\beta}}+\sum_{t=1}^TX_t^{\T}\Xi^{-1}\left(\bm{y}_t-\Lambda\bm{\eta}_t\right)\right]\right\rbrace \nonumber\\
			&\sim N_{p}\left( V_{\bm{\beta}}{\bm{\mu}}_{\bm{\beta}}, V_{\bm{\beta}}\right),\text{ where}  \nonumber\\
			&\;\; V_{\bm{\beta}} = \left[\Sigma_{0\bm{\beta}}^{-1} + \sum_{t=1}^TX_t^{\T}\Xi^{-1}X_t\right]^{-1} \text{ and }{\bm{\mu}}_{\bm{\beta}} = \Sigma_{0\bm{\beta}}^{-1}\bm{\mu}_{0\bm{\beta}}+\sum_{t=1}^TX_t^{\T}\Xi^{-1}\left(\bm{y}_t-\Lambda\bm{\eta}_t\right).
			\nonumber
		\end{align}
		\item Sampling from the full conditional distributions of $(\sigma^2)^o(\bm{s}_{i})$'s when not all observations are from the distribution family \texttt{`binomial'} (\Cref{appenJ}):\par
		Fix any arbitrary $(i,o)\in\{1,\dots,m\}\times\{1,\dots,O\}$. Then
		\begin{align}\label{postSigma2}
			&f\left((\sigma^2)^o\left(\bm{s}_{i}\right)|\cdot\right)\propto f_0\left((\sigma^2)^o\left(\bm{s}_{i}\right)\right)\times \prod_{t=1}^T f\left({y}_t^o\left(\bm{s}_{i}\right)|\Lambda^o\left(\bm{s}_{i}\right),\bm{\eta}_t,\bm{\beta},(\sigma^2)^o\left(\bm{s}_{i}\right)\right)\\
			&\quad\propto \left[(\sigma^2)^o\left(\bm{s}_{i}\right)\right]^{-a-1}\exp\left\lbrace -\frac{b}{(\sigma^2)^o\left(\bm{s}_{i}\right)}\right\rbrace \times \nonumber\\
			&\quad\;\left[(\sigma^2)^o\left(\bm{s}_{i}\right)\right]^{-\frac{T}{2}}\exp\left\lbrace -\frac{1}{2(\sigma^2)^o\left(\bm{s}_{i}\right)}\sum_{t=1}^T\left[{y}_t^o\left(\bm{s}_{i}\right)-\bm{x}_t^o\left(\bm{s}_{i}\right)^{\T}\bm{\beta}-\bm{\lambda}^o\left(\bm{s}_{i}\right)^{\T}\bm{\eta}_t \right]^2\right\rbrace  \nonumber\\
			&\quad\propto \left[(\sigma^2)^o\left(\bm{s}_{i}\right)\right]^{-a-\frac{T}{2}-1}\exp\left\lbrace-\frac{1}{(\sigma^2)^o\left(\bm{s}_{i}\right)}\left(b+\frac{1}{2}\sum_{t=1}^T\left[{y}_t^o\left(\bm{s}_{i}\right)-\bm{x}_t^o\left(\bm{s}_{i}\right)^{\T}\bm{\beta}-\bm{\lambda}^o\left(\bm{s}_{i}\right)^{\T}\bm{\eta}_t \right]^2\right)\right\rbrace  \nonumber\\
			&\quad\sim \mathcal{IG}\left(a+\frac{T}{2},\; b+\frac{1}{2}\sum_{t=1}^T\left[{y}_t^o\left(\bm{s}_{i}\right)-\bm{x}_t^o\left(\bm{s}_{i}\right)^{\T}\bm{\beta}-\bm{\lambda}^o\left(\bm{s}_{i}\right)^{\T}\bm{\eta}_t \right]^2\right),\text{ where}\nonumber\\
			&\quad\bm{\lambda}^o\left(\bm{s}_{i}\right)^{\T}_{1\times k}\text{ denotes the row in $\Lambda_{mO\times k}$ corresponding to location $i$ and observation type }o. \nonumber
		\end{align}
	\end{enumerate}
	
	\subsection{Full Gibbs Sampling Details for the Basic Model Without Spatial Clustering Capabilities}\label{appenA1}
	
	\par When we do not consider the PSBP clustering mechanism and directly impose spatially correlated Gaussian Processes onto the columns $\{\bm{\lambda}_j\}_{j=1}^k$ of the factor loadings matrix $\Lambda_{mO\times k}$, i.e., assign a normal prior with Kronecker-type covariance structure $\bm{\lambda}_j\overset{\iid}{\sim}N_{Om}\left(\bm{0}, \kappa_{O\times O}\otimes F(\rho)_{m\times m}\right)$, $j\in\{1,\dots,k\} $, the corresponding Gibbs sampler steps are detailed in the following. The parameters $\bm{\lambda}_j$'s and $\bm{\sigma^2}$ (when not all observations are from the \texttt{`binomial'} family) are all ordered first spatially and then by observation type, i.e., the first $m$ components of each above $Om\times 1$ vector are from observation type 1, the next $m$ components are from observation type 2, and so on. $\kappa \in \RR^{O\times O}$ depicts the covariance between the $O$ observation types and is imposed an $\mathcal{IW}(\nu,\Theta)$ prior. For a continuous spatial domain, $F(\rho)$ denotes the covariance matrix from a Gaussian process over the $m$ spatial locations with an exponential covariance function tuning parameter $\rho$ that can be further assigned a uniform prior.
	\begin{enumerate}
		\item Sampling from the full conditional distributions of $\bm{\lambda}_{j}$'s:
		\par Fix any arbitrary $j\in\{1,\dots,k\}$. Let $\Xi_{Om\times Om}=\diag\left(\bm{\sigma^2}_1,\dots,\bm{\sigma^2}_O\right)$, where $\bm{\sigma^2}_o=\left((\sigma^2)^o(\bm{s}_{1}),\dots,(\sigma^2)^o(\bm{s}_{m})\right)^{\T}\;\forall\;o\in\{1,\dots,O\}$. Denote $\Lambda_{-j}$ as $\left(\bm{\lambda}_1,\dots,\bm{\lambda}_{j-1},\bm{\lambda}_{j+1},\dots,\bm{\lambda}_k\right)$, the sub-matrix obtained by removing the $j^{\text{th}}$ column from $\Lambda_{Om\times k}$. For each $t$, let $\bm{\eta}_t^{-j}$ represent the length-$(k-1)$ sub-vector of $\bm{\eta}_t$ obtained by removing its $j^{\text{th}}$ element. Then
		\begin{align}
			&f\left(\bm{\lambda}_{j}|\cdot\right)
			\propto \prod_{t=1}^T f\left(\bm{y}_t|\Lambda,\bm{\eta}_t,\bm{\beta},\Xi\right)\times \pi\left(\bm{\lambda}_{j}|\kappa,\rho\right)\\
			&\:\propto\exp\{-\frac{1}{2}\bm{\lambda}_j^{\T}\left[\kappa^{-1}\otimes F(\rho)^{-1}\right]\bm{\lambda}_j\}\times\nonumber\\
			&\quad\prod_{t=1}^T\exp\left\lbrace-\frac{1}{2}\left(\bm{y}_t-X_t\bm{\beta}-\Lambda_{-j}\bm{\eta}_t^{-j}-\eta_{tj}\bm{\lambda}_j\right)^{\T}\Xi^{-1}\left(\bm{y}_t-X_t\bm{\beta}-\Lambda_{-j}\bm{\eta}_t^{-j}-\eta_{tj}\bm{\lambda}_j\right)\right\rbrace \nonumber \\
			&\:\propto\exp\left\lbrace-\frac{1}{2}\bm{\lambda}_j^{\T}\left(\kappa^{-1}\otimes F(\rho)^{-1}+\sum_{t=1}^T \eta_{tj}^2\Xi^{-1}\right)\bm{\lambda}_j+\bm{\lambda}_j^{\T}\Xi^{-1}\left[\sum_{t=1}^T\eta_{tj}\left(\bm{y}_t-\Lambda_{-j}\bm{\eta}_t^{-j}-X_t\bm{\beta} \right)\right]\right\rbrace\nonumber\\
			&\;\sim N_{Om}\left(V_{\bm{\lambda}_j}\bm{\mu}_{\bm{\lambda}_j}, V_{\bm{\lambda}_j}\right),\text{ where } \nonumber\\
			&\quad V_{\bm{\lambda}_j}=\left[\kappa^{-1}\otimes F(\rho)^{-1}+\sum_{t=1}^T \eta_{tj}^2\Xi^{-1} \right]^{-1}\text{ and }\bm{\mu}_{\bm{\lambda}_j}=\Xi^{-1}\left[ \sum_{t=1}^T\eta_{tj}\left(\bm{y}_t-\Lambda_{-j}\bm{\eta}_t^{-j}-X_t\bm{\beta} \right)\right]. \nonumber
		\end{align}
		
		When the NNGP instead of the full GP prior is adopted for the $\bm{\lambda}_{j}$'s, we replace $\pi\left(\bm{\lambda}_{j}|\kappa,\rho\right)$ and $F(\rho)$ by $\tilde{\pi}\left(\bm{\lambda}_{j}|\kappa,\rho\right)$ and $\tilde{F}(\rho)$ in the above equations.
		\item Sampling from the full conditional distribution of $\kappa_{O\times O}$:
		\begin{align}\label{postKappaFixedLnoCL}
			&f(\kappa|\cdot)\propto \prod_{j=1}^k\pi(\bm{\lambda}_{j}|\kappa,\rho)\times f_0(\kappa)
			\sim \mathcal{IW}\left(mk+\nu , \sum_{j=1}^kS_{\lambda^{j}_{\rho}}+\Theta \right),\\ &\text{where }S_{\lambda^{j}_{\rho}} = A_{j}^{\T} F(\rho)^{-1}A_{j}\text{ with }(A_{j})_{m\times O}=(\bm{\lambda}_{j1},\dots,\bm{\lambda}_{jO})\:\forall\:j.\nonumber
		\end{align}
		In \Cref{postKappaFixedLnoCL}, we have written the $Om\times 1$ vector $\bm{\lambda}_j$ as $(\bm{\lambda}_{j1}^{\T},\dots,\bm{\lambda}_{jO}^{\T})^{\T}$ for all $j\in\{1,\dots,k\}$. \\When $O=1$, \Cref{postKappaFixedLnoCL} reduces to $\mathcal{IG}\left(\frac{mk+\nu}{2}, \frac{\sum_{j=1}^kS_{\lambda^{j}_{\rho}}+\Theta}{2}\right) $.
		\par When the NNGP prior rather than the full GP prior is adopted for the $\bm{\lambda}_{j}$'s, we replace $\pi\left(\bm{\lambda}_{j}|\kappa,\rho\right)$ and $F(\rho)$ in \Cref{postKappaFixedLnoCL} by $\tilde{\pi}\left(\bm{\lambda}_{j}|\kappa,\rho\right)$ and $\tilde{F}(\rho)$ respectively.
		\item Sampling $\rho$ via a Metropolis step when \texttt{include.space = TRUE}:
		\begin{align}
			f(\rho|\cdot)\propto\prod_{j=1}^k&\:\pi(\bm{\lambda}_{j}|\kappa,\rho)\times f_0(\rho)\times\left| \frac{\partial}{\partial\Delta}h^{-1}(\Delta) \right| \label{postRhoFixedLnoCL}\\
			&\text{with }\bm{\lambda}_{j}|\kappa,\rho\sim N_{Om}\left(\bm{0},  \kappa\otimes F(\rho)\right)\qquad\forall\;j.\nonumber
		\end{align}
		When the NNGP prior rather than the full GP prior is adopted for the $\bm{\lambda}_{j}$'s, we replace $\pi\left(\bm{\lambda}_{j}|\kappa,\rho\right)$ and $F(\rho)$ in \Cref{postRhoFixedLnoCL} by $\tilde{\pi}\left(\bm{\lambda}_{j}|\kappa,\rho\right)$ and $\tilde{F}(\rho)$ respectively.
		\par In \Cref{postRhoFixedLnoCL}, $f_0(\rho)$ refers to the uniform prior $\unif\left(a_{\rho}, b_{\rho}\right)$, where $a_{\rho}=0,b_{\rho}=1$ when \texttt{spatial.structure = `discrete'} and $a_{\rho},b_{\rho}\in\mathbb{R}^+$ when \texttt{spatial.structure = `continuous'}. $\Delta$ is a new parameter defined to be $h(\rho)=\ln\left(\frac{\rho-a_{\rho}}{b_{\rho}-\rho}\right)$. Hence $\rho=h^{-1}(\Delta)=\frac{b_{\rho}\exp(\Delta)+a_{\rho}}{1+\exp(\Delta)}$ and $\lvert\frac{\partial}{\partial\Delta}h^{-1}(\Delta)\rvert\propto\frac{\exp(\Delta)}{\left[1+\exp(\Delta)\right]^2}$. 
		\par At each MCMC iteration $s\in\mathbb{N}$, let the current parameter estimate for $\rho$ be $\rho^{(s)}$. We propose a new parameter value $\rho^*=h^{-1}(\Delta^*)$ with $\Delta^*$ sampled from a symmetric kernel $N\left(\Delta^{(s)},\delta\right)$, where $\Delta^{(s)}=h\left(\rho^{(s)}\right)$ and $\delta>0$ is a tuning parameter decided after the burn-in MCMC iterations based on the acceptance rate for $\rho^*$, and set
		\begin{align*}
			{\rho}^{(s+1)}=
			\begin{cases}
				{\rho}^*,&\text{ w.p. }\alpha\left({\rho}^{(s)},{\rho}^*\right)=\min\left\lbrace\frac{f({\rho}^*|\cdot)}{f({\rho}^{(s)}|\cdot)},1\right\rbrace \\ {\rho}^{(s)},&\text{ w.p. }1-\alpha\left({\rho}^{(s)},{\rho}^*\right)
			\end{cases}.
		\end{align*}
		\item Same as Step 1 given at the start of \Cref{appenA}.
		\item Same as Step 2 given at the start of \Cref{appenA}.
		\item Same as Step 3 given at the start of \Cref{appenA}.
		\item Same as Step 4 given at the start of \Cref{appenA}.
		\item Same as Step 5 given at the start of \Cref{appenA}.
	\end{enumerate}
	
	\subsection{Full Gibbs Sampling Details for Our Model in \Cref{sect2}}\label{appenA2}   
	\par When we adopt the spatial PSBP clustering mechanism with a pre-determined fixed common $L\in\mathbb{N}, L>1$ for all factors and assign the prior $\bm{\alpha}_{jl_j}\overset{\iid}{\sim}N_{mO}\left(\bm{0}, F(\rho)_{m\times m}\otimes \kappa_{O\times O}\right)$, $j\in\{1,\dots,k\}$, $l_j\in\{1,\dots,L-1\}$, the corresponding Gibbs sampler steps are detailed in the following. We take note that only $\bm{z}_{jl_j}$'s and $\bm{\alpha}_{jl_j}$'s are ordered first by observation type and then spatially, i.e., the first $O$ components of each above $mO\times 1$ vector are from location point 1, the next $O$ components are from location point 2, and so on. All other concerned parameters, i.e., $\bm{w}_{jl_j}$'s, $\bm{\xi}_j$'s, $\bm{\lambda}_j$'s, and $\bm{\sigma^2}$ when not all observations are from the \texttt{`binomial'} family (\Cref{appenJ}), are ordered first spatially and then by observation type, i.e., the first $m$ components of each above $Om\times 1$ vector are from observation type 1, the next $m$ components are from observation type 2, and so on.
	\begin{enumerate}
		\item 
		Sampling from the marginal conditional distributions of $\xi_j^o(\bm{s}_{i})$'s with the introduced latent normal $z_{jl_j}^o(\bm{s}_{i})$'s integrated out:
		\par Fix any arbitrary $j\in\{1,\dots,k\}$.
		\begin{align}\label{postXiFixedL}
			&\forall\:(o,i)\in\{1,\dots,O\}\times\{1,\dots,m \},\;\forall\; l_j\in\{1,\dots,L\},\; \mathbb{P}\left(\xi_j^o(\bm{s}_{i})=l_j|\cdot\right)\\
			& \propto \mathbb{P}\left(\xi_j^o(\bm{s}_{i})=l_j|\alpha_{jr_j}^o(\bm{s}_{i}),r_j\leq L-1\right) \prod_{t=1}^T f\left(y_t^o(\bm{s}_{i})|\bm{\beta},\bm{\eta}_t,(\sigma^2)^o(\bm{s}_{i}),\xi_j^o(\bm{s}_{i})=l_j,\bm{\xi}_{-j}^o(\bm{s}_{i}),\bm{\theta}_{\bm{\xi}^o(\bm{s}_{i})}\right) \nonumber\\
			& \propto w_{jl_j}^o(\bm{s}_{i})\times \exp\left\lbrace -\frac{1}{2}(\sigma^{-2})^o(\bm{s}_{i})\sum_{t=1}^T\left[y_t^o(\bm{s}_{i})-\bm{x}_t^o(\bm{s}_{i})^{\T}\bm{\beta}-\sum_{h\neq j}^{1\leq h\leq k}\theta_{h\xi_h^o(\bm{s}_{i})}\eta_{th}-\theta_{jl_j}\eta_{tj} \right]^2\right\rbrace.\nonumber
		\end{align}	
		\item Sampling from the full conditional distributions of $\theta_{jl_j}$'s for $l_j\leq L$:
		\par Fix any arbitrary $(j,l_j)\in\{1,\dots,k\}\times\{1,\dots,L\}$. Then 
		\begin{align}\label{postThetaFixedL}
			f(\theta_{jl_j}|\cdot )\propto f_0(\theta_{jl_j}|\tau_j) \times \prod_{(i,o):\xi_j^o(\bm{s}_{i})=l_j }\prod_{t=1}^T f\left(y_t^o(\bm{s}_{i})|\bm{\beta},\bm{\eta}_t,(\sigma^2)^o(\bm{s}_{i}),\bm{\xi}^o(\bm{s}_{i}),\bm{\theta}_{\bm{\xi}^o(\bm{s}_{i})}\right).
		\end{align}
		If $\nexists\; (i,o)$ s.t. $\xi_j^o(\bm{s}_{i})=l_j$, then $f(\theta_{jl_j}|\cdot )= f_0(\theta_{jl_j}|\tau_j)\sim N\left(0,\tau_j^{-1}\right)$.\\
		If $\exists\; (i,o)$ s.t. $\xi_j^o(\bm{s}_{i})=l_j$, then 
		\begin{align*}
			&f(\theta_{jl_j}|\cdot )\sim N\left(V_{\theta_{jl_j}}\mu_{\theta_{jl_j}}, V_{\theta_{jl_j}}\right) , \text{ where }V_{\theta_{jl_j}}=\left[\tau_j + \sum_{(i,o):\xi_j^o(\bm{s}_{i})=l_j}(\sigma^{-2})^o(\bm{s}_{i})\sum_{t=1}^T\eta_{tj}^2 \right]^{-1}\\
			&\text{and }\mu_{\theta_{jl_j}} = \sum_{(i,o):\xi_j^o(\bm{s}_{i})=l_j}(\sigma^{-2})^o(\bm{s}_{i})\sum_{t=1}^T\eta_{tj}\left[y_t^o(\bm{s}_{i})-\bm{x}_t^o(\bm{s}_{i})^{\T}\bm{\beta}-\sum_{h\neq j}^{1\leq h\leq k}\theta_{h\xi_h^o(\bm{s}_{i})}\eta_{th} \right]. 
		\end{align*}
		\item Sampling from the full conditional distributions of $\delta_{h}$'s:
		\par If a multiplicative gamma process shrinkage prior is placed on the atoms so that $\forall\:(j,l_j)$, $\theta_{jl_j}\overset{\text{ind}}{\sim}N\left(0, \tau_j^{-1}\right)$ with $\tau_j=\prod_{h=1}^j\delta_h,\;\delta_1\sim\Ga(a_1,1),\:\delta_h\sim\Ga(a_2,1)$ $\forall\:h\geq 2$, then for any arbitrary fixed $h\in\{1,\dots,k\}$,
		\begin{align}\label{postDeltaFixedL}
			f(\delta_h|\cdot)&\propto \prod_{j=h}^k\prod_{l_j=1}^{L}f(\theta_{jl_j}|\tau_j)\times f_0(\delta_h)\\
			&\sim \Ga\left(a_h+\frac{(k-h+1)L}{2}, 1+\frac{\sum_{j=h}^k \left[\prod_{x=1,x\neq h}^j\delta_x\right]\sum_{l_j=1}^{L}\theta_{jl_j}^2}{2} \right).\nonumber 
		\end{align}
		If we consider the non-shrinkage prior $\theta_{jl_j}\overset{\text{ind}}{\sim}N\left(0, \tau_j^{-1}=\delta_j^{-1}\right)\:\forall\:(j,l_j)$ with \\$\delta_j\overset{\iid}{\sim}\Ga(a_1,a_2)\:\forall\:j$, then for any arbitrary fixed $j\in\{1,\dots,k\}$,
		\begin{align}\label{postDeltaNoGSFixedL}
			f(\delta_j|\cdot)\propto \prod_{l_j=1}^{L}f(\theta_{jl_j}|\delta_j)\times f_0(\delta_j)
			\sim \Ga\left(a_1+\frac{L}{2}, a_2+\frac{\sum_{l_j=1}^{L}\theta_{jl_j}^2}{2} \right). 
		\end{align}    	
		\item Sampling from the full conditional distributions of ${z}_{jl_j}^o(\bm{s}_{i})$'s for $l_j\in\{1,\dots,L-1\}$:
		\par For any arbitrary fixed $(j,i,o,l_j)\in\{1,\dots,k\}\times\{1,\dots,m\}\times\{1,\dots,O\}\times\{1,\dots,L-1\}$, 
		\begin{align}\label{postZ}
			& f(z_{jl_j}^o(\bm{s}_{i})|\cdot)\propto f(\xi_j^o(\bm{s}_{i})|z_{jl_j}^o(\bm{s}_{i}))\times f(z_{jl_j}^o(\bm{s}_{i})|\alpha_{jl_j}^o(\bm{s}_{i}))\\
			&\sim  \mathbbm{1}_{\{\xi_j^o(\bm{s}_{i})>l_j\}}N(\alpha_{jl_j}^o(\bm{s}_{i}),1)_{\mathbb{R}_-} + \mathbbm{1}_{\{\xi_j^o(\bm{s}_{i})=l_j\}}N(\alpha_{jl_j}^o(\bm{s}_{i}),1)_{\mathbb{R}_+} + 
			\mathbbm{1}_{\{\xi_j^o(\bm{s}_{i})<l_j\}}N(\alpha_{jl_j}^o(\bm{s}_{i}),1).\nonumber
		\end{align}
		\item Sampling from the full conditional distributions of $\bm{\alpha}_{jl_j}$'s for $l_j\in\{1,\dots,L-1\}$: 
		\par For any arbitrary fixed $(j,l_j)\in\{1,\dots,k\}\times\{1,\dots,L-1\}$,
		\begin{align}\label{postAlphaFixedL}
			f(\bm{\alpha}_{jl_j}|\cdot)&\propto \prod_{(i,o)}f\left(z_{jl_j}^o(\bm{s}_{i})|\alpha_{jl_j}^o(\bm{s}_{i})\right)\times \pi(\bm{\alpha}_{jl_j}|{\kappa},\rho)=  {f(\bm{z}_{jl_j}|\bm{\alpha}_{jl_j})}\times \pi(\bm{\alpha}_{jl_j}|{\kappa},\rho)\nonumber\\
			&\sim N_{mO}\left(\left[ {I}_{mO}+{F}(\rho)^{-1}\otimes {\kappa}^{-1}\right]^{-1}\bm{z}_{jl_j}, \left[ {I}_{mO}+{F}(\rho)^{-1}\otimes{\kappa}^{-1}\right] ^{-1}\right).
		\end{align}   
		When the NNGP prior rather than the full GP prior is adopted for the $\bm{\alpha}_{jl_j}$'s, we replace $\pi\left(\bm{\alpha}_{jl_j}|\kappa,\rho\right)$ and $F(\rho)$ in \Cref{postAlphaFixedL} by $\tilde{\pi}\left(\bm{\alpha}_{jl_j}|\kappa,\rho\right)$ and $\tilde{F}(\rho)$ respectively.
		\par After obtaining our new estimates for $\alpha_{jl_j}^o(\bm{s}_{i})$'s ($l_j\in\{1,\dots,L-1\}$), we calculate our new samples for the corresponding weights parameters $w_{jl_j}^o(\bm{s}_{i})$'s ($l_j\in\{1,\dots,L\}$) via the formula
		\begin{equation}\label{weightsFormulaFixedL}
			w_{jl_j}^o(\bm{s}_{i})=
			\begin{cases}
				\Phi\left(\alpha_{jl_j}^o(\bm{s}_{i})\right)\prod_{r<l}\left[1-\Phi\left(\alpha_{jr_j}^o(\bm{s}_{i})\right) \right],&\text{ if }l_j\in\{1,\dots,L-1\}\\
				\prod_{r<L}\left[1-\Phi\left(\alpha_{jr_j}^o(\bm{s}_{i})\right) \right],&\text{ if }l_j=L
			\end{cases}	  
		\end{equation}
		for any arbitrary $(j,o,i)\in\{1,\dots,k\}\times\{1,\dots,O\}\times\{1,\dots,m\}$ and $l_j\in\{1,\ldots,L\}$.
		
		{\centering \textbf{Sequentially Updating $\bm{\alpha}_{jl_j}=\left(\bm{\alpha}_{jl_j}(\bm{s}_1)^{\T}, \dots, \bm{\alpha}_{jl_j}(\bm{s}_m)^{\T}\right)^{\T}\;\forall\;(j,l_j) $}\par}
		Under the circumstance when \texttt{include.space = TRUE}, \texttt{spatial.structure = "continuous"}, and \texttt{spatApprox = TRUE}, we can opt to sequentially update the $\bm{\alpha}_{jl_j}$'s instead:
		\par Under our NNGP prior $\tilde{\pi}\left(\bm{\alpha}_{jl_j}|\kappa,\rho\right)=N_{mO}\left(\bm{0},\tilde{F}(\rho)\otimes\kappa\right)$ for all $(j,l_j)$, for any arbitrary fixed $(j,l_j,i)\in\{1,\dots,k\}\times\{1,\dots,L-1\}\times\{1,\dots,m\}$,
		\begin{align}
			f(\bm{\alpha}_{jl_j}(\bm{s}_i)|\cdot)&\propto f(\bm{z}_{jl_j}(\bm{s}_i)|\bm{\alpha}_{jl_j}(\bm{s}_i))\times \tilde f(\bm{\alpha}_{jl_j}|\cdot)
			= f(\bm{z}_{jl_j}(\bm{s}_i)|\bm{\alpha}_{jl_j}(\bm{s}_i)) \prod_{r=1}^m {f}(\bm{\alpha}_{jl_j}(\bm{s}_r)|\bm{\alpha}_{jl_j,N(\bm{s}_r)},\kappa,\rho)\nonumber\\
			&\propto f(\bm{z}_{jl_j}(\bm{s}_i)|\bm{\alpha}_{jl_j}(\bm{s}_i))\times  {f}(\bm{\alpha}_{jl_j}(\bm{s}_i)|\bm{\alpha}_{jl_j,N(\bm{s}_i)},\kappa,\rho)\prod_{r:\bm{s}_i\in N(\bm{s}_r)}{f}(\bm{\alpha}_{jl_j}(\bm{s}_r)|\bm{\alpha}_{jl_j,N(\bm{s}_r)},\kappa,\rho) \nonumber\\
			&\propto \exp\left\lbrace -\frac{1}{2}\left(\bm{z}_{jl_j}(\bm{s}_i)-\bm{\alpha}_{jl_j}(\bm{s}_i)\right)^{\T}I_{O}\left(\bm{z}_{jl_j}(\bm{s}_i)-\bm{\alpha}_{jl_j}(\bm{s}_i)\right) \right\rbrace \times \label{postAlphaFixedLseq}\\ 
			&\quad\; \exp\left\lbrace -\frac{1}{2}\left(\bm{\alpha}_{jl_j}(\bm{s}_i)-B_{\bm{s}_i}\bm{\alpha}_{jl_j,N(\bm{s}_i)} \right)^{\T}F_{\bm{s}_i}^{-1} \left(\bm{\alpha}_{jl_j}(\bm{s}_i)-B_{\bm{s}_i}\bm{\alpha}_{jl_j,N(\bm{s}_i)}\right)\right\rbrace\times \nonumber\\
			&\quad\; \exp\left\lbrace -\frac{1}{2}\sum_{r:\bm{s}_i\in N(\bm{s}_r)}\left(\bm{\alpha}_{jl_j}(\bm{s}_r)-B_{\bm{s}_r}\bm{\alpha}_{jl_j,N(\bm{s}_r)} \right)^{\T}F_{\bm{s}_r}^{-1} \left(\bm{\alpha}_{jl_j}(\bm{s}_r)-B_{\bm{s}_r}\bm{\alpha}_{jl_j,N(\bm{s}_r)}\right)\right\rbrace \nonumber
		\end{align}
		\vspace{-5mm}
		\begin{align}
			&\sim N_O\left(V_{\bm{s}_i}\bm{\mu}_{jl_j,\bm{s}_i}, V_{\bm{s}_i} \right),\text{ where }V_{\bm{s}_i} = \left[I_O+F^{-1}_{\bm{s}_i} + \sum_{r:\bm{s}_i\in N(\bm{s}_r)}B_{\bm{s}_r,k_r(\bm{s}_i)}^{\T}F^{-1}_{\bm{s}_r}B_{\bm{s}_r,k_r(\bm{s}_i)} \right]^{-1}\text{ and} \label{seqAlphaIndPostDensFixedL} \\
			&\bm{\mu}_{jl_j,\bm{s}_i} = \bm{z}_{jl_j}(\bm{s}_i) + F^{-1}_{\bm{s}_i}B_{\bm{s}_i}\bm{\alpha}_{jl_j,N(\bm{s}_i)} + \sum_{r:\bm{s}_i\in N(\bm{s}_r)}B_{\bm{s}_r,k_r(\bm{s}_i)}^{\T}F^{-1}_{\bm{s}_r}\left\lbrace \bm{\alpha}_{jl_j}(\bm{s}_r)-\sum^{1\leq k\leq |N(\bm{s}_r)|}_{k\neq k_r(\bm{s}_i)}B_{\bm{s}_r,k}\bm{\alpha}_{jl_j,N(\bm{s}_r)[k]} \right\rbrace. \nonumber
		\end{align}
		For each $r$ such that  $\bm{s}_i\in N(\bm{s}_r) $, $k_r(\bm{s}_i)$ in the above refers to the positive integer less than or equal to $|N(\bm{s}_r)|$ such that $\bm{s}_i=N(\bm{s}_r)[k_r(\bm{s}_i)]$. We have written the $O\times (|N(\bm{s}_r)|\cdot O)$ matrix $B_{\bm{s}_r}$ as $(B_{\bm{s}_r,1},\ldots,B_{\bm{s}_r,|N(\bm{s}_r)|})$, where each sub-matrix is $O\times O$, and written the $\left(|N(\bm{s}_r)|\cdot O\right)\times 1$ vector $\bm{\alpha}_{jl_j,N(\bm{s}_r)}$ as 
		$\left(\bm{\alpha}_{jl_j,N(\bm{s}_r)[1]}^\T, \ldots,\bm{\alpha}_{jl_j,N(\bm{s}_r)[|N(\bm{s}_r)|]}^\T\right)^\T$ so that $B_{\bm{s}_r}\bm{\alpha}_{jl_j,N(\bm{s}_r)}=\sum_{k=1}^{|N(\bm{s}_r)|}B_{\bm{s}_r,k}\bm{\alpha}_{jl_j,N(\bm{s}_r)[k]}$. Note that we only need to consider $r=i+1,\dots,m$ for each $i<m$ in order to find out the set $\{r\in\{1,\dots,m\}:\bm{s}_i\in N(\bm{s}_r)\}$. Also note that $F_{\bm{s}_1}=\kappa_{O\times O}$, $N(\bm{s}_1)=\emptyset\Rightarrow\text{ we can let } F^{-1}_{\bm{s}_1}B_{\bm{s}_1}\bm{\alpha}_{jl_j,N(\bm{s}_1)}=\bm{0}_{O\times 1}$. 
		
		\item Sampling from the full conditional distribution of $\kappa_{O\times O}$:
		\begin{align}\label{postKappaFixedLCL}
			&f(\kappa|\cdot)\propto \prod_{j=1}^k\prod_{l_j=1}^{L-1}\pi(\bm{\alpha}_{jl_j}|\kappa,\rho)\times f_0(\kappa)
			\sim \mathcal{IW}\left(mk(L-1)+\nu , \sum_{j=1}^k\sum_{l_j=1}^{L-1}S_{\alpha^{jl_j}_{\rho}}+\Theta \right),\\ &\text{where }S_{\alpha^{jl_j}_{\rho}} = A_{jl_j} F(\rho)^{-1}A_{jl_j}^{\T}\text{ with }(A_{jl_j})_{O\times m}=(\bm{\alpha}_{jl_j}(\bm{s}_1),\dots,\bm{\alpha}_{jl_j}(\bm{s}_m))\:\forall\:(j,l_j).\nonumber
		\end{align}
		When $O=1$, \Cref{postKappaFixedLCL} reduces to $\mathcal{IG}\left(\frac{mk(L-1)+\nu}{2}, \frac{\sum_{j=1}^k\sum_{l_j=1}^{L-1}S_{\alpha^{jl_j}_{\rho}}+\Theta}{2}\right) $.\\
		When the NNGP prior rather than the full GP prior is adopted for the $\bm{\alpha}_{jl_j}$'s, we replace $\pi\left(\bm{\alpha}_{jl_j}|\kappa,\rho\right)$ and $F(\rho)$ in \Cref{postKappaFixedLCL} by $\tilde{\pi}\left(\bm{\alpha}_{jl_j}|\kappa,\rho\right)$ and $\tilde{F}(\rho)$ respectively.
		\item Sampling $\rho$ via a Metropolis step when \texttt{include.space = TRUE}:
		\begin{align}
			f(\rho|\cdot)\propto\prod_{j=1}^k\prod_{l_j=1}^{L-1}&\;\pi(\bm{\alpha}_{jl_j}|\kappa,\rho)\times f_0(\rho)\times\left| \frac{\partial}{\partial\Delta}h^{-1}(\Delta) \right| \label{postRhoFixedLCL}\\
			&\text{with }\bm{\alpha}_{jl_j}|\kappa,\rho\sim N_{mO}\left(\bm{0},  F(\rho)\otimes\kappa\right)\qquad\forall\;(j,l_j).\nonumber
		\end{align}
		When the NNGP prior rather than the full GP prior is adopted for the $\bm{\alpha}_{jl_j}$'s, we replace $\pi\left(\bm{\alpha}_{jl_j}|\kappa,\rho\right)$ and $F(\rho)$ in \Cref{postRhoFixedLCL} by $\tilde{\pi}\left(\bm{\alpha}_{jl_j}|\kappa,\rho\right)$ and $\tilde{F}(\rho)$ respectively.
		\par In \Cref{postRhoFixedLCL}, $f_0(\rho)$ refers to the uniform prior $\unif\left(a_{\rho}, b_{\rho}\right)$, where $a_{\rho}=0,b_{\rho}=1$ when \texttt{spatial.structure = `discrete'} and $a_{\rho},b_{\rho}\in\mathbb{R}^+$ when \texttt{spatial.structure = `continuous'}. $\Delta$ is a new parameter defined to be $h(\rho)=\ln\left(\frac{\rho-a_{\rho}}{b_{\rho}-\rho}\right)$. Hence $\rho=h^{-1}(\Delta)=\frac{b_{\rho}\exp(\Delta)+a_{\rho}}{1+\exp(\Delta)}$ and $\lvert\frac{\partial}{\partial\Delta}h^{-1}(\Delta)\rvert\propto\frac{\exp(\Delta)}{\left[1+\exp(\Delta)\right]^2}$. 
		\par At each MCMC iteration $s\in\mathbb{N}$, let the current parameter estimate for $\rho$ be $\rho^{(s)}$. We propose a new parameter value $\rho^*=h^{-1}(\Delta^*)$ with $\Delta^*$ sampled from a symmetric kernel $N\left(\Delta^{(s)},\delta\right)$, where $\Delta^{(s)}=h\left(\rho^{(s)}\right)$ and $\delta>0$ is a tuning parameter decided after the burn-in MCMC iterations based on the acceptance rate for $\rho^*$, and set
		\begin{align*}
			{\rho}^{(s+1)}=
			\begin{cases}
				{\rho}^*,&\text{ w.p. }\alpha\left({\rho}^{(s)},{\rho}^*\right)=\min\left\lbrace\frac{f({\rho}^*|\cdot)}{f({\rho}^{(s)}|\cdot)},1\right\rbrace \\ 
				{\rho}^{(s)},&\text{ w.p. }1-\alpha\left({\rho}^{(s)},{\rho}^*\right)
			\end{cases}.
		\end{align*}
		\item Same as Step 1 given at the start of \Cref{appenA}.
		\item Same as Step 2 given at the start of \Cref{appenA}.
		\item Same as Step 3 given at the start of \Cref{appenA}.
		\item Same as Step 4 given at the start of \Cref{appenA}.
		\item Same as Step 5 given at the start of \Cref{appenA}.
	\end{enumerate}

	\subsection{Full Gibbs Sampling Details for Our Model in \Cref{sect3}}\label{appenA3}
	\par When we adopt the spatial PSBP clustering mechanism and further integrate pertinent ideas of slice sampling \parencite{Walker2007} by introducing new parameters $L_j\in\mathbb{N}\backslash\{0\}$, $j\in\{1,\dots,k\}$ and $ u_j^o\left(\bm{s}_{i}\right)$, $(j,i,o)\in\{1,\dots,k\}\times\{1,\dots,m\}\times\{1,\dots,O\}$ so that the spatial prior becomes $\bm{\alpha}_{jl_j}\overset{\iid}{\sim}N_{mO}\left(\bm{0}, F(\rho)_{m\times m}\otimes \kappa_{O\times O}\right)$, $j\in\{1,\dots,k\}$, $l_j\in\{1,\dots,L_j-1\}$, the corresponding Gibbs sampler steps are detailed in the following. Gibbs sampling steps for $\theta_{jl_j}$ ($j\in\{1,\ldots,k\}$, $l_j\in\{1,\ldots,L_j\}$), $\delta_h$ ($h\in\{1,\dots,k\}$), $\kappa_{O\times O}$, and $\rho$ are almost the same as their counterparts in \Cref{appenA2}. We simply sample $L_j$ instead of $L$ many $\theta_{jl_j}$'s for each $j$ and replace $L$ by $L_j$'s in the three full conditional densities \Cref{postDeltaFixedL,postKappaFixedLCL,postRhoFixedLCL} to obtain \Cref{postDelta,postKappa,postRhoVaryLj}. We take note that only $\bm{\alpha}_{jl_j}$'s are ordered first by observation type and then spatially, i.e., the first $O$ components of each above $mO\times 1$ vector are from location point 1, the next $O$ components are from location point 2, and so on. All other concerned parameters, i.e., $\bm{u}_j$'s, $\bm{w}_{jl_j}$'s, $\bm{\xi}_j$'s, $\bm{\lambda}_j$'s, and $\bm{\sigma^2}$ when not all observations are from the \texttt{`binomial'} family (\Cref{appenJ}), are ordered first spatially and then by observation type, i.e., the first $m$ components of each above $Om\times 1$ vector are from observation type 1, the next $m$ components are from observation type 2, and so on. 
	
	\begin{enumerate}
		\item Sampling from the full conditional distributions of $u_j^o(\bm{s}_{i})$'s:
		\par For any arbitrary $(i,o,j)$, $f\left(u_j^o(\bm{s}_{i})|\cdot\right) \propto \mathbbm{1}_{\left\lbrace u_j^o(\bm{s}_{i})<w_{j\xi_j^o(\bm{s}_{i})}^o(\bm{s}_{i}) \right\rbrace}\sim\unif\left(0, w_{j\xi_j^o(\bm{s}_{i})}^o(\bm{s}_{i})\right) $.
		\item Updating the $L_j$'s and sampling from the full conditional distributions of $\xi_j^o(\bm{s}_{i})$'s:
		\par Fix any arbitrary $j\in\{1,\dots,k\}$.
		\begin{align}\label{LjNewFormula}
			\forall\;&(i,o),\text{ let }L_j^{i,o}=\text{ the smallest positive integer s.t. }
			\sum_{l_j=1}^{L_j^{i,o}}w_{jl_j}^o(\bm{s}_{i}) > 1-u_j^o(\bm{s}_{i})\nonumber\\
			&\text{ and let }L_j^{\text{new}}=\max\left\lbrace L_j^{i,o}:i=1,\dots,m,o=1,\dots,O\right\rbrace. 
		\end{align}
		After obtaining $\{L_j^{\text{new}}:j=1,\dots,k\}$ by \Cref{LjNewFormula}, we immediately update the parameter estimates objects $\bm{\alpha}_{jl_j}$'s (by only keeping the $mO\times 1$ vectors corresponding to $l_j\in\{1,2,\dots,L_j^{\text{new}}-1\}$ and discarding the ones corresponding to $l_j\in\{L_j^{\text{new}},L_j^{\text{new}}+1,\dots,L_j^{\text{old}}-1\}$) and $\bm{w}_{jl_j}$'s correspondingly. Note that 
		\begin{itemize}
			\item $\forall\;l_j\in\{L_j^{\text{new}}+1,L_j^{\text{new}}+2,\dots,L_j^{\text{old}}\}$, $w_{jl_j}^o\left(\bm{s}_{i}\right)^{\text{old}}<u_j^o\left(\bm{s}_{i}\right)$ for all $(i,o)$. Hence, $\{1,\dots,L_j^{\text{new}}\}$ comprises of all possible choices for $\xi_j^o\left(\bm{s}_{i}\right)\:\forall\:(i,o)$,
			\item $\forall\; j\in\{1,\dots,k\},1\leq L_j^{\text{new}}\leq L_j^{\text{old}}<\infty$, and
			\item $\forall\;(j,i,o)\in\{1,\dots,k\}\times\{1,\dots,m\}\times\{1,\dots,O\}$, $w_{jl_j}^o\left(\bm{s}_{i}\right)^{\text{old}}=w_{jl_j}^o\left(\bm{s}_{i}\right)^{\text{new}}\;\forall\;l_j\in\{1,\dots,L_j^{\text{new}}-1\}$ and $w_{jL_j^{\text{new}}}^o\left(\bm{s}_{i}\right)^{\text{old}}\leq w_{jL_j^{\text{new}}}^o\left(\bm{s}_{i}\right)^{\text{new}}$. Hence, $\xi_j^o\left(\bm{s}_{i}\right)^{\text{old}}$ is still in the support for $\xi_j^o\left(\bm{s}_{i}\right)^{\text{new}}$.
		\end{itemize}	
		Hence, we can update our parameter estimates for $\xi_j^o\left(\bm{s}_{i}\right)$'s via the following:   
		\begin{align}\label{postXi}
			\forall\; l_j\in\{1,\dots,L_j^{\text{new}}\},\;\forall\:(i,o),&\; \mathbb{P}\left(\xi_j^o(\bm{s}_{i})=l_j|\cdot\right) \propto \mathbbm{1}_{\left\lbrace w_{jl_j}^o(\bm{s}_{i})^{\text{new}}>u_j^o(\bm{s}_{i})\right\rbrace }\\ \times& \prod_{t=1}^T f\left(y_t^o(\bm{s}_{i})|\bm{\beta},\bm{\eta}_t,(\sigma^2)^o(\bm{s}_{i}),\xi_j^o(\bm{s}_{i})=l_j,\bm{\xi}_{-j}^o(\bm{s}_{i}),\bm{\theta}_{\bm{\xi}^o(\bm{s}_{i})}\right), \nonumber
		\end{align}	
		where $\forall\:(i,o,j,l_j)$ s.t. $w_{jl_j}^o(\bm{s}_{i})^{\text{new}}>u_j^o(\bm{s}_{i})$, $\forall\;t\in\{1,\dots,T\}$,
		\begin{align} &f\left(y_t^o(\bm{s}_{i})|\bm{\beta},\bm{\eta}_t,(\sigma^2)^o(\bm{s}_{i}),\xi_j^o(\bm{s}_{i})=l_j,\bm{\xi}_{-j}^o(\bm{s}_{i}),\bm{\theta}_{\bm{\xi}^o(\bm{s}_{i})}\right)\\
			=\;&(2\pi)^{-\frac{1}{2}}(\sigma^{-1})^o(\bm{s}_{i})\exp\left\lbrace -\frac{1}{2}(\sigma^{-2})^o(\bm{s}_{i})\left[y_t^o(\bm{s}_{i})-\bm{x}_t^o(\bm{s}_{i})^{\T}\bm{\beta}-\sum_{h\neq j}^{1\leq h\leq k}\theta_{h\xi_h^o(\bm{s}_{i})}\eta_{th}-\theta_{jl_j}\eta_{tj} \right]^2\right\rbrace. \nonumber		
		\end{align}		 
		\item 
		Sampling from the full conditional distributions of $\theta_{jl_j}$'s for $l_j\leq L_j$, as we have assumed a finite mixture model with $L_j$'s as the spatial cluster numbers in this MCMC iteration :
		\par Fix any arbitrary $j\in\{1,\dots,k\}$ and $l_j\in\{1,\dots,L_j\}$. Then 
		\begin{align}\label{postTheta}
			f(\theta_{jl_j}|\cdot )\propto f_0(\theta_{jl_j}|\tau_j) \times \prod_{(i,o):\xi_j^o(\bm{s}_{i})=l_j}\prod_{t=1}^T f\left(y_t^o(\bm{s}_{i})|\bm{\beta},\bm{\eta}_t,(\sigma^2)^o(\bm{s}_{i}),\bm{\xi}^o(\bm{s}_{i}),\bm{\theta}_{\bm{\xi}^o(\bm{s}_{i})}\right).
		\end{align}
		If $\nexists\; (i,o)$ s.t. $\xi_j^o(\bm{s}_{i})=l_j$, then $f(\theta_{jl_j}|\cdot )= f_0(\theta_{jl_j}|\tau_j)\sim N\left(0,\tau_j^{-1}\right)$.\\
		If $\exists\; (i,o)$ s.t. $\xi_j^o(\bm{s}_{i})=l_j$, then 
		\begin{align*}
			&f(\theta_{jl_j}|\cdot )\sim N\left(V_{\theta_{jl_j}}\mu_{\theta_{jl_j}}, V_{\theta_{jl_j}}\right) , \text{ where }V_{\theta_{jl_j}}=\left[\tau_j + \sum_{(i,o):\xi_j^o(\bm{s}_{i})=l_j}(\sigma^{-2})^o(\bm{s}_{i})\sum_{t=1}^T\eta_{tj}^2 \right]^{-1}\\
			&\text{and }\mu_{\theta_{jl_j}} = \sum_{(i,o):\xi_j^o(\bm{s}_{i})=l_j}(\sigma^{-2})^o(\bm{s}_{i})\sum_{t=1}^T\eta_{tj}\left[y_t^o(\bm{s}_{i})-\bm{x}_t^o(\bm{s}_{i})^{\T}\bm{\beta}-\sum_{h\neq j}^{1\leq h\leq k}\theta_{h\xi_h^o(\bm{s}_{i})}\eta_{th} \right]. 
		\end{align*}
		\item Sampling from the full conditional distributions of $\delta_{h}$'s:
		\par If a multiplicative gamma process shrinkage prior is placed on the atoms so that $\forall\:(j,l_j)$, $\theta_{jl_j}\overset{\text{ind}}{\sim}N\left(0, \tau_j^{-1}\right)$ with $\tau_j=\prod_{h=1}^j\delta_h,\;\delta_1\sim\Ga(a_1,1),\:\delta_h\sim\Ga(a_2,1)$ $\forall\:h\geq 2$, then for any arbitrary fixed $h\in\{1,\dots,k\}$,
		\begin{align}\label{postDelta}
			f(\delta_h|\cdot)&\propto \prod_{j=h}^k\prod_{l_j=1}^{L_j}f(\theta_{jl_j}|\tau_j)\times f_0(\delta_h)\\
			&\sim \Ga\left(a_h+\frac{\sum_{j=h}^k L_j}{2}, 1+\frac{\sum_{j=h}^k \left[\prod_{x=1,x\neq h}^j\delta_x\right]\sum_{l_j=1}^{L_j}\theta_{jl_j}^2}{2} \right).\nonumber 
		\end{align}
		If we consider the non-shrinkage prior $\theta_{jl_j}\overset{\text{ind}}{\sim}N\left(0, \tau_j^{-1}=\delta_j^{-1}\right)\:\forall\:(j,l_j)$ with \\$\delta_j\overset{\iid}{\sim}\Ga(a_1,a_2)\;\forall\;j$, then for any arbitrary fixed $j\in\{1,\dots,k\}$,
		\begin{align}\label{postDeltaNoGS}
			f(\delta_j|\cdot)\propto \prod_{l_j=1}^{L_j}f(\theta_{jl_j}|\delta_j)\times f_0(\delta_j)
			\sim \Ga\left(a_1+\frac{L_j}{2}, a_2+\frac{\sum_{l_j=1}^{L_j}\theta_{jl_j}^2}{2} \right). 
		\end{align}
		\item Sampling from the full conditional distributions of $\bm{\alpha}_{jl_j}$'s ($l_j\in\{1,\dots,L_j-1\}$ for each $j$) and calculating new weights $w_{jl_j}^o(\bm{s}_{i})$'s ($l_j\in\{1,\dots,L_j\}$ for each $j$) when $L_j>1$:
		\par If $L_j=1$, then we do not need to sample any $\bm{\alpha}_{jl_j}$'s for this $j$. The corresponding $w_{j1}^o(\bm{s}_{i})$'s have already been set to $1$ in the previous Gibbs sampler step for $\xi_j^o(\bm{s}_{i})$'s in this same MCMC iteration.
		\par For any arbitrary fixed $(j, l_j)$ where $j\in\{1,\dots,k\}$ and $l_j\in\{1,\dots,L_j-1\}$,
		\begin{align}
			&f\left(\bm{\alpha}_{jl_j}|\cdot\right)
			\propto \pi\left(\bm{\alpha}_{jl_j}|\kappa,\rho\right)\times\prod_{i=1}^m\prod_{o=1}^O \mathbbm{1}_{\left\lbrace l_j>\xi_j^o(\bm{s}_{i})\text{ or }  u_j^o(\bm{s}_{i})<w_{j\xi_j^o(\bm{s}_{i})}^o(\bm{s}_{i})\right\rbrace}\label{postAlphaVaryLj01}\\
			&\quad\quad\sim N_{mO}\left(\bm{0}, F(\rho)\otimes\kappa\right)\times\prod_{i=1}^m\prod_{o=1}^O \mathbbm{1}_{\left\lbrace l_j>\xi_j^o(\bm{s}_{i})\text{ or }  u_j^o(\bm{s}_{i})<w_{j\xi_j^o(\bm{s}_{i})}^o(\bm{s}_{i})\right\rbrace}.\nonumber\\
			&\quad\quad\sim N_{mO}\left(\bm{0}, F(\rho)\otimes\kappa\right)\times\prod_{i=1}^m\prod_{o=1}^O
			\begin{cases}
				1,&\text{ if }l_j>\xi_j^o(\bm{s}_{i})\\
				\mathbbm{1}_{\left\lbrace {\alpha}_{jl_j}^o(\bm{s}_{i})^{\text{new}}>\text{lowerBound}_{(j,l_j,i,o)} \right\rbrace},&\text{ if }l_j=\xi_j^o(\bm{s}_{i})\\
				\mathbbm{1}_{\left\lbrace {\alpha}_{jl_j}^o(\bm{s}_{i})^{\text{new}}<\text{upperBound}_{(j,l_j,i,o)}\right\rbrace},&\text{ if }l_j<\xi_j^o(\bm{s}_{i})\\
			\end{cases},\label{postAlphaVaryLj02}
		\end{align} 
		where $\forall\;(j,l_j,i,o)$,
		\begin{align}
			&\text{lowerBound}_{(j,l_j,i,o)}=\Phi^{-1}\left( \frac{u_j^o(\bm{s}_{i})\cdot\Phi\left({\alpha}_{jl_j}^o(\bm{s}_{i})^{\text{old}}\right) }{w_{j\xi_j^o(\bm{s}_{i})}^o(\bm{s}_{i})} \right) \text{ and }\\ 
			&\text{upperBound}_{(j,l_j,i,o)}=\Phi^{-1}\left(1 - \frac{u_j^o(\bm{s}_{i})\cdot\left[ 1-\Phi\left({\alpha}_{jl_j}^o(\bm{s}_{i})^{\text{old}}\right)\right]  }{w_{j\xi_j^o(\bm{s}_{i})}^o(\bm{s}_{i})} \right).
		\end{align}
		
		When the NNGP instead of the full GP prior is adopted for the $\bm{\alpha}_{jl_j}$'s, we replace $\pi\left(\bm{\alpha}_{jl_j}|\kappa,\rho\right)$ and $F(\rho)$ by $\tilde{\pi}\left(\bm{\alpha}_{jl_j}|\kappa,\rho\right)$ and $\tilde{F}(\rho)$ in the above equations.		
		\par After obtaining our new estimates for $\alpha_{jl_j}^o(\bm{s}_{i})$'s ($l_j\in\{1,\dots,L_j-1\}$), we calculate our new samples for the corresponding weights parameters $w_{jl_j}^o(\bm{s}_{i})$'s ($l_j\in\{1,\dots,L_j\}$) via 
		\begin{equation}\label{weightsFormulaVaryLj}
			w_{jl_j}^o(\bm{s}_{i})=
			\begin{cases}
				\Phi\left(\alpha_{jl_j}^o(\bm{s}_{i})\right)\prod_{r_j<l_j}\left[1-\Phi\left(\alpha_{jr_j}^o(\bm{s}_{i})\right) \right],&\text{ if }l_j\in\{1,\dots,L_j-1\}\\
				\prod_{r_j<L_j}\left[1-\Phi\left(\alpha_{jr_j}^o(\bm{s}_{i})\right) \right],&\text{ if }l_j=L_j
			\end{cases}	  
		\end{equation}
		for any arbitrary $(j,o,i)\in\{1,\dots,k\}\times\{1,\dots,O\}\times\{1,\dots,m\}$ and $l_j\in\{1,\dots,L_j\}$.
		\par The truncated multivariate normal posterior distributions of $\bm{\alpha}_{jl_j}$'s (\Cref{postAlphaVaryLj02}) can be directly sampled by multivariate rejection sampling via for instance the \texttt{rtmvnorm()} function in the R package \texttt{tmvtnorm}. One noteworthy problem with multivariate rejection sampling, in this case, is that it typically gets very inefficient when $m$ is large due to an excessively small acceptance rate associated with sampling from a high-dimensional truncated $mO$-variate normal distribution. This issue can be successfully addressed by integrating spatial scalability via assuming an NNGP prior and sequentially sampling $\bm{\alpha}_{jl_j}$'s instead. 
		
		{\centering \textbf{Sequentially Updating $\bm{\alpha}_{jl_j}=\left(\bm{\alpha}_{jl_j}(\bm{s}_1)^{\T}, \dots, \bm{\alpha}_{jl_j}(\bm{s}_m)^{\T}\right)^{\T}\;\forall\;(j,l_j) $}\par}    
		Under the circumstance when \texttt{include.space = TRUE}, \texttt{spatial.structure = "continuous"}, and \texttt{spatApprox = TRUE}, we can opt to sequentially update the $\bm{\alpha}_{jl_j}$'s instead:
		\par Under our NNGP prior $\tilde{\pi}\left(\bm{\alpha}_{jl_j}|\kappa,\rho\right)=N_{mO}\left(\bm{0}, \tilde{F}(\rho)\otimes\kappa\right)$ for all $(j,l_j)$, for any arbitrary fixed $(j,l_j,i)$ where $j\in\{1,\dots,k\}$, $l_j\in\{1,\dots,L_j-1\}$, and $i\in\{1,\dots,m\}$,
		\begin{align}\label{varyLjAlphaSequen}
			f&\left(\bm{\alpha}_{jl_j}(\bm{s}_i)|\cdot\right) \propto\tilde{f}\left(\bm{\alpha}_{jl_j}|\cdot\right) \times \prod_{o=1}^O \mathbbm{1}_{\left\lbrace l_j>\xi_j^o(\bm{s}_{i})\text{ or }  u_j^o(\bm{s}_{i})<w_{j\xi_j^o(\bm{s}_{i})}^o(\bm{s}_{i})\right\rbrace}\\
			&=
			\prod_{r=1}^m {f}(\bm{\alpha}_{jl_j}(\bm{s}_r)|\bm{\alpha}_{jl_j,N(\bm{s}_r)},\kappa,\rho)\times \prod_{o=1}^O \mathbbm{1}_{\left\lbrace l_j>\xi_j^o(\bm{s}_{i})\text{ or }  u_j^o(\bm{s}_{i})<w_{j\xi_j^o(\bm{s}_{i})}^o(\bm{s}_{i})\right\rbrace}\nonumber\\
			&\propto  \prod_{r:\;r=i\text{ or }\bm{s}_i\in N(\bm{s}_r)}{f}(\bm{\alpha}_{jl_j}(\bm{s}_r)|\bm{\alpha}_{jl_j,N(\bm{s}_r)},\kappa,\rho)\times \prod_{o=1}^O \mathbbm{1}_{\left\lbrace l_j>\xi_j^o(\bm{s}_{i})\text{ or }  u_j^o(\bm{s}_{i})<w_{j\xi_j^o(\bm{s}_{i})}^o(\bm{s}_{i})\right\rbrace} \nonumber\\
			&\propto \prod_{o=1}^O \mathbbm{1}_{\left\lbrace l_j>\xi_j^o(\bm{s}_{i})\text{ or }  u_j^o(\bm{s}_{i})<w_{j\xi_j^o(\bm{s}_{i})}^o(\bm{s}_{i})\right\rbrace}\times\label{seqAlphaPostDensVaryLj}\\ 
			&\quad\; \exp\left\lbrace -\frac{1}{2}\left(\bm{\alpha}_{jl_j}(\bm{s}_i)-B_{\bm{s}_i}\bm{\alpha}_{jl_j,N(\bm{s}_i)} \right)^{\T}F_{\bm{s}_i}^{-1} \left(\bm{\alpha}_{jl_j}(\bm{s}_i)-B_{\bm{s}_i}\bm{\alpha}_{jl_j,N(\bm{s}_i)}\right)\right\rbrace\times \nonumber\\
			&\quad\; \exp\left\lbrace -\frac{1}{2}\sum_{r:\bm{s}_i\in N(\bm{s}_r)}\left(\bm{\alpha}_{jl_j}(\bm{s}_r)-B_{\bm{s}_r}\bm{\alpha}_{jl_j,N(\bm{s}_r)} \right)^{\T}F_{\bm{s}_r}^{-1} \left(\bm{\alpha}_{jl_j}(\bm{s}_r)-B_{\bm{s}_r}\bm{\alpha}_{jl_j,N(\bm{s}_r)}\right)\right\rbrace.\nonumber \\
			&\sim N_O\left(V_{\bm{s}_i}\bm{\mu}_{jl_j,\bm{s}_i}, V_{\bm{s}_i} \right)\times\prod_{o=1}^O \mathbbm{1}_{\left\lbrace l_j>\xi_j^o(\bm{s}_{i})\text{ or }  u_j^o(\bm{s}_{i})<w_{j\xi_j^o(\bm{s}_{i})}^o(\bm{s}_{i})\right\rbrace},\label{seqAlphaIndPostDens1}\\
			&\text{ where }V_{\bm{s}_i} = \left[F^{-1}_{\bm{s}_i} + \sum_{r:\bm{s}_i\in N(\bm{s}_r)}B_{\bm{s}_r,k_r(\bm{s}_i)}^{\T}F^{-1}_{\bm{s}_r}B_{\bm{s}_r,k_r(\bm{s}_i)} \right]^{-1}\text{and } \nonumber \\
			&\bm{\mu}_{jl_j,\bm{s}_i} =  F^{-1}_{\bm{s}_i}B_{\bm{s}_i}\bm{\alpha}_{jl_j,N(\bm{s}_i)} + \sum_{r:\bm{s}_i\in N(\bm{s}_r)}B_{\bm{s}_r,k_r(\bm{s}_i)}^{\T}F^{-1}_{\bm{s}_r}\left\lbrace \bm{\alpha}_{jl_j}(\bm{s}_r)-\sum^{1\leq k\leq |N(\bm{s}_r)|}_{k\neq k_r(\bm{s}_i)}B_{\bm{s}_r,k}\bm{\alpha}_{jl_j,N(\bm{s}_r)[k]} \right\rbrace. \nonumber
		\end{align}
		For each $r$ such that  $\bm{s}_i\in N(\bm{s}_r) $, $k_r(\bm{s}_i)$ in the above refers to the positive integer less than or equal to $|N(\bm{s}_r)|$ such that $\bm{s}_i=N(\bm{s}_r)[k_r(\bm{s}_i)]$. We have written the $O\times (|N(\bm{s}_r)|\cdot O)$ matrix $B_{\bm{s}_r}$ as $(B_{\bm{s}_r,1},\ldots,B_{\bm{s}_r,|N(\bm{s}_r)|})$, where each sub-matrix is $O\times O$, and written the $\left(|N(\bm{s}_r)|\cdot O\right)\times 1$ vector $\bm{\alpha}_{jl_j,N(\bm{s}_r)}$ as $\left(\bm{\alpha}_{jl_j,N(\bm{s}_r)[1]}^\T, \ldots,\bm{\alpha}_{jl_j,N(\bm{s}_r)[|N(\bm{s}_r)|]}^\T\right)^\T$ so that $B_{\bm{s}_r}\bm{\alpha}_{jl_j,N(\bm{s}_r)}=\sum_{k=1}^{|N(\bm{s}_r)|}B_{\bm{s}_r,k}\bm{\alpha}_{jl_j,N(\bm{s}_r)[k]}$. Note that we only need to consider $r=i+1,\dots,m$ for each $i<m$ in order to find out the set $\{r\in\{1,\dots,m\}:\bm{s}_i\in N(\bm{s}_r)\}$. Also note that $F_{\bm{s}_1}=\kappa_{O\times O}$, $N(\bm{s}_1)=\emptyset\Rightarrow\text{ we can let } F^{-1}_{\bm{s}_1}B_{\bm{s}_1}\bm{\alpha}_{jl_j,N(\bm{s}_1)}=\bm{0}_{O\times 1}$. 
		\item Sampling from the full conditional distribution of $\kappa_{O\times O}$:
		\begin{align}\label{postKappa}
			&f(\kappa|\cdot)\propto \prod_{j=1}^k\prod_{l_j=1}^{L_j-1}\pi(\bm{\alpha}_{jl_j}|\kappa,\rho)\times f_0(\kappa)
			\sim \mathcal{IW}\left(m\sum_{j=1}^k(L_j-1)+\nu , \sum_{j=1}^k\sum_{l_j=1}^{L_j-1}S_{\alpha^{jl_j}_{\rho}}+\Theta \right),\\ &\text{where }S_{\alpha^{jl_j}_{\rho}} = A_{jl_j} F(\rho)^{-1}A_{jl_j}^{\T}\text{ with }(A_{jl_j})_{O\times m}=(\bm{\alpha}_{jl_j}(\bm{s}_1),\dots,\bm{\alpha}_{jl_j}(\bm{s}_m))\:\forall\:(j,l_j).\nonumber
		\end{align}
		When $O=1$, \Cref{postKappa} reduces to $\mathcal{IG}\left(\frac{m\sum_{j=1}^k(L_j-1)+\nu}{2}, \frac{\sum_{j=1}^k\sum_{l_j=1}^{L_j-1}S_{\alpha^{jl_j}_{\rho}}+\Theta}{2}\right) $.\\
		When the NNGP prior rather than the full GP prior is adopted for the $\bm{\alpha}_{jl_j}$'s, we replace $\pi\left(\bm{\alpha}_{jl_j}|\kappa,\rho\right)$ and $F(\rho)$ in \Cref{postKappa} by $\tilde{\pi}\left(\bm{\alpha}_{jl_j}|\kappa,\rho\right)$ and $\tilde{F}(\rho)$ respectively.
		\item Sampling $\rho$ via a Metropolis step when \texttt{include.space = TRUE}:
		\begin{align}
			f(\rho|\cdot)\propto\prod_{j=1}^k\prod_{l_j=1}^{L_j-1}&\;\pi(\bm{\alpha}_{jl_j}|\kappa,\rho)\times f_0(\rho)\times\left| \frac{\partial}{\partial\Delta}h^{-1}(\Delta) \right| \label{postRhoVaryLj}\\
			&\text{with }\bm{\alpha}_{jl_j}|\kappa,\rho\sim N_{mO}\left(\bm{0},  F(\rho)\otimes\kappa\right)\qquad\forall\;(j,l_j).\nonumber
		\end{align}
		When the NNGP prior rather than the full GP prior is adopted for the $\bm{\alpha}_{jl_j}$'s, we replace $\pi\left(\bm{\alpha}_{jl_j}|\kappa,\rho\right)$ and $F(\rho)$ in \Cref{postRhoVaryLj} by $\tilde{\pi}\left(\bm{\alpha}_{jl_j}|\kappa,\rho\right)$ and $\tilde{F}(\rho)$ respectively.
		\par In \Cref{postRhoVaryLj}, $f_0(\rho)$ refers to the uniform prior $\unif\left(a_{\rho}, b_{\rho}\right)$, where $a_{\rho}=0,b_{\rho}=1$ when \texttt{spatial.structure = `discrete'} and $a_{\rho},b_{\rho}\in\mathbb{R}^+$ when \texttt{spatial.structure = `continuous'}. $\Delta$ is a new parameter defined to be $h(\rho)=\ln\left(\frac{\rho-a_{\rho}}{b_{\rho}-\rho}\right)$. Hence $\rho=h^{-1}(\Delta)=\frac{b_{\rho}\exp(\Delta)+a_{\rho}}{1+\exp(\Delta)}$ and $\lvert\frac{\partial}{\partial\Delta}h^{-1}(\Delta)\rvert\propto\frac{\exp(\Delta)}{\left[1+\exp(\Delta)\right]^2}$. 
		\par At each MCMC iteration $s\in\mathbb{N}$, let the current parameter estimate for $\rho$ be $\rho^{(s)}$. We propose a new parameter value $\rho^*=h^{-1}(\Delta^*)$ with $\Delta^*$ sampled from a symmetric kernel $N\left(\Delta^{(s)},\delta\right)$, where $\Delta^{(s)}=h\left(\rho^{(s)}\right)$ and $\delta>0$ is a tuning parameter decided after the burn-in MCMC iterations based on the acceptance rate for $\rho^*$, and set
		\begin{align*}
			{\rho}^{(s+1)}=
			\begin{cases}
				{\rho}^*,&\text{ w.p. }\alpha\left({\rho}^{(s)},{\rho}^*\right)=\min\left\lbrace\frac{f({\rho}^*|\cdot)}{f({\rho}^{(s)}|\cdot)},1\right\rbrace \\ 
				{\rho}^{(s)},&\text{ w.p. }1-\alpha\left({\rho}^{(s)},{\rho}^*\right)
			\end{cases}.
		\end{align*}
		\item Same as Step 1 given at the start of \Cref{appenA}.
		\item Same as Step 2 given at the start of \Cref{appenA}.
		\item Same as Step 3 given at the start of \Cref{appenA}.
		\item Same as Step 4 given at the start of \Cref{appenA}.
		\item Same as Step 5 given at the start of \Cref{appenA}.
	\end{enumerate}

	\section{Temporal Posterior Sampling and Prediction Acceleration for Evenly Dispersed Time Points}\label{appenB}
	
	\par Throughout this section, we sometimes write $H(\psi)$ as $H$ to simplify notations.
	\subsection{Temporal Part Computational Burdens When $T$ is Large}\label{appenB1}
	\par Our Gibbs sampling steps with respect to temporal-related parameters $\bm{\eta}_t$'s, $\Upsilon_{k\times k}$, and $\psi$ given in \Cref{appenA} can get very slow when $T$ is large due to the primary reasons listed below, especially the last one. In each MCMC iteration in any of our Gibbs samplers (\Cref{appenA}),
	\begin{itemize}
		\item The full conditional distribution of $\Upsilon_{k\times k}$ (\Cref{postUpsilon}) concerns $H(\psi)^{-1}$, whose computation requires $\Ocal(T^3)$ flops via standard methods;
		\item To sample $\psi$ via a Metropolis step, we need to calculate $\log f(\psi|\cdot)  $ for both the current parameter estimate and the newly proposed value of $\psi$. As the full conditional density of $\psi$ is given by \Cref{postPsi}, we have to evaluate $H(\psi)^{-1}$ and $\log\left[ \determ\left(H(\psi) \right)\right]$,\footnote{Note that these two quantities suffice for the $\psi$ part since $\left[H(\psi)\otimes\Upsilon\right]^{-1}=H(\psi)^{-1}\otimes\Upsilon^{-1}$ and $\log\left[\determ\left(H(\psi)\otimes\Upsilon\right)\right] = k\cdot \log\left[\determ\left(H(\psi)\right)\right] + T\cdot \log\left[\determ\left(\Upsilon\right)\right]$.} which both require $\Ocal(T^3)$ flops under standard methods. 
		Let $\rooti(H)=\left[\Chol(H)\right]^{-1}$ be the upper triangular matrix with positive diagonal entries such that $H(\psi)^{-1}=\rooti(H)\left[\rooti(H)\right]^{\T}$. Then $\determ\left(H(\psi)\right)$ can also be conveniently calculated as the inverse of the square of the product of all diagonal entries in $\rooti(H)$. Adopting $\rooti(H)$ instead of $H(\psi)^{-1}$ and $\determ\left(H(\psi)\right)$ for this step is usually faster, but $\rooti(H)$ also has a computational complexity of $\Ocal(T^3)$ for a general $H(\psi)_{T\times T}$ without any special structures.
		\item When sampling $\bm{\eta}_t$'s from their full conditional distributions (\Cref{postEta}), we would need to loop over $t\in\{1,2,\dots,T\}$ and calculate $H^+_t=H_{t,(1:T)_{-t}}H^{-1}_{(1:T)_{-t},(1:T)_{-t}}, H^*_t=H_{t,t}-H^+_tH_{(1:T)_{-t},t}=H_{t,t}-H_{t,(1:T)_{-t}}H^{-1}_{(1:T)_{-t},(1:T)_{-t}}H_{(1:T)_{-t},t}$ for each $t$. This can be computationally intimidating when $T$ is large given that we need to invert $T$ many $(T-1)\times (T-1)$ principal sub-matrices of $H(\psi)_{T\times T}$, which results in an overall computational complexity of $\Ocal(T^4)$. Also, we would have to compute and store the entire $H(\psi)_{T\times T}$ matrix to extract out its aforementioned sub-matrices.
	\end{itemize}
	When making predictions at $q\in\mathbb{N}$ future time points $(T+1):(T+q)$, $H(\psi)^{-1}$ is also needed to obtain a predicted $\hat{\bm{\eta}}_{(T+1):(T+q)}$ given $\bm{\eta}_{1:T},\,\Upsilon,\,\psi$ (\Cref{appenG1}) in each MCMC iteration. 
	\par The above points do not exist as problems when temporal independence is assumed, i.e., when the temporal correlation structure matrix $H(\psi)_{T\times T}$ is fixed as the $T\times T$ identity matrix. In that case,
	\begin{itemize}
		\item When sampling $\Upsilon_{k\times k}$ from its full conditional distribution and when obtaining $\hat{\bm{\eta}}_{(T+1):(T+q)}$ given $\bm{\eta}_{1:T},\,\Upsilon,\,\psi$, we can directly take $H(\psi)^{-1}_{T\times T}$ as $I_{T\times T}$ throughout;
		\item We no longer need to sample $\psi$ anymore in any MCMC iteration;
		\item The step sampling $\bm{\eta}_t$'s can be greatly simplified and accelerated:
		\begin{align*}
			&H=H(\psi)=I_{T\times T}\\
			\Rightarrow &H^+_t=(0,0,\dots,0)_{1\times (T-1)}I_{(T-1)\times (T-1)}=(0,0,\dots,0)_{1\times (T-1)}\text{ and }
			\\ &H^*_t=1-(0,0,\dots,0)_{1\times (T-1)}(0,0,\dots,0)_{(T-1)\times 1}^{\T}=1-0=1\quad\forall\;t\in\{1,2,\dots,T\}.\\
			\text{Hence }&\mathbb{C}_{0\bm{\eta}_t}=\Upsilon_{k\times k} \text{ and } \mathbb{E}_{0\bm{\eta}_t}=(0,0,\dots\dots,0)^{\T}_{k\times 1}\;\forall\;t\\
			\Rightarrow & \;\mathbb{C}_{\bm{\eta}_t}=\left( \Lambda^{\T}\Xi^{-1}\Lambda + \Upsilon^{-1} \right)^{-1}\text{ and }\mathbb{\mu}_{\bm{\eta}_t}= \Lambda^{\T}\Xi^{-1}(\bm{y}_t-X_t\bm{\beta}) \;\forall\;t\text{ in \Cref{postEta}}.
		\end{align*}	
	\end{itemize}
	\par Under most circumstances, however, temporal correlation should indeed be taken into consideration. Fortunately, under a widely encountered special application scenario when all our time points are equally spaced, substantial computational acceleration is indeed achievable by exploiting the corresponding temporal correlation matrix's special structure.
	
	\subsection{Some Set-Up and Notations}\label{appenB2} 
	
	\par Throughout the rest of this section, we assume the scenario when temporal correlation is included and all time points are equispaced with the adjacent distances normalized to 1. Four temporal structures -- \texttt{ar1}, \texttt{exponential}, \texttt{sar1}, \texttt{sexponential}, which denote $\AR(1)$, the exponential process, and their seasonal counterparts (\Cref{appenA}), respectively, are considered. 
	{\begin{align*}
			\text{Let }\rho=
			\begin{cases}
				\psi\in(0,1), &\text{ if \texttt{temporal.structure = `ar1'} or \texttt{`sar1'}}\\
				e^{-\psi}\in(0,1), &\text{ if \texttt{temporal.structure = `exponential'} or \texttt{`sexponential'}}
			\end{cases}.
		\end{align*}
	}	
	Then for $T\geq 3$, the $T\times T$ temporal correlation structure matrix, which is ensured positive definite (\Cref{appenB3}), is 
	$$
	H=H(\psi)=
	\begin{pmatrix}
		1 & \rho & \rho^2 & \dots & \rho^{T-1} \\
		\rho & 1 & \rho & \dots & \rho^{T-2} \\
		\rho^2 & \rho & 1 & \dots & \rho^{T-3} \\
		\vdots & \vdots & \vdots & \ddots & \vdots \\
		\rho^{T-1} & \rho^{T-2} & \rho^{T-3} & \dots & 1 \\
	\end{pmatrix}
	$$
	if \texttt{temporal.structure = "ar1"} or \texttt{"exponential"},
	and $H=H(\psi, d)=$
	\begin{adjustwidth}{-1.5cm}{-0.5cm}
		\begin{center}
			\scalebox{1}{
				$\begin{blockarray}{ccccccccccccccc}
					& \bm{1} & & & \bm{d+1} & & & \bm{2d+1}  & & & & \bm{kd+1} & & & \\
					\begin{block}{c(ccccccccccccc)c}
						\bm{1}& 1 &  &  & \rho &  &  & \rho^2 &  & \dots & \dots & \rho^k & & &\\
						&  & 1 &  &  & \rho &  &  & \rho^2 & \dots & \dots &  & \ddots & &\\
						&  &  & \ddots &  &  & \ddots &  &  &  \ddots & & \vdots & & \rho^k & \bm{T-kd}\\
						\bm{d+1}& \rho &  &  & 1 &  &  & \rho &  &  & \rho^2 & \vdots &  & \vdots & \\
						&  & \rho &  &  & 1 &  &  & \rho &  &  & \rho^2 &  & \vdots & \\
						&  &  & \ddots &  &  & \ddots &  &  & \ddots &  &  & \ddots & & \\
						\bm{2d+1}& \rho^2 &  &  & \rho &  &  & 1 &  &  & \rho &  &  & \rho^2 & \bm{T-2d}\\
						&  &  \rho^2 &  &  & \rho &  &  & 1 &  &  & \rho &  & & \\ 
						& \vdots & \vdots & \ddots &  &  & \ddots &  &  & \ddots &  &  & \ddots & & \\
						& \vdots & \vdots &  & \ddots &  &  & \ddots &  &  & \ddots &  &  & \rho &\bm{T-d}\\
						\bm{kd+1}& \rho^k &  & \dots & \dots & \rho^2 &  &  & \rho &  &  & 1 &  & & \\
						&  & \ddots &  &  &  & \ddots &  &  & \ddots &  &  & \ddots &  & \\
						&  &  & \rho^k & \dots & \dots &  & \rho^2 &  &  & \rho &  &  & 1 & \bm{T} \\
					\end{block}	
					& & & \bm{T-kd} &  & & & \bm{T-2d} & & & \bm{T-d} & & & \bm{T} &
				\end{blockarray}$
			}
		\end{center}
	\end{adjustwidth}
	if \texttt{temporal.structure = "sar1"} or \texttt{"sexponential"}, where $d>1,\,d\in\mathbb{N}$ is an arbitrary temporal seasonal period and $k=\lfloor \frac{T-1}{d} \rfloor$.

	\subsection{Simple Sparse Closed-Form Representations for $H(\psi)^{-1}_{T\times T}$ and $\rooti(H)_{T\times T}$}\label{appenB3}
	\subsubsection{When \texttt{temporal.structure = `ar1'} or \texttt{`exponential'}}\label{appenB3a}
	Let $D_{T\times T}=\diag\left(1,1-\rho^2,1-\rho^2,\ldots,1-\rho^2\right)$ and 
	$$
	L_{T\times T}=
	\begin{pmatrix}
		1 & 0 & 0 & \dots & 0 \\
		\rho & 1 & 0 & \dots & 0 \\
		\rho^2 & \rho & 1 & \dots & 0 \\
		\vdots & \vdots & \vdots & \ddots & \vdots \\
		\rho^{T-1} & \rho^{T-2} & \rho^{T-3} & \dots & 1 
	\end{pmatrix}.
	\text{ Note that }
	L^{-1}=\begin{pmatrix}
		1 &  &  &  &  &  \\
		-\rho & 1&  &  &  &  \\
		& -\rho  & 1 &  &  &  \\
		&  & \ddots & \ddots &  &  \\
		&  &  & -\rho & 1 &  \\
		&  &  &  & -\rho & 1 
	\end{pmatrix}_{T\times T}.
	$$
	Then $H=LDL^{\T}=\Chol(H)^{\T}\Chol(H)$, where  
	\begin{align*}
		\Chol(H)=D^{\frac{1}{2}}L^{\T}
		&=
		\begin{pmatrix}
			1 & 0 & 0 & \dots & 0 \\
			0 & \sqrt{1-\rho^2} & 0 & \dots & 0 \\
			0 & 0 & \sqrt{1-\rho^2} & \dots & 0 \\
			\vdots & \vdots & \vdots & \ddots & \vdots \\
			0 & 0 & 0 & \dots & \sqrt{1-\rho^2}
		\end{pmatrix}
		\begin{pmatrix}
			1 & \rho & \rho^2 & \dots & \rho^{T-1} \\
			0 & 1 & \rho & \dots & \rho^{T-2} \\
			0 & 0 & 1 & \dots & \rho^{T-3} \\
			\vdots & \vdots & \vdots & \ddots & \vdots \\
			0 & 0 & 0 & \dots & 1 
		\end{pmatrix}\\
		&=\begin{pmatrix}
			1 & \rho & \rho^2 & \dots & \rho^{T-1} \\
			0 & \sqrt{1-\rho^2} & \rho \sqrt{1-\rho^2}& \dots & \rho^{T-2} \sqrt{1-\rho^2}\\
			0 & 0 & \sqrt{1-\rho^2} & \dots & \rho^{T-3} \sqrt{1-\rho^2} \\
			\vdots & \vdots & \vdots & \ddots & \vdots \\
			0 & 0 & 0 & \dots & \sqrt{1-\rho^2}
		\end{pmatrix}\text{ is upper triangular}.
		\\
		\text{Hence }\rooti(H)&=[\Chol(H)]^{-1}=(L^{-1})^{\T}D^{-\frac{1}{2}}\\
		&=
		\begin{pmatrix}
			1 & -\rho & 0 & 0 & \dots & 0 \\
			0 & 1& -\rho & 0 & \dots & 0 \\
			0 & 0 & 1 & -\rho & \dots & 0 \\
			\vdots & \vdots & \vdots & \ddots & \ddots & \vdots \\
			0 & 0 & 0 & \dots & 1 & -\rho \\
			0 & 0 & 0 &\dots  &0 & 1 
		\end{pmatrix}
		\begin{pmatrix}
			1 & 0 & 0 & \dots & 0 \\
			0 & \frac{1}{\sqrt{1-\rho^2}} & 0 & \dots & 0 \\
			0 & 0 & \frac{1}{\sqrt{1-\rho^2}} & \dots & 0 \\
			\vdots & \vdots & \vdots & \ddots & \vdots \\
			0 & 0 & 0 & \dots & \frac{1}{\sqrt{1-\rho^2}} 
		\end{pmatrix}\\
		&=\begin{pmatrix}
			1 & -\frac{\rho}{\sqrt{1-\rho^2}} & 0 & 0 & \dots & 0 \\
			0 & \frac{1}{\sqrt{1-\rho^2}}& -\frac{\rho}{\sqrt{1-\rho^2}} & 0 & \dots & 0 \\
			0 & 0 & \frac{1}{\sqrt{1-\rho^2}} & -\frac{\rho}{\sqrt{1-\rho^2}} & \dots & 0 \\
			\vdots & \vdots & \vdots & \ddots & \ddots & \vdots \\
			0 & 0 & 0 & \dots & \frac{1}{\sqrt{1-\rho^2}} & -\frac{\rho}{\sqrt{1-\rho^2}} \\
			0 & 0 & 0 &\dots  &0 & \frac{1}{\sqrt{1-\rho^2}} 
		\end{pmatrix}_{T\times T},\text{ and }\\
		H^{-1}&=(L^{-1})^{\T}D^{-1}L^{-1}=\rooti(H)\left[\rooti(H)\right]^{\T}\\
		&=\frac{1}{1-\rho^2}
		\begin{pmatrix}
			1 & -\rho & &  &  & & \\
			-\rho & 1+\rho^2 & -\rho &  &  &  &\\
			& -\rho & 1+\rho^2 & \ddots &  &  &\\
			&  & -\rho & \ddots & -\rho & & \\
			&  &  & \ddots & 1+\rho^2 & -\rho & \\
			&  &  &  & -\rho & 1+\rho^2 & -\rho \\
			&  &  &  & & -\rho & 1
		\end{pmatrix}_{T\times T}.
	\end{align*}
	
	\subsubsection{When \texttt{temporal.structure = `sar1'} or \texttt{`sexponential'}}\label{appenB3b}
	{\begin{align*}
			& \text{Let } L_{T\times T}=\\
			& \begin{blockarray}{cccccccccccccc}
				& \bm{1} & & & \bm{d+1} & & & \bm{2d+1}  & & & & \bm{kd+1} & &\\
				\begin{block}{c(ccccccccccccc)}
					\bm{1}& 1 &  &  &  &  &  &  &  &  &  &  & & \\
					&  & 1 &  &  &  &  &  &  & & &  & & \\
					&  &  & \ddots &  &  &  &  &  &   & &  & &  \\
					\bm{d+1}& \rho &  &  & 1 &  &  & &  &  & &  &  & \\
					&  & \rho &  &  & 1 &  &  &  &  &  &  &  &   \\
					&  &  & \ddots &  &  & \ddots &  &  &  &  &  &  &  \\
					\bm{2d+1}& \rho^2 &  &  & \rho &  &  & 1 &  &  &  &  &  &  \\
					&  &  \rho^2 &  &  & \rho &  &  & 1 &  &  &  &  &  \\ 
					& \vdots & \vdots & \ddots &  &  & \ddots &  &  & \ddots &  &  &  &  \\
					& \vdots & \vdots &  & \ddots &  &  & \ddots &  &  & \ddots &  &  &  \\
					\bm{kd+1}& \rho^k &  & \dots & \dots & \rho^2 &  &  & \rho &  &  & 1 &  &  \\
					&  & \ddots &  &  &  & \ddots &  &  & \ddots &  &  & \ddots &  \\
					&  &  & \rho^k &  & \dots & \dots & \rho^2 &  &  & \rho &  &  & 1 \\
				\end{block}	
				& & & \bm{T-kd} &  & & & \bm{T-2d} & & & \bm{T-d} & & & \bm{T}
			\end{blockarray}\; ,\\
			& D_{T\times T} = \begin{blockarray}{cccccccccc}
				& \bm{1} & & & \bm{d} & \bm{d+1} & & & & \bm{T} \\
				\begin{block}{c(ccccccccc)}
					\bm{1}  & 1 & & & & & & & & \\
					& & 1 & & & & & & &  \\
					& & & \ddots & & & & & & \\
					\bm{d}  & & & & 1 & & & & & \\
					\bm{d+1}& & & & & 1-\rho^2 & & & & \\
					& & & & & & 1-\rho^2 & & & \\
					& & & & & & & \ddots & & \\
					& & & & & & & & \ddots & \\
					\bm{T}  & & & & & & & & & 1-\rho^2 \\
				\end{block}
			\end{blockarray}\;.	\\
			& \text{Then } H=LDL^{\T}=\Chol(H)^{\T}\Chol(H),\text{ where }\Chol(H)=D^{\frac{1}{2}}L^{\T}\text{ is upper triangular}. \text{ Note that}\\ 
			& L^{-1} = 
			\begin{blockarray}{cccccccccccc}
				& \bm{1} & & & \bm{d+1} & & & & & & & \\
				\begin{block}{c(ccccccccccc)}
					\bm{1} & 1 & & & & & & & & & & \\
					& & 1 & & & & & & & & & \\
					& & & \ddots & & & & & & & & \\
					\bm{d+1} & -\rho & & & 1 & & & & & & & \\
					& & -\rho & & & 1 & & & & & & \\
					& & & \ddots & & & \ddots & & & & & \\
					\bm{2d+1}& & & & -\rho & & & \ddots & & & & \\
					& & & & & -\rho & & & 1 & & & \\
					& & & & & & \ddots & & & 1 & & \\
					& & & & & & & \ddots & & & \ddots & \\
					& & & & & & & & -\rho & & & 1\\
				\end{block}
				& & & & & & & & \bm{T-d} & & & \bm{T} 
			\end{blockarray}_{T\times T}\;.	 \\
			& \text{Hence } \rooti(H)=[\Chol(H)]^{-1}=(L^{-1})^{\T}D^{-\frac{1}{2}}=\\ 
			&	\begin{blockarray}{cccccccccccc}
				& \bm{1} & \bm{2} & & \bm{d} & \bm{d+1} & \bm{d+2} & & & & \bm{T} & \\
				\begin{block}{c(cccccccccc)c}
					\bm{1} & 1 & & & & \frac{-\rho}{\sqrt{1-\rho^2}} & & & & & & \\
					\bm{2}	& & 1 & & & & \frac{-\rho}{\sqrt{1-\rho^2}} & & & & & \\
					& & & \ddots & & & & \ddots & & & &\\
					\bm{d}& & & & 1 & & & & \ddots & & &\\
					\bm{d+1}& & & & & \frac{1}{\sqrt{1-\rho^2}}& & & & \ddots & &   \\
					& & & & & & \ddots & & & & \frac{-\rho}{\sqrt{1-\rho^2}} & \bm{T-d}\\
					& & & & & & & & \ddots & & &\\
					& & & & & & & &  & \ddots & &\\
					\bm{T}& & & & & & & & & &\frac{1}{\sqrt{1-\rho^2}} & \bm{T} \\
				\end{block}
			\end{blockarray}_{T\times T},
	\end{align*}}
	and $H^{-1}=(L^{-1})^{\T}D^{-1}L^{-1}=\rooti(H)\left[\rooti(H)\right] ^{\T}={\frac{1}{1-\rho^2}\times}$
	\begin{adjustwidth}{-1cm}{-0.5cm}
		\begin{center}
			\scalebox{0.98}{$
				\begin{blockarray}{ccccccccccccccc}
					& \bm{1} & \bm{2} & & \bm{d} & \bm{d+1} & \bm{d+2} & & & & & & & \bm{T} & \\
					\begin{block}{c(ccccccccccccc)c}
						\bm{1}  & 1 & & & & -\rho & & & & & & & & & \\
						\bm{2}  & & 1 & & & & -\rho & & & & & & & & \\
						& & & \ddots & & & & \ddots & & & & & & & \\
						\bm{d}  & & & & 1 & & & & -\rho & & & & & & \\
						\bm{d+1}& -\rho & & & & 1+\rho^2 & & & & -\rho & & & & & \\
						\bm{d+2}& & -\rho & & & & 1+\rho^2 & & & & -\rho & & & & \\
						& & & \ddots & & & & \ddots & & & & \ddots & & & \\
						& & & & -\rho & & & & \ddots & & & & \ddots & & \\
						& & & & & -\rho & & & & 1+\rho^2 & & & & -\rho & \bm{T-d} \\
						& & & & & & -\rho & & & & 1 & & & & \bm{{T-d+1}}\\
						& & & & & & & \ddots & & & & 1 & & & \\
						& & & & & & & & \ddots & & & & \ddots & & \\
						\bm{T}& & & & & & & & & -\rho & & & & 1 & \bm{T} \\
					\end{block}
					& & & & & & & & & \bm{T-d} & \bm{T-d+1} & & & \bm{T} & \\
				\end{blockarray}.$}
		\end{center}
	\end{adjustwidth}
	
	\subsubsection{Attained Acceleration Outcomes Pertaining to Sampling $\Upsilon_{k\times k}$ and $\psi$ \& Predicting at Future Time Points}\label{appenB3c}
	\begin{enumerate}
		\item We have seen in \Cref{appenB1} that $H(\psi)^{-1}$ and $\rooti(H)$ are the only temporal correlation quantities required in each MCMC iteration for the posterior sampling of $\Upsilon_{k\times k}$ and $\psi$ as well as obtaining a predicted $\hat{\bm{\eta}}_{(T+1):(T+q)}$ given $\bm{\eta}_{1:T},\,\Upsilon,\,\psi$ when making future-time predictions. Since both $T\times T$ matrices have simple sparse closed-form solutions (\Cref{appenB3a,appenB3b}), we can now directly specify these two matrices in each MCMC iteration instead of calculating them and thus save the corresponding $\Ocal(T^3)$ flops.
		\item We have shown in  \Cref{appenB3a,appenB3b} that the symmetric matrix $H=LDL^{\T}=\left[D^{\frac{1}{2}}L^{\T}\right]^{\T}D^{\frac{1}{2}}L^{\T}=\Chol(H)^{\T}\Chol(H)$, where  $\Chol(H) = D^{\frac{1}{2}}L^{\T}$ is upper triangular whose diagonal entries are all strictly positive. Hence, the temporal correlation structure matrix $H(\psi)_{T\times T}$ is guaranteed positive definite. 
	\end{enumerate}
	
	\subsection{Simple Sparse Closed-Form Representations for the $T$ Vectors $H_t^+$'s of Length $T-1$ \& Repetitive Closed-Form Solutions for the $T$ Scalars $H_t^*$'s}\label{appenB4}
	\par Throughout this subsection, we sometimes write the subscript $(1:T)_{-t}$ as $-t$ to simplify notations.
	
	\subsubsection{When \texttt{temporal.structure = `ar1'} or \texttt{`exponential'}}\label{appenB4a}
	\begin{center}
		\textbf{\color{black}{$(a)$ For $t\in\{1,T\}$}}
	\end{center}
	As deduced in \Cref{appenB3a}, for $t\in\{1,T\}$, $$H^{-1}_{(1:T)_{-t},(1:T)_{-t}}=\frac{1}{1-\rho^2}\begin{pmatrix}
		1 & -\rho & &  &  & & \\
		-\rho & 1+\rho^2 & -\rho &  &  &  &\\
		& -\rho & 1+\rho^2 & \ddots &  &  &\\
		&  & -\rho & \ddots & -\rho & & \\
		&  &  & \ddots & 1+\rho^2 & -\rho & \\
		&  &  &  & -\rho & 1+\rho^2 & -\rho \\
		&  &  &  & & -\rho & 1
	\end{pmatrix}_{(T-1)\times (T-1)}.
	$$
	\begin{align*}
		&\text{Hence, } H^+_1=(\rho, \rho^2,\dots,\rho^{T-1})H^{-1}_{(2:T),(2:T)}=\color{black}{(\rho,0,0,\dots,0,0)_{1\times(T-1)}}\color{black}\text{ and }\\
		&H^*_1 = H_{1,1}-H^+_1H_{(2:T),1} = 1-(\rho,0,\dots,0)_{1\times(T-1)}\begin{pmatrix}
			\rho\\
			\rho^2\\
			\vdots\\
			\rho^{T-1}
		\end{pmatrix}_{(T-1)\times 1} = \color{black}{1-\rho^2}\color{black}.\\
		&\text{Similarly, }H^+_T=(\rho^{T-1}, \rho^{T-2},\dots,\rho)H^{-1}_{(1:(T-1)),(1:(T-1))}=\color{black}{(0,0,\dots,0,0,\rho)_{1\times(T-1)}}\color{black}\text{ and }\\
		&H^*_T = H_{T,T}-H^+_TH_{(1:(T-1)),T} = 1-(0,\dots,0,\rho)_{1\times(T-1)}\begin{pmatrix}
			\rho^{T-1}\\
			\rho^{T-2}\\
			\vdots\\
			\rho
		\end{pmatrix}_{(T-1)\times 1} = \color{black}{1-\rho^2}\color{black}.
	\end{align*}
	\begin{center}
		\textbf{\color{black}{$(b)$ For $t\in\{2,3,\dots,T-1\}$}}
	\end{center}
	\par $H^*_t=\color{black}{\frac{1-\rho^2}{1+\rho^2}}$ and $H^+_t$ is the vector of length $(T-1)$ whose $(t-1)^{\supth}$, $t^{\supth}$ positions are $\frac{\rho}{1+\rho^2}$ and all other positions are $0$, i.e., $H^+_t=\color{black}{\left(0,\dots,0,\overset{\bm{t-1}}{\frac{\rho}{1+\rho^2}}, \overset{\bm{t}}{\frac{\rho}{1+\rho^2}}, 0,\dots,0\right)_{1\times(T-1)}}$.
	\begin{proof}
		\begin{adjustwidth}{-0.7cm}{}
			\begin{align*}
				&\text{For any arbitrary }t\in(2,3,\dots,T-1),\\
				H_{-t,-t}&=\begin{blockarray}{ccccccccc}
					& & & & & \bm{t-1} & \bm{t} &  &\\
					\begin{block}{c(cccccccc)}
						& 1 & \rho & \rho^2 & \dots & \rho^{t-2} & \rho^{t} & \dots &\rho^{T-1} \\
						& \rho & 1 & \rho & \dots & \rho^{t-3} & \rho^{t-1} & \dots &\rho^{T-2} \\
						& \rho^2 & \rho & 1 & \dots & \rho^{t-4} & \rho^{t-2} & \dots & \rho^{T-3}  \\
						& \vdots & \vdots & \vdots & \ddots & \vdots & \vdots &  & \vdots \\
						\bm{t-1} & \rho^{t-2} & \rho^{t-3} & \rho^{t-4} & \dots & 1 & \rho^2 & \dots & \rho^{T-t+1}  \\
						\bm{t} & \rho^{t} & \rho^{t-1} & \rho^{t-2 }& \dots & \rho^2 & 1 & \dots & \rho^{T-t-1} \\
						& \vdots & \vdots & \vdots &  & \vdots & \vdots & \ddots & \vdots  \\
						& \rho^{T-1} & \rho^{T-2} & \rho^{T-3} & \dots & \rho^{T-t+1} & \rho^{T-t-1} & \dots & 1 \\
					\end{block}	
				\end{blockarray}_{(T-1)\times(T-1)}.\\
				\text{Let }L_{-t} & = \begin{blockarray}{ccccccccc}
					& & & & & \bm{t-1} & \bm{t} &  &\\
					\begin{block}{c(cccccccc)}
						& 1 & 0 & 0 & \dots & 0 & 0 & \dots & 0 \\
						& \rho & 1 & 0 & \dots & 0 & 0 & \dots & 0 \\
						& \rho^2 & \rho & 1 & \dots & 0 & 0 & \dots & 0  \\
						& \vdots & \vdots & \vdots & \ddots & \vdots & \vdots &  & \vdots \\
						\bm{t-1} & \rho^{t-2} & \rho^{t-3} & \rho^{t-4} & \dots & 1 & 0 & \dots & 0  \\
						\bm{t} & \rho^{t} & \rho^{t-1} & \rho^{t-2 }& \dots & \rho^2 & 1 & \dots & 0\\
						& \vdots & \vdots & \vdots &  & \vdots & \vdots & \ddots & \vdots  \\
						& \rho^{T-1} & \rho^{T-2} & \rho^{T-3} & \dots & \rho^{T-t+1} & \rho^{T-t-1} & \dots & 1 \\
					\end{block}
				\end{blockarray}_{(T-1)\times(T-1)}\\
				\text{and }D_{-t}&=\begin{blockarray}{cccccccccc}
					& & & & & \bm{t-1} & {\bm{t}} & \bm{t+1} & &\\
					\begin{block}{c(ccccccccc)}
						& 1 & 0 & 0 & \dots & 0 & 0 & 0 & \dots & 0 \\
						& 0 & 1-\rho^2 & 0 & \dots & 0 & 0 & 0 & \dots & 0 \\
						& 0 & 0 & 1-\rho^2 & \dots & 0 & 0 & 0 & \dots & 0  \\
						& \vdots & \vdots & \vdots & \ddots & \vdots & \vdots & \vdots &  & \vdots \\
						\bm{t-1} & 0 & 0 & 0 & \dots & 1-\rho^2 & 0 & 0 & \dots & 0  \\
						{\bm{t}} & 0 & 0 & 0 & \dots & 0 & {1-\rho^4} & 0 &  \dots & 0\\
						\bm{t+1} & 0 & 0 & 0 & \dots & 0 & 0 & 1-\rho^2 & \dots & 0\\
						& \vdots & \vdots & \vdots &  & \vdots & \vdots & \vdots & \ddots & \vdots  \\
						& 0 & 0 & 0 & \dots & 0 & 0 & 0 & \dots & 1-\rho^2 \\
					\end{block}	
				\end{blockarray}_{(T-1)\times(T-1)}.\\
				\text{Then }& H_{-t,-t}=L_{-t}D_{-t}L_{-t}^{\T}. \text{ Note that}\\
				L_{-t}^{-1} &=\begin{blockarray}{cccccccccc}
					& & & & \bm{t-2} & {\bm{t-1}} & {\bm{t}} & & &\\
					\begin{block}{c(ccccccccc)}
						& 1 &  &  &  &  &  &  &  &  \\
						& -\rho & 1 &  &  &  &  &  &  &  \\
						&  & -\rho & \ddots &  &  &  &  &  &   \\
						&  &  & \ddots & 1 &  &  & &  &   \\
						{\bm{t-1}} &  &  &  & -\rho & 1 & &  &  & \\
						{\bm{t}} &  &  &  &  & {-\rho^2} & 1 &  &  &   \\
						\bm{t+1} &  &  &  &  &  & -\rho & \ddots & &\\
						&  &  &  &  &  &  & \ddots & 1 & \\
						&  &  &  &  &  &  &  & -\rho & 1 \\
					\end{block}	
				\end{blockarray}_{(T-1)\times(T-1)}.
			\end{align*}
		\end{adjustwidth}
		\begin{align*}
			&\text{Hence, the }(T-1)\times(T-1)\text{ matrix } H_{-t,-t}^{-1}=(L_{-t}^{-1})^{\T}D_{-t}^{-1}L_{-t}^{-1}\\
			&=\begin{blockarray}{cccccccccccc}
				& & & & & \bm{t-2} & {\bm{t-1}} & {\bm{t}} & \bm{t+1} & & &\\
				\begin{block}{c(ccccccccccc)}
					& \frac{1}{1-\rho^2} & \frac{-\rho}{1-\rho^2} &  &  &  &  &  &  &  & & \\
					& \frac{-\rho}{1-\rho^2} & \frac{1+\rho^2}{1-\rho^2} & \frac{-\rho}{1-\rho^2} &  &  &  &  &  & & &  \\
					&  & \frac{-\rho}{1-\rho^2} & \frac{1+\rho^2}{1-\rho^2} & \ddots &  &  &  &  & & &  \\
					&  &  &  \frac{-\rho}{1-\rho^2} & \ddots & \frac{-\rho}{1-\rho^2} & & &  & & &  \\
					\bm{t-2} &  &  &  & \ddots & \frac{1+\rho^2}{1-\rho^2} & \frac{-\rho}{1-\rho^2} & &  & & &\\
					{\bm{t-1}} &  &  &  &  & \frac{-\rho}{1-\rho^2} & {+} & {\frac{-\rho^2}{1-\rho^4}} &  & & &\\
					{\bm{t}} &  &  &  &  &  & {\frac{-\rho^2}{1-\rho^4}} & {*} & \frac{-\rho}{1-\rho^2} & & & \\
					\bm{t+1} &  &  &  &  &  &  & \frac{-\rho}{1-\rho^2} & \frac{1+\rho^2}{1-\rho^2} & \ddots &  & \\
					&  &  &  &  &  & & &  \frac{-\rho}{1-\rho^2} &  \ddots & \frac{-\rho}{1-\rho^2} & \\
					&  &  &  &  &  &  & & & \ddots  & \frac{1+\rho^2}{1-\rho^2} & \frac{-\rho}{1-\rho^2} \\
					&  &  &  &  &  &  & & &  & \frac{-\rho}{1-\rho^2} & \frac{1}{1-\rho^2} \\
				\end{block}	
			\end{blockarray}\;,\\
			&\text{where }{+}=
			\begin{cases}
				(1,-\rho^2)\begin{pmatrix}
					1 & 0\\
					0 & \frac{1}{1-\rho^4}
				\end{pmatrix}
				\begin{pmatrix}
					1 \\
					-\rho^2
				\end{pmatrix}=1+\frac{\rho^4}{1-\rho^4}=\frac{1}{1-\rho^4},&\color{black}\text{ if }t=2\color{black} \\              
				(1,-\rho^2)\begin{pmatrix}
					\frac{1}{1-\rho^2} & 0\\
					0 & \frac{1}{1-\rho^4}
				\end{pmatrix}
				\begin{pmatrix}
					1 \\
					-\rho^2
				\end{pmatrix}=\frac{1}{1-\rho^2}+\frac{\rho^4}{1-\rho^4}=\frac{1+\rho^2+\rho^4}{1-\rho^4},&\color{black}\text{ if }t\in\{3,\dots,T-1\}\color{black}\\
			\end{cases},\\
			&\text{and }{*}=
			\begin{cases}
				(1,-\rho)\begin{pmatrix}
					\frac{1}{1-\rho^4} & 0\\
					0 & \frac{1}{1-\rho^2}
				\end{pmatrix}
				\begin{pmatrix}
					1 \\
					-\rho
				\end{pmatrix}=\frac{1}{1-\rho^4}+\frac{\rho^2}{1-\rho^2}=\frac{1+\rho^2+\rho^4}{1-\rho^4},&\color{black}\text{ if }t\in\{2,\dots,T-2\}\color{black} \\  
				1\cdot\frac{1}{1-\rho^4}\cdot 1 = \frac{1}{1-\rho^4},&\color{black}\text{ if }t=T-1\color{black}\\
			\end{cases}.\\
			&\text{Since for }t\in\{2,3,\dots,T-1\}, H_{t,(1:T)_{-t}}=\left(\rho^{t-1},\dots,\rho^2,\overset{\bm{t-1}}{\rho}, \overset{\bm{t}}{\rho}, \rho^2,\dots,\rho^{T-t}\right)_{1\times(T-1)},\\
			&\text{we have }H^+_t=H_{t,(1:T)_{-t}}H^{-1}_{(1:T)_{-t},(1:T)_{-t}}=\color{black}{\left(0,\dots,0,\overset{\bm{t-1}}{\frac{\rho}{1+\rho^2}}, \overset{\bm{t}}{\frac{\rho}{1+\rho^2}}, 0,\dots,0\right)_{1\times(T-1)}}\color{black}\text{ and }\\
			&H^*_t = H_{t,t}-H^+_tH_{(1:T)_{-t},t}=1-2*\frac{\rho^2}{1+\rho^2}=\color{black}{\frac{1-\rho^2}{1+\rho^2}}\color{black}.
		\end{align*}
	\end{proof}
	\textbf{Notes:} $\forall\; t\in\{2,3,\dots,T-1\}, H_t^+[t]=H_t^+[t-1]=\frac{\rho}{1+\rho^2}$ since
	\begin{itemize}
		\item
		$H_{T-1}^+[T-1]=(\rho,\rho)\cdot\left(\frac{-\rho^2}{1-\rho^4},\frac{1}{1-\rho^4}\right) =\rho\cdot\frac{1-\rho^2}{1-\rho^4}=\frac{\rho}{1+\rho^2}$ \color{black}for $t=T-1$ \color{black}and \\
		$H_t^+[t]=(\rho,\rho,\rho^2)\cdot\left(\frac{-\rho^2}{1-\rho^4},\frac{1}{1-\rho^4}+\frac{\rho^2}{1-\rho^2}, \frac{-\rho}{1-\rho^2}\right)=\rho\times\left(\frac{-\rho^2+1}{1-\rho^4} \right) + (\rho,\rho^2)\cdot \left(\frac{\rho^2}{1-\rho^2}, \frac{-\rho}{1-\rho^2} \right) =\frac{\rho}{1+\rho^2}\color{black}\;\forall\;t\in\{2,\dots,T-2\}$;
		\item $H_2^+[2-1]=(\rho,\rho)\cdot\left(\frac{1}{1-\rho^4},\frac{-\rho^2}{1-\rho^4}\right)=\rho\cdot\left(\frac{1-\rho^2}{1-\rho^4} \right) =\frac{\rho}{1+\rho^2}$ \color{black}for $t=2$ \color{black} and\\$H_t^+[t-1]=(\rho^2,\rho,\rho)\cdot\left(\frac{-\rho}{1-\rho^2},\frac{1}{1-\rho^2}+\frac{\rho^4}{1-\rho^4}, \frac{-\rho^2}{1-\rho^4}\right)=\rho\times\left(\frac{-\rho^2+1}{1-\rho^2}+\frac{\rho^2(\rho^2-1)}{1-\rho^4}\right) =\rho\times\left(1-\frac{\rho^2}{1+\rho^2} \right)=\frac{\rho}{1+\rho^2}\;\color{black}\forall\; t\in\{3,\dots,T-1\}$.
	\end{itemize}
	
	\subsubsection{When \texttt{temporal.structure = `sar1'} or \texttt{`sexponential'}}\label{appenB4b}
	\par We have 
	\begin{equation}\label{HplusSeason}
		H^{+}_{t}=
		\begin{cases}
			\color{black}{\left(0,\dots,0,\overset{\bm{t+d-1}}{\rho},0,\dots \dots,0\right)}\color{black}, & \text{if }t\in\{1,2,\dots, d\} \\
			\color{black}{\left(0,\dots\dots,0,\overset{\bm{t-d}}{\frac{\rho}{1+\rho^2}},0,\dots,0,\overset{\bm{t+d-1}}{\frac{\rho}{1+\rho^2}}, 0,\dots \dots,0\right)}\color{black}, & \text{if }t\in\{d+1,d+2,\dots\dots, T-d\} \\
			\color{black}{\left(0,\dots\dots, 0,\overset{\bm{t-d}}{\rho},0,\dots,0\right)}\color{black}, & \text{if }t\in\{T-d+1,\dots, T-1,T\}
		\end{cases}
	\end{equation}
	$\Rightarrow$
	\begin{equation}\label{HstarSeason}
		H^*_t=
		\begin{cases}
			\color{black}{1-\rho^2}\color{black}\color{black}, & \text{if }t\in\{1,2,\dots,d,T-d+1,\dots, T-1,T\}\\
			1-2\cdot\rho\cdot\frac{\rho}{1+\rho^2}=\color{black}{\frac{1-\rho^2}{1+\rho^2}}\color{black}, & \text{if }t\in\{d+1,d+2,\dots\dots,T-d\}
		\end{cases}
	\end{equation}	
	since the $t^{\supth}$ position in the $t^{\supth}$ column of $H_{T\times T}$ is $1$ and thus
	\begin{itemize}
		\item The $(t+d-1)^{\supth}$ position in the ${(T-1)\times 1}$ vector $H_{(1:T)_{-t},t}$ is $\rho$ for $t\in\{1,2,\dots, d\}$;
		\item The $(t-d)^{\supth}$ and $(t+d-1)^{\supth}$ positions in the ${(T-1)\times 1}$ vector $H_{(1:T)_{-t},t}$ are both ${\rho}$ for $t\in\{d+1,d+2,\dots\dots, T-d\}$;
		\item The $(t-d)^{\supth}$ position in the ${(T-1)\times 1}$ vector $H_{(1:T)_{-t},t}$ is $\rho$  for $t\in\{T-d+1,\dots, T-1,T\}$.
	\end{itemize}
	\vspace{2mm}
	\par We leave the proof of \Cref{HplusSeason} to \Cref{appenB5}.
	
	\subsubsection{Attained Acceleration Outcomes Pertaining to Sampling \boldmath{$\eta$}$_t$'s}\label{appenB4c}
	\par As shown by \Cref{appenB4a,appenB4b}, we have simple sparse closed-form representations for the $T$ vectors $H^+_t=H_{t,{-t}}H^{-1}_{{-t},{-t}}$ of length $T-1$ and repetitive closed-form solutions for the $T$ scalars $H^*_t=H_{t,t}-H^+_tH_{{-t},t}=H_{t,t}-H_{t,{-t}}H^{-1}_{{-t},{-t}}H_{{-t},t},\,t\in\{1,\dots,T\}$. Hence, in each MCMC iteration,
	\begin{enumerate}
		\item We take $d=1$ when there is no temporal seasonality. Then
		\begin{align*}
			H_t^* &= 
			\begin{cases}
				1-\rho^2 & \forall\;t\in\{1,2,\dots,d,T-d+1,\dots, T-1,T\}\\
				\frac{1-\rho^2}{1+\rho^2} & \forall\;t\in\{d+1,d+2,\dots\dots,T-d\}
			\end{cases}\text{ and thus }\\
			\mathbb{C}_{0\bm{\eta}_t} = H_t^* \otimes \Upsilon_{k\times k}&= 
			\begin{cases}
				(1-\rho^2)\cdot\Upsilon & \forall\;t\in\{1,2,\dots,d,T-d+1,\dots, T-1,T\}\\
				\frac{1-\rho^2}{1+\rho^2}\cdot\Upsilon & \forall\;t\in\{d+1,d+2,\dots\dots,T-d\}
			\end{cases}.
		\end{align*} 
		Hence, the values of these quantities can be directly specified rather than calculated. Furthermore, we can opt to first specify $H^*_t=\frac{1-\rho^2}{1+\rho^2}$ and $\mathbb{C}_{0\bm{\eta}_t}=\frac{1-\rho^2}{1+\rho^2}\cdot\Upsilon\;\forall\;t\in\{1,\dots,T\}$ outside of the loop over $t$ and then only modify these two values inside the loop over $t$ for $t\in\{1,2,\dots,d,T-d+1,\dots, T-1,T\}$. 
		\item For each $t\in\{1,\dots,T\}$ inside the loop over $t$, we can specify $H^+_t$ by modifying one or two entries on the zero vector of length $T-1$ instead of calculating $H^+_t$ from $H^{-1}_{(1:T)_{-t},(1:T)_{-t}}$.
		\item We no longer need to compute and store $H(\psi)_{T\times T}$ itself, since $H(\psi)$ is only used when sampling $\bm{\eta}_t$'s in the entire Gibbs sampler, where we extract out its sub-matrices $H_{-t,-t}$'s, $H_{t,-t}$'s $H_{-t,t}$'s, $H_{t,t}$'s to calculate $H_t^+$'s and $H_t^*$'s. 
	\end{enumerate}

	\subsection{Proof of \Cref{HplusSeason}}\label{appenB5}
	Let $k=\lfloor \frac{T-1}{d} \rfloor$, where $d\in\mathbb{N},\,d>1$ is the temporal season period.
	\begin{center}
		\textbf{\color{black}{$(a)$ For $t\in\{1,T\}$}}
	\end{center}
	As deduced in \Cref{appenB3b}, for $t\in\{1,T\}$, the $(T-1)\times (T-1)$ matrix $	H^{-1}_{-t,-t}={\frac{1}{1-\rho^2}\times}$ 
	\begin{adjustwidth}{-1.8cm}{-0.5cm}
		\begin{center}
			\scalebox{0.95}{$
				\begin{blockarray}{ccccccccccccccc}
					& \bm{1} & & & \bm{d} & \bm{d+1} & \bm{d+2} & & & & & & & \bm{T-1} & \\
					\begin{block}{c(ccccccccccccc)c}
						\bm{1}  & 1 & & & & -\rho & & & & & & & & & \\
						& & 1 & & & & -\rho & & & & & & & & \\
						& & & \ddots & & & & \ddots & & & & & & & \\
						\bm{d}  & & & & 1 & & & & -\rho & & & & & & \\
						\bm{d+1}& -\rho & & & & 1+\rho^2 & & & & -\rho & & & & & \\
						\bm{d+2}& & -\rho & & & & 1+\rho^2 & & & & -\rho & & & & \\
						& & & \ddots & & & & \ddots & & & & \ddots & & & \\
						& & & & -\rho & & & & \ddots & & & & \ddots & & \\
						& & & & & -\rho & & & & 1+\rho^2 & & & & -\rho & \bm{(T-1)-d} \\
						& & & & & & -\rho & & & & 1 & & & & \bm{T-d}\\
						& & & & & & & \ddots & & & & 1 & & & \\
						& & & & & & & & \ddots & & & & \ddots & & \\
						\bm{T-1}& & & & & & & & & -\rho & & & & 1 & \bm{T-1} \\
					\end{block}
					& & & & & & & & & \bm{(T-1)-d} & \bm{T-d} & & & \bm{T-1} & \\
				\end{blockarray}$}.
		\end{center}
	\end{adjustwidth}
	\begin{align*}
		&\text{Hence, } H^+_1=\left(0,\dots,0,\overset{\bm{d}}{\rho^1},0,\dots,0,\overset{\bm{2\cdot d}}{\rho^2}, 0,\dots \dots,0, \overset{\bm{k\cdot d}}{\rho^k},0,\dots,0 \right)_{1\times(T-1)}H^{-1}_{(2:T),(2:T)}\\
		&=\color{black}{\left(0,\dots,0,\overset{\bm{d}}{\rho},0,\dots \dots,0\right)_{1\times(T-1)}}\color{black}.\text{ Similarly, }\\
		&H^+_T=\left(0,\dots,0,\overset{\bm{T-k\cdot d}}{\rho^k},0,\dots\dots,0,\overset{\bm{T-2\cdot d}}{\rho^2}, 0,\dots, 0, \overset{\bm{T- d}}{\rho^1},0,\dots,0 \right)_{1\times(T-1)}H^{-1}_{(1:(T-1)),(1:(T-1))}\\
		&=\color{black}{\left(0,\dots \dots,0,\overset{\bm{T-d}}{\rho},0,\dots,0\right)_{1\times(T-1)}}\color{black}.
	\end{align*}
	\begin{center}
		\textbf{\color{black}{$(b)$ For $t\in\{2,\dots,T-1\}$}}
	\end{center}
	\par Denote $L_{-t}$ as the $(T-1)\times(T-1)$ matrix whose upper-triangular entries are all $0$ and lower-triangular (including the main diagonal) entries all equal their counterparts in $H_{-t,-t}$.
	\begin{center}
		\color{black}{$(i)$\textbf{ If} $t\in\{2,\dots,d\}$,}
	\end{center}
	\color{black}
	then each one of the $d$ sets $\{1,d,2d,\dots\dots\},\,\{2,d+1,2d+1,\dots\dots\},\dots,\,\{\color{black}{t-1}\color{black},\color{black}{t+d-2}\color{black},t+2d-2,\dots\dots\}$, $\{\color{black}{t}\color{black},\color{black}t+d\color{black},t+2d,\dots\dots\},\,\{t+1,t+1+d,t+1+2d,\dots\dots\},\dots,\,\{d-1,2d-1,3d-1,\dots\dots\}$ and $\{{t+d-1},t+2d-1,t+3d-1,\dots\dots\}$ contains time points \textbf{in $H_{-t,-t}$ (after deducting the original $t^{\supth}$ time point)} with a same temporal season period component.
	\par Hence, $L_{-t}^{-1}=$
	\begin{adjustwidth}{-2cm}{-1.2cm}
		\begin{center}
			\scalebox{0.75}{
				$\begin{blockarray}{cccccccccccccccccccccc}
					& \bm{1} & \bm{2} &  & {\bm{t-1}} & {\bm{t}} & \bm{t+1} &  & \bm{d} &  &  &  & \bm{T-1-d} &  &  &  & & & & & \bm{T-1} &  \\
					\begin{block}{c(cccccccccccccccccccc)c}
						\bm{1}    & 1 &  &  &  &  &  &  &  &  &  &  &  &  &  &  &  &  &  &  &  &  \\
						\bm{2}    &  & 1 &  &  &  &  &  &  &  &  &  &  &  &  &  &  &  &  &  &  &  \\
						&  &  & \ddots &  &  &  &  &  &  &  &  &  &  &  &  &  &  &  &  &  &  \\
						\bm{t-1}  &  &  &  & 1 &  &  &  &  &  &  &  &  &  &  &  &  &  &  &  &  &  \\
						\bm{t}    &  &  &  &  & 1 &  &  &  &  &  &  &  &  &  &  &  &  &  &  &  &  \\
						\bm{t+1}  &  &  &  &  &  & 1 &  &  &  &  &  &  &  &  &  &  &  &  &  &  &  \\
						&  &  &  &  &  &  & \ddots &  &  &  &  &  &  &  &  &  &  &  &  &  &  \\
						\bm{d}    & -\rho &  &  &  &  &  &  & 1 &  &  &  &  &  &  &  &  &  &  &  &  &  \\
						\bm{d+1}  &  & -\rho &  &  &  &  &  &  & \ddots &  &  &  &  &  &  &  &  &  &  &  &  \\
						&  &  & \ddots &  &  &  &  &  &  & \ddots &  &  &  &  &  &  &  &  &  &  &  \\
						{\bm{t+d-2}}&  &  &  & -\rho &  &  &  &  &  &  & \ddots &  &  &  &  &  &  &  &  &  &  \\
						&  &  &  &  &  &  &  &  &  &  &  & 1 &  &  &  &  &  &  &  &  & \bm{T-1-d} \\
						{\bm{t+d}}  &  &  &  &  & -\rho &  &  &  &  &  &  &  &  &  &  &  &  &  &  &  &  \\
						\bm{t+1+d}&  &  &  &  &  & -\rho &  &  &  &  &  &  &  &  &  &  &  &  &  &  &  \\
						&  &  &  &  &  &  & \ddots &  &  &  &  &  &  &  &  &  &  &  &  &  &  \\
						\bm{2d}   &  &  &  &  &  &  &  & -\rho &  &  &  &  &  &  &  & \ddots &  &  &  &  &  \\
						&  &  &  &  &  &  &  &  &  \ddots &  &  &  &  &  &  &  &  &  &  &  &  \\
						&  &  &  &  &  &  &  &  &  & \ddots &  &  &  &  &  &  &  &  &  &  &  \\
						&  &  &  &  &  &  &  &  &  &  & \ddots &  &  &  &  &  &  &  &  &  &  \\
						\bm{T-1}  &  &  &  &  &  &  &  &  &  &  &  & -\rho  &  &  &  &  &  &  &  & 1 & \bm{T-1} \\
					\end{block}
					&  &  &  &  &  &  &  &  &  &  &  & \bm{T-1-d} &  &  &  &  &  &  &  & \bm{T-1} &  \\
				\end{blockarray}.
				$    		
			}
		\end{center}  
	\end{adjustwidth}
	Let $D_{-t}$ be the $(T-1)\times (T-1)$ diagonal matrix  
	$$\diag\left(1,1,\dots,\overset{{\bm{d-1}}}{1},\overset{{\bm{d}}}{1-\rho^2},1-\rho^2,\dots,1-\rho^2,\overset{{\bm{t+d-1}}}{{1}}, 1-\rho^2,\dots\dots, 1-\rho^2 \right).$$
	Then $H_{-t,-t}=L_{-t}D_{-t}L_{-t}^{\T}$ and hence $H_{-t,-t}^{-1}=\left(L_{-t}^{\T}\right)^{-1}D_{-t}^{-1}L_{-t}^{-1}={\frac{1}{1-\rho^2}\times}$ 
	\begin{adjustwidth}{-2cm}{-1.5cm}
		\begin{center}
			\scalebox{0.6}{
				$\begin{blockarray}{cccccccccccccccccccccccc}
					& \bm{1} &  & \color{black}{\bm{t-1}} & \color{black}{\bm{t}} &  & \color{black}{\bm{d-1}} & \color{black}{\bm{d}} &  & \color{black}{\bm{t+d-2}} & {\bm{t+d-1}} & \color{black}{\bm{t+d}} &  & \bm{2d-1} &  &  & {\bm{T-1-d}} & {\bm{T-d}} &  &  &  &  & \bm{T-1} &  \\
					\begin{block}{c(cccccccccccccccccccccc)c}
						\bm{1}     & 1 &  &  &  &  &  & -\rho &  &  &  &  &  &  &  &  &  &  &  &  &  &  &  &  \\
						&  & \ddots &  &  &  &  &  & \ddots &  &  &  &  &  &  &  &  &  &  &  &  &  &  &  \\
						\color{black}{\bm{t-1}}  &  &  & 1 &  &  &  &  &  & -\rho &  &  &  &  &  &  &  &  &  &  &  &  &  &  \\
						\color{black}{\bm{t}}     &  &  &  & 1 &  &  &  &  &  & 0 & -\rho &  &  &  &  &  &  &  &  &  &  &  &  \\
						&  &  &  &  & \ddots &  &  &  &  &  &  & \ddots &  &  &  &  &  &  &  &  &  &  &  \\
						\color{black}{\bm{d-1}}   &  &  &  &  &  & 1 &  &  &  &  &  &  & -\rho &  &  &  &  &  &  &  &  &  &  \\
						\color{black}{\bm{d}}     & -\rho &  &  &  &  &  & 1+\rho^2 &  &  &  &  &  &  &  &  &  &  &  &  &  &  &  &  \\
						&  & \ddots &  &  &  &  &  & \ddots &  &  &  &  &  &  &  &  &  &  &  &  &  &  &  \\
						\color{black}{\bm{t+d-2}} &  &  & -\rho &  &  &  &  &  & 1+\rho^2 &  &  &  &  &  &  &  &  &  &  &  &  &  &  \\
						{\bm{t+d-1}} &  &  &  & 0 &  &  &  &  &  & {1} &  &  &  &  &  & \ddots &  &  &  &  &  &  &  \\
						\color{black}{\bm{t+d}}   &  &  &  & -\rho &  &  &  &  &  &  & 1+\rho^2 &  &  &  &  &  & \ddots &  &  &  &  &  &  \\
						&  &  &  &  & \ddots &  &  &  &  &  &  & \ddots &  &  &  &  &  & \ddots &  &  &  &  &  \\
						\bm{2d-1}   &  &  &  &  &  & -\rho &  &  &  &  &  &  & 1+\rho^2 &  &  &  &  &  &  &  &  &  &  \\
						&  &  &  &  &  &  &  &  &  &  &  &  &  & \ddots &  &  &  &  &  &  &  &  &  \\
						&  &  &  &  &  &  &  &  &  &  &  &  &  &  & \ddots &  &  &  &  &  &  &  &  \\
						{\bm{T-d-1}}  &  &  &  &  &  &  &  &  &  & \ddots &  &  &  &  &  & 1+\rho^2 &  &  &  &  &  & -\rho &  \\
						{\bm{T-d}}    &  &  &  &  &  &  &  &  &  &  & \ddots &  &  &  &  &  & 1 &  &  &  &  &  & \bm{T-d} \\
						&  &  &  &  &  &  &  &  &  &  &  & \ddots &  &  &  &  &  &  &  &  &  &  &  \\
						&  &  &  &  &  &  &  &  &  &  &  &  &  &  &  &  &  &  &  &  &  &  &  \\
						&  &  &  &  &  &  &  &  &  &  &  &  &  &  &  &  &  &  &  & \ddots &  &  &  \\
						&  &  &  &  &  &  &  &  &  &  &  &  &  &  &  &  &  &  &  &  &  &  &  \\
						\bm{T-1}    &  &  &  &  &  &  &  &  &  &  &  &  &  &  &  & -\rho &  &  &  &  &  & 1 & \bm{T-1} \\                         
					\end{block}
					&  &  &  &  &  &  &  &  &  &  &  &  &  &  &  & \bm{T-1-d} &  &  &  &  &  & \bm{T-1} &  \\
				\end{blockarray}.	$
			}
		\end{center}
	\end{adjustwidth}
	\begin{align*} \text{Hence, }H^+_t&=\left[H_{t,(1:T)_{-t}}\right]_{1\times(T-1)} H^{-1}_{(1:T)_{-t},(1:T)_{-t}}\\
		&=\left(0,\dots,0,\overset{\bm{t+d-1}}{\rho^1},0,\dots,0,\overset{\bm{t+2\cdot d-1}}{\rho^2}, 0,\dots,0, \overset{\bm{t+3\cdot d-1}}{\rho^3},\dots\dots \right)H^{-1}_{(1:T)_{-t},(1:T)_{-t}}\\
		&=\color{black}{\left(0,\dots,0,\overset{\bm{t+d-1}}{\rho},0,\dots \dots,0\right)_{1\times(T-1)}}\color{black}.
	\end{align*}
	\begin{center}
		\color{black}{$(ii)$\textbf{ If} $t\in\{T-d+1,\dots,T-1\}$,}
	\end{center}
	\color{black}
	then each one of the $d$ sets $\{T-1,T-d,T-2d,\dots\dots\},\,\{T-2,T-d-1,T-2d-1,\dots\dots\},\dots$, $\{\color{black}{t}\color{black},\color{black}{t-d+1}\color{black},t-2d+1,\dots\dots\}$, $\{\color{black}{t-1}\color{black},\color{black}{t-1-d}\color{black},t-1-2d,\dots\dots\},\,\{t-2,t-2-d,t-2-2d,\dots\dots\},\dots,\,\{T-d+1,T-2d+1,T-3d+1,\dots\dots\}$ and $\{{t-d},t-2d,t-3d,\dots\dots\}$ contains time points, which include exactly one from $\{1,2,\ldots,d\}$, \textbf{in $H_{-t,-t}$ (after deducting the original $t^{\supth}$ time point)} with a same temporal season period component.
	\par Hence, $L_{-t}^{-1}=$
	\begin{adjustwidth}{-2cm}{-1.2cm}
		\begin{center}
			\scalebox{0.63}{
				$\begin{blockarray}{cccccccccccccccccccccc}
					& \bm{1} &  &  &  &  &  &  &  & \bm{d+1} &  &  &  &  &  &  &  &  &  &  &  &  \\
					\begin{block}{c(cccccccccccccccccccc)c}
						\bm{1} & 1 &  &  &  &  &  &  &  &  &  &  &  &  &  &  &  &  &  &  &  &  \bm{1}\\
						&  &  &  &  &  &  &  &  &  &  &  &  &  &  &  &  &  &  &  &  &  \\
						&  &  &  &  &  &  &  &  &  &  &  &  &  &  &  &  &  &  &  &  &  \\
						&  &  &  &  &  &  &  &  &  &  &  &  &  &  &  &  &  &  &  &  &  \\
						&  &  &  &  & \ddots &  &  &  &  &  &  &  &  &  &  &  &  &  &  &  &  \\
						&  &  &  &  &  &  &  &  &  &  &  &  &  &  &  &  &  &  &  &  &  \\
						&  &  &  &  &  &  &  &  &  &  &  &  &  &  &  &  &  &  &  &  &  \\
						&  &  &  &  &  &  &  &  &  &  &  &  &  &  &  &  &  &  &  &  &  \\
						\bm{d+1}& -\rho &  &  &  &  &  &  &  & 1 &  &  &  &  &  &  &  &  &  &  &  & \bm{d+1} \\
						&  & \ddots &  &  &  &  &  &  &  & \ddots &  &  &  &  &  &  &  &  &  &  &  \\
						&  &  & \ddots &  &  &  &  &  &  &  & \ddots &  &  &  &  &  &  &  &  &  &  \\
						&  &  &  & \ddots &  &  &  &  &  &  &  & \ddots &  &  &  &  &  &  &  &  &  \\
						&  &  &  &  & -\rho &  &  &  &  &  &  &  & 1 &  &  &  &  &  &  &  &  \bm{T-d}\\
						&  &  &  &  &  & \ddots &  &  &  &  &  &  &  & \ddots &  &  &  &  &  &  &  \\
						&  &  &  &  &  &  & -\rho &  &  &  &  &  &  &  & 1 &  &  &  &  &  &  \bm{t-2}\\
						&  &  &  &  &  &  &  & -\rho &  &  &  &  &  &  &  & 1 &  &  &  &  &  {\bm{t-1}} \\
						&  &  &  &  &  &  &  &  &  & -\rho &  &  &  &  &  &  & 1 &  &  &  &  {\bm{t}} \\
						&  &  &  &  &  &  &  &  &  &  & \ddots &  &  &  &  &  &  & \ddots &  &  &  \\
						&  &  &  &  &  &  &  &  &  &  &  & -\rho &  &  &  &  &  &  & 1 &  &  \bm{T-2}  \\
						&  &  &  &  &  &  &  &  &  &  &  &  & -\rho &  &  &  &  &  &  & 1 & \bm{T-1} \\
					\end{block}
					& \bm{1} &  &  &  & \bm{T-2d} &  &  & {\bm{t-d-1}} &  & {\bm{t-d+1}} &  & \bm{T-d-1} & \bm{T-d} &  & \bm{t-2} & \bm{t-1} & \bm{t} &  & \bm{T-2} & \bm{T-1} &  \\
				\end{blockarray}.
				$    		
			}
		\end{center} 
	\end{adjustwidth} 
	Let $D_{-t}$ be the $(T-1)\times (T-1)$ diagonal matrix  
	$$\diag\left(1,1,\dots,\overset{{\bm{d}}}{1},\overset{{\bm{d+1}}}{1-\rho^2},1-\rho^2,\dots\dots, 1-\rho^2 \right).$$
	Then $H_{-t,-t}=L_{-t}D_{-t}L_{-t}^{\T}$ and hence $H_{-t,-t}^{-1}=\left(L_{-t}^{\T}\right)^{-1}D_{-t}^{-1}L_{-t}^{-1}={\frac{1}{1-\rho^2}\times}$ 
	\begin{adjustwidth}{-2cm}{-1.7cm}
		\begin{center}
			\scalebox{0.72}{
				$\begin{blockarray}{cccccccccccccccccccc}
					& \bm{1} &  & \bm{d} & \bm{d+1} &  &  &  &  &  &  & \bm{2d+1} &  &  &  &  &  &  &  &  \\
					\begin{block}{c(cccccccccccccccccc)c}
						\bm{1}	& 1 &  &  & -\rho &  &  &  &  &  &  &  &  &  &  &   &  &  &  & \bm{1} \\
						&  & \ddots &  &  &  &  & \ddots &  &  &  &  &  &  &  &   &  &  &  &  \\
						\bm{d}	&  &  & 1 &  &  &  &  &  &  &  &  &  &  &  &  &  &  &  & \bm{d} \\
						\bm{d+1} & -\rho &  &  & 1+\rho^2 &  &  &  &  &  &  &  -\rho &  &  &   &  &  &  &  & \bm{d+1} \\
						&  &  &  &  & \ddots &  &  &  &  &  &  & \ddots &  &  &  &   &  &  &  \\
						&  &  &  &  &  & \ddots &  &  &  &  &  &  &  \ddots &  &  &   &  &  &  \\
						&  & \ddots &  &  &  &  & \ddots &  &  &  &  &  &  &  \ddots &  &   &  &  &  \\
						&  &  &  &  &  &  &  & 1+\rho^2 &  &  &  &  &  &  & -\rho &  &  &  & \color{black}{\bm{t-d-1}} \\
						&  &  &  &  &  &  &  &  & {1} &  &  &  &  &  &  0 &  &  &  & {\bm{t-d}} \\
						&  &  &  &  &  &  &  &  &  & 1+\rho^2 &  &  &  &  &  &  -\rho &  &  & \color{black}{\bm{t-d+1}} \\
						\bm{2d+1} &  &  &  & -\rho &  &  &  &  &  &  & \ddots &  &  &  &  &  &  \ddots &  &  \\
						&  &  &  &  & \ddots &  &  &  &  &  &  & 1+\rho^2 &  &  &  &  &  & -\rho & \color{black}{\bm{T-d}} \\
						&  &  &  &  &  & \ddots &  &  &  &  &  &  & 1 &  &  &  &  &  & \color{black}{\bm{T-d+1}} \\
						&  &  &  &  &  &  & \ddots &  &  &  &  &  &  & \ddots &  &  &  &  &  \\
						&  &  &  &  &  &  &  & -\rho & 0 &  &  &  &  &  & 1 &  &  &  & \color{black}{\bm{t-1}} \\
						&  &  &  &  &  &  &  &  &  & -\rho &  &  &  &  &   & 1 &  &  & \color{black}{\bm{t}} \\
						&  &  &  &  &  &  &  &  &  &  & \ddots &  &  &  &   &  & \ddots &  &  \\
						&  &  &  &  &  &  &  &  &  &  &  & -\rho &  &  &  &  &  & 1 &\bm{T-1}  \\                    
					\end{block}
					& \bm{1} &  & \bm{d} & \bm{d+1} &  &  &  & \color{black}{\bm{t-d-1}} & {\bm{t-d}} & \color{black}{\bm{t-d+1}} &  & \color{black}{\bm{T-d}} & \color{black}{\bm{T-d+1}} &  & \color{black}{\bm{t-1}} & \color{black}{\bm{t}} &  & \bm{T-1} &  \\
				\end{blockarray}.	$
			}
		\end{center}
	\end{adjustwidth}
	\begin{align*}
		\text{Hence, }H^+_t & = \left[H_{t,(1:T)_{-t}} \right]_{1\times(T-1)}H^{-1}_{(1:T)_{-t},(1:T)_{-t}} \\
		& =\left(\dots\dots,\overset{\bm{t-3\cdot d}}{\rho^3},0,\dots,0,\overset{\bm{t-2\cdot d}}{\rho^2}, 0,\dots, 0, \overset{\bm{t-d}}{\rho^1},0,\dots,0 \right)H^{-1}_{(1:T)_{-t},(1:T)_{-t}}\\
		&=\color{black}{\left(0,\dots \dots,0,\overset{\bm{t-d}}{\rho},0,\dots,0\right)_{1\times(T-1)}}\color{black}.
	\end{align*}
	\begin{center}	
		\color{black}{$(iii)$\textbf{ If} $t\in\{d+1,\dots\dots,T-d\}$,}
	\end{center}
	\color{black}
	then each one of the $d$ sets $\{\dots\dots, t-2d-1,\color{black}t-d-1\color{black}, \color{black}t-1\color{black}, \color{black}t+d-2\color{black}, t+2d-2, \dots\dots\},\,\{\dots\dots,t-2d, {t-d}, {t+d-1}, t+2d-1,\dots\dots\},\, \{\dots\dots, t-2d+1, \color{black}t-d+1\color{black}, \color{black}t\color{black}, \color{black}t+d\color{black}, t+2d, \dots\dots\},\, \{\dots\dots, t-2d+2, t-d+2, t+1, t+d+1, t+2d+1, \dots\dots\}, \dots,\, \{\dots\dots, t-2, t+d-3, t+2d-3, \dots\dots\}$ contains time points, which include exactly one from $\{1,2,\ldots,d\}$, \textbf{in $H_{-t,-t}$ (after deducting the original $t^{\supth}$ time point)} with a same temporal season period component.
	\par Hence, $L_{-t}^{-1}=$
	\begin{adjustwidth}{-2.2cm}{-1cm}
		\begin{center}
			\scalebox{0.7}{$
				\begin{blockarray}{cccccccccccccccccccccc}
					& \bm{1} & & \bm{d+1} & & & & & \color{black}{\bm{t-d-1}} & {\bm{t-d}} & \color{black}{\bm{t-d+1}} & & \color{black}{\bm{t-1}} & \color{black}{\bm{t}} & & & & & \bm{T-d-1} & & \bm{T-1} & \\
					\begin{block}{c(cccccccccccccccccccc)c}
						\bm{1}     & 1& & & & & & & & & & & & & & & & & & & & \\
						& & \ddots& & & & & & & & & & & & & & & & & & & \\
						\bm{d+1}   & -\rho& & 1& & & & & & & & & & & & & & & & & & \\
						& & & & \ddots& & & & & & & & & & & & & & & & & \\
						& & & & & \ddots& & & & & & & & & & & & & & & & \\
						& & & & & & \ddots& & & & & & & & & & & & & & & \\
						& & & &\ddots & & & \ddots& & & & & & & & & & & & & & \\
						\bm{t-d-1} & & & & &\ddots & & & 1& & & & & & & & & & & & & \\
						\bm{t-d}   & & & & & & & & & 1& & & & & & & & & & & & \\
						\bm{t-d+1} & & & & & & & & & & 1& & & & & & & & & & & \\
						& & & & & & & & & & & \ddots& & & & & & & & & & \\
						\color{black}{\bm{t-1}}& & & & & & & & -\rho& & & & 1& & & & & & & & & \\
						\color{black}{\bm{t}}& & & & & & & & & & -\rho& & & 1& & & & & & & & \\
						& & & & & & & & & & & \ddots& & & \ddots& & & & & & & \\
						\color{black}{\bm{t+d-2}}& & & & & & & & & & & & -\rho& & & \ddots& & & & & & \\
						{\bm{t+d-1}}  & & & & & & & & & {-\rho^2}& & & & & & & \ddots& & & & & \\
						\color{black}{\bm{t+d}}  & & & & & & & & & & & & & -\rho& & & & \ddots& & & & \\
						& & & & & & & & & & & & & & & \ddots& & & 1& & & \bm{T-d-1}\\
						& & & & & & & & & & & & & & & & \ddots& & & \ddots& & \\
						& & & & & & & & & & & & & & & & & & -\rho& & 1& \bm{T-1}\\
					\end{block}
					& & & & & & & & & & & & & & & & & & \bm{T-d-1} & & \bm{T-1} & \\
				\end{blockarray}
				$}
		\end{center}
	\end{adjustwidth}
	Let $D_{-t}$ be the $(T-1)\times(T-1)$ diagonal matrix
	$$
	\diag\left(1,1,\dots,\overset{{\bm{d}}}{1},\overset{{\bm{d+1}}}{1-\rho^2},1-\rho^2,\dots,1-\rho^2,\overset{{\bm{t+d-1}}}{{1-\rho^4}},1-\rho^2,\dots\dots, 1-\rho^2 \right). 
	$$
	Then $H_{-t,-t}=L_{-t}D_{-t}L_{-t}^{\T}$ and hence $H_{-t,-t}^{-1}=\left(L_{-t}^{\T}\right)^{-1}D_{-t}^{-1}L_{-t}^{-1}=$
	\begin{adjustwidth}{-2.2cm}{-1.2cm}
		\begin{center}
			\scalebox{0.6}{$
				\begin{blockarray}{ccccccccccccccccccccccc}
					& \bm{1} &  & \bm{d} & \bm{d+1} &  &  & \color{black}{\bm{t-d-1}} & {\bm{t-d}} & \color{black}{\bm{t-d+1}} &  & \color{black}{\bm{t-1}} & \color{black}{\bm{t}} &  & \color{black}{\bm{t+d-2}} & {\bm{t+d-1}} & \color{black}{\bm{t+d}} &  &  & \bm{T-d-1} & \bm{T-d} &  & \bm{T-1}\\
					\begin{block}{c(cccccccccccccccccccccc)}
						\bm{1}         & \frac{1}{1-\rho^2} &  &  & \frac{-\rho}{1-\rho^2} &  &  &  &  &  &  &  &  &  &  &  &  &  &  &  &  &  & \\
						&  & \ddots &  &  &  &  &  &  &  &  &  &  &  &  &  &  &  &  &  &  &  & \\
						\bm{d}         &  &  & \frac{1}{1-\rho^2} &  &  &  &  &  &  &  &  &  &  &  &  &  &  &  &  &  &  & \\
						\bm{d+1}       & \frac{-\rho}{1-\rho^2} &  &  & \frac{1+\rho^2}{1-\rho^2} &  &  & \ddots &  &  &  &  &  &  &  &  &  &  &  &  &  &  & \\
						&  &  &  &  & \ddots &  &  & \ddots &  &  &  &  &  &  &  &  &  &  &  &  &  & \\
						&  &  &  &  &  & \ddots &  &  &  &  &  &  &  &  &  &  &  &  &  &  &  & \\
						\color{black}{\bm{t-d-1}}&  &  &  & \ddots &  &  & \frac{1+\rho^2}{1-\rho^2} &  &  &  & \frac{-\rho}{1-\rho^2} &  &  &  &  &  &  &  &  &  &  & \\
						{\bm{t-d}}     &  &  &  &  & \ddots &  &  & {+} &  &  &  &  &  &  & {-\frac{\rho^2}{1-\rho^4}} &  &  &  &  &  &  & \\
						\color{black}{\bm{t-d+1}}&  &  &  &  &  &  &  &  & \frac{1+\rho^2}{1-\rho^2} &  &  & \frac{-\rho}{1-\rho^2} &  &  &  &  &  &  &  &  &  & \\
						&  &  &  &  &  &  &  &  &  & \ddots &  &  & \ddots &  &  &  &  &  &  &  &  & \\
						\color{black}{\bm{t-1}}  &  &  &  &  &  &  & \frac{-\rho}{1-\rho^2} &  &  &  & \frac{1+\rho^2}{1-\rho^2} &  &  & \frac{-\rho}{1-\rho^2} &  &  &  &  &  &  &  & \\
						\color{black}{\bm{t}}	   &  &  &  &  &  &  &  &  & \frac{-\rho}{1-\rho^2} &  &  & \frac{1+\rho^2}{1-\rho^2} &  &  &  & \frac{-\rho}{1-\rho^2} &  &  &  &  &  & \\
						&  &  &  &  &  &  &  &  &  & \ddots &  &  & \ddots &  &  &  &  &  &  &  &  & \\
						\color{black}{\bm{t+d-2}}&  &  &  &  &  &  &  &  &  &  & \frac{-\rho}{1-\rho^2} &  &  & \frac{1+\rho^2}{1-\rho^2} &  &  &  &  &  &  &  & \\
						{\bm{t+d-1}}		   &  &  &  &  &  &  &  & {-\frac{\rho^2}{1-\rho^4}} &  &  &  &  &  &  & {*} &  &  &  & \ddots &  &  &  \\
						\color{black}{\bm{t+d}}  &  &  &  &  &  &  &  &  &  &  &  & \frac{-\rho}{1-\rho^2} &  &  &  & \frac{1+\rho^2}{1-\rho^2} &  &  &  & \ddots &  &  \\
						&  &  &  &  &  &  &  &  &  &  &  &  &  &  &  &  & \ddots &  &  &  &  &  \\
						&  &  &  &  &  &  &  &  &  &  &  &  &  &  &  &  &  & \ddots &  &  &  &  \\
						\bm{T-d-1}		   &  &  &  &  &  &  &  &  &  &  &  &  &  &  & \ddots &  &  &  & \frac{1+\rho^2}{1-\rho^2} &  &  & \frac{-\rho}{1-\rho^2} \\
						\bm{T-d}	       &  &  &  &  &  &  &  &  &  &  &  &  &  &  &  & \ddots &  &  &  & \frac{1}{1-\rho^2} &  &  \\
						&  &  &  &  &  &  &  &  &  &  &  &  &  &  &  &  &  &  &  &  & \ddots &  \\
						\bm{T-1}	       &  &  &  &  &  &  &  &  &  &  &  &  &  &  &  &  &  &  & \frac{-\rho}{1-\rho^2} &  &  & \frac{1}{1-\rho^2} \\
					\end{block}			
				\end{blockarray}\:$},
		\end{center}
	\end{adjustwidth}
	where 
	\begin{equation}\label{HtInvSeasonDiagtminusd}
		{+}=
		\begin{cases}
			\frac{1^2}{1-\rho^2} + \frac{\left(-\rho^2 \right)^2 }{1-\rho^4} = {\frac{\rho^4+\rho^2+1}{1-\rho^4}}, & \color{black}\text{if }t-d>d\iff t\in\{2d+1,2d+2,\dots\dots,T-d\}\\
			\frac{1^2}{1}+\frac{\left(-\rho^2 \right)^2 }{1-\rho^4} = {\frac{1}{1-\rho^4}}, & \color{black}\text{if }t-d\leq d\iff t\in\{d+1,d+2,\dots,2d\}
		\end{cases}
	\end{equation}
	and 	
	\begin{equation}\label{HtInvSeasonDiagtplusdminus1}
		{*}=
		\begin{cases}
			\frac{1^2}{1-\rho^4} + \frac{\left(-\rho \right)^2 }{1-\rho^2} = {\frac{\rho^4+\rho^2+1}{1-\rho^4}}, & \color{black}\text{if }t+d-1<T-d\iff t\in\{d+1,d+2,\dots\dots,T-2d\}\\
			\frac{1^2}{1-\rho^4}={\frac{1}{1-\rho^4}}, & \color{black}\text{if }t+d-1\geq T-d\iff t\in\{T-2d+1,T-2d+2,\dots,T-d\}
		\end{cases}\color{black}.
	\end{equation}
	\begin{align*}
		\text{Hence, }H^+_t & = \left[H_{t,(1:T)_{-t}} \right]_{1\times(T-1)}H^{-1}_{(1:T)_{-t},(1:T)_{-t}} \\
		& =\left(\dots\dots,\overset{\bm{t-2d}}{\rho^2},\dots,0,\overset{{\bm{t-d}}}{\rho^1},0,\dots,0,\overset{{\bm{t+d-1}}}{\rho^1}, 0,\dots,\overset{\bm{t+2d-1}}{\rho^2}, \dots\dots\right)H^{-1}_{(1:T)_{-t},(1:T)_{-t}}\\
		&=\color{black}\left(0,\dots\dots,0,\overset{{\bm{t-d}}}{\frac{\rho}{1+\rho^2}},0,\dots,0,\overset{{\bm{t+d-1}}}{\frac{\rho}{1+\rho^2}}, 0,\dots \dots,0\right)\color{black}\;\forall\; t\in\{d+1,d+2,\dots\dots,T-d\}.
	\end{align*}
	\qed \\
	\textbf{Notes:} $\forall\; t\in\{d+1,d+2,\dots\dots,T-d\},$
	\begin{itemize}
		\item The $(t-d)^{\supth}$ entry in the length-$(T-1)$ vector $H_t^+$ is (by \Cref{HtInvSeasonDiagtminusd})
		\begin{align*}
			&\rho\times\left({+} - \frac{\rho^2}{1-\rho^4}\right) + \mathbbm{1}_{\{t>2d\}}\times \rho^2\times \left(-\frac{\rho}{1-\rho^2} \right) \\
			= &\begin{cases}
				\quad\rho\times\left(\frac{\rho^4+\rho^2+1}{1-\rho^4}- \frac{\rho^2}{1-\rho^4}\right) + 	\rho^2\times \left(-\frac{\rho}{1-\rho^2}\right)&=\rho\times\left(\frac{\rho^4+1}{1-\rho^4}-\frac{\rho^2}{1-\rho^2} \right)=\rho\times\frac{\rho^4+1-\rho^2(1+\rho^2)}{1-\rho^4}\\  =\rho\times\frac{1-\rho^2}{1-\rho^4}=\frac{\rho}{1+\rho^2},&\color{black} \text{ if }t>2d\iff t\in\{2d+1,\dots\dots,T-d\} \\
				\quad\rho\times\left(\frac{1}{1-\rho^4}- \frac{\rho^2}{1-\rho^4}\right)=\rho\times\frac{1}{1+\rho^2}=\frac{\rho}{1+\rho^2}, &\color{black}\text{ if }t\leq 2d\iff t\in\{d+1,\dots,2d\}
			\end{cases}\\
			=&\frac{\rho}{1+\rho^2}\;\forall\; t\in\{d+1,d+2,\dots\dots,T-d\}.
		\end{align*}
		\item Similarly, $(t+d-1)^{\supth}$ entry in the length-$(T-1)$ vector $H_t^+$ is (by \Cref{HtInvSeasonDiagtplusdminus1})
		\begin{align*}
			&\rho\times\left({*} - \frac{\rho^2}{1-\rho^4}\right) + \mathbbm{1}_{\{t\leq T-2d\}}\times \rho^2\times \left(-\frac{\rho}{1-\rho^2} \right) \\
			= &\begin{cases}
				\quad\rho\times\left(\frac{\rho^4+\rho^2+1}{1-\rho^4}- \frac{\rho^2}{1-\rho^4}\right) + 	\rho^2\times \left(-\frac{\rho}{1-\rho^2}\right)&=\rho\times\left(\frac{\rho^4+1}{1-\rho^4}-\frac{\rho^2}{1-\rho^2} \right)=\rho\times\frac{\rho^4+1-\rho^2(1+\rho^2)}{1-\rho^4}\\  =\rho\times\frac{1-\rho^2}{1-\rho^4}=\frac{\rho}{1+\rho^2},&\color{black} \text{if }t\leq T-2d\iff t\in\{d+1,\dots\dots,T-2d\} \\
				\quad\rho\times\left(\frac{1}{1-\rho^4}- \frac{\rho^2}{1-\rho^4}\right)=\rho\times\frac{1}{1+\rho^2}=\frac{\rho}{1+\rho^2}, &\color{black}\text{if }t> T-2d\iff t\in\{T-2d+1,\dots,T-d\}
			\end{cases}\\
			=&\frac{\rho}{1+\rho^2}\;\forall\; t\in\{d+1,d+2,\dots\dots,T-d\}.
		\end{align*}
		\item The $\{(t-d),(t-d)\}^{\supth}$ entry in $H_{-t,-t}\left[H_{-t,-t}\right]^{-1} = 1$ because
		\begin{itemize}
			\item \color{black}if $t>2d$\color{black}, then this entry equals
			\begin{align*}
				&(\rho,1,\rho^2)\left(-\frac{\rho}{1-\rho^2}, \frac{1}{1-\rho^2}+\frac{\rho^4}{1-\rho^4}, -\frac{\rho^2}{1-\rho^4} \right)^{\T}\\
				=&\frac{1}{1-\rho^2}\left[\rho\cdot(-\rho)+1\cdot 1\right]+\frac{1}{1-\rho^4}\left[1\cdot\rho^4+\rho^2\cdot(-\rho^2) \right] =1+0=1;
			\end{align*}
			\item \color{black}if $t\leq2d$\color{black}, then this entry equals
			$$
			(1,\rho^2)\left(\frac{1}{1-\rho^4}, -\frac{\rho^2}{1-\rho^4} \right)^{\T}
			=1.
			$$
		\end{itemize}
		\item Similarly, the $\{(t+d-1),(t+d-1)\}^{\supth}$ entry in $H_{-t,-t}\left[H_{-t,-t}\right]^{-1} = 1$ because
		\begin{itemize}
			\item \color{black}if $t\leq T-2d$\color{black}, then this entry equals
			\begin{align*}
				&(\rho^2, 1, \rho)\left(-\frac{\rho^2}{1-\rho^4}, \frac{1}{1-\rho^4}+\frac{\rho^2}{1-\rho^2}, -\frac{\rho}{1-\rho^2}  \right)^{\T}\\
				=&\frac{1}{1-\rho^4}\left[\rho^2\cdot(-\rho^2)+1\cdot1 \right]+\frac{1}{1-\rho^2}\left[1\cdot\rho^2-\rho\cdot\rho\right]
				=1+0=1;
			\end{align*}
			\item \color{black}if $t>T-2d$\color{black}, then this entry equals
			$$
			(\rho^2, 1)\left(-\frac{\rho^2}{1-\rho^4}, \frac{1}{1-\rho^4} \right)^{\T}
			=1.
			$$
		\end{itemize}
	\end{itemize}

	\section{A $\VAR(1)$ Process for the Latent Temporal Factors \boldmath{$\eta$}$_{1:T}$}\label{appenC}  
	\par Throughout this section, we assume evenly dispersed time points corresponding to our observed data. To model the zero-mean $k\times 1$ latent temporal factors $\bm{\eta}_t,\:t\in\{1,2,\dots,T\}$, we consider a $\VAR(1)$ model in reduced form
	\begin{equation}\label{var1}
		\bm{\eta}_t = A \bm{\eta}_{t-1} + \bm{\epsilon}_t,\: \bm{\epsilon}_t \sim N_k(\bm{0},\Upsilon), \qquad t=1,2,\dots,T,
	\end{equation}
	where ${\Upsilon}_{k\times k}$ is positive definite, $A_{k\times k}$ is invertible, and \underline{$\bm{\eta}_0$ is assumed to be $\bm{0}_{k\times 1}$}. \Cref{var1} simplifies to an $\AR(1)$ model if we take $A_{k\times k}={\psi}I_{k\times k}$ for some $\psi\in(0,1)$. We assign priors $A\sim{\MN}_{k,k}\left(M,\Upsilon,V\right)$ and $\Upsilon\sim \mathcal{IW}_k\left(\zeta, \Omega\right)$ (the Inverse Wishart distribution with degrees of freedom $\zeta>k-1$ and $k\times k$ positive definite scale matrix $\Omega$) to parameters $A$ and $\Upsilon$ respectively. By modifying the original Gibbs sampling steps for temporal-related parameters, we can form integrated spatiotemporal Bayesian Gaussian factor models that permit convenient adequate predictions at any arbitrary future time points. 
	
	\subsection{New Gibbs Sampling Steps for Temporal-Related Parameters}\label{appenC1}
	\par The temporal-related parameters in our new holistic modeling frameworks are now $\{\bm{\eta}_{1:T},\Upsilon,A\}$ instead of $\{\bm{\eta}_{1:T},\Upsilon,\psi\}$. We detail the corresponding new Gibbs sampler steps in what ensues.
	\begin{center}
		\textbf{Sampling From the Full Conditional Distributions of the $k\times 1$ Vectors $\bm{\eta}_t$'s}
	\end{center}	
	Fix any arbitrary $t\in\{1,\dots,T-1\}$. Then
	\begin{align}\label{etaGibbs}
		&\, f(\bm{\eta}_t|\cdot) \propto f\left(\bm{y}_t|\bm{\beta},\Lambda,\bm{\eta}_t,\Xi \right) \times f\left(\bm{\eta}_{1:T}|A,\Upsilon\right) \propto f\left(\bm{y}_t|\bm{\beta},\Lambda,\bm{\eta}_t,\Xi \right) \times f\left(\bm{\eta}_t|\bm{\eta}_{t-1},A,\Upsilon\right) \times f\left(\bm{\eta}_{t+1}|\bm{\eta}_{t},A,\Upsilon\right)\nonumber\\
		&\quad \propto \exp\left\lbrace -\frac{1}{2}\left(\bm{y}_t-X_t\bm{\beta}-\Lambda\bm{\eta}_t\right)^{\T}\Xi^{-1}\left(\bm{y}_t-X_t\bm{\beta}-\Lambda\bm{\eta}_t\right)  \right\rbrace  \times\nonumber\\ &\quad\quad\: \exp\left\lbrace-\frac{1}{2}\left(\bm{\eta}_t-A\bm{\eta}_{t-1}\right)^{\T}{\Upsilon}^{-1}\left(\bm{\eta}_t-A\bm{\eta}_{t-1}\right)\right\rbrace \times \exp\left\lbrace-\frac{1}{2}\left(\bm{\eta}_{t+1}-A\bm{\eta}_{t}\right)^{\T}{\Upsilon}^{-1}\left(\bm{\eta}_{t+1}-A\bm{\eta}_{t}\right)\right\rbrace           \nonumber\\
		&\quad \propto \exp\left\lbrace-\frac{1}{2}\left[\bm{\eta}_t^{\T}V_{\bm{\eta}_t}^{-1}\bm{\eta}_t - 2\cdot\bm{\eta}_t^{\T}\left(\Lambda^{\T}\Xi^{-1}(\bm{y}_t-X_t\bm{\beta})+{\Upsilon}^{-1}A\bm{\eta}_{t-1}+A^{\T}{\Upsilon}^{-1}\bm{\eta}_{t+1} \right)\right]\right\rbrace              \nonumber\\
		& \Rightarrow\;  \bm{\eta}_t~|~\cdot\sim N_k\left(V_{\bm{\eta}_t}\left[\Lambda^{\T}\Xi^{-1}(\bm{y}_t-X_t\bm{\beta})+{\Upsilon}^{-1}A\bm{\eta}_{t-1}+A^{\T}{\Upsilon}^{-1}\bm{\eta}_{t+1} \right], V_{\bm{\eta}_t}\right), \nonumber\\
		&\quad\; \text{ where }V_{\bm{\eta}_t} = \left[\Lambda^{\T}\Xi^{-1}\Lambda+{\Upsilon}^{-1}+A^{\T}{\Upsilon}^{-1}A\right]^{-1}.\\
		&\text{For }t=T,\nonumber\\
		&\,f(\bm{\eta}_T|\cdot) \propto f\left(\bm{y}_T|\bm{\beta},\Lambda,\bm{\eta}_T,\Xi \right) \times f\left(\bm{\eta}_{1:T}|A,{\Upsilon}\right) \propto f\left(\bm{y}_T|\bm{\beta},\Lambda,\bm{\eta}_T,\Xi \right) \times f\left(\bm{\eta}_T|\bm{\eta}_{T-1},A,{\Upsilon}\right)\nonumber\\
		&\qquad\quad\; \propto \exp\left\lbrace -\frac{1}{2}\left(\bm{y}_T-X_T\bm{\beta}-\Lambda\bm{\eta}_T\right)^{\T}\Xi^{-1}\left(\bm{y}_T-X_T\bm{\beta}-\Lambda\bm{\eta}_T\right)  \right\rbrace  \times\nonumber\\ &\quad\;\:\qquad\quad \exp\left\lbrace-\frac{1}{2}\left(\bm{\eta}_T-A\bm{\eta}_{T-1}\right)^{\T}{\Upsilon}^{-1}\left(\bm{\eta}_T-A\bm{\eta}_{T-1}\right)\right\rbrace           \nonumber\\
		&\qquad\quad\;\propto \exp\left\lbrace-\frac{1}{2}\left[\bm{\eta}_T^{\T}\left( \Lambda^{\T}\Xi^{-1}\Lambda+{\Upsilon}^{-1}\right)\bm{\eta}_T - 2\cdot\bm{\eta}_T^{\T}\left(\Lambda^{\T}\Xi^{-1}(\bm{y}_T-X_T\bm{\beta})+{\Upsilon}^{-1}A\bm{\eta}_{T-1} \right)\right]\right\rbrace              \nonumber\\
		& \Rightarrow\;  \bm{\eta}_T ~|~\cdot\sim N_k\left(\left[\Lambda^{\T}\Xi^{-1}\Lambda+{\Upsilon}^{-1}\right]^{-1}\left[\Lambda^{\T}\Xi^{-1}(\bm{y}_T-X_T\bm{\beta})+{\Upsilon}^{-1}A\bm{\eta}_{T-1} \right], \left[\Lambda^{\T}\Xi^{-1}\Lambda+{\Upsilon}^{-1}\right]^{-1}\right).
	\end{align}
	\begin{center}
		\textbf{Sampling From the Full Conditional Distribution of the Positive Definite Matrix ${\Upsilon}_{k\times k}$}
	\end{center}
	\begin{align}\label{UGibbs}
		f({\Upsilon}|\cdot) &\propto f(\bm{\eta}_{1:T}|A,{\Upsilon},\bm{\eta}_0=\bm{0}_{k\times 1})\times f_0({\Upsilon}),\text{ where }f_0=\mathcal{IW}_k\left(\zeta, \Omega\right)\nonumber\\
		&\propto |\Upsilon|^{-\frac{T}{2}}\prod_{t=1}^T \exp\left\lbrace-\frac{1}{2}\left(\bm{\eta}_t-A\bm{\eta}_{t-1}\right)^{\T}{\Upsilon}^{-1}\left(\bm{\eta}_t-A\bm{\eta}_{t-1}\right)  \right\rbrace \times |\Upsilon|^{-\frac{\zeta+k+1}{2}}\exp\left\lbrace-\frac{1}{2}\trace(\Omega {\Upsilon}^{-1}) \right\rbrace     \nonumber\\
		&=|\Upsilon|^{-\frac{T+\zeta+k+1}{2}}\exp\left\lbrace-\frac{1}{2}\sum_{t=1}^T\trace\left(\left(\bm{\eta}_t-A\bm{\eta}_{t-1}\right)^{\T} {\Upsilon}^{-1}\left(\bm{\eta}_t-A\bm{\eta}_{t-1}\right)\right)\right\rbrace \cdot \exp\left\lbrace-\frac{1}{2}\trace(\Omega {\Upsilon}^{-1}) \right\rbrace  \nonumber\\
		&=|\Upsilon|^{-\frac{T+\zeta+k+1}{2}}\exp\left\lbrace-\frac{1}{2}\left[ \sum_{t=1}^T\trace\left(\left(\bm{\eta}_t-A\bm{\eta}_{t-1}\right)\left(\bm{\eta}_t-A\bm{\eta}_{t-1}\right)^{\T} {\Upsilon}^{-1}\right)+\trace\left(\Omega {\Upsilon}^{-1}\right)\right] \right\rbrace   \nonumber\\
		&=|\Upsilon|^{-\frac{T+\zeta+k+1}{2}}\exp\left\lbrace-\frac{1}{2} \trace\left(\left[ \sum_{t=1}^T\left(\bm{\eta}_t-A\bm{\eta}_{t-1}\right)\left(\bm{\eta}_t-A\bm{\eta}_{t-1}\right)^{\T}+\Omega\right]  {\Upsilon}^{-1}\right) \right\rbrace   \nonumber\\
		\Rightarrow\;& \Upsilon~|~\cdot \sim \mathcal{IW}_k\left(T+\zeta, \; \sum_{t=1}^{T}\left(\bm{\eta}_t-A\bm{\eta}_{t-1}\right)\left(\bm{\eta}_t-A\bm{\eta}_{t-1}\right)^{\T}+\Omega\right). 
	\end{align}
	\begin{center}
		\textbf{Sampling From the Full Conditional Distribution of the Invertible Matrix $A_{k\times k}$}
	\end{center}
	Write $A=\left(\bm{a}_1,\bm{a}_2,\ldots,\bm{a}_k\right)$, where $\bm{a}_j$ denotes the $j^{\text{th}}$ column of $A$ for $j=1,2,\ldots,k$, and let $f_0(A)={\MN}_{k,k}\left(M,\Upsilon,V\right)$, where $\Upsilon_{k\times k}$ and $V_{k\times k}$ are both positive definite scale matrices and $M_{k\times k}$ is a location matrix. Since $A\sim {\MN}_{k, k}\left(M,\Upsilon,V\right)\iff \vect(A)\sim N_{k^2}\left(\vect(M),V\otimes \Upsilon\right)$, 
	\begin{align}\label{AGibbs}
		f(A|\cdot) &\propto f(\bm{\eta}_{1:T}|A,{\Upsilon},\bm{\eta}_0=\bm{0}_{k\times 1})\times f_0(A) \nonumber\\
		&\propto \prod_{t=1}^{T} \exp\left\lbrace-\frac{1}{2}\left(\bm{\eta}_t-A\bm{\eta}_{t-1}\right)^{\T}{\Upsilon}^{-1}\left(\bm{\eta}_t-A\bm{\eta}_{t-1}\right)\right\rbrace \times f_0(A) \nonumber\\
		&\propto \exp\left\lbrace-\frac{1}{2}[\vect(A)-\vect(M)]^{\T}\left(V^{-1}\otimes\Upsilon^{-1}\right)[\vect(A)-\vect(M)]\right\rbrace\; \times\nonumber\\
		&\;\;\;\prod_{t=1}^{T} \exp\left\lbrace-\frac{1}{2}\left({\eta}_{t-1,1}\bm{a}_1+\ldots+{\eta}_{t-1,k}\bm{a}_k-\bm{\eta}_t\right)^{\T}{\Upsilon}^{-1}\left({\eta}_{t-1,1}\bm{a}_1+\ldots+{\eta}_{t-1,k}\bm{a}_k-\bm{\eta}_t\right)\right\rbrace \nonumber\\
		& = \exp\left\lbrace-\frac{1}{2}[\vect(A)-\vect(M)]^{\T}\left(V^{-1}\otimes\Upsilon^{-1}\right)[\vect(A)-\vect(M)]\right\rbrace\; \times \nonumber\\
		&\;\;\;\prod_{t=1}^{T}\exp\left[-\frac{1}{2}\vect(A)^{\T}\left\lbrace\left[\bm{\eta}_{t-1}\bm{\eta}_{t-1}^{\T}\right]\otimes \Upsilon^{-1}\right\rbrace\vect(A) + \vect(A)^{\T}(\bm{\eta}_{t-1}\otimes\Upsilon^{-1})\bm{\eta}_t\right]\nonumber\\
		\propto \exp&\left\lbrace-\frac{1}{2}\vect(A)^{\T}\left[V_A^{-1}\otimes \Upsilon^{-1}\right]\vect(A)+\vect(A)^{\T}\left[\left(V^{-1}\otimes\Upsilon^{-1}\right)\vect(M) + \sum_{t=1}^T(\bm{\eta}_{t-1}\otimes\Upsilon^{-1})\bm{\eta}_t\right]\right\rbrace \nonumber\\
		\quad\quad\qquad\;\text{with} &\;V_A=\left(\sum_{t=1}^T\left[\bm{\eta}_{t-1}\bm{\eta}_{t-1}^{\T}\right]+V^{-1}\right)^{-1}\nonumber\\
		\Rightarrow\vect(A) &~|~ \cdot \sim N_{k^2}\left([V_AV^{-1}\otimes I_{k\times k}]\vect(M)+\sum_{t=1}^T[V_A\bm{\eta}_{t-1}\otimes I_{k\times k}]\bm{\eta}_t,\;V_A\otimes \Upsilon\right) \nonumber \\
		\iff \qquad A &~|~ \cdot  \sim {\MN}_{k,k}\left(M_A,\Upsilon,V_A\right), \\
		\text{where }&M_A\text{ is }k\times k\text{ with vectorized form }[V_AV^{-1}\otimes I_{k\times k}]\vect(M)+\sum_{t=1}^T[V_A\bm{\eta}_{t-1}\otimes I_{k\times k}]\bm{\eta}_t.\nonumber
	\end{align}

	\subsection{Predictions at Desired Future Time Points}\label{appenC2}   
	\par Suppose we have $q\geq 1$, $q\in\mathbb{N}$ new time points $T+1,\dots,T+q$. Then for predicting $\mathbf{\hat{y}}_{(T+1):(T+q)}$ given $\mathbf{y}_{1:T}$ and covariates matrix $\left[ X_{(T+1):(T+q)}\right] _{qOm\times p}$ (if $p\geq 1$) under our integrated frameworks, we can write the posterior predictive distribution (PPD) as
	\begin{align}\label{PPDtempVAR1}
		&f\left(\bm{y}_{(T+1):(T+q)}|\bm{y}_{1:T},X_{(T+1):(T+q)}\right)=\int_{\Theta}f\left(\bm{y}_{(T+1):(T+q)}|\Theta,\bm{y}_{1:T},X_{(T+1):(T+q)}\right)\pi\left(\Theta|\bm{y}_{1:T}\right)d\Theta, \nonumber\\ &\text{where }\Theta=
		\begin{cases}
			(\Lambda,\bm{\beta},\bm{\sigma^2},\bm{\eta}_{(T+1):(T+q)},\bm{\eta},\Upsilon,A),&\text{ if \texttt{clustering = FALSE}}\\(\bm{\theta},\bm{\xi},\bm{\beta},\bm{\sigma^2},\bm{\eta}_{(T+1):(T+q)},\bm{\eta},\Upsilon,A),&\text{ if \texttt{clustering = TRUE}}
		\end{cases}.
	\end{align}
    $\bm{\eta}$ denotes $\bm{\eta}_{1:T}$. \texttt{clustering = FALSE} refers to the scenario when no spatial clustering mechanism is integrated, and \texttt{clustering = TRUE} refers to the case
	when a spatial PSBP clustering mechanism is incorporated. We can then partition the integral in \Cref{PPDtempVAR1} into
	\begin{align}
		&\int_{\Theta}\underbrace{f\left(\bm{y}_{(T+1):(T+q)}|\Lambda,\bm{\eta}_{(T+1):(T+q)},\bm{\beta},\bm{\sigma^2},X_{(T+1):(T+q)}\right)}_{T_1}\underbrace{f\left(\bm{\eta}_{(T+1):(T+q)}|\bm{\eta},\Upsilon,A\right)}_{T_2}\nonumber\\
		&\qquad\times\underbrace{\pi\left(\Lambda,\bm{\beta},\bm{\sigma^2},\bm{\eta},\Upsilon,A|\bm{y}_{1:T}\right)}_{T_3}d\Theta \text{ if \texttt{clustering = FALSE}, and}\label{PPDnewTimeNoCLVAR1}\\
		&\int_{\Theta}\underbrace{f\left(\bm{y}_{(T+1):(T+q)}|\bm{\theta},\bm{\xi},\bm{\eta}_{(T+1):(T+q)},\bm{\beta},\bm{\sigma^2},X_{(T+1):(T+q)}\right)}_{T_1}\underbrace{f\left(\bm{\eta}_{(T+1):(T+q)}|\bm{\eta},\Upsilon,A\right)}_{T_2}\nonumber\\
		&\qquad\times\underbrace{\pi\left(\bm{\theta},\bm{\xi},\bm{\beta},\bm{\sigma^2},\bm{\eta},\Upsilon,A|\bm{y}_{1:T}\right)}_{T_3}d\Theta\text{ if \texttt{clustering = TRUE}}.\label{PPDnewTimeCLVAR1}
	\end{align}
	In \Cref{PPDnewTimeNoCLVAR1,PPDnewTimeCLVAR1}, $T_1$ is the observed likelihood and $T_3$ is the parameters' posterior distribution obtained from the original model fit's MCMC sampler. For $T_2$, 
	\begin{align}\label{VARpredEta}
		f\left({\bm{\eta}}_{(T+1):(T+q)}|\bm{\eta}_{1:T},\Upsilon,A\right) 
		&= \prod_{r=1}^q f\left(\bm{\eta}_{T+r}|{\bm{\eta}}_{(T+1):(T+r-1)}, \bm{\eta}_{1:T},\Upsilon, A\right) \\
		&= \prod_{r=1}^q N_k\left(\bm{\eta}_{T+r}|A{\bm{\eta}}_{T+r-1}, \Upsilon\right),\nonumber
	\end{align}
	where for each $r$, $N_k\left(\bm{\eta}_{T+r}|A{\bm{\eta}}_{T+r-1}, \Upsilon\right)$ denotes the density of a $k$-variate normal distribution with mean $A{\bm{\eta}}_{T+r-1}$ and covariance matrix $\Upsilon$ evaluated at $\bm{\eta}_{T+r}$. Hence, we can obtain a predicted $\hat{\bm{\eta}}_{(T+1):(T+q)} | \bm{\eta}_{1:T},A,\Upsilon$ by first sampling $\hat{\bm{\eta}}_{T+1}$ from $N_k\left(A\bm{\eta}_{T},\Upsilon\right)$ and then sampling $\hat{\bm{\eta}}_{T+r}$ from $N_k\left(A\hat{\bm{\eta}}_{T+r-1},\Upsilon\right)$ one by one for $r=2,3,\dots,q$.

	\section{Facilitating Efficient Computation for Spatial PSBP}\label{appenD} 
	One major computational issue pertaining to our model in \Cref{sect2} lies in sampling $\bm{\alpha}_{jl_j}$'s from their full conditionals $f(\bm{\alpha}_{jl_j}|\cdot)$'s.
	We notice that $\{\bm{\alpha}_{jl_j}:j=1,\dots,k,\;l_j=1,\dots,L-1\}$ affects the fitted response $\bm{\hat{y}}$ via $\{\bm{w}_{jl_j}:j=1,\dots,k,\;l_j=1,\dots,L\}$, the set of probability parameters for the multinomial distributions of $\xi_j^o(\bm{s}_{i})$'s, and that 
	\begin{align}
		\forall\;&(i,o,l_j)\in\{1,\dots,m\}\times\{1,\dots,O\}\times\{1,\dots,L\},
		\\
		&\mathbb{P}\left(\xi_j^o(\bm{s}_{i})=l_j|{\alpha}_{jr_j}^o(\bm{s}_{i}),1\leq r_j\leq L-1\right)= \mathbb{P}\left(\xi_j^o(\bm{s}_{i})=l_j|{\alpha}_{jr_j}^o(\bm{s}_{i}),1\leq r_j\leq \min\{l_j,L-1\}\right)\nonumber\\ 
		=\; & w_{jl_j}^o(\bm{s}_{i}) =
		\begin{cases}
			\Phi\left(\alpha_{jl_j}^o(\bm{s}_{i})\right)\prod_{r_j<l_j}\left[1-\Phi\left(\alpha_{jr_j}^o(\bm{s}_{i})\right) \right],&\text{ if }l_j\in\{1,\dots,L-1\}\\
			\prod_{r_j<L}\left[1-\Phi\left(\alpha_{jr_j}^o(\bm{s}_{i})\right) \right],&\text{ if }l_j=L
		\end{cases}.\nonumber 
	\end{align} 	
	Hence for any arbitrary $(j,l_j)\in\{1,\dots,k\}\times\{1,\dots,L-1\}$, $f(\bm{\alpha}_{jl_j}|\cdot)\propto{f(\bm{\xi}_j|\bm{\alpha}_{jr_j},1\leq r_j\leq L-1)}\times f_0(\bm{\alpha}_{jl_j}|\kappa,\rho)$, which is burdensome to sample from due to loss of conjugacy.  
	
	This issue can be tackled by adopting ideas put forward by \textcite{Rodriguez2011}. In this context, we introduce a set of latent variables $z_{jl_j}^o(\bm{s}_{i})\overset{\ind}{\sim}N(\alpha_{jl_j}^o(\bm{s}_{i}),1)$ and define $\xi_j^o(\bm{s}_{i})$'s values deterministically based on $z_{jl_j}^o(\bm{s}_{i})$'s as follows:
	\begin{align}\label{xiz}
		&\forall\;(j,i,o)\in\{1,\dots,k\}\times\{1,\dots,m\}\times\{1,\dots,O\},\\
		&\forall\;l_j\in\{1,\dots,L\},\;\xi_j^o(\bm{s}_{i})=l_j\iff
		\begin{cases}
			z_{jl_j}^o(\bm{s}_{i})>0\text{ and } z_{jr_j}^o(\bm{s}_{i})<0\;\forall\; r_j<l_j,&\text{ if }l_j<L\\
			z_{jr_j}^o(\bm{s}_{i})<0\;\forall\; r_j<L,&\text{ if }l_j=L
		\end{cases}.\nonumber
	\end{align}       
	Fix any arbitrary $(j,i,o)$. With this set of introduced latent variables $\{z_{jl_j}^o(\bm{s}_{i}):1\leq l_j\leq L-1\}$ and value of the clustering label parameter $\xi_j^o(\bm{s}_{i})$ defined according to \Cref{xiz}, for any arbitrary $l_j\in\{1,\dots,L\}$, 
	\begin{align*}
		&\mathbb{P}\left(\xi_j^o(\bm{s}_{i})=l_j|{z}_{jr_j}^o(\bm{s}_{i}),{\alpha}_{jr_j}^o(\bm{s}_{i}),1\leq r_j\leq L-1\right) = \mathbb{P}\left(\xi_j^o(\bm{s}_{i})=l_j|{z}_{jr_j}^o(\bm{s}_{i}),1\leq r_j\leq L-1\right),\\
		&\text{and the expression }\int\dots\int \mathbb{P}\left(\xi_j^o(\bm{s}_{i})=l_j|{z}_{jr_j}^o(\bm{s}_{i}),{\alpha}_{jr_j}^o(\bm{s}_{i}),1\leq r_j\leq L-1\right)\times\\
		&\qquad \mathbb{P}\left({z}_{jr_j}^o(\bm{s}_{i}),1\leq r_j\leq L-1|{\alpha}_{jr_j}^o(\bm{s}_{i}),1\leq r_j\leq L-1\right)dz_{j1}^o(\bm{s}_{i})\dots dz_{j(L-1)}^o(\bm{s}_{i})\\
		&=
		\begin{cases}
			\mathbb{P}\left( z_{jl_j}^o(\bm{s}_{i})>0\text{ and } z_{jr_j}^o(\bm{s}_{i})<0\;\forall\; r_j<l_j | {\alpha}_{jr_j}^o(\bm{s}_{i}),1\leq r_j\leq L-1 \right) ,&\text{ if }l_j<L\\
			\mathbb{P}\left( z_{jr_j}^o(\bm{s}_{i})<0\;\forall\; r_j<L | {\alpha}_{jr_j}^o(\bm{s}_{i}),1\leq r_j\leq L-1 \right) ,&\text{ if }l_j=L
		\end{cases}\\
		&=
		\begin{cases}
			\mathbb{P}\left(z_{jl_j}^o(\bm{s}_{i})>0|{\alpha}_{jl_j}^o(\bm{s}_{i})\right)\prod_{r_j=1}^{l-1} \mathbb{P}\left(z_{jr_j}^o(\bm{s}_{i})<0|{\alpha}_{jr_j}^o(\bm{s}_{i})\right)   ,&\text{ if }l_j<L\\
			\prod_{r_j=1}^{L-1} \mathbb{P}\left(z_{jr_j}^o(\bm{s}_{i})<0|{\alpha}_{jr_j}^o(\bm{s}_{i})\right)   ,&\text{ if }l_j=L
		\end{cases}\\
		&=
		\begin{cases}
			\Phi\left(\alpha_{jl_j}^o(\bm{s}_{i})\right)\prod_{r_j<l_j}\left[1-\Phi\left(\alpha_{jr_j}^o(\bm{s}_{i})\right) \right],&\text{ if }l_j\in\{1,\dots,L-1\}\\
			\prod_{r_j<L}\left[1-\Phi\left(\alpha_{jr_j}^o(\bm{s}_{i})\right) \right],&\text{ if }l_j=L
		\end{cases}
	\end{align*}
	does turn out to be the same as $w_{jl_j}^o(\bm{s}_{i})=\mathbb{P}\left(\xi_j^o(\bm{s}_{i})=l_j|{\alpha}_{jr_j}^o(\bm{s}_{i}),1\leq r_j\leq L-1\right)$, where $\xi_j^o(\bm{s}_{i})$'s value is obtained probabilistically conditioning on ${\alpha}_{jr_j}^o(\bm{s}_{i}),1\leq r_j\leq L-1$, as in our original model without the introduced latent normal variables $z_{jr_j}^o(\bm{s}_{i})$'s.
	
	Hence, we can opt to sample $\xi_j^o(\bm{s}_{i})$'s from their marginal distributions with the latent variables $z_{jl_j}^o(\bm{s}_{i})$'s integrated out as above. The corresponding Gibbs sampler step for $\xi_{j}^o(\bm{s}_{i})$'s is thus unaffected by the introduction of $z_{jl_j}^o(\bm{s}_{i})$'s:
	\begin{align}\label{postXiFixedL1}
		&\text{Fix any arbitrary }j\in\{1,\dots,k\}.\quad\forall\:(o,i)\in\{1,\dots,O\}\times\{1,\dots,m \},\nonumber\\
		&\forall\; l_j\in\{1,\dots,L\},\; \mathbb{P}\left(\xi_j^o(\bm{s}_{i})=l_j|\cdot\right)\\
		& \propto \mathbb{P}\left(\xi_j^o(\bm{s}_{i})=l_j|\alpha_{jr_j}^o(\bm{s}_{i}),r_j\leq L-1\right) \prod_{t=1}^T f\left(y_t^o(\bm{s}_{i})|\bm{\beta},\bm{\eta}_t,(\sigma^2)^o(\bm{s}_{i}),\xi_j^o(\bm{s}_{i})=l_j,\bm{\xi}_{-j}^o(\bm{s}_{i}),\bm{\theta}_{\bm{\xi}^o(\bm{s}_{i})}\right) \nonumber\\
		& \propto w_{jl_j}^o(\bm{s}_{i})\times \exp\left\lbrace -\frac{1}{2}(\sigma^{-2})^o(\bm{s}_{i})\sum_{t=1}^T\left[y_t^o(\bm{s}_{i})-\bm{x}_t^o(\bm{s}_{i})^{\T}\bm{\beta}-\sum_{h\neq j}\theta_{h\xi_h^o(\bm{s}_{i})}\eta_{th}-\theta_{jl_j}\eta_{tj} \right]^2\right\rbrace.\nonumber
	\end{align}	
	
	It is also straightforward to sample from the full conditional distributions of ${z}_{jl_j}^o(\bm{s}_{i})$'s:
	\begin{align}\label{postZ1}
		&\text{For any arbitrary fixed } (j,i,o,l_j)\in\{1,\dots,k\}\times\{1,\dots,m\}\times\{1,\dots,O\}\times\{1,\dots,L-1\}, \nonumber\\
		& f(z_{jl_j}^o(\bm{s}_{i})|\cdot)\propto f(\xi_j^o(\bm{s}_{i})|z_{jl_j}^o(\bm{s}_{i}))\times f(z_{jl_j}^o(\bm{s}_{i})|\alpha_{jl_j}^o(\bm{s}_{i}))\\
		&\sim  \mathbbm{1}_{\{\xi_j^o(\bm{s}_{i})>l_j\}}N(\alpha_{jl_j}^o(\bm{s}_{i}),1)_{\mathbb{R}_-} + \mathbbm{1}_{\{\xi_j^o(\bm{s}_{i})=l_j\}}N(\alpha_{jl_j}^o(\bm{s}_{i}),1)_{\mathbb{R}_+} + 
		\mathbbm{1}_{\{\xi_j^o(\bm{s}_{i})<l_j\}}N(\alpha_{jl_j}^o(\bm{s}_{i}),1).\nonumber
	\end{align}
	
	The essence of this latent variables introduction approach lies in bringing about conjugacy for $\bm{\alpha}_{jl_j}$'s. With our introduced $z_{jl_j}^o(\bm{s}_{i})$'s and $\xi_j^o(\bm{s}_{i})$'s defined deterministically by \Cref{xiz}, for any arbitrary fixed $(j,l_j)\in\{1,\dots,k\}\times\{1,\dots,L-1\}$,
	\begin{align}
		&f(\bm{\alpha}_{jl_j}|\cdot)\propto{f(\bm{\xi}_j,\bm{z}_{jr_j},1\leq r_j\leq L-1|\bm{\alpha}_{jr_j},1\leq r_j\leq L-1)}\times f_0(\bm{\alpha}_{jl_j}|\kappa,\rho),\nonumber\\
		&\text{where }f(\bm{\xi}_j,\bm{z}_{jr_j},1\leq r_j\leq L-1|\bm{\alpha}_{jr_j},1\leq r_j\leq L-1) \nonumber\\
		&\quad = f(\bm{\xi}_j|\bm{z}_{jr_j}, \bm{\alpha}_{jr_j},1\leq r_j\leq L-1) \times f(\bm{z}_{jr_j},1\leq r_j\leq L-1|\bm{\alpha}_{jr_j},1\leq r_j\leq L-1)\nonumber\\
		&\quad = f(\bm{\xi}_j|\bm{z}_{jr_j}, 1\leq r_j\leq L-1) \times \prod\nolimits_{(i,o,r_j)\in\{1,\dots,m\}\times\{1,\dots,O\}\times\{1,\dots,L-1\}}f\left(z_{jr_j}^o(\bm{s}_{i})|\alpha_{jr_j}^o(\bm{s}_{i})\right) \nonumber\\
		&\quad = f(\bm{\xi}_j|\bm{z}_{jr_j}, 1\leq r_j\leq L-1) \times \prod\nolimits_{1\leq r_j\leq L-1}f\left(\bm{z}_{jr_j}|\bm{\alpha}_{jr_j}\right).\nonumber\\
		&\text{Hence, }f(\bm{\alpha}_{jl_j}|\cdot)\propto{\prod\nolimits_{1\leq r_j\leq L-1}f\left(\bm{z}_{jr_j}|\bm{\alpha}_{jr_j}\right)}\times f_0(\bm{\alpha}_{jl_j}|\kappa,\rho)\propto{f(\bm{z}_{jl_j}|\nonumber\bm{\alpha}_{jl_j})}\times f_0(\bm{\alpha}_{jl_j}|{\kappa},\rho)\nonumber\\
		&\propto \exp\left\lbrace (\bm{z}_{jl_j}-\bm{\alpha}_{jl_j})^{\T}I_{mO}(\bm{z}_{jl_j}-\bm{\alpha}_{jl_j}) + \bm{\alpha}_{jl_j}^{\T} \left[ F(\rho)^{-1}\otimes\kappa^{-1}\right] \bm{\alpha}_{jl_j} \right\rbrace \nonumber\\
		&\propto \exp\left\lbrace \bm{\alpha}_{jl_j}^{\T} \left[I_{mO}+ F(\rho)^{-1}\otimes\kappa^{-1}\right] \bm{\alpha}_{jl_j} - 2\cdot\bm{z}_{jl_j}^{\T}\bm{\alpha}_{jl_j} \right\rbrace \nonumber\\
		&\sim N_{mO}\left(\left[ {I}_{mO}+{F}(\rho)^{-1}\otimes {\kappa}^{-1}\right]^{-1}\bm{z}_{jl_j}, \left[ {I}_{mO}+{F}(\rho)^{-1}\otimes{\kappa}^{-1}\right] ^{-1}\right).\label{postAlphaFixedL1}    
	\end{align}
	
	Under our model in \Cref{sect2}, $\bm{z}_{jl_j}$'s and $\bm{\alpha}_{jl_j}$'s are ordered first by observation type and then spatially, whereas $\bm{w}_{jl_j}$'s, $\bm{\xi}_j$'s, $\bm{\lambda}_j$'s and $\bm{\sigma^2}$ are ordered first spatially and then by observation type. We also take note that \textcite{Berchuck2021}'s finite-mixture-version implementation of \Cref{PSBP} with a common fixed number of spatial mixture components $L\in\mathbb{N}, L>1$ for all factors contains some mistakes, in that they omitted the term $\mathbbm{1}_{\{\xi_j^o(\bm{s}_{i})<l_j\}}N(\alpha_{jl_j}^o(\bm{s}_{i}),1)$ in \Cref{postZ1} for posterior sampling of $z_{jl_j}^o\left(\bm{s}_{i}\right)$'s, calculated the last mixture weights $w_{jL}^o\left(\bm{s}_{i}\right)=\prod_{r=1}^{L-1}[1-\Phi(\alpha_{jr_j}^o(\bm{s}_{i}))]\;\forall\;(j,i,o)$ wrongly, and included redundant posterior updating steps for ${\alpha}_{jL}^o\left(\bm{s}_{i}\right)$'s and $z_{jL}^o\left(\bm{s}_{i}\right)$'s.

	\section{Some Notes on Slice Sampling Adapted in Our Context }\label{appenE}
	While it might be theoretically more adequate to assume independent infinite mixture models specified by \Cref{PSBP} $\forall\;(j,i,o)$ to impose a spatial structure permitting convenient clustering on $\Lambda_{mO\times k}$, we instead opt to introduce new parameters $L_j\in\mathbb{N}\backslash\{0\},j\in\{1,\dots,k\}$ and proceed with a finite mixture model with the unknown $L_j$'s as spatial cluster numbers primarily due to the following two reasons:
	\begin{enumerate}
		\item Since we do have a spatial cluster number upper bound $mO$, it would indeed be reasonable to introduce unknown cluster number parameters $L_j\in\mathbb{N}\backslash\{0\}$ for all latent factors $j=1,\dots,k$ and proceed with the corresponding finite mixture models. 
		\item Fix any arbitrary $j\in\{1,\dots,k\}$.    
		Under a baseline infinite mixture model, problems pertaining to the normalizing constants emerge when sampling $\xi_j^o\left(\bm{s}_{i}\right)$'s, as the support for any $\xi_j^o\left(\bm{s}_{i}\right)$ is the countably infinite set $\mathbb{Z}^+$. This issue can be tackled by introducing latent variables $u_j^o\left(\bm{s}_{i}\right)$'s with uniform full conditional distributions and thus making the possible values each $\xi_j^o\left(\bm{s}_{i}\right)$ can take finite.  
		However, when deciding the new maximum cluster number estimate $L_j$ in the Gibbs sampling step for $\xi_j^o(\bm{s}_{i})$'s, we may well need more $w_{jl_j}^o\left(\bm{s}_{i}\right)$'s for each $(i,o)$ to ensure the criterion in \Cref{LjNewFormula1} due to the slight differences calculating $w_{jl_j}^o\left(\bm{s}_{i}\right)$'s between a finite and an infinite mixture model. We can thus end up with an $L_j$ larger than its counterpart in the previous MCMC iteration. 
		Since $L_j$'s are not ensured non-increasing through the MCMC iterations under an infinite mixture model, we likely need to sample much more $\bm{\alpha}_{jl_j}$'s and $\theta_{jl_j}$'s, especially when the cluster number upper bound $mO$ is large. Computational and storage burdens can thus ensue.  
	\end{enumerate}
	
	Our approach elaborated in \Cref{sect3} successfully bypasses the above concerns. We shall also note that \textcite{Berchuck2021}'s implementation attempt regarding infinite mixture model via slice sampling in their supplementary material and R package \texttt{spBFA} actually does not follow \textcite{Walker2007} and is erroneous. One major accomplishment brought about by the introduced latent variables $u_j^o\left(\bm{s}_{i}\right)$'s (\Cref{sect3}) is that we no longer need to enable feasible posterior sampling of $\bm{\alpha}_{jl_j}^o\left(\bm{s}_{i}\right)$'s by introducing latent normal parameters $z_{jl_j}^o\left(\bm{s}_{i}\right)$'s (\Cref{appenD}). Hence \textcite{Berchuck2021}'s Gibbs sampler step for $z_{jl_j}^o\left(\bm{s}_{i}\right)$'s is redundant when slice sampling is adopted. Their procedure updating $L_j^*$'s is also incorrect and placed wrongly.

	\section{Some Further Comments Regarding the Latent NNGP}\label{appenF}
	The NNGP is a highly scalable alternative to the full parent GP while producing comparably good inferences, and has numerous advantages over other approaches for modeling large geostatistical data sets.
	
	In terms of scalability, NNGP typically reduces computational complexity from cubic to linear in $m$ in our context setting $\mathcal{S}=\mathcal{T}=\{\bm{s}_1,\dots,\bm{s}_m\}$ and significantly outperforms alternative methods like low rank approximation. Low rank models typically require approximately $\Ocal(mr^2)$ flops, where $r\in\mathbb{N}, r\ll m$ is the number of knots that by empirical experiments must be quite large to approximate the GP well when $m$ is large, thereby making $mr^2$ computationally prohibitive. NNGP, on the other hand, needs $\Ocal\left(|\mathcal{S}\cup\mathcal{T}|\cdot h^3\right)$ flops with $|\mathcal{T}|=m$, where $\mathcal{S}=\mathcal{T}$ and a quite small $h$, e.g., between 10 to 15, have been shown to perform well. NNGP's capabilities in significantly easing storage (\Cref{sect4b,sect4c}) may also make it the only feasible candidate under some requirements, e.g., delivering process-based inferences, for certain super large geostatistical data sets (Section 5.2 in \cite{Datta2016}).
	
	From the inferential perspective, the NNGP incorporates parameter estimation, outcome prediction, and latent process interpolation into a single fully process-based framework, which had not been explored by any other methods in the literature. A legitimate GP with sparse precision matrices on the entire geographical domain $\mathcal{D}$, which we shall denote as $\NNGP\big(\bm{0}, \tilde C(\cdot,\cdot|\bm{\theta}) \big) $ derived from the parent Gaussian process $\GP\left(\bm{0}, C(\cdot,\cdot|\bm{\theta}) \right)$, the latent NNGP needs not be conceived as an approximation to its parent GP when modeling the latent or observed spatial surface and permits fast and adequate predictions of the latent and outcome variables at arbitrary new locations, a main issue covariance tapering methods are unable to handle.
	Various simulation experiments \parencite{Datta2016} suggest that NNGP produces estimation \& kriging closely resembling those from the parent GP models for diverse covariance functions $C(\cdot,\cdot|\bm{\theta})$.
	
	Under our spatiotemporal Bayesian Gaussian factor analysis modeling framework, other NNGP methods like Response NNGP and Conjugate NNGP \parencite{Finley2019} are not applicable. A few points to note in our adaptations of the NNGP compared to the framework presented in \textcite{Datta2016} not mentioned earlier are as follows:
	\setlist{nolistsep}
	\begin{itemize}[noitemsep] 
		\item The parametric covariance function $C\left(\cdot,\cdot|\bm{\theta}\right)$ is over $\mathcal{D}\times \mathcal{D}$, not just over $\mathcal{S}\times \mathcal{S}$. $\forall\;\bm{s},\bm{t}\in\mathcal{D}$, $ C(\bm{s},\bm{t}|\bm{\theta})_{q\times q}=\Cov\left\lbrace \bm{w}(\bm{s}), \bm{w}(\bm{t}) \right\rbrace $. $q$ corresponds to $O$ in our model.
		\item $C_{\mathcal{S}}(\bm{\theta})$, the covariance function over the reference set $\mathcal{S}$, corresponds to $F(\rho)_{m\times m}\otimes\kappa_{O\times O}$ in our model. 
		\item We have let $\mathcal{S}=\mathcal{T}$. Thus, $\mathcal{S}^*=\mathcal{T}\cap\mathcal{S}=\mathcal{T}\text{ and } \mathcal{U}= \mathcal{T}\backslash\mathcal{S}^*=\emptyset$. Hence the likelihood in Equation (10) in \textcite{Datta2016} is simplified to only the first portion there, as $k=n=r=m$ in their notations. 
		\item Some software features to further accelerate computation are presented in Section 2.5 of \textcite{Finley2022}.
	\end{itemize}

	\section{More on Predictions at Future Time Points or New Spatial Locations (Complementary to \Cref{sect5a})}\label{appenG}
	
	\subsection{Predicting $\mathbf{\hat{y}}_{(T+1):(T+q)}$ Given $\mathbf{y}_{1:T}$ and $\left[ X_{(T+1):(T+q)}\right] _{qOm\times p}$ (if $p\geq 1$)}\label{appenG1}
	Suppose we have $q\geq 1,q\in\mathbb{N}$ new time points $T+1,\dots,T+q$ with corresponding time ${\nu}_{T+1},\dots,{\nu}_{T+q}$ (distances standardized to 1 if \texttt{include.time = equalTimeDist = TRUE}). Then under all our modeling frameworks, for Bayesian predictions at ${\nu}_{(T+1):(T+q)}$ given their corresponding covariates matrix $\left[ X_{(T+1):(T+q)}\right] _{qOm\times p}$, if any, we can write the PPD as
	\begin{align}\label{PPDtemp}
		&f\left(\bm{y}_{(T+1):(T+q)}|\bm{y}_{1:T},X_{(T+1):(T+q)}\right)=\int_{\Theta}f\left(\bm{y}_{(T+1):(T+q)}|\Theta,\bm{y}_{1:T},X_{(T+1):(T+q)}\right)\pi\left(\Theta|\bm{y}_{1:T}\right)d\Theta, \nonumber\\ &\text{where }\Theta=
		\begin{cases}
			(\Lambda,\bm{\beta},\bm{\sigma^2},\bm{\eta}_{(T+1):(T+q)},\bm{\eta},{\Upsilon},\psi),&\text{ if \texttt{clustering = FALSE}}\\
			(\bm{\theta},\bm{\xi},\bm{\beta},\bm{\sigma^2},\bm{\eta}_{(T+1):(T+q)},\bm{\eta},{\Upsilon},\psi),&\text{ if \texttt{clustering = TRUE}}
		\end{cases}.
	\end{align}
	$\bm{\eta}$ denotes $\bm{\eta}_{1:T}$. \texttt{clustering = FALSE} refers to the scenario when no spatial clustering mechanism is integrated, i.e., when our basic model in \Cref{appenA1} is adopted, and \texttt{clustering = TRUE} refers to the case
	when a spatial PSBP clustering mechanism is incorporated, i.e., when our model in \Cref{sect2} or \labelcref{sect3} is adopted. We can then partition the integral in \Cref{PPDtemp} into
	\begin{align}
		&\int_{\Theta}\underbrace{f\left(\bm{y}_{(T+1):(T+q)}|\Lambda,\bm{\eta}_{(T+1):(T+q)},\bm{\beta},\bm{\sigma^2},X_{(T+1):(T+q)}\right)}_{T_1}\underbrace{f\left(\bm{\eta}_{(T+1):(T+q)}|\bm{\eta},\Upsilon,\psi\right)}_{T_2}\nonumber\\
		&\qquad\times\underbrace{\pi\left(\Lambda,\bm{\beta},\bm{\sigma^2},\bm{\eta},\Upsilon,\psi|\bm{y}_{1:T}\right)}_{T_3}d\Theta \text{ if \texttt{clustering = FALSE}, and}\label{PPDnewTimeNoCL}\\
		&\int_{\Theta}\underbrace{f\left(\bm{y}_{(T+1):(T+q)}|\bm{\theta},\bm{\xi},\bm{\eta}_{(T+1):(T+q)},\bm{\beta},\bm{\sigma^2},X_{(T+1):(T+q)}\right)}_{T_1}\underbrace{f\left(\bm{\eta}_{(T+1):(T+q)}|\bm{\eta},\Upsilon,\psi\right)}_{T_2}\nonumber\\
		&\qquad\times\underbrace{\pi\left(\bm{\theta},\bm{\xi},\bm{\beta},\bm{\sigma^2},\bm{\eta},\Upsilon,\psi|\bm{y}_{1:T}\right)}_{T_3}d\Theta\text{ if \texttt{clustering = TRUE}}.\label{PPDnewTimeCL}
	\end{align}
	In \Cref{PPDnewTimeNoCL,PPDnewTimeCL}, $T_1$ is the likelihood, $T_3$ is the parameters' posterior distribution obtained from the original model fit's MCMC sampler, and $T_2$ can be calculated by properties of conditional distributions of jointly multivariate normals as follows. Depending on whether a temporal structure is included and whether the distances between adjacent time points are equal, we can get a predicted value $\hat{\bm{\eta}}_{(T+1):(T+q)}$ conditioning on $\bm{\eta}_{1:T},\Upsilon,\psi$: 
	\begin{itemize}
		\item When \texttt{include.time = FALSE}, i.e., when temporal independence is assumed and the temporal correlation structure matrix $H(\psi)$ is set to the identity matrix
		\begin{equation}\label{temppred1}
			f\left({\bm{\eta}}_{(T+1):(T+q)}|\bm{\eta}_{1:T},\Upsilon,\psi\right) = \pi\left({\bm{\eta}}_{(T+1):(T+q)}|\Upsilon,\psi\right) \sim N_{qk}\left(\bm{0},I_{q\times q}\otimes \Upsilon\right). 
		\end{equation}
		\item When \texttt{include.time = TRUE}, i.e., when temporal correlation is taken into consideration
		\begin{align}\label{temppred2}
			&\text{By properties of conditional distributions of jointly multivariate normals, }\nonumber\\
			&{\bm{\eta}}_{(T+1):(T+q)}|\bm{\eta}_{1:T},\Upsilon,\psi \sim N_{qk}\left(\left[H^+_{(T+1):(T+q)}\otimes I_{k\times k}\right]\bm{\eta}_{1:T}, \left[H^*_{(T+1):(T+q)}\otimes {\Upsilon}\right]\right),\\
			&\text{where }H_{(T+1):(T+q)}^+=H_{(T+1):(T+q),(1:T)}H_{(1:T),(1:T)}^{-1}\text{ and }\nonumber\\
			&H_{(T+1):(T+q)}^*=H_{(T+1):(T+q),(T+1):(T+q)}-H_{(T+1):(T+q)}^+H_{(1:T), (T+1):(T+q)}\nonumber\\
			&\qquad\qquad\quad\;=H_{(T+1):(T+q),(T+1):(T+q)}-H_{(T+1):(T+q),(1:T)}H_{(1:T),(1:T)}^{-1}H_{(1:T), (T+1):(T+q)} \nonumber\\
			&\text{with }H_{(1:T),(1:T)}, H_{(T+1):(T+q),(1:T)}, H_{(1:T), (T+1):(T+q)}, H_{(T+1):(T+q),(T+1):(T+q)}\text{ being the}\nonumber\\
			&\text{corresponding }T\times T, q\times T, T\times q, q\times q \text{ sub-matrices of the }(T+q)\times (T+q)\nonumber\\
			&\text{temporal correlation structure matrix }H(\psi)_{1:(T+q)}\text{ for all }(T+q) \text{ time points.}\nonumber
		\end{align}
		\begin{itemize}
			\item When \texttt{equalTimeDist = FALSE}, i.e., when the distances between adjacent time points are unequal, we may only get $H^{-1}_{(1:T),(1:T)}$ by straightforwardly inverting $H_{(1:T),(1:T)}$.
			\item When \texttt{equalTimeDist = TRUE}, i.e., when the time $v_{1},\dots,v_{T},v_{T+1},\dots,v_{T+q}$ are equispaced with distances normalized to 1, we can bypass this matrix inversion by directly specifying $H^{-1}_{(1:T),(1:T)}$ according to our formula presented in \Cref{appenB}.
		\end{itemize}
	\end{itemize}
	
	\subsection{Predicting $\mathbf{\hat{y}}(\mathbf{s}_{(m+1):(m+r)})$ Given $\mathbf{y}(\mathbf{s}_{1:m})$ and $X(\mathbf{s}_{(m+1):(m+r)})_{rTO\times p}$ (if $p\geq 1$)} \label{appenG2}
	\subsubsection{Under the Basic Modeling Framework in \Cref{appenA1}}\label{appenG2a}
	When spatially correlated Gaussian Processes are directly imposed onto the columns $\{\bm{\lambda}_j\}_{j=1}^k$ of the factor loadings matrix $\Lambda_{mO\times k}$, for Bayesian predictions at $r\in\mathbb{N},r\geq 1$ new locations $\bm{s}_{(m+1):(m+r)}$ given their covariates matrix $X(\bm{s}_{(m+1):(m+r)})_{rTO\times p}$, if any, the PPD becomes
	\begin{align}\label{PPDspatNoCL}
		&f(\bm{y}(\bm{s}_{(m+1):(m+r)})|\bm{y}(\bm{s}_{1:m}),X(\bm{s}_{(m+1):(m+r)}))\\
		=&\int_{\Theta}f\left(\bm{y}(\bm{s}_{(m+1):(m+r)})|\Theta,\bm{y}(\bm{s}_{1:m}),X(\bm{s}_{(m+1):(m+r)}) \right)\pi\left(\Theta|\bm{y}(\bm{s}_{1:m})\right)d\Theta,\text{ where}\nonumber\\ \Theta&=\left(\bm{\eta},\bm{\beta},\bm{\sigma^2}(\bm{s}_{(m+1):(m+r)}),{\Lambda}(\bm{s}_{(m+1):(m+r)}),\Lambda,\kappa,\rho
		\right).\nonumber    	
	\end{align} 
	In the above, $\Lambda_{mO\times k}=\left(\bm{\lambda}_1,\dots,\bm{\lambda}_k\right)\text{ denotes } \Lambda\left(\bm{s}_{1:m}\right),\text{ with each }mO\times 1\text{ vector }\bm{\lambda}_j = \bm{\lambda}_j\left(\bm{s}_{1:m}\right),\\ j\in\{1,\dots,k\}$ ordered first by observation type and then spatially (different from the ordering of our original factor loadings matrix), i.e., the first $O$ entries correspond to the first location point, the next $O$ entries correspond to the second location point and so on. ${\Lambda}(\bm{s}_{(m+1):(m+r)})_{rO\times k}=\left(\bm{\lambda}_1\left(\bm{s}_{(m+1):(m+r)}\right) ,\dots,\bm{\lambda}_k\left(\bm{s}_{(m+1):(m+r)}\right)\right)$, where for each $
	j\in\{1,\dots,k\}$, the $rO\times 1$ vector $\bm{\lambda}_j\left(\bm{s}_{(m+1):(m+r)}\right)$ is ordered first by observation type and then spatially. The integral in \Cref{PPDspatNoCL} can then be partitioned into 
	\begin{align}\label{PPDnewLocNoCL}
		&\int_{\Theta}\underbrace{f\left(\bm{y}(\bm{s}_{(m+1):(m+r)})|\Lambda\left(\bm{s}_{(m+1):(m+r)}\right),\bm{\eta},\bm{\beta},\bm{\sigma^2}(\bm{s}_{(m+1):(m+r)}),X(\bm{s}_{(m+1):(m+r)})\right) }_{T_1}\nonumber\\
		&\qquad\times\underbrace{f\left(\Lambda\left(\bm{s}_{(m+1):(m+r)}\right)|{\Lambda},\kappa,\rho\right) }_{T_2}\underbrace{\pi\left(\bm{\eta},\bm{\beta},\Lambda,\kappa,\rho |\bm{y}(\bm{s}_{1:m})\right)}_{T_3}\underbrace{\pi\left(\bm{\sigma^2}(\bm{s}_{(m+1):(m+r)})\right)}_{T_4}d\Theta\\
		&\text{since the posterior density }\pi\left(\bm{\sigma^2}(\bm{s}_{(m+1):(m+r)})|\bm{y}(\bm{s}_{1:m})\right)\text{ equals the prior }\pi\left(\bm{\sigma^2}(\bm{s}_{(m+1):(m+r)})\right).\nonumber
	\end{align}
	
	In \Cref{PPDnewLocNoCL}, $T_1$ is the likelihood, $T_3$ is the parameters' posterior distribution obtained from the original model fit’s MCMC sampler, $T_4$ denotes the prior density for $\bm{\sigma^2}(\bm{s}_{(m+1):(m+r)})=\left( (\sigma^2)^1\left(\bm{s}_{m+1}\right),\dots,(\sigma^2)^O\left(\bm{s}_{m+1}\right),\dots\dots,(\sigma^2)^1\left(\bm{s}_{m+r}\right),\dots,(\sigma^2)^O\left(\bm{s}_{m+r}\right) \right)$ with\\ $(\sigma^2)^o\left(\bm{s}_{m+i_r}\right)\overset{\iid}{\sim}\mathcal{IG}(a,b),\,i_r\in\{1,\dots,r\},\,o\in\{1,\dots,O\}$, and $T_2$ can be written as $$f\left(\Lambda\left(\bm{s}_{(m+1):(m+r)}\right)|{\Lambda}\left(\bm{s}_{1:m}\right) ,\kappa,\rho\right)=\prod_{j=1}^kf\left(\bm{\lambda}_{j}(\bm{s}_{(m+1):(m+r)})|\bm{\lambda}_{j}(\bm{s}_{1:m}),\kappa,\rho\right).$$ Hence, we can get a predicted value $\hat{\Lambda}\left(\bm{s}_{(m+1):(m+r)}\right)$ conditioning on $\Lambda(\bm{s}_{1:m}),\kappa,\rho$ depending on the following 3 scenarios regarding the spatially extended GP prior on $\bm{\lambda}_{j}(\bm{s}_{1:(m+r)})$'s (ordered first by observation type and then spatially here):
	\begin{itemize}
		\item When \texttt{include.space = FALSE}, i.e., when spatial independence is assumed and the corresponding spatially extended full GP prior $\bm{\lambda}_j\left(\bm{s}_{1:(m+r)}\right)\overset{\iid}{\sim} N_{(m+r)O}\left(\bm{0},I_{(m+r)\times (m+r)}\otimes \kappa_{O\times O} \right),\\j\in\{1,\dots,k\}$ is imposed  
		\begin{align}\label{spatpredFixedLnoCL1}
			\text{For any arbitrary }j\in\{1,\dots,k\},\quad &f\left( \bm{\lambda}_{j}(\bm{s}_{(m+1):(m+r)})|\bm{\lambda}_{j}(\bm{s}_{1:m}),\kappa,\rho\right) 
			\nonumber\\ 
			=\;& \pi\left(\bm{\lambda}_{j}(\bm{s}_{(m+1):(m+r)})|\kappa,\rho\right)\sim N_{rO}\left(\bm{0},I_{r\times r}\otimes\kappa\right). 
		\end{align}  	
		\item When \texttt{include.space = TRUE} and \texttt{spatApprox = FALSE}, i.e., when spatial dependence is taken into consideration and the spatially extended full GP prior $\bm{\lambda}_{j}(\bm{s}_{1:(m+r)})|\kappa,\rho\overset{\iid}{\sim}N_{(m+r)O}\left(\bm{0},  F(\rho)_{(m+r)\times (m+r)} \otimes \kappa_{O\times O}\right),\,j\in\{1,\dots,k\}$ is placed
		\begin{align}\label{spatpredFixedLnoCL2}
			&\text{By properties of conditional distributions of jointly multivariate normals, } \forall\text{ arbitrary } j,\nonumber\\
			&\bm{\lambda}_{j}(\bm{s}_{(m+1):(m+r)})|\bm{\lambda}_{j}(\bm{s}_{1:m}),\kappa,\rho \sim N_{rO}\left(B_{\bm{s}_{(m+1):(m+r)}}\bm{\lambda}_{j}(\bm{s}_{1:m}), F_{\bm{s}_{(m+1):(m+r)}}\right),\\
			&\text{where the }rO\times mO\text{ matrix }B_{\bm{s}_{(m+1):(m+r)}}=\left\lbrace F(\rho)_{\bm{s}_{(m+1):(m+r)}, \bm{s}_{1:m}}\left[F(\rho)_{ \bm{s}_{1:m}}\right]^{-1}\right\rbrace \otimes I_{O\times O} \nonumber\\
			&\text{and the }rO\times rO \text{ matrix }F_{\bm{s}_{(m+1):(m+r)}}\nonumber\\ & = \left\lbrace F(\rho)_{\bm{s}_{(m+1):(m+r)}} - F(\rho)_{\bm{s}_{(m+1):(m+r)}, \bm{s}_{1:m}}\left[F(\rho)_{ \bm{s}_{1:m}}\right]^{-1}F(\rho)_{\bm{s}_{1:m}, \bm{s}_{(m+1):(m+r)}}\right\rbrace \otimes \kappa_{O\times O} \nonumber\\
			&\text{with }F(\rho)_{ \bm{s}_{1:m}}, F(\rho)_{\bm{s}_{(m+1):(m+r)}, \bm{s}_{1:m}}, F(\rho)_{\bm{s}_{1:m}, \bm{s}_{(m+1):(m+r)}}, F(\rho)_{\bm{s}_{(m+1):(m+r)}}\text{ being the}\nonumber\\
			&\text{corresponding }m\times m, r\times m, m\times r, r\times r \text{ sub-matrices of }\left[F(\rho)_{\bm{s}_{1:(m+r)}}\right]_{(m+r)\times (m+r)}.\nonumber
		\end{align}
		\item When \texttt{include.space = spatApprox = TRUE}, i.e., when spatial dependence is taken into consideration and the spatially extended NNGP prior $\tilde{\pi}\left( \bm{\lambda}_{j}(\bm{s}_{1:(m+r)})|\kappa,\rho\right)\;\forall\;j$ is placed
		\begin{align}\label{spatpredFixedLnoCL3}
			&\text{Under the spatially extended NNGP prior applied to $\bm{\lambda}_j$'s instead of $\bm{\alpha}_{jl_j}$'s} \nonumber\\
			&\text{and by properties of conditional distributions of jointly multivariate normals,}\nonumber\\ &\text{for any arbitrary fixed }j\in\{1,\dots,k\},\quad 
			\tilde{f}\left(\bm{\lambda}_{j}(\bm{s}_{(m+1):(m+r)})|\bm{\lambda}_{j}(\bm{s}_{1:m}),\kappa,\rho\right)\\
			&= \prod_{i_r=1}^r \tilde{f}\left(\bm{\lambda}_{j}(\bm{s}_{(m+i_r)})|\bm{\lambda}_{j}(\bm{s}_{1:m}),\kappa,\rho\right) = \prod_{i_r=1}^r f\left(\bm{\lambda}_{j}(\bm{s}_{(m+i_r)})|\bm{\lambda}_{j,N(\bm{s}_{(m+i_r)})},\kappa,\rho\right), \nonumber\\
			&\quad\text{where for each }i_r\in\{1,\dots,r\},\nonumber\\ 
			&\quad\bm{\lambda}_{j}(\bm{s}_{(m+i_r)})|\bm{\lambda}_{j,N(\bm{s}_{(m+i_r)})},\kappa,\rho \sim N_O\left(B_{\bm{s}_{(m+i_r)}}\bm{\lambda}_{j,N(\bm{s}_{(m+i_r)})}, F_{\bm{s}_{(m+i_r)}}\right)\text{ with} \nonumber\\
			&\text{the }O\times hO\text{ matrix }B_{\bm{s}_{(m+i_r)}} = \left\lbrace F(\rho)_{\bm{s}_{(m+i_r)}, N(\bm{s}_{(m+i_r)})}\left[F(\rho)_{ N(\bm{s}_{(m+i_r)})}\right]^{-1}\right\rbrace \otimes I_{O\times O} \nonumber\\
			&\text{and the }O\times O \text{ matrix }F_{\bm{s}_{(m+i_r)}}\nonumber\\ & = \left\lbrace F(\rho)_{\bm{s}_{(m+i_r)}} - F(\rho)_{\bm{s}_{(m+i_r)}, N(\bm{s}_{(m+i_r)})}\left[F(\rho)_{ N(\bm{s}_{(m+i_r)})}\right]^{-1}F(\rho)_{N(\bm{s}_{(m+i_r)}), \bm{s}_{(m+i_r)}}\right\rbrace \otimes \kappa_{O\times O}. \nonumber\\
			&\text{with }F(\rho)_{\bm{s}_{(m+i_r)}}=1, F(\rho)_{\bm{s}_{(m+i_r)}, N(\bm{s}_{(m+i_r)})}, F(\rho)_{N(\bm{s}_{(m+i_r)}), \bm{s}_{(m+i_r)}}, F(\rho)_{ N(\bm{s}_{(m+i_r)})}\text{ being the}\nonumber\\
			&\text{corresponding }1\times 1, 1\times h, h\times 1, h\times h \text{ sub-matrices of }
			\left[F(\rho)_{\bm{s}_{1:(m+r)}}\right]_{(m+r)\times (m+r)}.\nonumber
		\end{align}
	\end{itemize} 
	
	\subsubsection{Under Our Modeling Framework in \Cref{sect2}}\label{appenG2b}
	When the prior $\bm{\alpha}_{jl_j}\overset{\iid}{\sim}N_{mO}\left(\bm{0},F(\rho)_{m\times m}\otimes\kappa_{O\times O}\right)$ for all $(j,l_j)\in\{1,\ldots,k\}\times\{1,\ldots,L-1\}$ is assigned, for Bayesian predictions at $r\in\mathbb{N},r\geq 1$ new locations $\bm{s}_{(m+1):(m+r)}$ given their corresponding covariates matrix $X(\bm{s}_{(m+1):(m+r)})_{rTO\times p}$, if any, we can write the PPD as
	\begin{align*}
		&f(\bm{y}(\bm{s}_{(m+1):(m+r)})|\bm{y}(\bm{s}_{1:m}),X(\bm{s}_{(m+1):(m+r)}))\\
		=&\int_{\Theta}f\left(\bm{y}(\bm{s}_{(m+1):(m+r)})|\Theta,\bm{y}(\bm{s}_{1:m}),X(\bm{s}_{(m+1):(m+r)}) \right)\pi\left(\Theta|\bm{y}(\bm{s}_{1:m})\right)d\Theta,
	\end{align*} 
	where $\Theta=\left(\bm{\eta},\bm{\beta},\bm{\sigma^2}(\bm{s}_{(m+1):(m+r)}),\bm{\theta},\bm{\xi}(\bm{s}_{(m+1):(m+r)}),\bm{\alpha}(\bm{s}_{(m+1):(m+r)}),\bm{\alpha},\kappa,\rho
	\right)$ and $\bm{\alpha}$ denotes $\bm{\alpha}(\bm{s}_{1:m})$, and then partition the integral into 
	\begin{align}\label{PPDnewLocAppen}
		&\int_{\Theta}\underbrace{f\left(\bm{y}(\bm{s}_{(m+1):(m+r)})|\bm{\theta},\bm{\xi}(\bm{s}_{(m+1):(m+r)}),\bm{\eta},\bm{\beta},\bm{\sigma^2}(\bm{s}_{(m+1):(m+r)}),X(\bm{s}_{(m+1):(m+r)})\right) }_{T_1}\nonumber\\
		&\qquad\times\underbrace{f\left(\bm{\xi}(\bm{s}_{(m+1):(m+r)})|\bm{\alpha}(\bm{s}_{(m+1):(m+r)})\right) }_{T_2}\underbrace{f\left(\bm{\alpha}(\bm{s}_{(m+1):(m+r)})|\bm{\alpha},\kappa,\rho\right) }_{T_3}\nonumber\\
		&\qquad\times\underbrace{\pi\left(\bm{\eta},\bm{\beta},\bm{\theta},\bm{\alpha},\kappa,\rho |\bm{y}(\bm{s}_{1:m})\right)}_{T_4}\underbrace{\pi\left(\bm{\sigma^2}(\bm{s}_{(m+1):(m+r)})\right)}_{T_5}d\Theta\\
		&\text{since the posterior density }\pi\left(\bm{\sigma^2}(\bm{s}_{(m+1):(m+r)})|\bm{y}(\bm{s}_{1:m})\right)\text{ equals the prior }\pi\left(\bm{\sigma^2}(\bm{s}_{(m+1):(m+r)})\right).\nonumber	
	\end{align}
	In \Cref{PPDnewLocAppen}, $T_1$ is the likelihood, $T_4$ is the parameters' posterior distribution obtained from the original model fit’s MCMC sampler, $T_5$ denotes the prior density for $\bm{\sigma^2}(\bm{s}_{(m+1):(m+r)})=\left( (\sigma^2)^1\left(\bm{s}_{m+1}\right),\dots,(\sigma^2)^O\left(\bm{s}_{m+1}\right),\dots\dots,(\sigma^2)^1\left(\bm{s}_{m+r}\right),\dots,(\sigma^2)^O\left(\bm{s}_{m+r}\right) \right)$ with\\ $(\sigma^2)^o\left(\bm{s}_{m+i_r}\right)\overset{\iid}{\sim}\mathcal{IG}(a,b),i_r\in\{1,\dots,r\},o\in\{1,\dots,O\}$, $T_2$ is the density of the multinomial distribution described in \Cref{sect2}, and $T_3$ can be written as  
	\begin{equation}\label{spatpredCompoFixedL}
		f\left(\bm{\alpha}(\bm{s}_{(m+1):(m+r)})|\bm{\alpha}(\bm{s}_{1:m}),\kappa,\rho\right)=\prod_{j=1}^k\prod_{l_j=1}^{L-1} f\left(\bm{\alpha}_{jl_j}(\bm{s}_{(m+1):(m+r)})|\bm{\alpha}_{jl_j}(\bm{s}_{1:m}),\kappa,\rho\right). 
	\end{equation}
	Hence, we can directly get a predicted value $\hat{\bm{\alpha}}(\bm{s}_{(m+1):(m+r)})$ conditioning on $\bm{\alpha}(\bm{s}_{1:m}),\kappa,\rho$ depending on the following 3 scenarios regarding the spatially extended GP prior on $\bm{\alpha}_{jl_j}(\bm{s}_{1:(m+r)})$'s:
	\begin{itemize}
		\item When \texttt{include.space = FALSE}, i.e., when spatial independence is assumed and the spatial neighborhood structure matrix $F(\rho)$ is set to the identity matrix
		\begin{align}\label{spatpredFixedLCL1appen}
			&\forall\; (j,l_j)\in\{1,\ldots,k\}\times\{1,\ldots,L-1\},\\ 
			& f\left(\bm{\alpha}_{jl_j}(\bm{s}_{(m+1):(m+r)})|\bm{\alpha}_{jl_j}(\bm{s}_{1:m}),\kappa,\rho\right) = \pi\left(\bm{\alpha}_{jl_j}(\bm{s}_{(m+1):(m+r)})|\kappa,\rho\right)\sim N_{rO}\left(\bm{0},I_{r\times r}\otimes\kappa\right). \nonumber
		\end{align}
		\item When \texttt{include.space = TRUE} and \texttt{spatApprox = FALSE}, i.e., when spatial dependence is taken into consideration and the spatially extended full GP prior $\pi(\bm{\alpha}_{jl_j}(\bm{s}_{1:(m+r)})|\kappa,\rho)=N_{(m+r)O}\left(\bm{0},F(\rho)_{(m+r)\times (m+r)}\otimes\kappa_{O\times O} \right)\;\forall\;(j,l_j)\in\{1,\ldots,k\}\times\{1,\ldots,L-1\}$ is placed
		\begin{align}\label{spatpredFixedLCL2appen}
			&\text{By properties of conditional distributions of jointly multivariate normals, } \forall\text{ arbitrary } (j,l_j),\nonumber\\
			&\bm{\alpha}_{jl_j}(\bm{s}_{(m+1):(m+r)})|\bm{\alpha}_{jl_j}(\bm{s}_{1:m}),\kappa,\rho \sim N_{rO}\left(B_{\bm{s}_{(m+1):(m+r)}}\bm{\alpha}_{jl_j}(\bm{s}_{1:m}), F_{\bm{s}_{(m+1):(m+r)}}\right),\\
			&\text{where the }rO\times mO\text{ matrix }B_{\bm{s}_{(m+1):(m+r)}}=\left\lbrace F(\rho)_{\bm{s}_{(m+1):(m+r)}, \bm{s}_{1:m}}\left[F(\rho)_{ \bm{s}_{1:m}}\right]^{-1}\right\rbrace \otimes I_{O\times O} \nonumber\\
			&\text{and the }rO\times rO \text{ matrix }F_{\bm{s}_{(m+1):(m+r)}}\nonumber\\ & = \left\lbrace F(\rho)_{\bm{s}_{(m+1):(m+r)}} - F(\rho)_{\bm{s}_{(m+1):(m+r)}, \bm{s}_{1:m}}\left[F(\rho)_{ \bm{s}_{1:m}}\right]^{-1}F(\rho)_{\bm{s}_{1:m}, \bm{s}_{(m+1):(m+r)}}\right\rbrace \otimes \kappa_{O\times O} \nonumber\\
			&\text{with }F(\rho)_{ \bm{s}_{1:m}}, F(\rho)_{\bm{s}_{(m+1):(m+r)}, \bm{s}_{1:m}}, F(\rho)_{\bm{s}_{1:m}, \bm{s}_{(m+1):(m+r)}}, F(\rho)_{\bm{s}_{(m+1):(m+r)}}\text{ being the}\nonumber\\
			&\text{corresponding }m\times m, r\times m, m\times r, r\times r \text{ sub-matrices of }\left[F(\rho)_{\bm{s}_{1:(m+r)}}\right]_{(m+r)\times (m+r)}.\nonumber
		\end{align}
		\item When \texttt{include.space = spatApprox = TRUE}, i.e., when spatial dependence is taken into consideration and the spatially extended NNGP prior $\tilde{\pi}(\bm{\alpha}_{jl_j}(\bm{s}_{1:(m+r)})|\kappa,\rho)\;\forall\;(j,l_j)$ is placed, under the spatially extended NNGP prior and by properties of conditional distributions of jointly multivariate normals, for any arbitrary fixed $(j,l_j)$,
		\begin{align}\label{spatpredFixedLCL3appen}
			&			\tilde{f}\left(\bm{\alpha}_{jl_j}(\bm{s}_{(m+1):(m+r)})|\bm{\alpha}_{jl_j}(\bm{s}_{1:m}),\kappa,\rho\right)\\
			&= \prod_{i_r=1}^r \tilde{f}\left(\bm{\alpha}_{jl_j}(\bm{s}_{(m+i_r)})|\bm{\alpha}_{jl_j}(\bm{s}_{1:m}),\kappa,\rho\right) = \prod_{i_r=1}^r f\left(\bm{\alpha}_{jl_j}(\bm{s}_{(m+i_r)})|\bm{\alpha}_{jl_j,N(\bm{s}_{(m+i_r)})},\kappa,\rho\right), \nonumber\\
			&\quad\text{where for each }i_r\in\{1,\dots,r\},\nonumber\\ 
			&\quad\bm{\alpha}_{jl_j}(\bm{s}_{(m+i_r)})|\bm{\alpha}_{jl_j,N(\bm{s}_{(m+i_r)})},\kappa,\rho \sim N_O\left(B_{\bm{s}_{(m+i_r)}}\bm{\alpha}_{jl_j,N(\bm{s}_{(m+i_r)})}, F_{\bm{s}_{(m+i_r)}}\right)\text{ with} \nonumber\\
			&\text{the }O\times hO\text{ matrix }B_{\bm{s}_{(m+i_r)}}=\left\lbrace F(\rho)_{\bm{s}_{(m+i_r)}, N(\bm{s}_{(m+i_r)})}\left[F(\rho)_{ N(\bm{s}_{(m+i_r)})}\right]^{-1}\right\rbrace \otimes I_{O\times O} \nonumber\\
			&\text{and the }O\times O \text{ matrix }F_{\bm{s}_{(m+i_r)}}\nonumber\\ & =\left\lbrace F(\rho)_{\bm{s}_{(m+i_r)}} - F(\rho)_{\bm{s}_{(m+i_r)}, N(\bm{s}_{(m+i_r)})}\left[F(\rho)_{ N(\bm{s}_{(m+i_r)})}\right]^{-1}F(\rho)_{N(\bm{s}_{(m+i_r)}), \bm{s}_{(m+i_r)}}\right\rbrace \otimes \kappa_{O\times O}. \nonumber\\
			&F(\rho)_{\bm{s}_{(m+i_r)}}=1, F(\rho)_{\bm{s}_{(m+i_r)}, N(\bm{s}_{(m+i_r)})}, F(\rho)_{N(\bm{s}_{(m+i_r)}), \bm{s}_{(m+i_r)}}, F(\rho)_{ N(\bm{s}_{(m+i_r)})}\text{ denote the}\nonumber\\
			&\text{corresponding }1\times 1, 1\times h, h\times 1, h\times h \text{ sub-matrices of }
			\left[F(\rho)_{\bm{s}_{1:(m+r)}}\right]_{(m+r)\times (m+r)}.\nonumber
		\end{align}
	\end{itemize}

	\section{Scalability Improvements Corresponding to Our 3 Novelties -- A Summary of Computational Complexity and Memory}\label{appenH}
	\par Our three novelties for enhancing scalability are \textbf{slice sampling}, \textbf{spatial NNGP prior}, and \textbf{sequential updating algorithms}. We first list some important points to note. 
	\begin{itemize}
		\item $O\in\mathbb{N}$ is assumed very small; $h,\,k,\,p\ll m$; $1\leq L$ or $L_j$'s $\leq m$; $T \leq m$.
		\item Slice sampling is useful when $L$ has to be at least moderately large. Moreover, it resolves intimidating storage concerns regarding $\theta_{jl_j}$'s, $\bm{\alpha}_{jl_j}$'s, and $\bm{w}_{jl_j}$'s and thus enables practicable implementation of spatial prediction and clustering when $m$ and $L$ are large.
		\item Spatial NNGP prior and sequential updates are helpful for a large $m$. Sequential updating algorithms can only be adopted when our spatial NNGP prior is specified. Hence, when slice sampling is not adopted, we have 3 methods, which we denote as \texttt{fullGPfixedL}, \texttt{NNGPblockFixedL}, and \texttt{NNGPsequenFixedL}.
		\item When slice sampling is adopted and $m$ is at least moderately large, we may only also specify our spatial NNGP prior with sequential updates for a computationally feasible model, which we shall denote as \texttt{NNGPsequenVaryLj}, as multivariate rejection sampling from high-dimensional truncated $mO$-variate normal distributions of the $\bm{\alpha}_{jl_j}$'s (\Cref{postAlphaVaryLj12,blockAlphaVaryLj}) is unbearably slow.  
	\end{itemize}
	\par Hence, we only implement the four methods \texttt{fullGPfixedL}, \texttt{NNGPblockFixedL}, \texttt{NNGPsequenFixedL}, and \texttt{NNGPsequenVaryLj} for our simulation experiments. To verify what we have obtained theoretically regarding scalability, we may compare 
	\begin{itemize}
		\item \texttt{fullGPfixedL} and \texttt{NNGPblockFixedL} for effects caused by spatial NNGP prior (without slice sampling); 
		\item \texttt{NNGPblockFixedL} and \texttt{NNGPsequenFixedL} for effects caused by sequential updates (without slice sampling);
		\item \texttt{NNGPsequenFixedL} and \texttt{NNGPsequenVaryLj} for effects caused by slice sampling (with our spatial NNGP prior and sequential updating algorithms adopted).
	\end{itemize}
	\par We now elaborate on \textbf{theoretical} scalability improvements attributable to each of our three novelties. Computational complexity reduction for certain involved major steps in posterior sampling and spatial prediction, as well as storage requirements alleviation for some concerned key quantities in main model fitting and spatial prediction \& clustering, are considered measures of scalability improvements. {All computational complexity orders in the subsections below are for only one typical MCMC iteration.}
	
	\subsection{Flexibility, Computation, and Storage Improvements Brought About by Our {Slice Sampling Approach} in \Cref{sect3} with Spatial NNGP \& Sequential Updates}\label{appenH1}
	\par We assume that $\Ocal(T)\leq \Ocal(m)$. Recall that $1\leq L$ or $L_j$'s $\leq m$ and that we have assumed $O, h, k, p\ll m$. It is likely restrictive to pre-specify a fixed common $L\in\mathbb{N},\,L>1$, as the numbers of spatial mixtures are unknown and may well differ across the $k$ latent factors. Moreover, an at least moderately large $L$ would typically be required to guarantee satisfactory model fitting, resulting in computational burdens and storage issues. Our slice sampling approach (\Cref{sect3}) not only brings about desired modeling flexibility and adequacy but also decently addresses the involved scalability issues (\Cref{{sectH1summaryTable}}) by ensuring non-increasing $L_j$ estimates through the MCMC iterations.
	\par Let $N\in\mathbb{N}$ be the number of kept post-burn-in MCMC iterations. We denote the corresponding estimated number(s) of spatial mixture components as $L_j^{(t_i)}$'s, $i\in\{1,\ldots,N\},\:j\in\{1,\ldots,k\}$. In \Cref{sectH1summaryTable}, overall storage orders correspond to all kept MCMC iterations, whereas overall computational complexity orders are only for one typical MCMC iteration.
	
	\begin{table}[h]
		{
			\begin{adjustwidth}{-2.4cm}{-1cm}
				\begin{center}
					\scalebox{0.608}{\begin{tabular}{|*{6}{c|}}
							\hline
							\multicolumn{2}{|c|}{\multirow{2}{*}{\textbf{Domain of Scalability Improvements}}} & \multicolumn{2}{|c|}{\textbf{Computational Acceleration}} &  \multicolumn{2}{|c|}{\textbf{Storage Alleviation}} \\
							\cline{3-6}
							\multicolumn{2}{|c|}{} & \textbf{Posterior Sampling} & \multicolumn{2}{|c|}{\textbf{Spatial Prediction} at $r\in\mathbb{N}\:(r<m)$ new locations} &  \textbf{Spatial Clustering} \\
							\hline
							\multicolumn{2}{|c|}{\textbf{Key Parameters / Quantities Involved}} & $u_j^o\left(\bm{s}_{i}\right)$'s / $z_{jl_j}^o\left(\bm{s}_{i}\right)$'s, ${\xi}_j^o\left(\bm{s}_{i}\right)$'s, $\bm{\alpha}_{jl_j}$'s, $\theta_{jl_j}$'s, $\delta_{1:k}$, $\kappa$, $\rho$  & $\bm{\alpha}_{jl_j}\big(\bm{s}_{(m+1):(m+r)}\big)$'s & $\bm{\alpha}_{jl_j}$'s and $\theta_{jl_j}$'s& $\bm{w}_{jl_j}$'s \\
							\hline
							{\textbf{{Overall Standard Computational}}} & \textbf{Before} & $\bm{\Ocal\left(kLmO\left\lbrace h^2O + T(p+k)\right \rbrace + mh^3\right)}$  & $\bm{\Ocal\left(r\big[h^3+k(L-1)O(h+O)\big]+O^3\right)}$ &  \multicolumn{2}{|c|}{$\bm{\Ocal(NmOkL)}$} \\
							\cline{2-6}
							\multirow{2}{*}{{\textbf{{Complexity or Storage Order}}}} & \multirow{2}{*}{\textbf{After}}  & $\bm{\Ocal\left(mO\sum_{j=1}^k L_j \left\lbrace h^2O + T(p+k)\right \rbrace + mh^3\right)}$ & \multirow{2}{*}{$\bm{\Ocal\left(r\left[h^3+\sum_{j=1}^{k}(L_j-1)\cdot O(h+O)\right]+O^3\right)}$} &  \multicolumn{2}{|c|}{\multirow{2}{*}{$\bm{\Ocal\left(mO\sum_{i=1}^N\sum_{j=1}^kL_j^{(t_i)}\right)}$}}\\
							& & $\bm{+ \Ocal\left(O^2\sum_{j=1}^k\sum_{l_j=1}^{L_j-1}\sum_{i=1}^m [k_{jl_ji}^* / O +n_{jl_ji}^*]\right)}$ &  &  \multicolumn{2}{|c|}{} \\
							\hline
							\multirow{2}{*}{\textbf{References}} & \textbf{Section(s)} & \multirow{2}{*}{\Cref{sect3} and \Cref{appenA2,appenA3}} &  \multicolumn{2}{|c|}{\Cref{sect5a} and \Cref{appenG2b}}  &  \multirow{2}{*}{\Cref{sect5b}}\\
							\cline{2-2}\cline{4-5}
							\multirow{2}{*}{} & \textbf{Equations}  &  & \multicolumn{2}{|c|}{\Cref{spatpredFixedLCL3,spatpredFixedLCL3appen}}  & \\
							\hline
					\end{tabular}}
				\end{center}
			\end{adjustwidth}
			\caption{A theoretical summary table of posterior sampling \& spatial prediction computation acceleration and spatial prediction \& clustering storage alleviation brought about by our slice sampling approach in \Cref{sect3} with our spatial NNGP prior and sequential updates (\Cref{sect4}). Our simulation experiments comparing \texttt{NNGPsequenFixedL} and \texttt{NNGPsequenVaryLj} should lead to results corresponding well to the table in terms of main model fitting time for the involved parameters and spatial prediction time for $\bm{\alpha}_{jl_j}(\bm{s}_{(m+1):(m+r)})$'s. }
			\label{sectH1summaryTable}	
		}
	\end{table}
	
	\begin{center}
		\textbf{Detailed Posterior Sampling Computational Complexity Order Derivations}
	\end{center}
	\begin{center}
		\textbf{\small\color{black}{$(a)$ Without Slice Sampling (Before)}}
	\end{center}
	\begin{align}\label{noSliceSampling}
		&\Ocal\left(\left\lbrace k(L-1)O^2[h^2+h+1]+h^3\right\rbrace m\right) + \Ocal\big(m\big[k(L-1)hO(h+O)+O^3\big]\big) + \Ocal(kLmOT(p+k))\nonumber\\
		=\;& \bm{\Ocal\left(kLmO\left\lbrace h^2O + T(p+k)\right \rbrace + mh^3\right)} =\Ocal(mT)\leq \Ocal(m^2) \text{ if }L\ll m
	\end{align}
	\begin{center}
		\textbf{\small\color{black}{$(b)$ With Slice Sampling (After)}}
	\end{center}
	\begin{align}\label{withSliceSampling}
		&\Ocal\left(\left\lbrace\sum_{j=1}^k(L_j-1)\cdot O^2[h^2+h+1]+h^3\right\rbrace m\right) + \Ocal\left(\sum_{j=1}^k L_j\cdot mOT(p+k)\right)\nonumber\\
		& + \Ocal\left(O\left\lbrace \sum_{j=1}^k \Bigg[(L_j-1)mh(h+O) + \sum_{l_j=1}^{L_j-1}\sum_{i=1}^m \left(k_{jl_ji}^*+n_{jl_ji}^*\cdot O\right)\Bigg] + O^2 \right\rbrace  \right) \nonumber\\
		=\;& \bm{\Ocal\left(mO\sum_{j=1}^k L_j \left\lbrace h^2O + T(p+k)\right \rbrace + mh^3 + O^2\sum_{j=1}^k\sum_{l_j=1}^{L_j-1}\sum_{i=1}^m \left[\frac{k_{jl_ji}^*}{O}+n_{jl_ji}^*\right]\right)} 
	\end{align}
	
	\par With our slicing sampling approach (\Cref{sect3}) adopted on top of our spatial NNGP prior (\Cref{sect4}) coupled with sequential updates (\Cref{sect4d}), rejection sampling from low-dimensional truncated $O$-variate normal distributions would have adequate acceptance rates and thus small required numbers $n_{jl_ji}^*$'s of draws from $N_O\left(V_{\bm{s}_i}\bm{\mu}_{jl_j,\bm{s}_i}, V_{\bm{s}_i} \right)$'s (\Cref{{seqAlphaIndPostDens2}} in \Cref{sect4d}) per sample of $\bm{\alpha}_{jl_j}(\bm{s}_i)$'s. If $\Ocal\left(\underset{(j,l_j,i)}{\max}\;\left[\frac{k_{jl_ji}^*}{O}+n_{jl_ji}^*\right]\right) \leq \Ocal(T)$ and $\underset{j\in\{1,\ldots,k\}}{\max}L_j\ll m$, then $$\Ocal\left(mO\sum_{j=1}^k L_j \left\lbrace h^2O + T(p+k)\right \rbrace + mh^3 + O^2\sum_{j=1}^k\sum_{l_j=1}^{L_j-1}\sum_{i=1}^m \left[\frac{k_{jl_ji}^*}{O}+n_{jl_ji}^*\right]\right) = \Ocal(mT)\leq \Ocal(m^2). $$
	
	\subsection{Posterior Sampling Computational Acceleration \& Storage Alleviation and Faster Spatial Prediction Brought About by Our {Spatial NNGP Prior} in \Cref{sect4} with or without Slice Sampling}\label{appenH2}
	
	\begin{table}[h]
		{
			\begin{adjustwidth}{-2.45cm}{-0.4cm}
				\begin{center}
					\scalebox{0.7}{\begin{tabular}{|*{5}{c|}}
							\hline
							\multicolumn{2}{|c|}{\multirow{2}{*}{\textbf{Domain of Scalability Improvements}}} & \multicolumn{2}{|c|}{\textbf{Posterior Sampling}} &  \textbf{Spatial Prediction} \\
							\cline{3-5}
							\multicolumn{2}{|c|}{} & \textbf{Acceleration} & \textbf{Storage Alleviation} &  \textbf{Acceleration} \\
							\hline
							\multicolumn{2}{|c|}{\textbf{Parameters Involved}} & $\kappa$ and $\rho$ & $\kappa$ and $\rho$ (and $\bm{\alpha}_{jl_j}$'s) & $\bm{\alpha}_{jl_j}\big(\bm{s}_{(m+1):(m+r)}\big)\big|\bm{\alpha}_{jl_j}\big(\bm{s}_{1:m}\big),\kappa,\rho$'s for all $(j,l_j)$ \\
							\hline
							\multicolumn{2}{|c|}{\textbf{Key Quantities Concerned}} & $F(\rho)^{-1}$ and $\det(F(\rho))$ & $F(\rho)$ and $F(\rho)^{-1}$  & \scriptsize{conditional normal variances $\&$ their Cholesky factors and mean vectors} \\
							\hline
							{\textbf{{Standard Computational Complexity}}} & \textbf{Before} & $\Ocal(m^3)$ & \bm{$\Ocal(m^2)$} & $\Ocal\big(rm^2\big) + \Ocal\big(r^3\big) + \Ocal\big(O^3\big)+ \Ocal\left(\sum_{j=1}^{k}(L_j-1)\cdot rmO\right)$\\
							\cline{2-5}
							{\textbf{{or Storage Order for the Key Quantities}}} & \textbf{After} & $\Ocal(mh^3)$ & \bm{$\Ocal\big(mh^2\big)$} & $\Ocal\big(rh^3\big) + \Ocal(r) + \Ocal\big(O^3\big) + \Ocal\left(\sum_{j=1}^{k}(L_j-1)\cdot r\cdot hO\right)$\\
							\hline
							{\textbf{{Overall Standard Order of Computational}}} & \textbf{Before} & \multicolumn{2}{|c|}{$\bm{\Ocal(m^3)}$} & $\bm{\Ocal\big(r\left[m^2+r^2+k(L-1)mO\right]\big)}$\\
							\cline{2-5}
							{\textbf{{Complexity Without Slice Sampling}}} & \textbf{After} & \multicolumn{2}{|c|}{$\bm{\Ocal\left(\left\lbrace k(L-1)O^2[h^2+h+1]+h^3\right\rbrace m\right)}$} & $\bm{\Ocal\left(r\big[h^3+k(L-1)O(h+O)\big]+O^3\right)}$\\
							\hline
							{\textbf{\color{black}{Overall Standard Order of Computational}}} & \textbf{Before} & \multicolumn{2}{|c|}{$\bm{\Ocal(m^3)}$} & $\bm{\Ocal\left(r\left[m^2+r^2+\sum_{j=1}^{k}(L_j-1)\cdot mO\right]\right)}$\\
							\cline{2-5}
							{\textbf{\color{black}{Complexity With Slice Sampling}}} & \textbf{After} & \multicolumn{2}{|c|}{$\bm{\Ocal\left(\left\lbrace\sum_{j=1}^k(L_j-1)\cdot O^2[h^2+h+1]+h^3\right\rbrace m\right)}$ } & $\bm{\Ocal\left(r\left[h^3+\sum_{j=1}^{k}(L_j-1)\cdot O(h+O)\right]+O^3\right)}$\\
							\hline
							\multirow{2}{*}{\textbf{References}} & \textbf{Section(s)} & \Cref{sect4c} & \Cref{sect4b,sect4c} & \Cref{sect5a} and \Cref{appenG2b} \\
							\cline{2-5}
							\multirow{2}{*}{} & \textbf{Equations} & \multicolumn{2}{|c|}{\Cref{postKappaFixedLCL,postKappa,postRhoFixedLCL,postRhoVaryLj}} & \Cref{spatpredFixedLCL2,spatpredFixedLCL3,spatpredFixedLCL2appen,spatpredFixedLCL3appen}\\
							\hline
					\end{tabular}}
				\end{center}
			\end{adjustwidth}
			\caption{A theoretical summary table of posterior sampling computation acceleration \& storage alleviation and spatial prediction acceleration brought about by our spatial NNGP prior in \Cref{sect4} with or without slice sampling (\Cref{sect3}). Our simulation experiments comparing \texttt{fullGPfixedL} and \texttt{NNGPblockFixedL} should lead to results corresponding well to the table in terms of main model fitting time for $\kappa$ and $\rho$ and spatial prediction time for $\bm{\alpha}_{jl_j}(\bm{s}_{(m+1):(m+r)})$'s. }
			\label{sectH2summaryTable}	
		}
	\end{table}
	Suppose we have a large $m$. Recall that $1\leq L$ or $L_j$'s $\leq m$ and that we have assumed $O, h, k\ll m$. We also assume $rO<m$ for spatial prediction at $r\in\mathbb{N}$ new locations $\bm{s}_{(m+1):(m+r)}$. Note that $\left[ F(\rho)\otimes\kappa\right]^{-1}= F(\rho)^{-1}\otimes \kappa^{-1}$ and that $\det\left( F(\rho)\otimes\kappa\right)=\det\left( F(\rho) \right)^O\cdot \det\left( \kappa\right)^m$.
	
	\begin{center}
		\textbf{Detailed Derivations Regarding the Overall Computational Complexity Orders}
	\end{center}

	\begin{center}
		\textbf{\color{black}{1. Computational Complexity for the Posterior Sampling Steps of $\rho$ and $\kappa_{O\times O}$}}
	\end{center}
	
	\begin{center}
		\textbf{\small\color{black}{$(a)$ The Full Spatial GP Prior Adopted (Before)}}
	\end{center}
	\begin{center}
		\footnotesize{\color{black}{\textbf{Without Slice Sampling}}}
		\begin{align}\label{postsamplingGPfixedL}
			\left[\Ocal\big(m^3\big) + \Ocal\big(k(L-1)(Om)^2\big)\right] + \Ocal\big(k(L-1)\cdot Om^2\big) = \bm{\Ocal\big(m^3\big)}
		\end{align}
	\end{center}
	\begin{center}
		\textbf{\footnotesize\color{black}{With Slice Sampling}}
		\begin{align}\label{postsamplingGPvaryLj}
			\left[\Ocal\big(m^3\big) + \Ocal\left(\sum_{j=1}^k(L_j-1)\cdot (Om)^2\right)\right] + \Ocal\left(\sum_{j=1}^k(L_j-1)\cdot Om^2\right) = \bm{\Ocal\big(m^3\big)}
		\end{align}
	\end{center}
	
	\begin{center}
		\textbf{\small\color{black}{$(b)$ Our Spatial NNGP Prior in \Cref{sect4} Adopted (After)}}
	\end{center}
	\begin{center}
		\footnotesize{\color{black}{\textbf{Without Slice Sampling}}}
		\begin{align}\label{postsamplingNNGPfixedL}
			&\left[\left\lbrace \Ocal\big(mh^3\big)+\Ocal\big(O^3\big)+\Ocal\big(m(h+1)^2\big)+\Ocal(m)\right\rbrace+\Ocal\big(k(L-1)m[h^2+h+1]O^2\big)\right] \nonumber\\
			&  + \Ocal\big(k(L-1)\cdot Om[h^2+h+1]\big)\nonumber \\
			= \;&\bm{\Ocal\big(\left\lbrace k(L-1)O^2[h^2+h+1]+h^3\right\rbrace m\big)} = \Ocal(m)\text{ if }L\ll m
		\end{align}
	\end{center}
	\begin{center}
		\textbf{\footnotesize\color{black}{With Slice Sampling}}
		\begin{align}\label{postsamplingNNGPvaryLj}       &\left[\left\lbrace \Ocal\big(mh^3\big)+\Ocal\big(O^3\big)+\Ocal\big(m(h+1)^2\big)+\Ocal(m)\right\rbrace +\Ocal\left(\sum_{j=1}^k(L_j-1)\cdot m[h^2+h+1]O^2\right)\right] \nonumber\\
			&+ \Ocal\left(\sum_{j=1}^k(L_j-1)\cdot Om[h^2+h+1]\right)\nonumber\\
			=\; &\bm{\Ocal\left(\left\lbrace\sum_{j=1}^k(L_j-1)\cdot O^2[h^2+h+1]+h^3\right\rbrace m\right)} = \Ocal(m) \text{ if } \underset{j\in\{1,\ldots,k\}}{\max}L_j\ll m
		\end{align}
	\end{center}

	\begin{center}
		\textbf{\color{black}{2. Computational Complexity for the Spatial Prediction Step Obtaining $\hat{{\alpha}}_{jl_j}\big(\bm{s}_{(m+1):(m+r)}\big)$ for all $(j,l_j)$}}
	\end{center}
	
	\begin{center}
		\textbf{\small\color{black}{$(a)$ The Full Spatial GP Prior Adopted (Before)}}
	\end{center}
	\begin{center}
		\footnotesize{\color{black}{\textbf{Without Slice Sampling}}}
		\begin{equation}\label{spatpredGPfixedL}
			\Ocal\big(rm^2\big) + \Ocal\big(r^3\big) + \Ocal\big(O^3\big) + \Ocal(k(L-1)\cdot rmO) + \Ocal\big(k(L-1)\cdot (rO)^2\big)= \bm{\Ocal\big(r\left[m^2+r^2+k(L-1)mO\right]\big)}
		\end{equation}
	\end{center}
	\begin{center}
		\textbf{\footnotesize\color{black}{With Slice Sampling}}
		\begin{align}\label{spatpredGPvaryLj}
			&\Ocal\big(rm^2\big) + \Ocal\big(r^3\big) + \Ocal\big(O^3\big)+ \Ocal\left(\sum_{j=1}^{k}(L_j-1)\cdot rmO\right) + \Ocal\left(\sum_{j=1}^{k}(L_j-1)\cdot (rO)^2\right) \nonumber\\
			= \; &\bm{\Ocal\left(r\left[m^2+r^2+\sum_{j=1}^{k}(L_j-1)\cdot mO\right]\right)}
		\end{align}
	\end{center}
	
	\begin{center}
		\textbf{\small\color{black}{$(b)$ Our Spatial NNGP Prior in \Cref{sect4} Adopted (After)}}
	\end{center}
	\begin{center}
		\footnotesize{\color{black}{\textbf{Without Slice Sampling}}}
		\begin{equation}\label{spatpredNNGPfixedL}
			\Ocal\big(rh^3\big)+\Ocal(r)+\Ocal\big(O^3\big)+\Ocal(k(L-1)\cdot r\cdot hO)  + \Ocal\big(k(L-1)\cdot r\cdot O^2\big)=\bm{\Ocal\left(r\big[h^3+k(L-1)O(h+O)\big]+O^3\right)} 
		\end{equation}
	\end{center}
	\begin{center}
		\textbf{\footnotesize\color{black}{With Slice Sampling}}
		\begin{align}\label{spatpredNNGPvaryLj}
			&\Ocal\big(rh^3\big) + \Ocal(r) + \Ocal\big(O^3\big) + \Ocal\left(\sum_{j=1}^{k}(L_j-1)\cdot r\cdot hO\right) + \Ocal\left(\sum_{j=1}^{k}(L_j-1)\cdot r\cdot O^2\right)\nonumber \\
			= \;& \bm{\Ocal\left(r\left[h^3+\sum_{j=1}^{k}(L_j-1)\cdot O(h+O)\right]+O^3\right)} 
		\end{align}
	\end{center}

	\subsection{Posterior Sampling Computational Acceleration Brought About by Our {Sequential Updating Methods} in \Cref{sect4d} (with spatial NNGP)}\label{appenH3}
	\begin{table}[ph!]
		{
			\begin{center}
				\begin{adjustwidth}{-2.45cm}{-0.5cm}
					\scalebox{0.655}{\begin{tabular}{|*{4}{c|}}
							\hline
							\multicolumn{2}{|c|}{\multirow{2}{*}{\textbf{Place of Scalability Improvements}}} & \multicolumn{2}{|c|}{\textbf{{Computational Acceleration for the Posterior Sampling Step of $\bm{\alpha}_{jl_j}$'s for all $(j,l_j)$} }} \\
							\cline{3-4}
							\multicolumn{2}{|c|}{} & \color{black}\textbf{Without Slice Sampling} \color{black} & \textbf{With Slice Sampling} \\
							\hline
							\multicolumn{2}{|c|}{\textbf{Key Quantities Concerned}} & $\big[ I_{mO}+\tilde F(\rho)^{-1}\otimes\kappa^{-1} \big]^{-1}$, $\chol\big( \big[ I_{mO}+\tilde F(\rho)^{-1}\otimes\kappa^{-1} \big]^{-1} \big) $ & $\chol\big(\tilde{F}(\rho)\otimes\kappa\big)$ and $n_{jl_j}$'s for $mO-$variate rejection sampling\\
							\hline
							{\textbf{{Overall 
										Order of Standard}}} & \textbf{Before} & $\Ocal(m^3)+\Ocal(k(L-1)\cdot (mO)^2)=\bm{\Ocal(m^3)}$ & $\bm{\Ocal(m^3)+\Ocal\left(\sum_{j=1}^k\sum_{l_j=1}^{L_j-1}\big[k_{jl_j}\cdot mO + n_{jl_j}\cdot (mO)^2\big]\right) > \Ocal(m^3)} $\\
							\cline{2-4}
							{\textbf{{Computational Complexity}}} & \textbf{After} & $\bm{\Ocal\big(m\big[k(L-1)hO(h+O)+O^3\big]\big)} = \Ocal(m)$ if $L\ll m$& $\bm{\Ocal\left(O\left\lbrace \sum\limits_{j=1}^k \Bigg[(L_j-1)mh(h+O) + \sum\limits_{l_j=1}^{L_j-1}\sum\limits_{i=1}^m \left(k_{jl_ji}^*+n_{jl_ji}^*\cdot O\right)\Bigg] + O^2 \right\rbrace  \right)} $\\
							\hline
							\multirow{2}{*}{\textbf{References}} & \textbf{Section(s)} & \multicolumn{2}{|c|}{\Cref{sect4d}} \\
							\cline{2-4}
							\multirow{2}{*}{} & \textbf{Equations} & \Cref{blockAlphaFixedL,seqAlphaIndPostDensFixedL1} & \Cref{blockAlphaVaryLj,seqAlphaIndPostDens2}\\
							\hline
					\end{tabular}}
				\end{adjustwidth}
				\caption{A theoretical summary table of posterior sampling computation acceleration further brought about by our sequential updating methods in \Cref{sect4d} on top of imposing our spatial NNGP prior (\Cref{sect4}) with or without slice sampling (\Cref{sect3}). Our simulation experiments comparing \texttt{NNGPblockFixedL} and \texttt{NNGPsequenFixedL} should lead to results corresponding well to the table in terms of main model fitting time for $\bm{\alpha}_{jl_j}$'s.}
				\label{sectH3summaryTable}	
			\end{center}
		}		
	\end{table}
    Suppose we have a large $m$. Recall that $1\leq L$ or $L_j$'s $\leq m$ and that we have assumed $O, h, k\ll m$. When slice sampling (\Cref{sect3}) is adopted without our sequential updating method, rejection sampling from a high-dimensional truncated $mO-$variate normal distribution (\Cref{blockAlphaVaryLj}) gets unbearably slow for feasible computation when $m$ gets large, as a very small acceptance rate leads to a huge number $n_{jl_j}\in \mathbb{N}$ of draws from $N_{mO}\left(\bm{0}, \tilde{F}(\rho)\otimes\kappa\right)$ required per sample of $\bm{\alpha}_{jl_j}$ for each $(j,l_j)$. When $m$ is large, we typically have $n_{jl_j}(m) \gg m$  for any $(j,l_j)$. With our spatial NNGP prior coupled with sequential updates, rejection sampling from low-dimensional truncated $O$-variate normal distributions would have adequate acceptance rates and thus small required numbers $n_{jl_ji}^*$'s of draws from $N_O\left(V_{\bm{s}_i}\bm{\mu}_{jl_j,\bm{s}_i}, V_{\bm{s}_i} \right)$'s (\Cref{{seqAlphaIndPostDens2}} in \Cref{sect4d}) per sample of $\bm{\alpha}_{jl_j}(\bm{s}_i)$'s. If $\underset{j\in\{1,\dots,k\}}{\max}L_j \ll m$, $\underset{(j,l_j,i)}{\max}\;k_{jl_ji}^* \ll m$, and $\underset{(j,l_j,i)}{\max}\;n_{jl_ji}^* \ll m$, then $$\Ocal\left(O\left\lbrace \sum_{j=1}^k \Bigg[(L_j-1)mh(h+O) + \sum_{l_j=1}^{L_j-1}\sum_{i=1}^m \left(k_{jl_ji}^*+n_{jl_ji}^*\cdot O\right)\Bigg] + O^2 \right\rbrace  \right) = \Ocal(m). $$

	\subsection{An Overall Scalability Improvements Summary for Our Three Novelties}\label{appenH4}
	Suppose we have a large $m$. Recall that $1\leq L$ or $L_j$'s $\leq m$ and that we have assumed $O, h, k,p\ll m$. We also assume $r<m$ for spatial prediction at $r\in\mathbb{N}$ new locations $\bm{s}_{(m+1):(m+r)}$. Let $N\in\mathbb{N}$ be the number of kept post-burn-in MCMC iterations. We denote the corresponding estimated number(s) of spatial mixture components as $L_j^{(t_i)}$'s, $j\in\{1,\ldots,k\},\:i\in\{1,\ldots,N\}$. In \Cref{sectH4summaryTable}, storage requirements for spatial prediction and clustering correspond to all kept MCMC iterations, whereas computational complexity orders are only for one typical MCMC iteration. 
	
	\begin{table}[h]
		{
			\begin{adjustwidth}{-2.53cm}{-0.5cm}
				\begin{center}
					\scalebox{0.58}{\begin{tabular}{|*{6}{c|}}
							\hline
							\multicolumn{2}{|c|}{\textbf{Simulation Method}} & \texttt{fullGPfixedL} & \texttt{NNGPblockFixedL} & \texttt{NNGPsequenFixedL} & \texttt{NNGPsequenVaryLj}\\
							\hline
							\multicolumn{2}{|c|}{\textbf{Spatial NNGP Prior}} & No & Yes & Yes & Yes\\
							\hline
							\multicolumn{2}{|c|}{\textbf{Sequential Updating Methods}} & No & No & Yes & Yes\\
							\hline
							\multicolumn{2}{|c|}{\textbf{Slice Sampling Approach}} & No & No & No & Yes\\
							\hline
							\multicolumn{2}{|c|}{\textbf{Spatial Prediction Speed for}} & \multirow{2}{*}{$\bm{\Ocal\big(r\left[m^2+r^2+k(L-1)mO\right]\big)}$} &  \multicolumn{2}{|c|}{\multirow{2}{*}{$\bm{\Ocal\left(r\big[h^3+k(L-1)O(h+O)\big]+O^3\right)}$}}  & \multirow{2}{*}{$\bm{\Ocal\left(r\left[h^3+\sum\limits_{j=1}^{k}(L_j-1)\cdot O(h+O)\right]+O^3\right)}$}\\
							\multicolumn{2}{|c|}{\textbf{$\hat{\bm{\alpha}}_{jl_j}(\bm{s}_{(m+1):(m+r)})|\bm{\alpha}_{jl_j}(\bm{s}_{1:m}),\kappa,\rho$'s}} &  &  \multicolumn{2}{|c|}{}  & \\
							\hline
							& \multirow{2}{*}{$\rho$ and $\kappa_{O\times O}$} & \multirow{2}{*}{$\bm{\Ocal(m^3)}$} &  \multicolumn{2}{|c|}{\multirow{2}{*}{$\bm{\Ocal\left(\left\lbrace k(L-1)O^2[h^2+h+1]+h^3\right\rbrace m\right)}$}}  & \multirow{2}{*}{$\bm{\Ocal\left(\left\lbrace\sum\limits_{j=1}^k(L_j-1)\cdot O^2[h^2+h+1]+h^3\right\rbrace m\right)}$}\\
							\textbf{Posterior} &  &  &  \multicolumn{2}{|c|}{}  & \\
							\cline{2-6}
							\textbf{Sampling} & \multirow{2}{*}{$\alpha_{jl_j}^o\big(\bm{s}_{i}\big)$'s}& \multicolumn{2}{|c|}{\multirow{2}{*}{$\bm{\Ocal(m^3)}$}} &  \multirow{2}{*}{$\bm{\Ocal\big(m\big[k(L-1)hO(h+O)+O^3\big]\big)}$}  & \multirow{2}{*}{$\bm{\Ocal\left(O\left\lbrace \sum\limits_{j=1}^k \Bigg[(L_j-1)mh(h+O) + \sum\limits_{l_j=1}^{L_j-1}\sum\limits_{i=1}^m \left(k_{jl_ji}^*+n_{jl_ji}^*\cdot O\right)\Bigg] + O^2 \right\rbrace  \right)} $}\\
							\textbf{Speed} & &\multicolumn{2}{|c|}{} &   & \\
							\cline{2-6}
							\textbf{for} & $z_{jl_j}^o\left(\bm{s}_{i}\right)$'s or $u_j^o\left(\bm{s}_{i}\right)$'s & \multicolumn{3}{|c|}{$\bm{\Ocal(kLmO)}$} & $\bm{\Ocal(kmO)}$\\
							\cline{2-6}
							& $\delta_{1:k}$, $\theta_{jl_j}$'s, & \multicolumn{3}{|c|}{\multirow{2}{*}{$\bm{\Ocal(kLmOT(p+k))}$ (absolute highest order from $\xi_j^o\left(\bm{s}_{i}\right)$'s)}} & \multirow{2}{*}{$\bm{\Ocal\left(\sum\limits_{j=1}^k L_j\cdot mOT(p+k)\right)}$ (absolute highest order from $\xi_j^o\left(\bm{s}_{i}\right)$'s)}\\
							& and $\xi_j^o\left(\bm{s}_{i}\right)$'s & \multicolumn{3}{|c|}{} & \\
							\hline
							\textbf{Storage} & $F(\rho)$ and $F(\rho)^{-1}$ & $\bm{\Ocal(m^2)}$ &  \multicolumn{3}{|c|}{$\bm{\Ocal\left(mh^2\right)}$}  \\
							\cline{2-6}
							\textbf{Require-} & Spatial Prediction & \multicolumn{3}{|c|}{\multirow{2}{*}{$\bm{\Ocal(NmOkL)}$}} & \multirow{2}{*}{$\bm{\Ocal\left(mO\sum\limits_{i=1}^N\sum\limits_{j=1}^kL_j^{(t_i)}\right)}$}\\
							\cline{2-2}
							\textbf{ments for} & Spatial Clustering & \multicolumn{3}{|c|}{} &    \\
							\hline
					\end{tabular}}
				\end{center}
			\end{adjustwidth}
			\caption{An overall theoretical summary table of scalability improvements brought about by our 3 novelties. For each of the four methods implemented in our simulation experiments, we list its computational complexity orders for involved Gibbs sampler steps in posterior sampling and a major spatial prediction procedure. We also present memory requirements for two key quantities and spatial prediction \& clustering under our four methods.}
			\label{sectH4summaryTable}	
		}
	\end{table}

	\section{Complementary Simulation Results and  Experiments}\label{appenI}

	\subsection{Further Details of the Toy Example in \Cref{sect2}} \label{appenIa}
	
\begin{figure}[h]
		\centering
		\includegraphics[width=\textwidth]{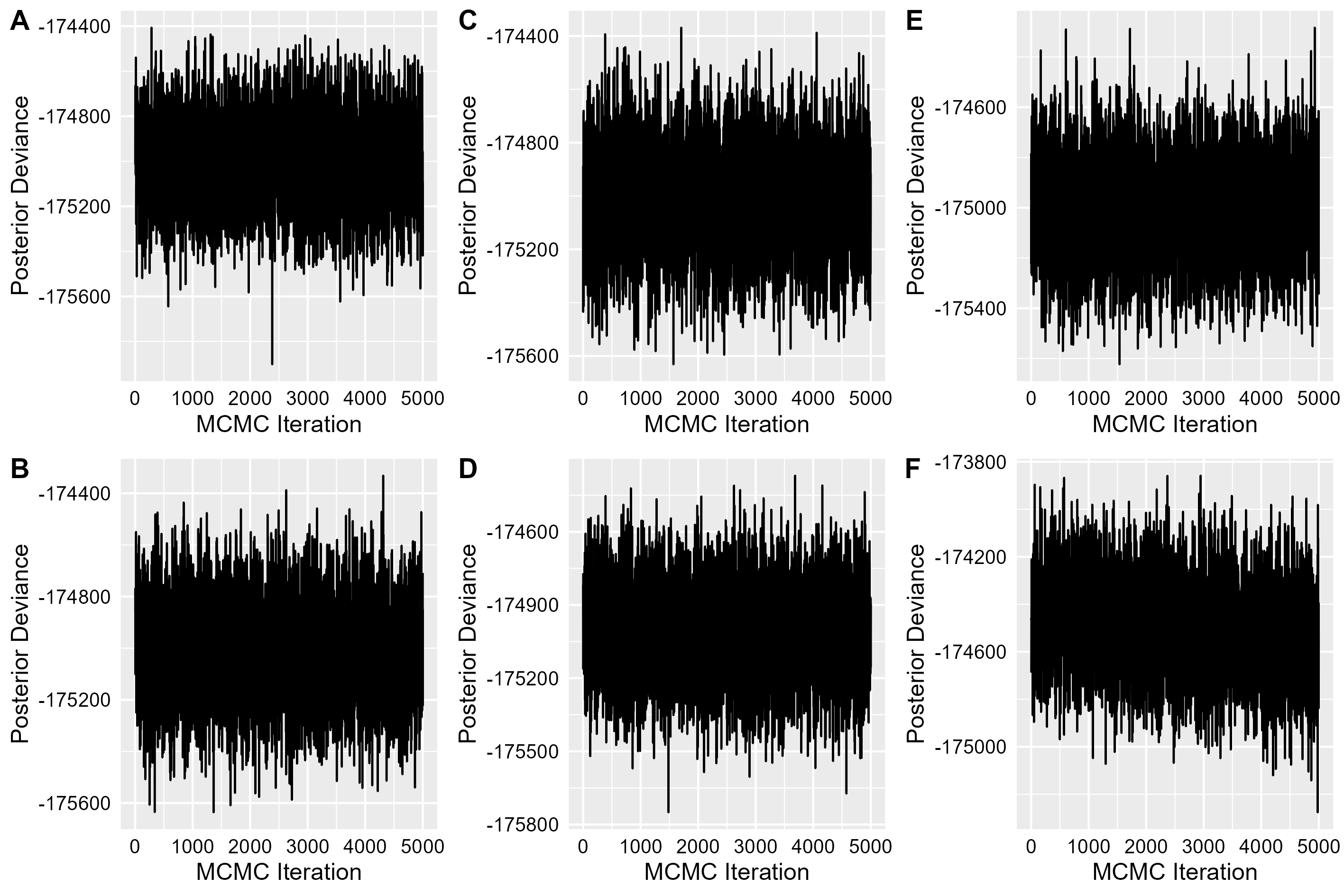}
		\caption{Posterior deviance plots for \texttt{fullGPfixedL} (\textbf{A}), \texttt{NNGPblockFixedL} (\textbf{B}), \texttt{NNGPsequenFixedL} (\textbf{C}), \texttt{NNGPsequenVaryLj} (\textbf{D}), and 
			\texttt{spBFAL10} (\textbf{E}), \texttt{spBFALInf} (\textbf{F}). For each of the six trace plots, the horizontal axis denotes the kept $20000\div 4=5000$ post-burn-in MCMC iterations, and the vertical axis presents the corresponding posterior deviance values.} 
		\label{toyExPosteriorDeviances}
	\end{figure}

	We present additional technical details and results for the toy example in \Cref{sect2}. All our four methods’ converged posterior deviances are comparable to that of \texttt{spBFAL10} and clearly better than that of \texttt{spBFALInf}; see \Cref{toyExPosteriorDeviances}.
	
	To get a more complete picture of all methods' diverse model fitting performances, we further consider some diagnostic statistics below. We first devise some diagnostic metrics based on in-sample prediction. Let $N=m_0\times O\times T_0$, where $m_0=352$, $O=1$, and $T_0=300$, and denote $\bm{y}_{\text{obs}}$ as our $N\times 1$ observed outcome vector. For each of the kept $W=5000$ post-burn-in MCMC iterations, we draw a predicted outcome vector $\bm{\hat{y}}^{(w)}_{N\times 1},w\in\{1,\dots,W\}$, based on the $w^{\text{th}}$ sampled posterior parameters. Let
	\begin{align} 
		\texttt{postMeanMSE} & = \frac{1}{m_0OT_0}\sum_{t=1}^{T_0}\sum_{o=1}^O\sum_{i=1}^{m_0} \left(\left[ \frac{1}{W}\sum_{w=1}^W\hat{y}^{(w)}(i,o,t)\right] -y_{\text{obs}}(i,o,t)\right)^2 , \nonumber\\
		\texttt{postMSE} & = \frac{1}{m_0OT_0}\sum_{t=1}^{T_0}\sum_{o=1}^O\sum_{i=1}^{m_0}\left\lbrace \frac{1}{W}\sum_{w=1}^W \left[\hat{y}^{(w)}(i,o,t)-y_{\text{obs}}(i,o,t)\right]^2 \right\rbrace , \nonumber\\
		\texttt{postVar} & = \frac{1}{m_0OT_0}\sum_{t=1}^{T_0}\sum_{o=1}^O\sum_{i=1}^{m_0} \Var\left(\bm{\hat{y}}^{(1:W)}(i,o,t)\right),\text{ and } \texttt{dinf} = \texttt{postMeanMSE} + \texttt{postVar}. \nonumber
	\end{align}
	Then \texttt{postMeanMSE} and \texttt{postMSE} adequately assess the model fitting accuracy, and \texttt{postVar} is an appropriate precision measure. Small values for the four statistics are desired. 
	We further consider two sets of diagnostic statistics based on the posterior mean estimates and all 5000 kept iterations' posterior parameter estimates, both of which stem from the posterior log-likelihoods and incorporate goodness-of-fit assessments as well as over-complexity penalty terms. The main deviance information criterion DIC statistic \texttt{dic} \parencite{Spiegelhalter2002} can be written as the sum of two values, one of which portrays model complexity and is denoted as \texttt{pD}. The Watanabe–Akaike information criterion WAIC \parencite{Vehtari2017} consists of two over-fit measurements \texttt{p\_waic\_1}, \texttt{p\_waic\_2}, a log predictive density \texttt{lppd}, and a main metric \texttt{waic = -2lppd + 2p\_waic\_2}. For all involved metrics except \texttt{lppd}, smaller values imply better model fitting. 
	
	\Cref{sect6dDiagnosticsTable} reassures us that our four methods deliver comparable satisfactory model fitting performances. \texttt{spBFAL10} results in similarly good diagnostic statistics, while \texttt{spBFALInf} produces the conspicuous worst metric values.
	
	\begin{table}[h]
		\centering
		\scalebox{0.7}{\begin{tabular}{|*{7}{c|}}
				\hline
				\textbf{Model} &{\texttt{fullGPfixedL}}&{\texttt{NNGPblockFixedL}}&{\texttt{NNGPsequenFixedL}}&{\texttt{NNGPsequenVaryLj}}&\texttt{spBFAL10}&\texttt{spBFALInf}\\ 
				\hline
				\textbf{\texttt{postMeanMSE}} &  0.009917465 & 0.009917465 & 0.009917531 & 0.00991737 & \bf0.009916747 & 0.009972942 \\
				\hline
				\textbf{\texttt{postMSE}} &  0.02668247 & 0.02668144 & 0.02668167 & {\bf 0.02668039} & 0.02668312 & 0.02679888 \\
				\hline
				\textbf{\texttt{postVar}} &  0.01676836 & 0.01676733 & 0.0167675 &  {\bf0.01676638} & 0.01677255 & 0.01683156 \\
				\hline
				\textbf{\texttt{dinf}} & 0.02668582 & 0.0266848 & 0.02668503 & {\bf 0.02668375} & 0.0266893 & 0.0268045 \\
				\hline
				\textbf{\texttt{pD}} & 621.2673 &  663.04 & 646.9392 & 662.9201 &  564.7672 & {\bf 557.425} \\
				\hline
				\textbf{\texttt{dic}} &  -174384.7 &  -174345.7 & -174361.8 & -174348 & {\bf -174440.4} &  -173945.4 \\
				\hline
				\textbf{\texttt{p\_waic\_1}} &   878.4693 & 611.0213 & 605.4927 & 721.9712 & {\bf600.7735} &  756.1147 \\
				\hline
				\textbf{\texttt{p\_waic\_2}} & 9000.328 & 8772.018 & {\bf8705.174} & 8917.304 &  9093.332 & 9281.774 \\
				\hline
				\textbf{\texttt{lppd}} & {\bf 87942.2} & 87809.87 & 87807.12 &  87866.43 & 87802.98 & 87629.49 \\
				\hline
				\textbf{\texttt{waic}} & -157883.7 &  -158075.7 & {\bf-158203.9} & -157898.2 & -157419.3 & -156695.4 \\
				\hline
		\end{tabular}}
		\caption{Values of ten diagnostic statistics in the toy example. Each method uses $W = 20000\div 4=5000$ kept post-burn-in MCMC iterations for analysis. } 
		\label{sect6dDiagnosticsTable}
	\end{table}
	
	We construct the following three temporal prediction metrics to assess the performance in predicting at the $q=10$ future time points $t = 301,302,\ldots,310$ for all $m_0=352$ training locations:
	\begin{align} 
		\texttt{postMeanMSE} & = \frac{1}{qOm_0}\sum_{t=T_0+1}^{T_0+q}\sum_{i=1}^{m_0}\sum_{o=1}^O \left(\left[ \frac{1}{W}\sum_{w=1}^W\hat{y}^{(w)}(t,o,i)\right] -y_{\text{actual}}(t,o,i)\right)^2,\nonumber\\
		\texttt{postMSE} & = \frac{1}{qOm_0}\sum_{t=T_0+1}^{T_0+q}\sum_{i=1}^{m_0}\sum_{o=1}^O\left\lbrace \frac{1}{W}\sum_{w=1}^W \left[\hat{y}^{(w)}(t,o,i)-y_{\text{actual}}(t,o,i)\right]^2 \right\rbrace,\text{ and} \nonumber\\
		\texttt{postVar} & = \frac{1}{qOm_0}\sum_{t=T_0+1}^{T_0+q}\sum_{i=1}^{m_0}\sum_{o=1}^O  \Var\left(\bm{\hat{y}}^{(1:W)}(t,o,i)\right).\nonumber
	\end{align}
	For each $(t,o,i,w)$, 
	$\hat{y}^{(w)}(t,o,i)$ represents our out-of-sample predicted outcome for observation type $o$ at location $\bm{s}_{i}$ and time $t$ based on the $w^{\text{th}}$ kept post-burn-in MCMC iteration's posterior parameter estimates, and
	$y_{\text{actual}}(t,o,i)$ is the corresponding actual observed outcome value in the testing data set not utilized when fitting the models. 
	As in \Cref{sect6dtemppredMetricsTable}, \texttt{spBFALInf} produces considerably worse future-time prediction than all other methods. 
	Among our 4 methods and \texttt{spBFAL10}, \texttt{NNGPsequenFixedL} appears to be the best, followed by \texttt{fullGPfixedL}, \texttt{NNGPblockFixedL}, \texttt{NNGPsequenVaryLj}, and finally \texttt{spBFAL10}.

	\begin{table}[h]
		\centering
		\scalebox{0.72}{
			\begin{tabular}{|*{7}{c|}}
				\hline
				\textbf{Model} & \texttt{fullGPfixedL} & \texttt{NNGPblockFixedL} & \texttt{NNGPsequenFixedL} & \texttt{NNGPsequenVaryLj} & \texttt{spBFAL10} & \texttt{spBFALInf}\\
				\hline
				\texttt{postMeanMSE} & 3.749685 & 3.753012 & 3.729849 & 3.728981
				&  4.009299  & 11.55326\\
				\hline
				\texttt{postMSE} & 18.518662 & 18.754345 & 18.299211 & 19.340736
				&  19.409493 & 164.42639\\
				\hline
				\texttt{postVar} & 14.771932 & 15.004334 & 14.572277 & 15.614878
				&  15.403274 & 152.90371\\
				\hline
		\end{tabular} }                                                          \caption{Temporal prediction metrics in the toy example. Smaller values for the three metrics imply better prediction capability at future time points.}
		\label{sect6dtemppredMetricsTable}
	\end{table}
	
	We utilize similar spatial prediction metrics to assess the prediction performance at the $r=9$ testing spatial locations for all training time points $t=1,2,\ldots,T_0=300$: 
	\begin{align} 
		\texttt{postMeanMSE} & = \frac{1}{rOT_0}\sum_{i_r=1}^r\sum_{t=1}^{T_0}\sum_{o=1}^O \left(\left[ \frac{1}{W}\sum_{w=1}^W\hat{y}^{(w)}(i_r,o,t)\right] -y_{\text{actual}}(i_r,o,t)\right)^2,\nonumber\\
		\texttt{postMSE} & = \frac{1}{rOT_0}\sum_{i_r=1}^r\sum_{t=1}^{T_0}\sum_{o=1}^O\left\lbrace \frac{1}{W}\sum_{w=1}^W \left[\hat{y}^{(w)}(i_r,o,t)-y_{\text{actual}}(i_r,o,t)\right]^2 \right\rbrace,\text{ and} \nonumber\\
		\texttt{postVar} & = \frac{1}{rOT_0}\sum_{i_r=1}^r\sum_{t=1}^{T_0}\sum_{o=1}^O  \Var\left(\bm{\hat{y}}^{(1:W)}(i_r,o,t)\right).\nonumber
	\end{align}
	For each $(i_r,o,t,w)$, 
	$\hat{y}^{(w)}(i_r,o,t)$ represents our out-of-sample predicted outcome for observation type $o$ at location $\bm{s}_{i_r}$ and
	time $t$ based on the $w^{\text{th}}$ kept post-burn-in MCMC iteration's posterior parameter estimates, and
	$y_{\text{actual}}(i_r,o,t)$ is the corresponding actual observed outcome value in the testing data set not utilized when fitting the models. All spatial prediction metrics are appropriately small (\Cref{sect6dspatpredMetricsStepTimeTable}). 
	\texttt{NNGPsequenVaryLj} appears to be the best, followed by \texttt{NNGPblockFixedL}, \texttt{fullGPfixedL}, and finally \texttt{NNGPsequenFixedL}. 
	
	We also recorded the time required to obtain $\hat{\bm{\alpha}}\left(\bm{s}_{(m_0+1):m}\right)$ given $\bm{\alpha}\left(\bm{s}_{1:m_0}\right),\,\kappa,\,\rho$ and to obtain $\hat{\bm{w}}\left(\bm{s}_{(m+1):(m+r)}\right)$, $\hat{\bm{\xi}}\left(\bm{s}_{(m+1):(m+r)}\right)$, and $\hat{{\Lambda}}\left(\bm{s}_{(m+1):(m+r)}\right)$ from $\hat{\bm{\alpha}}\left(\bm{s}_{(m_0+1):m}\right)$ and the kept posterior samples of $\theta_{jl_j}$'s (\texttt{alphaKrigTime} and \texttt{weightsXiLambdaKrigTime}). The results (\Cref{sect6dspatpredMetricsStepTimeTable}) correspond well to what we have derived in \Cref{appenH} and obtained in \Cref{sect6b} (\Cref{spatpredtimeBoxPlots}). \texttt{alphaKrigTime} corresponding to \texttt{NNGPblockFixedL} and \texttt{NNGPsequenFixedL} are close and more than 20 times smaller than that of \texttt{fullGPfixedL}.  \texttt{alphaKrigTime} corresponding to \texttt{NNGPsequenVaryLj}
	is even smaller. \texttt{weightsXiLambdaKrigTime} corresponding to \texttt{NNGPsequenVaryLj} is smaller than their counterparts from the other three methods.
	
	\begin{table}[h]
		\centering
		\scalebox{0.8}{
			\begin{tabular}{|*{5}{c|}}
				\hline
				\textbf{Model} & \texttt{fullGPfixedL} & \texttt{NNGPblockFixedL} & \texttt{NNGPsequenFixedL} & \texttt{NNGPsequenVaryLj}\\
				\hline
				\texttt{postMeanMSE} & 2.511976 & 2.478452 & 2.658194 & 2.476308\\
				\hline
				\texttt{postMSE} &  14.123149 &  14.003086 & 14.341286 & 12.764447\\
				\hline
				\texttt{postVar} & 11.613496 & 11.526939 & 11.685429 & 10.290197\\
				\hline
				\texttt{alphaKrigTime} & 63.988 & 2.838 & 2.888 & 2.492\\
				\hline
				\texttt{weightsXiLambdaKrigTime} & 0.118 & 0.129 & 0.139 & 0.076 \\
				\hline
		\end{tabular} }                                                          \caption{Spatial prediction metrics and time required (in seconds) for the two major spatial prediction steps in the toy example. Smaller values for the three metrics imply better prediction capability at new spatial locations.}
		\label{sect6dspatpredMetricsStepTimeTable}
	\end{table}
	
	For each of our four methods, we performed k-means clustering on three posterior weights matrices formulated from 10, 100, and 1000 selected equally dispersed post-burn-in MCMC iterations' parameter estimates, as described by the last paragraph in \Cref{sect5b}. Since we specified only 2 actual spatial groups and k-means clusters, there is no label-switching issue. Reassuringly, all clustering outcomes we obtained from all our four methods agree exactly with the actual spatial groupings and are thus completely accurate. 
	
	\newgeometry{top=2cm,left=1cm,right=1cm,bottom=2cm}
	\subsection{Additional Tables of Recorded Posterior Sampling Time for \Cref{sect6a}}\label{appenIb}
	\begin{table}[h]
		\centering
		\scalebox{0.82}{
			\centering
			\begin{tabular}{|*{4}{c|}}
				\hline
				\multicolumn{2}{|c|}{\backslashbox{\textbf{Setting}}{\textbf{Model}}} & \texttt{bfa\_sp()} from \texttt{spBFA} with a finite $L=50$ & \texttt{bfa\_sp()} from \texttt{spBFA} with \texttt{L = Inf} \\
				\hline
				\multirow{3}{*}{$T=30$} & $m=400$ & 1.41 days & 10.21 hours \\
				\cline{2-4}
				& $m=900$ &  18.5 days &  3.83 days\\
				\cline{2-4}
				& $m=1600$ & $>$ 3 months & 24.34 days \\
				\hline
				\multirow{3}{*}{$T=50$} & $m=400$ & 1.48 days & 11.21 hours \\
				\cline{2-4}
				& $m=900$ &  18.66 days &  4.63 days\\
				\cline{2-4}
				& $m=1600$ & $>$ 3 months &  25.72 days\\
				\hline
			\end{tabular}         
		}
		\caption{Total model fitting time by \texttt{spBFA}'s finite mixture and infinite mixture models with $2\times 10^4$ burn-in and $10^4$ post-burn-in MCMC iterations under 6 different pairs of $(m,T)$ values.}
		\label{spBFAsect6aModelRuntime}
	\end{table}
	\begin{table}[h]
		\centering
		\scalebox{0.82}{
			\centering
			\begin{tabular}{|*{5}{c|}}
				\hline
				\backslashbox{\textbf{Parameter}}{\textbf{Model}} & \texttt{fullGPfixedL} & \texttt{NNGPblockFixedL} & \texttt{NNGPsequenFixedL} & \texttt{NNGPsequenVaryLj}\\
				\hline
				$\rho$ & 48.8606 & 7.0378 & 7.1372 & 5.28 \\
				\hline
				$\kappa$ & 17.0116 & 2.9992 & 3.0006 & $\approx 1$ \\
				\hline
				$\alpha_{jl_j}^o(\bm{s}_{i})$'s & 67.0654 & 65.0538 & 87.1808 & 188.1246\\
				\hline
				$z_{jl_j}^o(\bm{s}_{i})$'s or $u_j^o(\bm{s}_{i})$'s & 219.643 & 147.0992 & 148.0586 & $\approx 0$\\
				\hline
				$\xi_j^o(\bm{s}_{i})$'s & 24.7242 & 24.4538 & 24.247 & 4.1488 \\
				\hline
				$\theta_{jl_j}$'s & 3.0336 & 3.1326 & 3.1088 & 2.297\\
				\hline
				$\delta_{1:k}$ & $4\times 10^{-4}$ & $\approx 0$ & $6\times 10^{-4}$ & $\approx 0$\\
				\hline
			\end{tabular}         
		}
		\caption{\footnotesize Average sampling time per MCMC iteration (in milliseconds) corresponding to the 7 spatial-related parameters $\rho$, $\kappa$, $\alpha_{jl_j}^o(\bm{s}_{i})$'s, $z_{jl_j}^o(\bm{s}_{i})$'s or $u_j^o(\bm{s}_{i})$'s, $\xi_j^o(\bm{s}_{i})$'s, $\theta_{jl_j}$'s, and $\delta_{1:k}$ across all 5000 post-burn-in MCMC iterations for our four methods with $(m,T)=(400, 30)$.}
		\label{m400T30sect6aGibbsStepTimeTable}
	\end{table}
	\begin{table}[h]
		\centering
		\scalebox{0.82}{
			\centering
			\begin{tabular}{|*{5}{c|}}
				\hline
				\backslashbox{\textbf{Parameter}}{\textbf{Model}} & \texttt{fullGPfixedL} & \texttt{NNGPblockFixedL} & \texttt{NNGPsequenFixedL} & \texttt{NNGPsequenVaryLj}\\
				\hline
				$\rho$ & 49.7274 & 7.1076 & 7.53 & 5.7782 \\
				\hline
				$\kappa$ & 17.0162 & 2.9856 & 3.0002 & $\approx 1$ \\
				\hline
				$\alpha_{jl_j}^o(\bm{s}_{i})$'s & 67.723 & 63.7966 & 85.1594 & 207.3494\\
				\hline
				$z_{jl_j}^o(\bm{s}_{i})$'s or $u_j^o(\bm{s}_{i})$'s & 211.5018 & 180.8608 & 183.3338 & $\approx 0$\\
				\hline
				$\xi_j^o(\bm{s}_{i})$'s & 35.9726 & 36.1242 & 35.6472 & 6.1232 \\
				\hline
				$\theta_{jl_j}$'s & 18.431 & 18.7772 & 18.6408 & 18.242\\
				\hline
				$\delta_{1:k}$ & $6\times 10^{-4}$ & $2\times 10^{-4}$ & $\approx 0$ & $\approx 0$\\
				\hline
			\end{tabular}         
		}
		\caption{\footnotesize Average sampling time per MCMC iteration (in milliseconds) corresponding to the 7 spatial-related parameters $\rho$, $\kappa$, $\alpha_{jl_j}^o(\bm{s}_{i})$'s, $z_{jl_j}^o(\bm{s}_{i})$'s or $u_j^o(\bm{s}_{i})$'s, $\xi_j^o(\bm{s}_{i})$'s, $\theta_{jl_j}$'s, and $\delta_{1:k}$ across all 5000 post-burn-in MCMC iterations for our four methods with $(m,T)=(400, 50)$.}
		\label{m400T50sect6aGibbsStepTimeTable}
		\captionsetup{font=footnotesize}
	\end{table}
	\begin{table}[h]
		\centering
		\scalebox{0.82}{
			\centering
			\begin{tabular}{|*{5}{c|}}
				\hline
				\backslashbox{\textbf{Parameter}}{\textbf{Model}} & \texttt{fullGPfixedL} & \texttt{NNGPblockFixedL} & \texttt{NNGPsequenFixedL} & \texttt{NNGPsequenVaryLj}\\
				\hline
				$\rho$ & 308.928 & 27.1156 & 26.7362 & 22.1778 \\
				\hline
				$\kappa$ & 93.9964 & 8.1494 & 8.8828 & 4.0032 \\
				\hline
				$\alpha_{jl_j}^o(\bm{s}_{i})$'s & 455.3278 & 423.2304 & 197.8944 & 282.4816\\
				\hline
				$z_{jl_j}^o(\bm{s}_{i})$'s or $u_j^o(\bm{s}_{i})$'s & 324.7298 & 335.7746 & 328.3926 & $\approx 0$\\
				\hline
				$\xi_j^o(\bm{s}_{i})$'s & 54.7586 & 57.1688 & 55.633 & 8.1984 \\
				\hline
				$\theta_{jl_j}$'s & 8.6664 & 8.8824 & 8.8414 & 6.9156\\
				\hline
				$\delta_{1:k}$ & 0.9486 & 0.9938 & 0.9842 & 0.0406\\
				\hline
			\end{tabular}         
		}
		\caption{\footnotesize Average sampling time per MCMC iteration (in milliseconds) corresponding to the 7 spatial-related parameters $\rho$, $\kappa$, $\alpha_{jl_j}^o(\bm{s}_{i})$'s, $z_{jl_j}^o(\bm{s}_{i})$'s or $u_j^o(\bm{s}_{i})$'s, $\xi_j^o(\bm{s}_{i})$'s, $\theta_{jl_j}$'s, and $\delta_{1:k}$ across all 5000 post-burn-in MCMC iterations for our four methods with $(m,T)=(900, 30)$.}
		\label{m900T30sect6aGibbsStepTimeTable}
		\captionsetup{font=footnotesize}
	\end{table}
	\begin{table}[h]
		\centering
		\scalebox{0.82}{
			\centering
			\begin{tabular}{|*{5}{c|}}
				\hline
				\backslashbox{\textbf{Parameter}}{\textbf{Model}} & \texttt{fullGPfixedL} & \texttt{NNGPblockFixedL} & \texttt{NNGPsequenFixedL} & \texttt{NNGPsequenVaryLj}\\
				\hline
				$\rho$ & 323.9814 & 28.1102 & 31.7456 & 26.6904 \\
				\hline
				$\kappa$ & 95.9408 & 8.467 & 9.0628 & 4.0716 \\
				\hline
				$\alpha_{jl_j}^o(\bm{s}_{i})$'s & 466.1494 & 423.1224 & 198.8446 & 406.3786 \\
				\hline
				$z_{jl_j}^o(\bm{s}_{i})$'s or $u_j^o(\bm{s}_{i})$'s & 389.276 & 533.31 & 541.5196 & $4\times 10^{-4}$\\
				\hline
				$\xi_j^o(\bm{s}_{i})$'s & 82.7366 & 84.2434 & 84.5774 & 12.1992 \\
				\hline
				$\theta_{jl_j}$'s & 43.8958 & 44.263 & 44.3988 & 42.588\\
				\hline
				$\delta_{1:k}$ & 1.008 & 1.0006 & 1.0018 & 0.8306\\
				\hline
			\end{tabular}         
		}
		\caption{\footnotesize Average sampling time per MCMC iteration (in milliseconds) corresponding to the 7 spatial-related parameters $\rho$, $\kappa$, $\alpha_{jl_j}^o(\bm{s}_{i})$'s, $z_{jl_j}^o(\bm{s}_{i})$'s or $u_j^o(\bm{s}_{i})$'s, $\xi_j^o(\bm{s}_{i})$'s, $\theta_{jl_j}$'s, and $\delta_{1:k}$ across all 5000 post-burn-in MCMC iterations for our four methods with $(m,T)=(900, 50)$.}
		\label{m900T50sect6aGibbsStepTimeTable}
		\captionsetup{font=footnotesize}
	\end{table}
	\begin{table}[h]
		\centering
		\scalebox{0.82}{
			\centering
			\begin{tabular}{|*{5}{c|}}
				\hline
				\backslashbox{\textbf{Parameter}}{\textbf{Model}} & \texttt{fullGPfixedL} & \texttt{NNGPblockFixedL} & \texttt{NNGPsequenFixedL} & \texttt{NNGPsequenVaryLj}\\
				\hline
				$\rho$ & 1477.033 & 93.8578 & 101.1084 & 95.6168 \\
				\hline
				$\kappa$ & 316.5146 & 20.362 & 20.5866 & 12.4996 \\
				\hline
				$\alpha_{jl_j}^o(\bm{s}_{i})$'s & 2358.661 & 2083.491 & 353.8104 & 604.7246\\
				\hline
				$z_{jl_j}^o(\bm{s}_{i})$'s or $u_j^o(\bm{s}_{i})$'s & 604.8164 & 582.528 & 574.3044 &  1.5622\\
				\hline
				$\xi_j^o(\bm{s}_{i})$'s & 101.5026 & 100.3484 & 98.4202 & 17.5796 \\
				\hline
				$\theta_{jl_j}$'s & 26.311 & 18.127 & 17.8028 & 15.6692\\
				\hline
				$\delta_{1:k}$ & 3.2974 & 3.2024 & 3.0554 & 2.307\\
				\hline
			\end{tabular}         
		}
		\caption{\footnotesize Average sampling time per MCMC iteration (in milliseconds) corresponding to the 7 spatial-related parameters $\rho$, $\kappa$, $\alpha_{jl_j}^o(\bm{s}_{i})$'s, $z_{jl_j}^o(\bm{s}_{i})$'s or $u_j^o(\bm{s}_{i})$'s, $\xi_j^o(\bm{s}_{i})$'s, $\theta_{jl_j}$'s, and $\delta_{1:k}$ across all 5000 post-burn-in MCMC iterations for our four methods with $(m,T)=(1600, 30)$.}
		\label{m1600T30sect6aGibbsStepTimeTable}
		\captionsetup{font=footnotesize}
	\end{table}
	\begin{table}[h]
		\centering
		\scalebox{0.82}{
			\centering
			\begin{tabular}{|*{5}{c|}}
				\hline
				\backslashbox{\textbf{Parameter}}{\textbf{Model}} & \texttt{fullGPfixedL} & \texttt{NNGPblockFixedL} & \texttt{NNGPsequenFixedL} & \texttt{NNGPsequenVaryLj}\\
				\hline
				$\rho$ & 1442.891 & 96.4844 & 105.1648 & 93.5652 \\
				\hline
				$\kappa$ & 315.5866 & 21.5504 & 21.8626 & 12.9298 \\
				\hline
				$\alpha_{jl_j}^o(\bm{s}_{i})$'s & 2306.249 & 2113.644 &  362.944 & 774.4342\\
				\hline
				$z_{jl_j}^o(\bm{s}_{i})$'s or $u_j^o(\bm{s}_{i})$'s & 723.9746 & 635.5834 & 714.0948 &  2.1136\\
				\hline
				$\xi_j^o(\bm{s}_{i})$'s & 151.4386 & 152.5778 & 152.0324 & 15.0362\\
				\hline
				$\theta_{jl_j}$'s & 82.9466 & 83.4584 & 82.454 & 77.637 \\
				\hline
				$\delta_{1:k}$ & 3.869 & 3.963 & 3.8754 & 2.9812\\
				\hline
			\end{tabular}         
		}
		\caption{\footnotesize Average sampling time per MCMC iteration (in milliseconds) corresponding to the 7 spatial-related parameters $\rho$, $\kappa$, $\alpha_{jl_j}^o(\bm{s}_{i})$'s, $z_{jl_j}^o(\bm{s}_{i})$'s or $u_j^o(\bm{s}_{i})$'s, $\xi_j^o(\bm{s}_{i})$'s, $\theta_{jl_j}$'s, and $\delta_{1:k}$ across all 5000 post-burn-in MCMC iterations for our four methods with $(m,T)=(1600, 50)$.}
		\label{m1600T50sect6aGibbsStepTimeTable}
		\captionsetup{font=footnotesize}
	\end{table}
	\begin{table}[h]
		\centering
		\scalebox{0.82}{
			\centering
			\begin{tabular}{|*{5}{c|}}
				\hline
				\backslashbox{\textbf{Parameter}}{\textbf{Model}} & \texttt{fullGPfixedL} & \texttt{NNGPblockFixedL} & \texttt{NNGPsequenFixedL} & \texttt{NNGPsequenVaryLj}\\
				\hline
				$\rho$ & 1675.367 & 97.143 & 106.6914 & 98.423 \\
				\hline
				$\kappa$ & 317.7624 & 21.976 & 22.5796 & 14.403 \\
				\hline
				$\alpha_{jl_j}^o(\bm{s}_{i})$'s & 2319.485 & 2124.528 & 365.9912 & 1002.139\\
				\hline
				$z_{jl_j}^o(\bm{s}_{i})$'s or $u_j^o(\bm{s}_{i})$'s & 1484.252 &  1155.679 & 1090.681 & 3.056 \\
				\hline
				$\xi_j^o(\bm{s}_{i})$'s & 286.3702 & 286.3756 & 286.3248 & 21.3178\\
				\hline
				$\theta_{jl_j}$'s & 317.6382 & 320.3912 & 321.6182 & 320.9168\\
				\hline
				$\delta_{1:k}$ & 4.032 & 4.0894 & 4.052 & 3.132 \\
				\hline
			\end{tabular}         
		}
		\caption{\footnotesize Average sampling time per MCMC iteration (in milliseconds) corresponding to the 7 spatial-related parameters $\rho$, $\kappa$, $\alpha_{jl_j}^o(\bm{s}_{i})$'s, $z_{jl_j}^o(\bm{s}_{i})$'s or $u_j^o(\bm{s}_{i})$'s, $\xi_j^o(\bm{s}_{i})$'s, $\theta_{jl_j}$'s, and $\delta_{1:k}$ across all 5000 post-burn-in MCMC iterations for our four methods with $(m,T)=(1600, 100)$.}
		\label{m1600T100sect6aGibbsStepTimeTable}
		\captionsetup{font=footnotesize}
	\end{table}
	\begin{table}[h]
		\centering
		\scalebox{0.82}{
			\centering
			\begin{tabular}{|*{5}{c|}}
				\hline
				\backslashbox{\textbf{Parameter}}{\textbf{Model}} & \texttt{fullGPfixedL} & \texttt{NNGPblockFixedL} & \texttt{NNGPsequenFixedL} & \texttt{NNGPsequenVaryLj}\\
				\hline
				$\rho$ & 25448.17 & 883.873 & 814.5748 & 793.9108 \\
				\hline
				$\kappa$ & 3683.99 & 170.5392 & 164.9444 & 143.551 \\
				\hline
				$\alpha_{jl_j}^o(\bm{s}_{i})$'s & 35893.15 & 22407.43 & 953.6206 & 2154.113\\
				\hline
				$z_{jl_j}^o(\bm{s}_{i})$'s or $u_j^o(\bm{s}_{i})$'s & 3819.395 &  2618.505 & 2692.092 & 55.5604 \\
				\hline
				$\xi_j^o(\bm{s}_{i})$'s & 542.3482 & 425.6026 & 403.9276 & 88.806\\
				\hline
				$\theta_{jl_j}$'s & 336.3604 & 242.8858 & 241.8212 & 234.8754\\
				\hline
				$\delta_{1:k}$ & 72.5452 & 58.4904 & 57.489 & 55.0086\\
				\hline
			\end{tabular}         
		}
		\caption{\footnotesize Average sampling time per MCMC iteration (in milliseconds) corresponding to the 7 spatial-related parameters $\rho$, $\kappa$, $\alpha_{jl_j}^o(\bm{s}_{i})$'s, $z_{jl_j}^o(\bm{s}_{i})$'s or $u_j^o(\bm{s}_{i})$'s, $\xi_j^o(\bm{s}_{i})$'s, $\theta_{jl_j}$'s, and $\delta_{1:k}$ across all 5000 post-burn-in MCMC iterations for our four methods with $(m,T)=(3600, 50)$.}
		\label{m3600T50sect6aGibbsStepTimeTable}
		\captionsetup{font=footnotesize}
	\end{table}
	\begin{table}[h]
		\centering
		\scalebox{0.82}{
			\centering
			\begin{tabular}{|*{5}{c|}}
				\hline
				\backslashbox{\textbf{Parameter}}{\textbf{Model}} & \texttt{fullGPfixedL} & \texttt{NNGPblockFixedL} & \texttt{NNGPsequenFixedL} & \texttt{NNGPsequenVaryLj}\\
				\hline
				$\rho$ & 24744.63 & 879.0884 & 805.6914 & 788.5346 \\
				\hline
				$\kappa$ & 3607.779 & 170.1582 & 162.2748 & 142.3264 \\
				\hline
				$\alpha_{jl_j}^o(\bm{s}_{i})$'s & 35275.15 & 22316.23 &  941.9378 &  2442.089\\
				\hline
				$z_{jl_j}^o(\bm{s}_{i})$'s or $u_j^o(\bm{s}_{i})$'s & 4460.977 &  2677.307 & 2687.017 &  56.1756\\
				\hline
				$\xi_j^o(\bm{s}_{i})$'s & 909.047 & 713.57 & 681.1242 & 99.2364\\
				\hline
				$\theta_{jl_j}$'s & 1359.953 & 779.6124 & 788.5526 & 787.6832\\
				\hline
				$\delta_{1:k}$ & 73.2364 & 59.2832 & 57.5628 & 55.5148\\
				\hline
			\end{tabular}         
		}
		\caption{\footnotesize Average sampling time per MCMC iteration (in milliseconds) corresponding to the 7 spatial-related parameters $\rho$, $\kappa$, $\alpha_{jl_j}^o(\bm{s}_{i})$'s, $z_{jl_j}^o(\bm{s}_{i})$'s or $u_j^o(\bm{s}_{i})$'s, $\xi_j^o(\bm{s}_{i})$'s, $\theta_{jl_j}$'s, and $\delta_{1:k}$ across all 5000 post-burn-in MCMC iterations for our four methods with $(m,T)=(3600, 100)$.}
		\label{m3600T100sect6aGibbsStepTimeTable}
		\captionsetup{font=footnotesize}
	\end{table}
	\restoregeometry
	
	\subsection{Accuracy for Spatial Clustering of Temporal Trends}\label{appenIc} 
	
	We design and implement a simulation experiment to demonstrate our four methods' satisfactory performances regarding our prime goal of clustering spatial locations into regions with similar temporal trajectories.
	
	We adopted a data generation mechanism specifically designed to assess clustering accuracy. The number of factors was set to $k=2$, and the actual numbers of clusters for both factors were also set to $2$. We specified $\theta_{11}=\theta_{12}=5$, $\theta_{21}=10$, and $\theta_{22}=-10$ for the atoms $\theta_{jl_j}$'s. $m=10^2=100$ spatial points $\bm{s}_i=(i_1,i_2)$ for $i=1,2,\ldots,100$ on an equispaced $2$-dimensional grid, where $i_1,i_2\in\{1,2,\ldots,10\}$ satisfy $10\cdot(i_1-1)+i_2=i$ for each $i$, were assigned to two actual spatial groups (one corresponding to $(\theta_{11},\theta_{21})=(5,10)$ and the other corresponding to $(\theta_{12},\theta_{22})=(5,-10)$) as depicted in \Cref{spatialPlot}. All other settings are almost identical to their counterparts in \Cref{sect6a}. We set $O=1$, $\psi=2.3$, $\sigma^2\left(\bm{s}_{i}\right)=0.01$ for all $i$ and specified $T=30$ time points $t=1,\ldots, 30$. When fitting the models, 10 is specified as both $L$ and an upper bound to all $L_j$'s. A burn-in length of 20000 and a post-burn-in length of 10000 (thinned to 5000) were used throughout.
	
	\begin{figure}[h]
		\centering
		\includegraphics[width=0.45\textwidth]{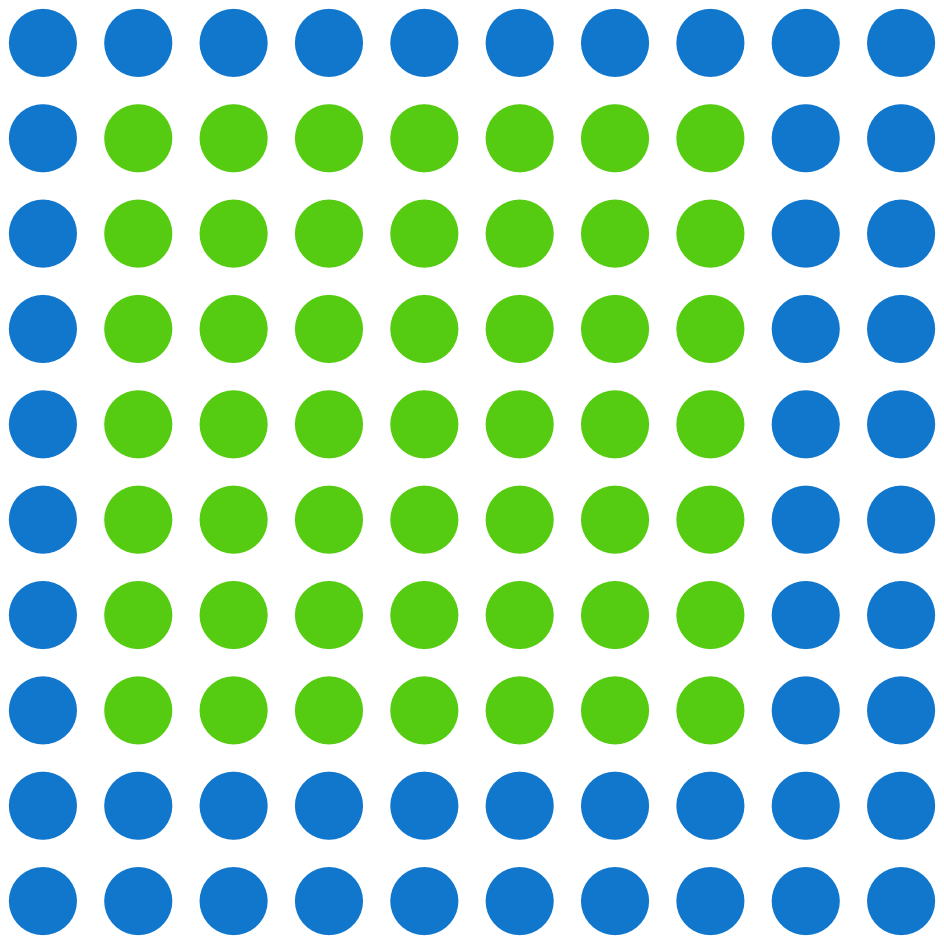}
		\caption{The actual spatial grouping for \Cref{appenIc}. The two colors represent the two groups for our $m=10^2=100$ spatial locations.} 
		\label{spatialPlot}
	\end{figure}
	
	For each of our $N=100$ simulated data sets and each of the four methods, we performed k-means clustering on three posterior weights matrices formulated from 10, 100, and 1000 selected equally dispersed kept post-burn-in MCMC iterations' parameter estimates, as described by the last paragraph in \Cref{sect5b}. Since we specified only 2 actual spatial groups and k-means clusters, there are no label-switching issues. Reassuringly, the $N\times 3\times 3$ clustering outcomes we obtained from three of our four methods, \texttt{fullGPfixedL}, \texttt{NNGPblockFixedL}, and \texttt{NNGPsequenFixedL}, all agree exactly with \Cref{spatialPlot} and are thus completely accurate. The fourth method, \texttt{NNGPsequenVaryLj}, also produces mean accuracy ratios of 0.9758, 0.9778, 0.9781 and mean rand indices of 0.9604242, 0.9621333, 0.9625717, which all indicate highly accurate clustering results. 
	
	\subsection{$\VAR(1)$ for the Latent Temporal Factors \boldmath{$\eta$}$_{1:T}$}\label{appenId}
	We fitted \texttt{fullGPfixedL}, \texttt{NNGPblockFixedL}, \texttt{NNGPsequenFixedL}, and \texttt{NNGPsequenVaryLj} under a $\VAR(1)$ structure for the latent temporal factors $\bm{\eta}_{1:T}$ on two sets of simulated data. The posterior sampling algorithm has been described in \Cref{appenC}.
	
\begin{figure}[H]
		\centering
		\includegraphics[width=0.6\textwidth]{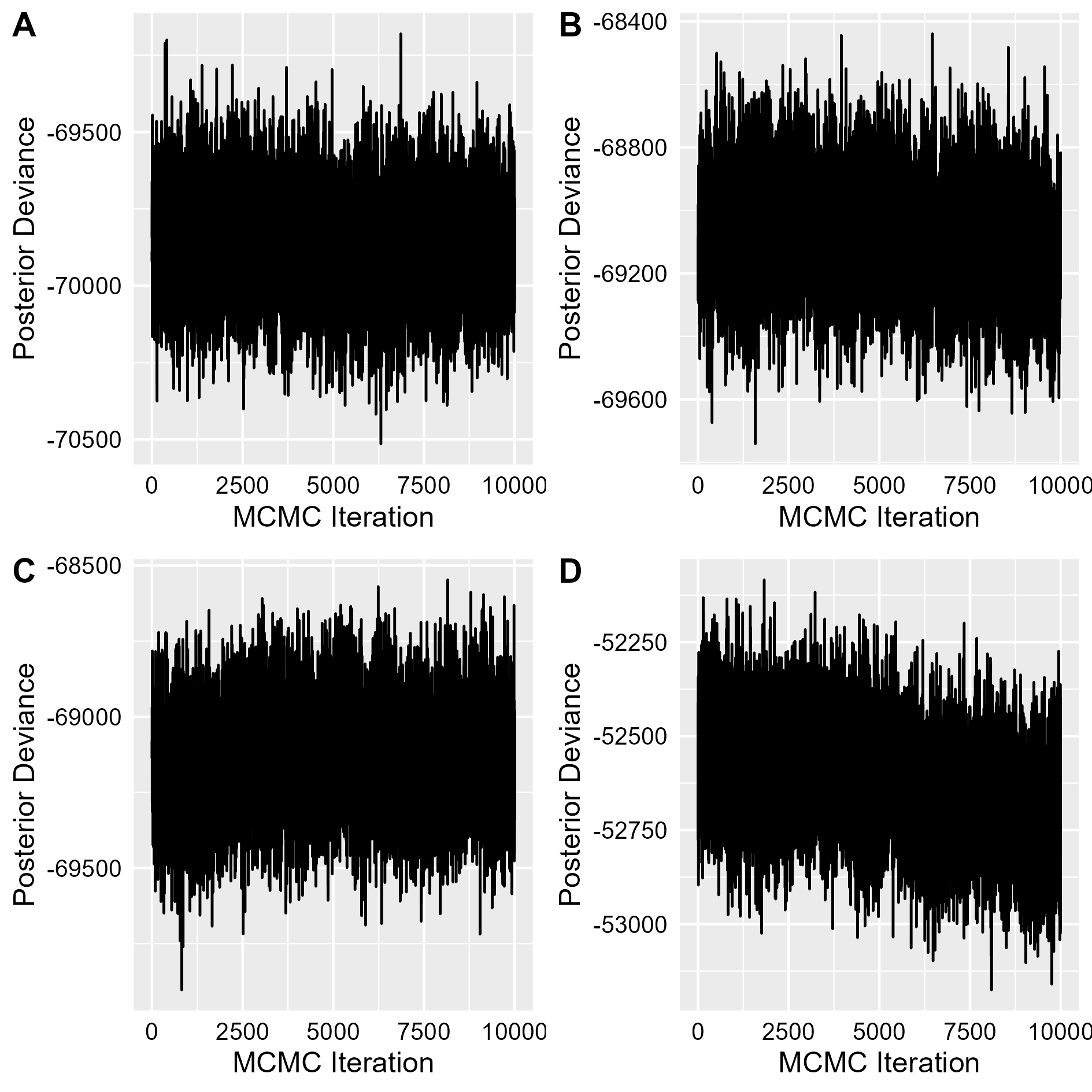}      
		\caption{\footnotesize $\VAR(1)$ temporal factors: Posterior deviances for \texttt{fullGPfixedL} (\textbf{A}), \texttt{NNGPblockFixedL} (\textbf{B}), \texttt{NNGPsequenFixedL} (\textbf{C}), and \texttt{NNGPsequenVaryLj} (\textbf{D}) under $(m,T)=(225,200)$, 80000 burn-in MCMC iterations, and 20000 post-burn-in MCMC iterations (thinned to 10000).}
		\label{VAR1deviances}
	\end{figure}

	We first adopted the data generation mechanism in \Cref{sect6a} with $(m,T)=(225,200)$. We ran 80000 burn-in and 20000 post-burn-in (thinned to 10000) MCMC iterations for our four methods, which took 23.99 hours, 1.05 days, 1.05 days, and 18.24 hours, respectively, and resulted in satisfactory diagnostics statistics (\Cref{VAR1diags}) and posterior deviances (\Cref{VAR1deviances}). 
	
	\begin{table}[H]
		\centering
		\scalebox{0.8}{\begin{tabular}{|*{5}{c|}}
				\hline
				\textbf{Model} &{\texttt{fullGPfixedL}}&{\texttt{NNGPblockFixedL}}&{\texttt{NNGPsequenFixedL}}&{\texttt{NNGPsequenVaryLj}}\\ 
				\hline
				\textbf{\texttt{postMeanMSE}} & 0.009766762 & 0.009880694 & 0.009925439 & 0.0165475\\
				\hline
				\textbf{\texttt{postMSE}} & 0.03064002 & 0.03112684 & 0.03108509 & 0.04475963\\
				\hline
				\textbf{\texttt{postVar}} & 0.02087535 & 0.02124827 & 0.02116177 & 0.02821495 \\
				\hline
				\textbf{\texttt{dinf}} & 0.03064211 & 0.03112897 & 0.0310872 & 0.04476245\\
				\hline
				\textbf{\texttt{pD}} & 985.3134 & 1081.714 & 1042.132 & 866.5728\\
				\hline
				\textbf{\texttt{dic}} & -68876.24 & -67995.26 & -68097.81 & -51755.75\\
				\hline
				\textbf{\texttt{p\_waic\_1}} & 634.8974 & 646.2995 & 744.9434 & 534.5163
				\\
				\hline
				\textbf{\texttt{p\_waic\_2}} & 7075.576 & 7234.089 & 7385.461 & 5357.02\\
				\hline
				\textbf{\texttt{lppd}} & 35248.22 & 34861.64 & 34942.44 & 26578.42 \\
				\hline
				\textbf{\texttt{waic}} & -56345.3 & -55255.1 & -55113.96 & -42442.8 \\
				\hline
		\end{tabular}}
		\caption{$\VAR(1)$ temporal factors: Diagnostic statistics for $(m,T)=(225,200)$, 80000 burn-in MCMC iterations, and 20000 post-burn-in MCMC iterations (thinned to 10000).} 
		\label{VAR1diags}
	\end{table}
	
	We then used a data generation mechanism specifically designed to assess clustering accuracy. The number of factors was set to $k=2$, and the actual numbers of clusters for both factors were also set to $2$. We specified $\theta_{11}=\theta_{12}=5$, $\theta_{21}=10$, and $\theta_{22}=-10$ for the atoms $\theta_{jl_j}$'s. $m=10^2=100$ spatial points $\bm{s}_i=(i_1,i_2)$ for $i=1,2,\ldots,100$ on an equispaced $2$-dimensional grid, where $i_1,i_2\in\{1,2,\ldots,10\}$ satisfy $10\cdot(i_1-1)+i_2=i$ for each $i$, were assigned to two actual spatial groups (one corresponding to $(\theta_{11},\theta_{21})=(5,10)$ and the other corresponding to $(\theta_{12},\theta_{22})=(5,-10)$) as depicted in \Cref{spatialPlot}. 
	All other settings are almost identical to their counterparts in \Cref{sect6a}. We set $O=1$, $\psi=2.3$, $\sigma^2\left(\bm{s}_{i}\right)=0.01$ for all $i$ and specified $T=30$ time points $t=1,\ldots, 30$. When fitting the models, 10 is specified as both $L$ and an upper bound to all $L_j$'s. A burn-in length of 20000 and a post-burn-in length of 10000 (thinned to 5000) were used. 
	For each of our $N=18$ simulated data sets and each of the four methods, we performed k-means clustering on three posterior weights matrices formulated from 10, 100, and 1000 selected equally dispersed kept post-burn-in MCMC iterations' parameter estimates, as described by the last paragraph in \Cref{sect5b}. Since we specified only 2 actual spatial groups and k-means clusters, we did not observe the label-switching issue. Reassuringly, the $N\times 3\times 3$ clustering outcomes we obtained from three of our four methods, \texttt{fullGPfixedL}, \texttt{NNGPblockFixedL}, and \texttt{NNGPsequenFixedL}, all agree exactly with \Cref{spatialPlot} and are thus completely accurate. The fourth method, \texttt{NNGPsequenVaryLj}, also produces mean accuracy ratios of 0.9538889, 0.9511111, 0.9722222 and mean rand indices of 0.9334119, 0.9288215, 0.9523008, which all indicate highly accurate clustering results. 
	
	\begin{figure}[H]
		\centering
		\includegraphics[width=0.32\textwidth]{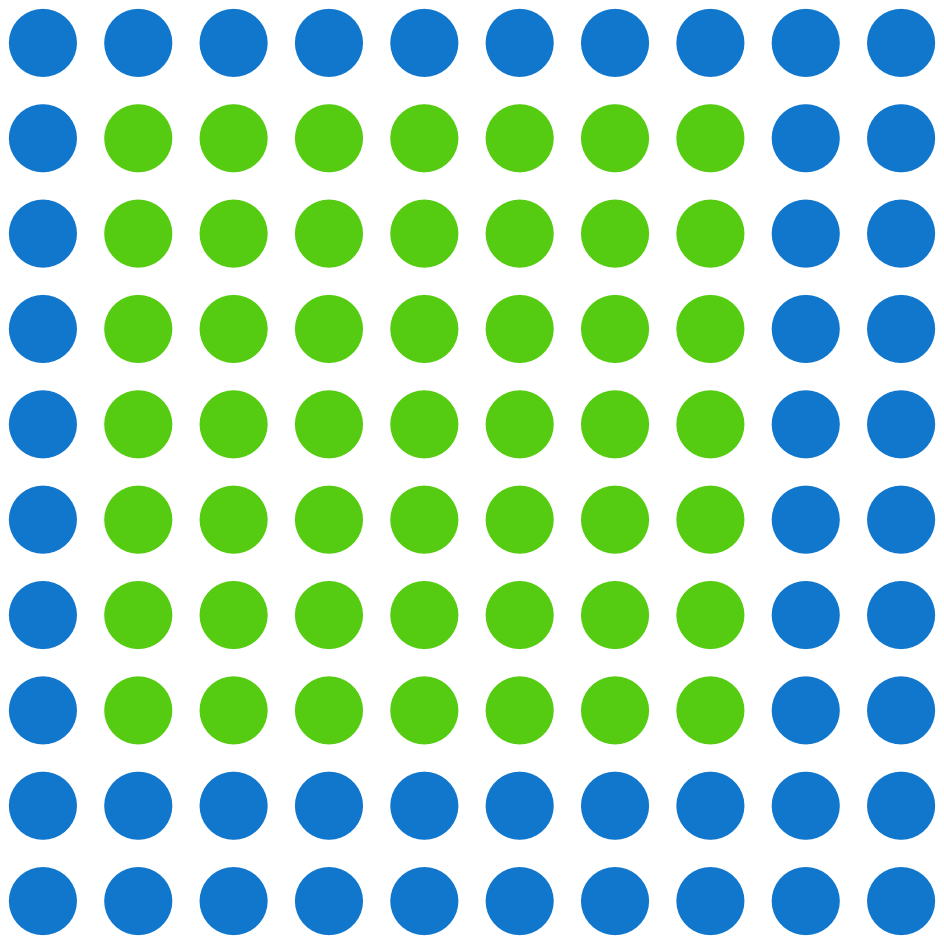}
		\caption{$\VAR(1)$ temporal factors: The actual spatial clusters. The two colors represent the two groups for our $m=10^2=100$ spatial locations.} 
		\label{spatialPlot}
	\end{figure}
	
	\subsection{Multiple Observation Types}\label{appenIe}
\begin{figure}[H]
		\centering
		\includegraphics[width=0.65\textwidth]{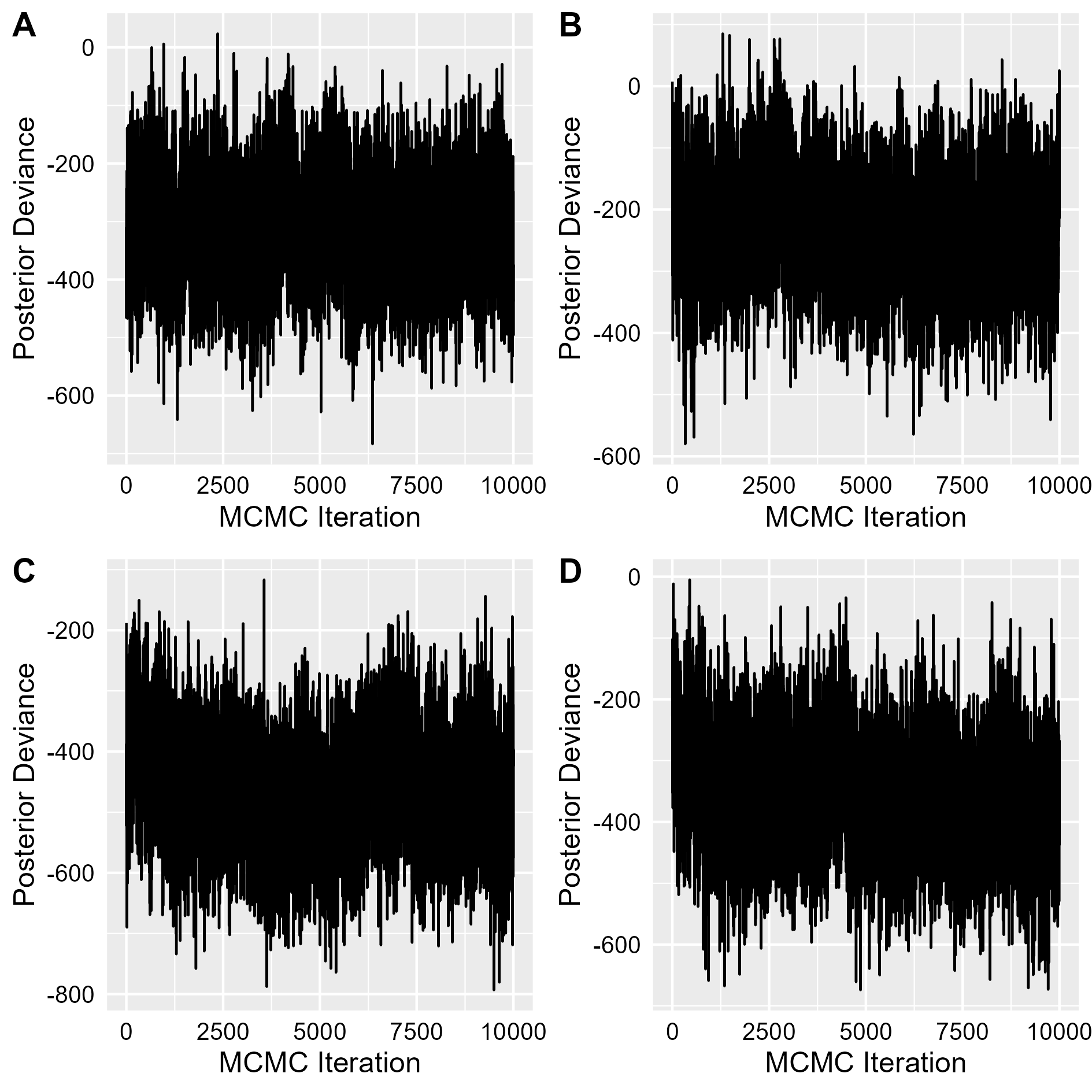}      
		\caption{Simulation for $O=2$: Posterior deviances for \texttt{fullGPfixedL} (\textbf{A}), \texttt{NNGPblockFixedL} (\textbf{B}), \texttt{NNGPsequenFixedL} (\textbf{C}), and \texttt{NNGPsequenVaryLj} (\textbf{D}) under $(m,T)=(100,20)$, 30000 burn-in MCMC iterations, and 20000 post-burn-in MCMC iterations (thinned to 10000).  \texttt{NNGPsequenFixedL} appears to be the best, followed by \texttt{NNGPsequenVaryLj}, \texttt{fullGPfixedL}, and \texttt{NNGPblockFixedL}.}
		\label{T20m100O2deviances}
	\end{figure}

\begin{figure}[H]
		\centering
		\includegraphics[width=0.67\textwidth]{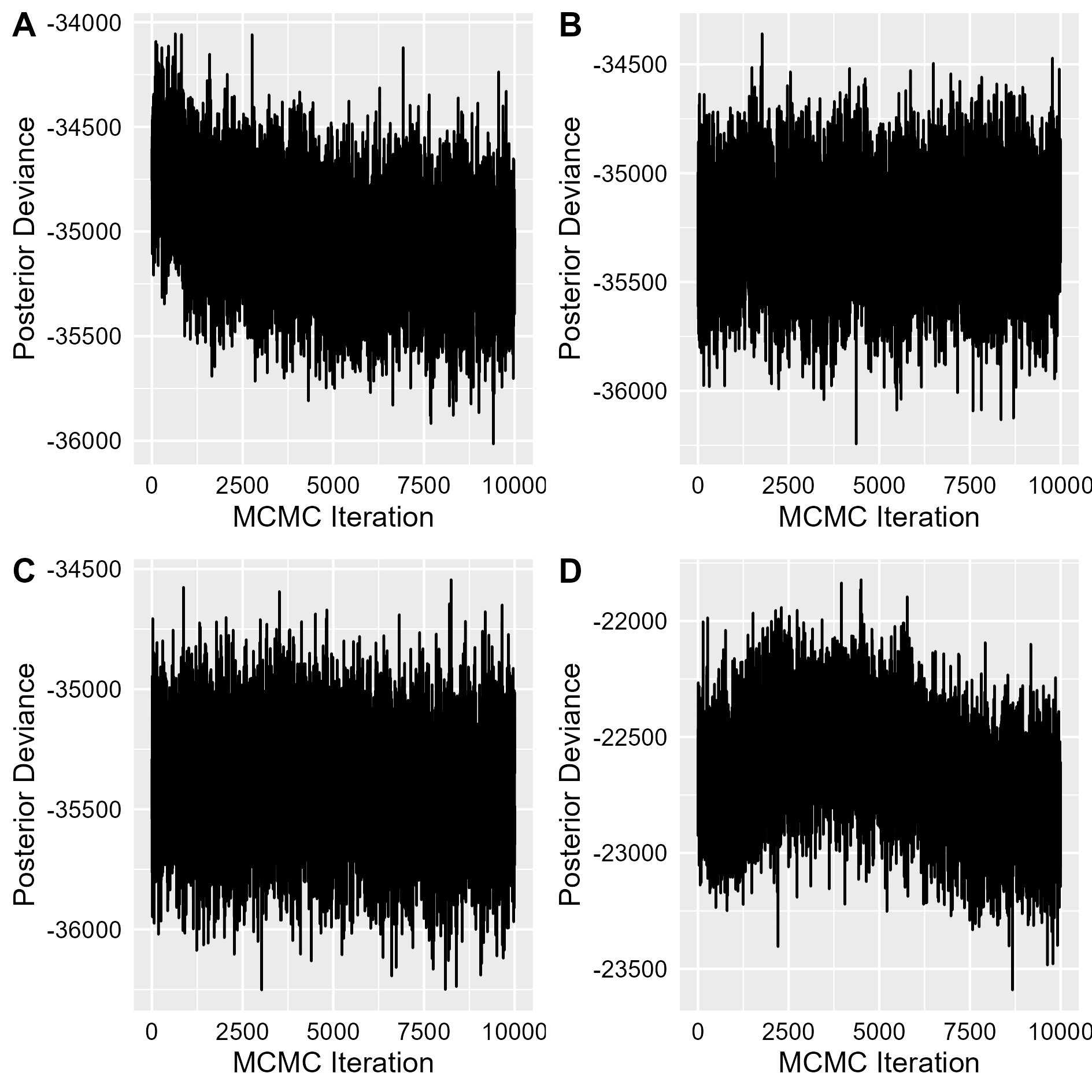}      
		\caption{Simulation for $O=2$: Posterior deviances for \texttt{fullGPfixedL} (\textbf{A}), \texttt{NNGPblockFixedL} (\textbf{B}), \texttt{NNGPsequenFixedL} (\textbf{C}), and \texttt{NNGPsequenVaryLj} (\textbf{D}) under $(m,T)=(400,50)$, 80000 burn-in MCMC iterations, and 20000 post-burn-in MCMC iterations (thinned to 10000).  \texttt{NNGPsequenFixedL} appears to be the best, followed by \texttt{NNGPblockFixedL}, \texttt{fullGPfixedL}, and \texttt{NNGPsequenVaryLj}.}
		\label{T50m400O2deviances}
	\end{figure}

	We performed extensive simulation experiments with 2 observation types, i.e., $O=2$, and the $O\times O$ matrix $\kappa$ set as $0.7$ times the $2\times 2$ identity matrix. We adopted the same data generation mechanism and model fitting specifications as in \Cref{sect6a} with four $(m,T)$ pairs - $(100,20)$, $(225,20)$, $(225,100)$, and $(400,50)$ and $2\times 10^4$ post-burn-in MCMC iterations (thinned to $10^4$). Our four methods \texttt{fullGPfixedL}, \texttt{NNGPblockFixedL}, \texttt{NNGPsequenFixedL}, and \texttt{NNGPsequenVaryLj} lead to comparable satisfactory diagnostics statistics and posterior deviances. For the cases $(m,T)=(100,20)$ and $(m,T)=(400,50)$, the posterior deviances are plotted in \Cref{T20m100O2deviances,T50m400O2deviances}, and summaries of diagnostic metrics are in \Cref{T20m100O2diags,T50m400O2diags}. 
	\begin{table}[H]
		\centering
		\scalebox{0.95}{\begin{tabular}{|*{5}{c|}}
				\hline
				\textbf{Model} &{\texttt{fullGPfixedL}}&{\texttt{NNGPblockFixedL}}&{\texttt{NNGPsequenFixedL}}&{\texttt{NNGPsequenVaryLj}}\\ 
				\hline
				\textbf{\texttt{postMeanMSE}} & 0.01718825 & 0.01856277 & 0.014993 & 0.01629108\\
				\hline
				\textbf{\texttt{postMSE}} & 0.1533105 & 0.1553441 & 0.1478036 & 0.1514525\\
				\hline
				\textbf{\texttt{postVar}} & 0.1361358 & 0.136795 & 0.1328239 & 0.1351749 \\
				\hline
				\textbf{\texttt{dinf}} & 0.1533241 & 0.1553578 & 0.1478169 & 0.151466\\
				\hline
				\textbf{\texttt{pD}} & -5516.225 & -406.5968 & -3493.642 & $-3.422094 \times 10^{36}$\\
				\hline
				\textbf{\texttt{dic}} & -5826.234 & -640.621 & -3957.715 & $-3.422094 \times 10^{36}$\\
				\hline
				\textbf{\texttt{p\_waic\_1}} & 354.9216 & 327.6262 & 310.7777 & 299.9709
				\\
				\hline
				\textbf{\texttt{p\_waic\_2}} & 1992.049 & 1910.587 & 2088.302 & 2091.322\\
				\hline
				\textbf{\texttt{lppd}} & 332.4655 & 280.8252 & 387.425 & 328.4316 \\
				\hline
				\textbf{\texttt{waic}} & 3319.167 & 3259.524 & 3401.753 & 3525.781 \\
				\hline
		\end{tabular}}
		\caption{Simulation for $O=2$: Diagnostic statistics for $(m,T)=(100,20)$, 30000 burn-in MCMC iterations, and 20000 post-burn-in MCMC iterations (thinned to 10000). The fixed number of clusters $L$ is specified to be $30$ for the first three methods, and a starting value and hence upper bound of $30$ is assigned to all $L_j$'s for \texttt{NNGPsequenVaryLj}. \texttt{NNGPsequenFixedL} appears to be the best, followed by \texttt{NNGPsequenVaryLj}, \texttt{fullGPfixedL}, and \texttt{NNGPblockFixedL}.} 
		\label{T20m100O2diags}
	\end{table}
	
	\begin{table}[H]
		\centering
		\scalebox{0.95}{\begin{tabular}{|*{5}{c|}}
				\hline
				\textbf{Model} &{\texttt{fullGPfixedL}}&{\texttt{NNGPblockFixedL}}&{\texttt{NNGPsequenFixedL}}&{\texttt{NNGPsequenVaryLj}}\\ 
				\hline
				\textbf{\texttt{postMeanMSE}} & 0.01156502 & 0.01163034 & 0.01151077 & 0.02350727\\
				\hline
				\textbf{\texttt{postMSE}} & 0.06627909 & 0.0659256 & 0.06565047 & 0.0902155\\
				\hline
				\textbf{\texttt{postVar}} & 0.05471954 & 0.0543007 & 0.05414511 & 0.0667149\\
				\hline
				\textbf{\texttt{dinf}} & 0.06628456 & 0.06593103 & 0.06565589 & 0.09022217\\
				\hline
				\textbf{\texttt{pD}} & -6991868 & 435.9175 & -1115973 & -73.88205\\
				\hline
				\textbf{\texttt{dic}} & -7026893 & -34833.96 & -1151389 & -22724.13\\
				\hline
				\textbf{\texttt{p\_waic\_1}} & 972.6698 & 955.6921 & 818.4227 & 922.2227\\
				\hline
				\textbf{\texttt{p\_waic\_2}} & 17411.46 & 13502.62 & 13483.7 & 12435.92 \\
				\hline
				\textbf{\texttt{lppd}} & 17998.47 & 18112.78 & 18117.06 & 11786.24  \\
				\hline
				\textbf{\texttt{waic}} & -1174.019 & -9220.335 & -9266.715 & 1299.361 \\
				\hline
		\end{tabular}}
		\caption{Simulation for $O=2$: Diagnostic statistics for $(m,T)=(400,50)$, 80000 burn-in MCMC iterations, and 20000 post-burn-in MCMC iterations (thinned to 10000). The fixed number of clusters $L$ is specified to be $50$ for the first three methods, and a starting value and hence upper bound of $50$ is assigned to all $L_j$'s for \texttt{NNGPsequenVaryLj}. \texttt{NNGPsequenFixedL} appears to be the best, followed by \texttt{NNGPblockFixedL}, \texttt{fullGPfixedL}, and \texttt{NNGPsequenVaryLj}.} 
		\label{T50m400O2diags}
	\end{table}
	
	\subsection{Sensitivity Analysis for the Number of Factors and Hyperparameters}\label{appenIf}
	We adopted the data generation mechanism and model fitting specifications in \Cref{sect6a} with $O=1$, $(m,T)=(225,100)$, 90000 burn-in and 10000 post-burn-in (thinned to 5000) MCMC iterations. We used the same hyper-parameter values as in \Cref{sect6a} (\texttt{setting 1}) with $k=3,4,5,6,7$. We also specified $k=5$ with five sets of different hyper-parameter combinations specified below, which we shall denote as \texttt{setting 1}, \texttt{setting 2}, \texttt{setting 3}, \texttt{setting 4}, and \texttt{setting 5}. 
	\begin{enumerate}
		\item $(a,b)=(1,1)$ for $\sigma^2\big(\bm{s}_{i}\big)$'s, $(a_1,a_2)=(1,1)$ for $\delta_{1:k}$, $(a_{\rho},b_{\rho})=(0.1,1)$ for $\rho$, $(a_{\psi},b_{\psi})=(0.1,4.5)$ for $\psi$, $(\nu,\Theta)=(2,1)$ for $\kappa$, and $(\zeta,\Omega)=(k+1,I_{k\times k})$ for $\Upsilon_{k\times k}$;
		\item $(a,b)=(1,1)$ for $\sigma^2\big(\bm{s}_{i}\big)$'s, $(a_1,a_2)=(1,1)$ for $\delta_{1:k}$, $(a_{\rho},b_{\rho})=(1,2)$ for $\rho$, $(a_{\psi},b_{\psi})=(1,2)$ for $\psi$, $(\nu,\Theta)=(2,1)$ for $\kappa$, and $(\zeta,\Omega)=(k+1,I_{k\times k})$ for $\Upsilon_{k\times k}$;
		\item $(a,b)=(1,1)$ for $\sigma^2\big(\bm{s}_{i}\big)$'s, $(a_1,a_2)=(1,1)$ for $\delta_{1:k}$, $(a_{\rho},b_{\rho})=(0.1,1)$ for $\rho$, $(a_{\psi},b_{\psi})=(0.1,4.5)$ for $\psi$, $(\nu,\Theta)=(2,0.1)$ for $\kappa$, and $(\zeta,\Omega)=(k+1,0.1\times I_{k\times k})$ for $\Upsilon_{k\times k}$;
		\item $(a,b)=(0.1,1)$ for $\sigma^2\big(\bm{s}_{i}\big)$'s, $(a_1,a_2)=(1,1)$ for $\delta_{1:k}$, $(a_{\rho},b_{\rho})=(0.1,1)$ for $\rho$, $(a_{\psi},b_{\psi})=(0.1,4.5)$ for $\psi$, $(\nu,\Theta)=(2,1)$ for $\kappa$, and $(\zeta,\Omega)=(k+1,I_{k\times k})$ for $\Upsilon_{k\times k}$;
		\item $(a,b)=(1,1)$ for $\sigma^2\big(\bm{s}_{i}\big)$'s, $(a_1,a_2)=(1,2)$ for $\delta_{1:k}$, $(a_{\rho},b_{\rho})=(0.1,1)$ for $\rho$, $(a_{\psi},b_{\psi})=(0.1,4.5)$ for $\psi$, $(\nu,\Theta)=(2,1)$ for $\kappa$, and $(\zeta,\Omega)=(k+1,I_{k\times k})$ for $\Upsilon_{k\times k}$.
	\end{enumerate}  
	Our four methods \texttt{fullGPfixedL}, \texttt{NNGPblockFixedL}, \texttt{NNGPsequenFixedL}, and \texttt{NNGPsequenVaryLj} lead to suitable diagnostics statistics and posterior deviances under diverse specifications of $k$ and certain hyper-parameters. \Cref{postMSEsensAnalysis} suggests that as long as $k$ is not too small (smaller than the true number of spatial clusters), a smaller $k$ summarizes information more efficiently and leads to better model fit compared to overly large $k$ values.
	\begin{table}[H]
		\centering
		\scalebox{0.9}{
			\begin{tabular}{|*{5}{c|}}
				\hline 
				& \texttt{fullGPfixedL} & \texttt{NNGPblockFixedL} & \texttt{NNGPsequenFixedL} & \texttt{NNGPsequenVaryLj} \\
				\hline
				$k = 3$ and \texttt{setting 1} & 0.04249498 & 0.04132792 & 0.04131845 & 0.04130782 \\
				\hline
				$k = 4$ and \texttt{setting 1} & 0.04169457 & 0.0458628 & 0.04170855 & 0.04228698 \\
				\hline
				$k = 5$ and \texttt{setting 1} & 0.04243256 & 0.04392605 & 0.04374525 & 0.07509951 \\
				\hline
				$k = 6$ and \texttt{setting 1} & 0.04654032 & 0.04463703 & 0.046456 & 0.07185972 \\
				\hline
				$k = 7$ and \texttt{setting 1} & 0.04590091 & 0.04622895 & 0.04577551 & 0.06927728 \\
				\hline
				$k = 5$ and \texttt{setting 2} & 0.05195494 & 0.05281261 & 0.05200683 & 0.08106986\\
				\hline
				$k = 5$ and \texttt{setting 3} & 0.04460182 & 0.0435225 & 0.04416449 & 0.07531454\\
				\hline
				$k = 5$ and \texttt{setting 4} & 0.04448045 & 0.04291583 & 0.04338732 & 0.08113479\\
				\hline
				$k = 5$ and \texttt{setting 5} & 0.04159748 & 0.04214983 & 0.04472333 & 0.07416565\\
				\hline
		\end{tabular} }                                         \caption{Sensitivity analysis: Posterior mean squared errors for our four methods to the number of latent factors $k$ and different choices of hyperparameters in the prior.}
		\label{postMSEsensAnalysis}
	\end{table}

	\section{Extension to Three Non-Gaussian Observed Data Types}\label{appenJ}
	
	Earlier on, we have only considered Gaussian observed data. As we shall see in this section, our models can be conveniently extended to accommodate three non-normal observed data types -- \texttt{`tobit'}, \texttt{`probit'}, and \texttt{`binomial'}, which can all still lead to Gaussian kernels. All we need to add is an extra \texttt{SampleY()} step sampling transformed normal response variables from our original observed non-normal data and based on our current posterior parameter estimates at the start of each MCMC iteration in the Gibbs samplers (\Cref{appenA}). After obtaining a model fit object, we shall also modify the corresponding portions calculating our diagnostics metrics and making both in-sample and out-of-sample predictions, which all turn out to be quite straightforward.
	
	The first two non-Gaussian observed data types are simply defined based on some original normal distributions. A \texttt{`tobit'} variable is obtained from a normally distributed instance that is set to 0 if negative and kept as it is otherwise. A \texttt{`probit'} instance is set to 0 if the original normal value is non-positive and set to 1 otherwise. The \texttt{SampleY()} steps corresponding to these two data types are hence quite straightforward. For an observed value of 0 corresponding to either data type and for a \texttt{`probit'} observed value of 1, we sample a transformed response variable from the truncated normal distribution with our current mean and variance posterior estimates left-censored and right-censored at 0 respectively. When making predictions for a certain outcome variable based on a fitted model, we first obtain a predicted transformed response as in the baseline \texttt{`normal'} case and then set the counterpart predicted original response to be 0 or 1 if needed. When calculating log likelihoods for certain diagnostics metrics, we take note that the \texttt{`tobit'} distribution is a rectified Gaussian distribution (a half-half mixture of a point mass at 0 and the positive portion of a normal distribution, which is neither discrete nor continuous), and that \texttt{`probit'} is a discrete distribution with discontinuity points 0 and 1.
	
	We now consider binomial observed data $y_t^o(\bm{s}_{i})\sim \Bin\left(n_t^o(\bm{s}_{i}),\pi_t^o(\bm{s}_{i}) \right),\;(i,o,t)\in\{1,\dots,m\}\times\{1,\dots,O\}\times\{1,\dots,T\} $, where $n_t^o(\bm{s}_{i})$'s are the pre-specified numbers of trials.     
	Following \cite{Berchuck2021}'s idea, we accommodate the binomial likelihood by extending \Cref{eq:single.obs} to the general modeling framework\footnote{Note that for the Gaussian specification, $g$ is the identity link and $f$ is Gaussian with mean $\mu_t^o(\bm{s}_{i})=g^{-1}(\vartheta_t^o(\bm{s}_{i}))=\vartheta_t^o(\bm{s}_{i})$ and variance (nuisance) $\zeta_t^o(\bm{s}_{i})=(\sigma^2)^o(\bm{s}_{i})$.}
	\begin{align}\label{BerchuckFramework}
		y_t^o(\bm{s}_{i})|\vartheta_t^o(\bm{s}_{i}),\zeta_t^o(\bm{s}_{i})&\overset{\ind}{\sim} f\left(y_t^o(\bm{s}_{i}); g^{-1}(\vartheta_t^o(\bm{s}_{i})),\zeta_t^o(\bm{s}_{i}) \right) \\
		\vartheta_t^o(\bm{s}_{i}) &= \bm{x}_t^o(\bm{s}_{i})^{\T}\bm{\beta} + \sum_{j=1}^k \lambda_j^o(\bm{s}_{i})\eta_{tj}\nonumber\\
		\forall\;(i,o,t)&\in\{1,\dots,m\}\times\{1,\dots,O\}\times\{1,\dots,T\},\nonumber    	
	\end{align}
	where $g$ is the logit link ($\zeta_t^o(\bm{s}_{i})$ is null) and $f$ is binomial with probability $\pi_t^o(\bm{s}_{i})=g^{-1}(\vartheta_t^o(\bm{s}_{i}))=\frac{e^{\vartheta_t^o(\bm{s}_{i})}}{1+e^{\vartheta_t^o(\bm{s}_{i})}}\in(0,1)$. Our joint data likelihood is hence
	\begin{align}\label{binLik}
		f(\bm{y}|\bm{\vartheta},\bm{n})=&\prod_{t=1}^T\prod_{o=1}^O\prod_{i=1}^m\left(\frac{e^{\vartheta_t^o(\bm{s}_{i})}}{1+e^{\vartheta_t^o(\bm{s}_{i})}}\right)^{y_t^o(\bm{s}_{i})}\left(\frac{1}{1+e^{\vartheta_t^o(\bm{s}_{i})}}\right)^{n_t^o(\bm{s}_{i})-y_t^o(\bm{s}_{i})}\nonumber\\
		=&\prod_{t=1}^T\prod_{o=1}^O\prod_{i=1}^m\frac{\exp\{\vartheta_t^o(\bm{s}_{i})y_t^o(\bm{s}_{i})\}}{(1+\exp\{\vartheta_t^o(\bm{s}_{i})\})^{n_t^o(\bm{s}_{i})}}.
	\end{align}    
	Directly proceeding with \Cref{binLik}, however, would be computationally prohibitive due to most parameters' loss of conjugacy. Hence, \textcite{Berchuck2021} introduced P\'olya-Gamma (PG) augmented parameters $\omega_t^o\left(\bm{s}_{i}\right)\overset{\ind}{\sim}\PG\left(n_t^o(\bm{s}_{i}),0\right)$ so that
	\begin{align}
		f(\bm{y},\bm{\omega}|\bm{\vartheta},\bm{n})=& \prod_{t=1}^T\prod_{o=1}^O\prod_{i=1}^m\left[\frac{\exp\{\vartheta_t^o(\bm{s}_{i})y_t^o(\bm{s}_{i})\}}{(1+\exp\{\vartheta_t^o(\bm{s}_{i})\})^{n_t^o(\bm{s}_{i})}}\right] \times f\left(\omega_t^o(\bm{s}_{i})|\vartheta_t^o(\bm{s}_{i}),n_t^o(\bm{s}_{i})\right)  \nonumber \\ 
		\propto &\prod_{t=1}^T\prod_{o=1}^O\prod_{i=1}^m\left[\frac{\exp\{\vartheta_t^o(\bm{s}_{i})y_t^o(\bm{s}_{i})\}}{(1+\exp\{\vartheta_t^o(\bm{s}_{i})\})^{n_t^o(\bm{s}_{i})}}\right] \exp\left\lbrace -\frac{[\vartheta_t^o(\bm{s}_{i})]^2\omega_t^o\left(\bm{s}_{i}\right)}{2}\right\rbrace \left(\frac{1+\exp\{\vartheta_t^o(\bm{s}_{i})\}}{2\exp\{\vartheta_t^o(\bm{s}_{i})/2\}}\right)^{n_t^o(\bm{s}_{i})} \nonumber\\
		\propto & \prod_{t=1}^T\prod_{o=1}^O\prod_{i=1}^m \exp\left\lbrace -\frac{\omega_t^o(\bm{s}_{i})}{2}\left[y^*_{t,o}(\bm{s}_{i})-\vartheta_t^o(\bm{s}_{i})\right]^2\right\rbrace, \label{binOmegaLikProp}
	\end{align}
	where $y^*_{t,o}\left(\bm{s}_{i}\right)=\frac{\chi_t^o(\bm{s}_{i})}{\omega_t^o(\bm{s}_{i})}$ with $\chi_t^o(\bm{s}_{i})=y_t^o(\bm{s}_{i})-\frac{n_t^o(\bm{s}_{i})}{2}\;\forall\;(i,o,t)$. This transformed kernel is now Gaussian and conjugacy has thus been introduced.
	
	\Cref{binOmegaLikProp} suggests that $(\sigma^2)^o\left(\bm{s}_{i}\right)$'s are no longer needed and the corresponding Gibbs sampler step can thus be removed. Instead, we add at the beginning of each MCMC iteration a step sampling the augmented PG parameters $\omega_t^o(\bm{s}_{i})$'s from their full conditional distributions $\omega_t^o\left(\bm{s}_{i}\right) \sim \PG\left(n_t^o\left(\bm{s}_{i}\right), \bm{x}_t^o\left(\bm{s}_{i}\right)^{\T}\bm{\beta}+\sum_{j=1}^k\lambda_j^o(\bm{s}_{i})\eta_{tj}\right)\;\forall\;(i,o,t)$. $y_t^o(\bm{s}_{i})$'s can then be transformed to $y_{t,o}^*(\bm{s}_{i})$'s accordingly in the \texttt{SampleY()} step. We should also replace $(\sigma^2)^o\left(\bm{s}_{i}\right)$'s by $\frac{1}{\omega_t^o(\bm{s}_{i})}$'s in all other concerned Gibbs sampler steps, i.e., the steps sampling $\bm{\eta}_t$'s, $\bm{\beta}$, and $\bm{\lambda}_j$'s or $\xi_j^o\left(\bm{s}_{i}\right)$'s, $\theta_{jl_j}$'s.
	When calculating a log likelihood for $y_t^o\left(\bm{s}_{i}\right)$ or obtaining an in-sample prediction $\hat{y}_t^o\left(\bm{s}_{i}\right)$ for any arbitrary $(i,o,t)\in\{1,\dots,m\}\times\{1,\dots,O\}\times\{1,\dots,T\}$ based on a fitted model, we first calculate $\hat{\pi}_t^o(\bm{s}_{i})=\frac{e^{\hat{\vartheta}_t^o(\bm{s}_{i})}}{1+e^{\hat{\vartheta}_t^o(\bm{s}_{i})}}\in(0,1)$ from our posterior mean estimate $\hat{\vartheta}_t^o(\bm{s}_{i})$ and then proceed according to $y_t^o\left(\bm{s}_{i}\right)$'s distribution $\Bin\left(n_t^o\left(\bm{s}_{i}\right), \hat{\pi}_t^o(\bm{s}_{i})\right)$. Out-of-sample predictions are similar to in-sample predictions in this case.
	
	\printbibliography
\end{document}